\documentclass[12pt,twoside]{report}

\usepackage{amsmath,latexsym,amssymb,slashed}
\usepackage{bbm}
\usepackage{pdfpages}
\usepackage{centernot}
\usepackage{multirow}
\usepackage{varwidth}
\usepackage{cite}
\usepackage{amsthm}
\usepackage{xypic}
\usepackage[hmarginratio=1:1]{geometry}
\usepackage{fancyhdr}
\usepackage{anonchap}
\usepackage{hyperref}

\fancypagestyle{main}{

     \fancyhead[RO]{\nouppercase{{\it \leftmark}}}
     \fancyfoot[C]{\thepage}
     }
\fancypagestyle{intro}{
             \fancyhead[RO]{Introduction}
             \fancyfoot[C]{\thepage}
             }
\fancypagestyle{conc}{
             \fancyhead[RO]{Conclusion}
             \fancyfoot[C]{\thepage}
             }

\fancypagestyle{ded}{\fancyfoot[C]{} \fancyhead[RO]{}}

\theoremstyle{definition}
\newtheorem{defn}{Definition}[section]
\newtheorem{propn}{Proposition}[section]
\newtheorem{thms}[propn]{Theorem}

\newtheorem{ber}[propn]{Theorem (Berger)}
\newtheorem{lef}[propn]{Proposition (Lefschetz decomposition)}

\newcommand{\eq}[1]{\begin{equation}\begin{split}#1\end{split}\end{equation}}
\newcommand{\spl}[1]{\begin{split}#1\end{split}}
\newcommand{\al}[1]{\begin{align}#1\end{align}}
\newcommand{\all}[1]{\begin{align*}#1\end{align*}}

\newcommand{\dfn}[1]{\begin{defn}#1\end{defn}}
\newcommand{\prop}[1]{\begin{propn}#1\end{propn}}
\newcommand{\thm}[1]{\begin{thms}#1\end{thms}}
\newcommand{\prf}[1]{\begin{proof}#1\end{proof}}

\newcommand{\arxth}[1]{\href{http://arxiv.org/abs/hep-th/#1}{[{\tt hep-th/#1}]}}
\newcommand{\arx}[1]{[\href{http://arxiv.org/abs/#1}{\tt #1}]}

\newcommand{\acal}{\mathcal{A}}
\newcommand{\ccal}{\mathcal{C}}
\newcommand{\ecal}{\mathcal{E}}
\newcommand{\fcal}{\mathcal{F}}
\newcommand{\gcal}{\mathcal{G}}
\newcommand{\hcal}{\mathcal{H}}
\newcommand{\ical}{\mathcal{I}}
\newcommand{\jcal}{\mathcal{J}}
\newcommand{\kcal}{\mathcal{K}}
\newcommand{\lcal}{\mathcal{L}}
\newcommand{\mcal}{\mathcal{M}}
\newcommand{\ncal}{\mathcal{N}}
\newcommand{\ocal}{\mathcal{O}}
\newcommand{\pcal}{\mathcal{P}}
\newcommand{\scal}{\mathcal{S}}
\newcommand{\tcal}{\mathcal{T}}
\newcommand{\vcal}{\mathcal{V}}
\newcommand{\wcal}{\mathcal{W}}

\newcommand{\nbb}{\mathbbm{N}}
\newcommand{\zbb}{\mathbbm{Z}}
\newcommand{\rbb}{\mathbbm{R}}
\newcommand{\cbb}{\mathbbm{C}}
\newcommand{\obb}{\mathbbm{1}}

\renewcommand{\a}{\alpha}
\renewcommand{\b}{\beta}
\newcommand{\G}{\Gamma}
\newcommand{\g}{\gamma}
\newcommand{\e}{\epsilon}
\newcommand{\m}{\mu}
\newcommand{\n}{\nu}
\renewcommand{\r}{\rho}
\newcommand{\s}{\sigma}
\renewcommand{\t}{\theta}
\renewcommand{\k}{\kappa}
\renewcommand{\l}{\lambda}
\renewcommand{\L}{\Lambda}
\renewcommand{\o}{\omega}
\renewcommand{\O}{\Omega}
\newcommand{\z}{\zeta}

\newcommand{\bs}{\left.\right|_{\Sigma}}
\newcommand{\ea}{\bigwedge\nolimits^{\!\bullet} T^*}
\newcommand{\bw}{\bigwedge\nolimits}
\newcommand{\p}{\partial}
\newcommand{\we}{\widetilde{\eta}}
\newcommand{\gs}{\left.g\right|_{\Sigma}}
\newcommand{\fwv}{{\mathcal{F}_{\text{wv}}}}
\newcommand{\tfwv}{\tilde{\fcal}_{\text{wv}}}
\newcommand{\vol}{\text{vol}}
\def\d{\text{d}}
\newcommand\nn{\nonumber}
\newcommand{\Op}{\Omega_{\text{p}}}
\renewcommand{\Re}{\text{Re}\;}
\renewcommand{\Im}{\text{Im}~}

\numberwithin{equation}{chapter}

\begin{document}

\begin{titlepage}
\begin{center}
\Huge \textbf{On flux vacua, $SU(n)$-structures and generalised complex geometry} \\
\vspace{25mm}
\Large Dani\"{e}l Prins\\
\vspace{10mm}
\vfill
\end{center}

\noindent\begin{varwidth}{0.5\linewidth} 
Ph.D. thesis \par
advisor: D. Tsimpis \par
September 2015\par
\par
\par
\end{varwidth} \hfill
\begin{varwidth}{0.5\linewidth}
Institut de Physique Nucl\'{e}aire de Lyon\par
\'{E}cole Doctorale de Physique et d'Astrophysique de Lyon \par
Universit\'{e} de Lyon 1\par
\noindent \par
\noindent  \par
\end{varwidth}

\end{titlepage}

\newpage
$\phantom{}$
\newpage

\section*{Abstract}
Understanding supersymmetric flux vacua is essential in order to connect string theory to observable physics. In this thesis, flux vacua are studied by making use of two mathematical frameworks: $SU(n)$-structures and generalised complex geometry. Manifolds with $SU(n)$-structure are generalisations of Calabi-Yau manifolds. Generalised complex geometry is a geometrical framework that simultaneously generalises complex and symplectic geometry. Classes of flux vacua of type II supergravity and M-theory are given on manifolds with $SU(4)$--structure. The $\ncal = (1,1)$ type IIA vacua uplift to $\ncal=1$ M-theory vacua, with four-flux that need not be (2,2) and primitive. Explicit vacua are given on Stenzel space, a non-compact Calabi-Yau. These are then generalised by constructing families of non-CY $SU(4)$-structures to find vacua on non-symplectic $SU(4)$-deformed Stenzel spaces.
It is shown that the supersymmetry conditions for $\ncal = (2,0)$ type IIB can be rephrased in the language of generalised complex geometry, partially in terms of integrability conditions of generalised almost complex structures. This rephrasing for $d=2$ goes beyond the calibration equations, in contrast to $d=4,6$ where the calibration equations are equivalent to supersymmetry.
Finally, Euclidean type II theory is examined on $SU(5)$-structure manifolds, where generalised equations are found which are necessary but not sufficient to satisfy the supersymmetry equations. Explicit classes of solutions are provided here as well. Contact with Lorentzian physics can be made by uplifting such solutions to $d=1$, $\ncal = 1 $ M-theory.

\setcounter{secnumdepth}{-1}

\chapter{Introduction}

\pagestyle{intro}
Flux vacua are solutions to the equations of motion of certain supergravity theories which describe low energy string theory.
String theory is an attempt to unify the Standard Model, which describes elementary particle physics, and general relativity, which describes classical gravity. Together, these two describe all known fundamental physics.

The Standard Model describes elementary particles, and hence is necessary at small scales. It works extremely well in cases where gravity can be disregarded, which is the case whenever one is not working with enormous amounts of mass, as gravity is far weaker than the electroweak and strong force described by the Standard Model. The Standard Model is a quantum field theory, which is itself a unification of the ideas of quantum mechanics and special relativity. The Standard Model formulates physical processes in terms of fields, with energetic states of these fields interpreted as particles. Although the formal mathematical framework of quantum field theory is incomplete, nevertheless the precision tests of the Standard Model (for example, the fine-structure constant of QED) are some of the most accurate empirical tests in all of physics. Furthermore, an array of predicted particles (the $W$ and $Z$ bosons, the top and charm quarks, and perhaps most well-known, the Higgs boson) have all been empirically confirmed to exist.

On the other hand, general relativity is based on differential geometry, and describes gravitational force not in terms of particles, but in terms of the curvature of spacetime due to the presence of energy. It is applicable in cases where one is dealing with large scales, where quantum effects do not play a role, such as cosmology and astronomy. General relativity has also undergone a vast number of experimental tests, some of the more well-known of which include the precession of Mercury and the gravitational lensing of galaxies.
\\
\\
General relativity is a classical theory, meaning that dynamics are derived by making use of the least action principle; minimising the action functional that defines the theory gives a set of equations of motions which describe the dynamics of the system. On the other hand, the Standard Model is a quantum theory, meaning that it is the partition function which describes the dynamics in terms of a path integral. This path integral yields probabilities for the system to end up in a certain state, given an initial state and an action. The {\it vacuum} is defined as the state of the theory with the lowest energy. In the classical limit $\hbar \rightarrow 0$, this is determined by the classical equations of motion $\delta S = 0$, by means of saddle point approximation. In this sense, the solution to the equations of motions yields the ground state or vacuum of the theory, with perturbations away from the vacuum considered quantum corrections. Of course, such local minima need not represent the global minimum, and hence, considering the enormous amount of vacua in string theory, are more likely to be false vacua.
Another difference between the Standard Model and general relativity can be found in terms of the symmetries. General relativity is a theory that is diffeomorphism invariant, and is defined for arbitrary (four-dimensional) manifolds. In fact, the geometry of the manifold is exactly the dynamical variable that one is interested in solving. On the other hand, the Standard Model is defined specifically on flat space, with a fixed Minkowski metric. As a consequence, its symmetries are global Poincar\'{e} invariance, which reflects the symmetries of the metric, and local gauge symmetries.
\\
\\
In order to unify the two, it is necessary to write down a  quantum theory of gravity. Unfortunately, the straightforward way of quantising a classical theory fails to yield a consistent theory, in the sense that general relativity is non-renormalisable. Very roughly speaking, string theory ameliorates this situation by `smoothing out' the divergences by replacing pointlike particles with one-dimensional strings, spanning a two-dimensional worldsheet rather than a worldline in the embedding target space (i.e., spacetime). However, this has some rather drastic consequences in addition. Firstly, in order to remove tachyonic states, it is natural to consider string theory which is supersymmetric, which leads to the existence of supersymmetric partner particles for all known elementary particles. Secondly, for superstring theory the target space is required to have ten dimensions, rather than four.
\\
\\
This thesis concerns itself not so much with the intricacies of string theory, but more so with supergravity. There are supergravities of various dimensions, with various supersymmetries and with various particle content. Some of them are related to others by means of dualities or Kaluza-Klein (KK) reduction, but there appear to be inequivalent kinds even taking such equivalences into account.
From the bottom-up point of view, a supergravity can be viewed as a supersymmetric extension of general relativity. Since the supersymmetry algebras are superalgebra extensions of the Poincar\'{e} algebra, any locally (i.e., gauged) supersymmetric field theory will also be invariant under local translation symmetry, or in other words, will be diffeomorphism invariant; this by definition describes classical gravity. From the top-down point of view, in the low energy limit $\a' \rightarrow 0$, $g_s \rightarrow 0$ where strings becomes particle-like, the worldsheet reduces to an ordinary worldline, and one is left with a description of the ten-dimensional target space. Although not obvious, supersymmetry of the worldsheet action implies supersymmetry of the target space action, hence the low energy theory will be a ten-dimensional supergravity theory.
\\
\\
In this thesis, we take the latter approach, and consider those supergravities related to type II (closed) string theory, respectively type IIA and type IIB supergravity. Furthermore, we also consider $D=11$ supergravity, which is the supergravity of maximal dimension that does not include massless particles of spin greather than two. In the same manner that type II string theory is the UV (i.e., high energy) completion to type II supergravity, it is believed that a UV completion to $D=11$ supergravity exists, which has been named M-theory. The field content of these supergravity theories consists of (at least) the graviton, its fermionic superpartner the gravitino, and certain other bosonic fields referred to as the {\it flux}. These fluxes are higher dimensional analogues of the electromagnetic field. They are sourced by branes, higher-dimensional supersymmetric objects which can be considered as stringy analogues of particles.
\\
\\
Although these supergravities, like any supergravity and general relativity itself, are non-renormalizable, if one views them as classical theories, the solutions to their equations of motions, that is, their vacua, can tell us how to connect string theory to low energy processes - and from the point of view of string theory, that includes pretty much everything and anything that is empirically accessible to us right now. In particular, if string theory describes nature, then it should be possible to obtain a description of four-dimensional physics from the higher-dimensional supergravities. This is accomplished by means of Kaluza-Klein reduction; by considering a target space of the form $\mcal_d \times \mcal_{D-d}$, under certain circumstances it is possible to find a way to write down a $d$-dimensional theory on $\mcal_d$ with vacua that can be uplifted to vacua of the $D$-dimensional theory.
The obvious cases of $d = 4$, $\ncal = 2$ type II vacua and $d=4$, $\ncal = 1$ type I or heterotic vacua with fluxes set to zero require the internal space $\mcal_{10-4}$ to be a Calabi-Yau manifold (CY). However, the KK reductions of these theories depend on the moduli space of the Calabi-Yau. More specifically, fluctuations of the Calabi-Yau metric are determined by fluctuations of the complex structure and K\"{a}hler form, which leads to additional scalar fields, called moduli fields, in the theory after KK reduction. Generically, these are massless, which is a problem since such massless particles are unobserved\footnote{Although the breaking of supersymmetry would give the moduli fields a mass at the scale of the breaking, it appears that this is phenomenologically not satisfactory.}. However, a closer investigation leads to the conclusion that the mass of the moduli fields is proportional to the value of the flux for the vacuum.
This has historically been the primary motivation to consider {\it flux vacua}\footnote{The study of flux vacua is oftentimes also referred to as string compactifications. Since the internal space need not be compact, nothing is being made compact, and there are generally no strings to be found anywhere, we will forgo this terminology.}: vacua with non-trivial fluxes lead to masses for the moduli fields, which can be heavy enough that such theories are in agreement with the empirical non-observation of moduli fields. Of course, another motivation might be the fact that, since fluxes are part of the theory, there is no a priori reason to discard them, other than simplification of the problem of solving the equations of motion.
\\
\\
The geometry that is associated to flux vacua tends to go beyond the Calabi-Yau (or more generally, special holonomy) scenario. A particularly useful geometrical framework that generalises the notion of special holonomy is that of {\it $G$-structures}. A $G$-structure can be viewed as some not necessarily integrable geometrical data that is associated to a manifold. Many familiar structures can be captured in this framework, such as product topologies, orientations, metrics, complex structures, K\"{a}hler forms and Calabi-Yau forms. Most conveniently, $G$-structures also give an algebraic description of such structures and obstructions to their existence, making explicit computations possible.
\\
\\
Although flux vacua, by definition, are solutions to the equations of motion, it turns out that it is not the equations of motions that play a central role. Instead, the central equations are the {\it supersymmetry equations}. These Killing spinor equations determine vacua which are supersymmetric. From a technical point of view, the integrability theorem shows that under mild conditions, solutions to the supersymmetry equations solve the equations of motion. Since solving first order Killing spinor equations is easier by far than solving the second order non-linear partial differential equations that furnish the equations of motion, this simplies life greatly. From a theoretical point of view, supersymmetric vacua are desirable since there are stability theorems for such vacua, and in addition to which, a number of cosmological and particle physics issues could be alleviated by the existence of (weak-scale) supersymmetry, for example the hierarchy problem and the problem of finding suitable dark matter candidates\footnote{At the time of this thesis, no weak-scale supersymmetry has been found at the LHC, however.}.
\\
\\
Remarkably, there appears to be a relation between the supersymmetry equations and another geometrical framework, that of {\it generalised complex geometry} (G$\mathbbm{C}$G). Generalised complex geometry is the study of the vector bundle $T \oplus T^*$ and endomorphisms of this bundle known as generalised almost complex structures. In some sense, generalised complex geometry simultaneously generalises symplectic and complex geometry. Under certain circumstances, the supersymmetry equations can be recast as integrability equations of generalised almost complex structures, and the relation between type IIA and type IIB on the supergravity side equates to the relation between symplectic and complex geometry in this framework. The appearance of generalised complex geometry is even more remarkable in that it also appears in similar but unrelated contexts in string theory, such as the reformulation of symmetries of certain NL$\s$M, or the reformulation of type II supergravity as a wholly geometric theory in light of the relation between T-duality and G$\mathbbm{C}$G.
\\
\\
This thesis centers on the study of aspects of the relation between geometry and flux vacua. The vacua under consideration will be of dimension other than the more well-studied four, in order to be able to compare aspects of vacua and find out whether or not certain features are generic or dimension-specific. In particular, classes of vacua on $SU(n)$-structure manifolds for $n\in \{4,5\}$ are constructed and an attempt is made to reformulate such vacua in terms of G$\mathbbm{C}$G.
\\
\\
In chapter \ref{gstruc}, $G$-structures will be introduced. This includes a brief reminder on the necessary concepts related to fibre bundles, a discussion on the geometrical interpretation of $SU(n)$-structures, and practical details for $SU(4)$ and $SU(5)$. In chapter \ref{sugra} the physics preliminaries are given. The relevant supergravities are discussed, their relations to one another, and what type of vacua are of interest and why. Furthermore, the integrability theorem that implies that solutions to the supersymmetry equations are solutions to the equations of motion will be reviewed. Following this, chapter \ref{III} contains original work \cite{pta} \cite{ptb} detailing the construction of flux vacua on $SU(4)$-structure manifolds. Specifically, these are $d=2$, $\ncal = (1,1)$ IIA, $d=2$, $\ncal = (2,0)$ IIB and $d = 3$, $\ncal = 1$ M-theory Minkowski vacua.
The chapter ends with explicit examples on the $SU(4)$-structure manifold $K3 \times K3$. What might be referred to as a more extensive example is worked out in chapter \ref{stnzl}, where flux vacua on the non-compact CY Stenzel space are discussed \cite{stnzl}. By making use of the coset structure that forms the base of the deformed cone that is Stenzel space, it is possible to construct families of $SU(4)$-structures on this space, leading to non-CY manifolds which are dubbed `$SU(4)$-deformed Stenzel spaces', on which more (classes of) vacua are given.
Next, generalised complex geometry is reviewed in chapter \ref{gcg}. The relation between G$\mathbbm{C}$G, the IIB $SU(4)$-structure supersymmetric solutions of chapter \ref{III}, and calibrated D-branes are discussed in chapter \ref{VI} \cite{ptb}. Pushing this idea further, in chapter \ref{su5} the case of type II supersymmetric solutions on $SU(5)$-structure manifolds is examined and reformulated to some extent in terms of G$\mathbbm{C}$G \cite{su5}. The price to pay for the generality of an $SU(5)$-structure manifold is that it is necessary to consider complexified supergravity. Physical results (and examples) are then obtained by lifting the type IIA solutions to $d=1$, $\ncal =1$ M-theory. Finally, the appendices contain conventions, a discussion on the intrinsic torsion that plays a key role in the $G$-structure story, and a slew of miscellaneous results.

\subsection*{About \& Acknowledgements}
The original contents of this thesis are (mostly) bijective to the papers \cite{pta, ptb, stnzl, su5}. They are not diffeomorphic: everything has been rewritten from scratch to present a more unified picture, with the exception of \cite{stnzl} which has been trivially embedded in chapter \ref{stnzl}. In addition, the preliminaries are more strongly emphasised (or in some cases, present) and I have attempted to give definitions for most of the terminology used throughout, in the hope of making this work understandable to a broad audience. In particular, it should be readable for beginning PhD students, stray geometers, and most anyone with a background in classical field theory, general relativity, a healthy dose of complex and differential geometry, some knowledge of supersymmetry and supergravity, quantum field theory and string theory (although I suspect the latter two are not even strictly necessary, merely recommended). In addition, the following table might be of use.
\begin{center}
\begin{tabular}{|l|l|}\hline
For more information on & the reader is referred to \\ \hline
complex geometry & \cite{voisin}, \cite{wells}, \cite{huybrechts} \\ \hline
string theory & \cite{bbs},  \cite{pol}, \cite{gsw} \\ \hline
flux vacua & \cite{grana}, \cite{denef}, \cite{koerber-gcg} \\ \hline
$G$-structures & \cite{joyce}, \cite{salamon} \\ \hline
G$\mathbbm{C}$G & \cite{gualtieri}, \cite{cavalcanti} \\ \hline
generalities & \cite{mirror},\cite{nakahara} \\ \hline
\end{tabular}
\end{center}
On the computational side, use has been made of the mathematica packages Gamma \cite{gran} and matrixEDC \cite{edc}.
 \newpage

 My gratitude to
 \begin{itemize}
\item[] first and foremost, my advisor Dimitrios Tsimpis, for being ever-supportive and ever-patient, always making time for me, always having good suggestions for reading material, problem-solving techniques, new projects, being willing to rephrase perfectly fine answers to my questions half a dozen times when I did not comprehend them, without whom this thesis would not have been possible and I would probably have abandoned
\item[] my fellow PhD students (and the accidental vagrant postdoc). For all of the coffeeless coffee breaks, the discussions (okay, maybe occasionally  monologues) on math and physics, proofs that 2=1, the topology of dinosaurs, and more, I am indebted to  Ahmad, Martin, Guillaume, Nicolas, and above all

\item[] my office mate Jean-Baptiste, who, in addition to the above, was always willing to listen to my ideas, unrelated to his field as they were, and often had useful insights, who helped me with the culture shock, and who made sure I survived the hardest part of the thesis, namely the bureaucracy. Though he's still to take me up on some rolling, possibly with

\item[] the Grapplers of team Team Icon Lyon, who have ensured that I stay sane. Their contribution is certainly not neglible, as without the balance provided by Laurent, Julien, Titou, Fabien and all the others, it would have been impossible for me to stay properly concentrated, as often emphasised by

\item[] my parents, who, besides everything else, ensured that in moving I did not become homeless, and in my absence looked after

\item[] Regina and Mop.
\end{itemize}

\newpage

\clearpage
\pagestyle{ded}
\vspace*{\stretch{2}}
\begin{center}
\emph{ Dedicated to all those who have taught me something of value.}
\end{center}
\vspace{\stretch{3}}
\clearpage

\pagestyle{main}

\newpage
$\phantom{a}$
\newpage
\tableofcontents
\newpage

\setcounter{secnumdepth}{2}
\setcounter{chapter}{0}

\chapter{$G$-structures}\label{gstruc}
In order to find solutions to the supersymmetry equations with non-vanishing fluxes, we will require some mathematical machinery. An extremely useful tool and one of the centerpieces of this thesis is the geometrical concept of a {\it $G$-structure}. $G$-structures are a way to encode the topology of vector bundles in group theoretic fashion, and allow us to encode connections on vector bundles algebraically through the notion of torsion classes. $G$-structures are closely related to holonomy, which have been exploited in the physics literature more frequently, as in many cases, the machinery of holonomy suffices to describe fluxless vacua. The notion of $G$-structures goes back to at least \cite{chern, kobnom}, but did not feature prominently in supergravity literature until \cite{strominger, chiossi, gkmw, gmpw, gmw}.
Standard references are \cite{joyce}, \cite{salamon}, both of which are highly recommended. An excellent overview from a physics point of view is given in \cite{koerber-gcg}.
\\
\\
In section \ref{fibre}, a short overview of fibre bundles is given, as these are essential to understanding $G$-structures and additionally, this will allow us to establish conventions. This is followed by \ref{g} where $G$-structures are defined, their use and examples are given, and torsion classes are discussed. We proceed to the $G$-structures of interest, $G=SU(n)$, in section \ref{su(n)}. After this formal introduction to $SU(n)$-structures, sections \ref{su(4)} and \ref{su(5)} offer a more utilitarian approach, in which necessary formulae for later use will be derived, both for $SU(4)$ and $SU(5)$. The latter two will hopefully clarify some of the more abstract statements made in the former sections.

\section{Fibre bundles}\label{fibre}
$G$-structures are intricately related with vector bundles and principal bundles. Although we assume the reader is familiar with these, let us briefly review and establish conventions.\\
\dfn{ A \textit{fibre bundle} is a tuple $ (E, F, M, \pi)$ with the following properties:
\begin{itemize}
\item[1)] $E,F,M$ are smooth manifolds. $E$ is known as the {\it total space}, $F$ the \textit{fibre}, $M$ the \textit{base space}, and $\pi : E \rightarrow M$ is a smooth projection. Often, the total space $E$ will be referred to as the fibre bundle itself.
\item[2)] $\forall p \in M$, $\pi^{-1} (\{p\}) \equiv E_p \simeq F$. $p$ is called the \textit{base point} of the fibre $E_p$. If $F$ has dimension $k$, the fibre bundle is said to be of \textit{rank $k$}.
\item[3)] $\forall p \in M$, there is an open neighbourhood $U \subset M$ containing $p$ such that the diagram
\all{
\xymatrix{
\pi^{-1} (U) \ar[d]_{\pi}  \ar[r]^{\simeq}_{\tau_U} & U \times F \ar[ld]^\Pi \\
U
}
}
commutes, with $\Pi$ the natural projection. The diffeomorphism $\tau_{U}$ is called a \textit{local trivialisation}.
\end{itemize}
}
Depending on the point of view we wish to stress, we will occasionally use the notation $E \rightarrow M$ or $F \hookrightarrow E$ for fibre bundles.
 \dfn{
A \textit{vector bundle} is a fibre bundle $(E, V, M, \pi)$ such that the fibre $V$ is a vector space, and the local trivialisations are linear maps. Generically we consider $V$ to be a vector space over $\rbb$. A \textit{complex vector bundle} is a vecor bundle where $V$ is a vector space over $\cbb$. A \textit{line bundle} is a vector bundle of rank $k=1$.
}
\dfn{
 A \textit{principal bundle} is a fibre bundle $(P, G, M, \pi)$ such that $G$ is a Lie group, and the local trivialisations are Lie group isomorphisms. The principal bundle comes equipped with an action $P \times G \rightarrow P$ that is smooth and free\footnote{That is, all stabilizers are trivial: $p \cdot g = p \implies g =e$.}. The local trivialisations are also required to be intertwiners\footnote{That is, $\forall p \in \pi^{-1}(U), g\in G$, the local trivialisation $\tau_U$ satisfies $\tau_U(pg) = \tau_U(p) g$} of the action.
}
The fibres of a principal bundle are the orbits of the action on the base point, so the action is fibre-preserving. The base space $M$ is diffeomorphic to $P/G$.

\dfn{
A \textit{subbundle} of $(E, F, M, \pi)$ is a fibre bundle $(E', F', M, \pi')$ such that $E' \subset E$, $F' \subset F$, $\pi'= \pi|_{E'}$. A vector subbundle is itself a vector bundle, hence $F'$ is a vector subspace of $F$, whereas a principal subbundle is itself a principal bundle, hence $F'$ is a subgroup of $F$.
}
An essential notion in the discussion of fibre bundles is that of {\it sections}.
\dfn{
Let $E$ be a fibre bundle. A {\it global section} $s$ is a smooth map $s : M \rightarrow E$ such that $s$ preserves fibres, i.e., $s (p) \in E_p$  $\forall p \in M$, or equivalently, $\pi \circ s = \obb_M$. Let $U \subset M$. A {\it local section} is a fibre-preserving smooth map $s : U \rightarrow E$. The set of global sections will be denoted either by $\G(M, E)$ or by $\G(E)$ if the base space is clear.
}

Fibre bundles can be wholly specified in terms of the fibre, bases space, and the transition data. Consider a set of coordinate patches $\{U_\a\}$ such that\footnote{The fact that we are taking the disjoint union is to be understood.} $M = \bigcup_\a U_\a$ with trivializations $\{\tau_\a\}$. Then one can define the atlas on $E$ by
\al{
\tau_{\a\b} \equiv \tau_\a \circ \tau_\b^{-1} : (U_\a \cap U_\b) \times F \rightarrow  (U_\a \cap U_\b) \times F\;,
}
which is base point preserving. In case of vector bundles, one has that for $p \in M, v \in V_p$,
\al{
\tau_{\a\b}\left( (p, v) \right) \equiv (p, t_{\a\b}(p) v)\;,\quad  t_{\a\b}(p) \in GL(k, \rbb)\;.
}
The maps $t_{\a\b} : U_\a \cap U_\b \rightarrow GL(k, \mathbbm{R})$ are referred to as {\it transition functions}. Hence transition functions describe the change of coordinates of the vector bundle considered as a manifold itself. In order for the transition functions to be sensible, they must satisfy $t_{\a\a} = \obb$, $t_{\a\b} t_{\b\a} = \obb$ and $t_{\a\b} t_{\b\g} t_{\g\a} = \obb$.
\\
\\
Every manifold naturally comes equipped with a vector bundle and a principal bundle. These are the tangent bundle and the (tangent) frame bundle. Let $M$ be a manifold. Then the tangent bundle $TM$ is the vector bundle whose fibres at each point $p \in M$ are given by the tangent plane to the manifold, $T_p M$. If $\{x_j\}$ are local coordinates, then the tangent space is given by
\eq{
T_p M = \text{span}_{\mathbbm{R}} \left\{ \frac{\p}{\p x^j}\right\}\;.
}
The (tangent) frame bundle is the principal fibre bundle whose fibre at $p$ consists of all
ordered bases of $T_p M$. Given an ordered basis, another ordered basis is found by acting on it with an element of $GL(n, \rbb)$. Hence the fibers of the frame bundle, which are called the \textit{structure group}, are isomorphic to $GL(n, \rbb)$. In fact, the transition functions $t_{ab}$ of the tangent bundle are local sections of the tangent frame bundle.
\\
\\
Understanding transition functions, a third kind of vector bundle can be defined.

\dfn{ A {\it holomorphic vector bundle} is a complex vector bundle over a complex base space such that the transition functions are biholomorphic.}

Coming back to the discussion on vector bundles and principal bundles, we note that associating the tangent frame bundle to the tangent bundle is a general feature of vector bundles: to every vector bundle $E$ a principal bundle can be associated, namely its frame bundle $FE$. Conversely, we can associate to each principal bundle a vector bundle. Let $(P, G, M)$ be a principal bundle. Let $\rho$ be a representation of $G$ on a vector space $V$, then there is an action of $G$ on $P \times V$ via the principal action on $P$ and $\rho$ on $V$. Then we define $E \equiv P \times V / \sim $, with the equivalence relation defined as follows: for $p, q \in P$, $p \sim q \iff \exists g \in G$, s.t. $p g = q$; that is to say, $p, q$ are both elements of the same fibre. In this way we get a vector bundle $(E, V, M)$ (with the projection induced from the principal bundle projection). These maps are inverse in the sense that $E \simeq FE \times V / \sim$ in terms of vector bundles. For principal bundles $P$ with fibres $G \neq GL(k, \rbb)$, $P \neq   F\left( P \times V / \sim \right)$, hence principal bundles are more general than vector bundles.
\\
\\

\section{$G$-structures \& Torsion Classes}\label{g}
We are now in a position to discuss the concepts that are of primary interest to us, namely $G$-structures and their intrinsic torsion, which is expressible in terms of torsion classes.
\dfn{
Let $E$ be a vector bundle with associated frame bundle $FE$. A {\it $G$-structure} is a principal subbundle $Q$ of a frame bundle $FE$ with fibres isomorphic to $G$.
}
A $G$-structure $Q$ exists if and only if all transition functions of $E$ take value in $G$. Existence of a $G$-structure also goes by the name of {\it reduction of the structure group}.
Generically, we will consider $E= TM$, and note that this induces $G$-structures on $T^*M$ and on the tensor bundle $\tcal^{(k,l)}$\footnote{See section \ref{gcg} for applications of $G$-structures to the generalised tangent bundle $T\oplus T^*$.}.
Although straightforward, this definition is not the most convenient way to work with $G$-structures. The point of a $G$-structure is as follows:
\prop{
Let $M$ be such that it admits a tensor $\phi$ that is nowhere vanishing and $G$-invariant. Then $M$ admits a $G$-structure.
}
\noindent This is easy to see: since $G$ is the stabiliser of $\phi$, if the transition functions would take value outside of $G$, $\phi$ would not be invariant. In all cases relevant for us, the converse is true as well; having a $G$-structure means only a certain set of frames on the tensor bundle is admitted, such that a certain tensor is stabilised.
\\
\\
Thus, we see that we can alternatively describe a $G$-structure algebraically in terms of tensors. Many geometrical structures, such as for example complex structures, symplectic structures, orientability, can be defined in terms of tensors, together with an integrability condition. In this sense, a $G$-structure can be thought of as a not necessarily integrable geometrical structure.  Suppose $M$ admits two tensors $\phi_1$, $\phi_2$ associated to $G$-structures with structure groups $H_1$, $H_2$. Then $M$ will also admit a $G$-structure with structure group $H_1 \cap H_2$. Some of the more useful examples are listed in the table below:
\begin{center}
\begin{tabular}{|c|c|}
\hline
Geometrical Structure &$G$-structure
\\ \hline
Orientability & $GL^+ (n, \rbb)$
\\ \hline
Volume form & $SL(n, \rbb)$
\\ \hline
Riemannian metric & $O(n)$
\\ \hline
Almost complex structure & $GL^+(n/2, \cbb)$
\\ \hline
Almost symplectic structure & $Sp(n, \rbb)$
\\ \hline
 \end{tabular}
 \end{center}
Let us give a proof for one of these as an example.
\propn{
Let $M$ be a manifold of dimension $2n$. $M$ admits an almost complex structure $J$ if and only if it admits a $GL(n, \cbb)$ structure.
}
\prf{
Suppose $M_{2n}$ admits an almost complex structure; a globally non-vanishing fibre-preserving map $I : TM \rightarrow TM$ such that $I^2 = - \obb$. Then $TM \otimes \cbb = T^{(1,0)} M \oplus T^{(0,1)}M$, with $T^{(1,0)}M$ the $+i$-eigenbundle of $I$. Consider a local frame  $\{ \z^j\}$ of the complex vector bundle $T^{(1,0)}$M. This local frame induces a basis for $T^{(1,0)}M$ viewed as a real vector bundle, namely $\{\Re \z^j, i \Im \z^j \}$. But then, this yields a local frame of $TM$, hence $TM$ and $T^{(1,0)}M$ are diffeomorphic as real vector bundles. As a consequence, they have the same structure group. Since the structure group of $T^{(1,0)}M$ is $GL(n, \cbb)$, it follows that $TM$ must also admit a reduction of its structure group to $GL(n, \cbb)$.

Conversely, suppose $M_{2n}$ has a $GL(n, \cbb)$-structure. Given a local frame $\{e_1, ..., e_{2n}\}$ of $TM$, $GL(n, \cbb)$ acts on the local frame of $2n$-dimensional vectors as $GL(n, \rbb^2)$. Thus, one can define a local frame $\{ \z^j, \bar{\z}^{\bar{\jmath}}\; |\; \zeta^j = e_{2j-1} + i e_{2j} \;, \bar{\z}^{\bar{\jmath}} =  e_{2j-1} - i e_{2j}   \}$ such that $GL(n, \cbb)$ acts canonically on
$\{ \z^j \}$ and via the complex conjugate representation on $\{\bar{\z}^{\bar{\jmath}} \}$. Thus, given a covering set of local coordinate patches $M = \bigcup_{\a} U_\a$ one can find local frames of the form $\{\z_\a^j, \bar{\z}_\a^{\bar{\jmath}}\}$ which are related to one another by $\z_\a^j = \left(t_{\a\b}\right)^j_{\phantom{j}k} \z_\b^k$. Define $T^{(1,0)}M$ to be spanned by local frames $\z^j$ and $I (\z) = i \z$ $\forall \z \in T^{(1,0)} M$, and analogously $T^{(0,1)}M$. Then $I$ is a globally well-defined invariant tensor satistying $I^2 = -\obb$, hence an almost complex structure.}

On the more practical side of things, part of $G$-structures' functionality is that they allow us to turn partial differential equations into algebraic ones. This comes about through {\it torsion classes}. Given a $G$-structure $Q$, torsion classes parametrise its {\it intrinsic torsion}. For Riemannian manifolds, intrinsic torsion can be viewed as the failure of the Levi-Civita connection to be compatible with the $G$-structure. Equivalently, it can be seen as a measure of the failure of the manifold to be special holonomy. See appendix \ref{torsion} for more details. Here, we shall restrict ourselves to the following more practical explanation in terms of the tensors that define the $G$-structure. Given a $G$-structure $Q$, globally well-defined tensors on $M$ can be decomposed into irreducible representations of $G$. If $Q$ is defined by a set of differential forms $\{\phi_j\}$, then the exterior derivatives of $\{\phi_j\}$ form a particular set of tensors, whose decomposition holds special meaning. The irreducible representations in their decomposition are expressible in terms of the $\{ \phi_j \}$ and in terms of a number of other differential forms $\{W_l\}$; these are the {\it torsion classes}. The number of torsion classes, as well as their type of representation, is determined by the decomposition of $T^* M \otimes \mathfrak{g}^\perp$, where $\mathfrak{g} \oplus \mathfrak{g}^\perp = \mathfrak{so}(n)$. The torsion classes can also be viewed as characterising obstructions to integrability of the geometrical objects $\{ \phi_j\}$. A $G$-structure with vanishing torsion classes is called {\it integrable} or {\it torsion-free}. This will be made more explicit in the explicit discussion in sections \ref{su(4)}, \ref{su(5)}.

\section{$SU(n)$-structures}\label{su(n)}
The $G$-structures we will be concerned with are $SU(n)$-structure, as manifolds with $SU(n)$-structures are closely related to Calabi-Yau manifolds. As fluxless $\ncal = 2$ vacua of type II supergravity on $\mathbbm{R}^{1,3} \times \mcal_6$ require $\mcal_6$ to be Calabi-Yau, $SU(n)$-structure manifolds are obvious candidates to examine for more general vacua, since a Calabi-Yau structure can be considered to be a torsion-free or integrable $SU(n)$-structure. Before discussing the relation with physics, let us first describe the mathematics of such manifolds, and explain the relation to Calabi-Yau manifolds in more detail. In this section, we will remain quite formal, and hope that the abstractions will be clarified by the explicit calculations given in the following sections where $n$ will be specified.
\\
\\
Let $M$ be a manifold of dimension $2n$. $U(n)$ is the intersection of $O(2n)$, $GL^+(n, \cbb)$ and $Sp(2n, \rbb)$. Hence, a manifold with a $U(n)$-structure will come equipped with an orientation,  an almost symplectic structure $J$, an almost complex structure $I$, and a Riemannian metric $g$. A further reduction of the structure group down to $SU(n)$ is then associated with a volume form $\O$
of $T^{*(1,0)}$ (i.e., a nowhere vanishing global section of the complex vector bundle known as the canonical bundle or determinant line bundle, $T^{*(n,0)} \equiv \bigwedge^n T^{* (1,0)}$ ) and thus, the vanishing of the first Chern class $c_1(T^{(1,0)})=0$. Let us stress that a priori, $T^{*(n,0)}$ is \textit{not} a holomorphic vector bundle, for the sole reason that the base space is equipped with an almost complex structure which need not be integrable. If the base space is complex, then $T^{*(n,0)}$ is holomorphic. Let us also stress that the volume form is a {\it smooth} global section of $T^{*(n,0)}$ rather than a {\it holomorphic} global section, even in the case where $T^{*(n,0)}$ is a holomorphic vector bundle. This will become more clear when discussing $SU(n)$-structures in terms of the associated forms.
\\
\\
We can be more concrete and reformulate this in terms of spinors and tensors. Let us start with the former\footnote{See appendix \ref{spinors} for spinor conventions.}. $M$ has an $SU(n)$-structure if and only if there exists a globally well-defined nowhere vanishing \textit{pure spinor}.

\dfn{
A {\it pure spinor} is a Weyl spinor that is annihilated by exactly half the Clifford algebra.
}
For $n\leq 3$, every Weyl spinor is pure. On the other hand, for $n> 3$, this is not guaranteed. In case $n=4$, there is a single scalar constraint placed on the spinor bilinears. For $n=5$, the constraint is vectorial. This will become clearer when considering the $SU(n)$-structure in terms of forms. Alternatively, see \cite{chevalley} for a far more comprehensive description.
\\
\\
In terms of tensors, an $SU(n)$-structure can be described as follows. Let $g$ be a Riemannian metric on $M$, and let $(J, \O)$ be respectively a real two-form and a complex $n$-form, satisfying the following:
\eq{\label{eq:sun}
J \wedge \O &= 0 \\
\frac{1}{n!} J^n &= \frac{1} {2^n} i^x \O \wedge \O^* = \text{vol}_{2n}\;,
}
for some $x \in \mathbb{N}$ depending on convention and dimension. The fact that $J^n \sim \vol_{2n}$ means that it is non-degenerate. A real non-degenerate two-form is known as an {\it almost symplectic form} or equivalently, {\it almost symplectic structure}. Using $J$, one can construct an almost complex structure $I$ by setting\footnote{See for example \cite{vandoren} for a proof.}
\al{
I_m^{\phantom{m}n} \equiv J_{mp} g^{pn} \;,
}
where the normalisation works out due to the prefactors chosen  in \eqref{eq:sun}. Unless it is necessary to refer explicitly to the fact that we are discussing the almost complex structure, we will abuse notation and simply denote the almost complex structure by $J$ as well. The almost complex structure induces a grading on $k$-forms, and with respect to this grading $\O$ is an $(n,0)$-form and is thus a volume form of $T^{*(1,0)}$.

The relation between the two descriptions is that the tensors can be constructed out of the spinors in terms of {\it spinor bilinears}, as will be made evident in the next section.
\\
\\
Any differential form  $\a \in \O^\bullet(M)$ on an oriented Riemannian manifold furnishes an irreducible $SO(2n)$ representation. The presence of an $SU(n)$-structure means that this representation can be decomposed further in terms of irreducible $SU(n)$ representations. Alternatively, there is a more geometrical way to view this. The $SU(n)$-structure defines an almost complex structure and an almost symplectic structure. An almost complex structure induces a grading on the exterior algebra via
\eq{
 \bigwedge\nolimits^k T^* &= \bigoplus_{p+q= k} \left(\bigwedge\nolimits^p T^{*(1,0)}\right) \wedge \left(\bigwedge\nolimits^q T^{*(0,1)}\right)\\
 &\equiv \bigoplus_{p+q= k} T^{*(p,q)}
}
and consequentially, a grading on sections of the exterior algebra,
\al{
\O^k(M) = \bigoplus_{p+q = k} \O^{(p,q)}(M)\;.
}
Similarly, the almost symplectic structure induces a decomposition of forms. Given an almost symplectic structure $J$, one has the following canonical maps on forms:
\eq{\label{eq:sympoperators}
L_{(k)} : \O^k(M) \rightarrow \O^{k+2}\;, \qquad \a \mapsto J \wedge \a \\
\L_{(k)} : \O^k(M) \rightarrow \O^{k-2}\;, \qquad \a \mapsto J \lrcorner \a \;.
}
The first map is sensible $\forall k \in \{0, ..., 2n\}$. In order to extend the second map to include $k < 2$, $\L_{(0)} \equiv \L_{(1)} \equiv 0$.
Note that $\L_{(k)} \sim \star L_{(2n-k)} \star$ by virtue of the fact that the volume form is expressible in terms of $J$.
The pedantic subscript will always be dropped, but is convenient for the following definition.
\dfn{
A $k$-form $\a$ is defined to be {\it primitive}\footnote{
The fact that a primitive two-form $\a$ satsifies $a_{mn} J^{mn} = 0$ has led some authors to refer to the condition as `tracelesness'. We will not do so.}
if $\a \in \Op^k(M)$, with
\al{
\O^k_{\text{p}}(M) \equiv \text{Ker}(\L_{(k)}) \;.
}
}
By non-degeneracy of the almost symplectic form and Hodge duality,
\eq{
\text{Ker}(\L_{(k)}) = \text{Ker}(L_{(k)}^{n+1-k}) \;,
}
hence this is another way to define primitive forms. Note in particular that $\O^k(M) = \Op^k(M)$ for $k \in \{0,1,2n-1, 2n\}$.

\begin{lef}
Existence of an almost symplectic structure induces a grading on $k$-forms
\eq{
\O^{2k}(M) &= \bigoplus_{l=0}^k L^{l}    \Op^{2(k-l)}(M) \\
\O^{2k+1}(M) &= \bigoplus_{l=0}^k L^{l}    \Op^{2(k-l)+1}(M)
}
and thus on sections of the exterior algebra as
\eq{
\O^\bullet(M) = \bigoplus_{k=0}^{2n} \bigoplus_{l=0}^{2n-k} L^l \Op^k(M) \;.
}
\end{lef}
In the simple cases that will be relevant for us, the proof consists of simply counting degrees of freedom. For a more general proof, one uses the fact that the operators $L, \L, [L, \L]$ form a representation of $sl(2, \cbb)$ with finite spectum, and that this is independent of closure of $J$. See \cite{yan, huybrechts, cavalcanti} for details, although closure of $J$ is assumed in all of these.
\\
\\
Note that in both the complex and symplectic case, decomposition of the forms does not require integrability of the almost complex structure or almost symplectic structure; on the other hand, demanding that this property descends to the level of cohomology is a far stronger constraint, requiring that $M$ is in fact compact and K\"{a}hler. This is because in both cases, the proof for the cohomology decomposition makes extensive use of Poincar\'{e} duality and the Hodge theorem, which states that on compact K\"{a}hler manifolds cohomology classes and harmonic forms are bijective. See for example \cite{voisin} for more details.

The bottom line is that decomposing $k$-forms into irreducible $SU(n)$-representations is equivalent to applying both Hodge and Lefschetz (i.e., complex and symplectic) decompostion to the forms.
Examples below will probably greatly clarify this procedure.

Let us also elaborate on the relation between Calabi-Yau manifolds and $SU(n)$-structure manifolds. There are various definitions for Calabi-Yau manifolds, most of which are inequivalent in one way or another. The one used here is as follows:
\dfn{
A {\it Calabi-Yau manifold} is a manifold $M$ together with a {\it Calabi-Yau structure} $(g, J, \O)$ such that $M$ is a connected complex manifold of complex dimension $n$, $g$ is a K\"{a}hler metric associated to the K\"{a}hler form $J$, and $\O$ is a holomorphic volume form, that is $\O \in \O^n_{\text{h}} (M)  \subset \O^{(n, 0)}(M)$ and $\O$ is nowhere vanishing.
}
Using this definition, a Calabi-Yau manifold has the following properties:
\begin{itemize}
\item $M$ is not necessarily compact.
\item The first Chern class satisfies $ c_1(T^{(1,0)})=0$.
\item $J^n \sim \O \wedge \O^* \sim \vol_{2n}$.
\item The Ricci curvature of $g$ is trivial.
\item The holonomy (see appendix \ref{holo}) satisfies $\text{Hol}(M) \subseteq SU(n)$. In case  $\text{Hol}(M) = SU(n)$, we refer to the Calabi-Yau as a {\it proper Calabi-Yau manifold}. Proper Calabi-Yau satisfy $h^{p,0} =  \delta_{p,0} + \delta_{p,n}$, with $h^{p,q} \equiv \text{dim}H^{p,q}$ the dimension of the Dolbeault cohomology.
\end{itemize}
Comparing this to an $SU(n)$-structure manifold, the following equivalence is found:
\propn{
A Calabi-Yau manifold is a manifold with an integrable $SU(n)$-structure.
}

The proof is to rather trivially check the relevant definitions. The reason we stress this point so heavily here is because understanding that $SU(n)$-structure manifolds are generalisations of Calabi-Yau manifolds is essential to understanding why they are relevant for finding supersymmetric flux vacua of supergravity. In this spirit, the complex $(n,0)$-form $\O$ defining the $SU(n)$-structure will occasionaly be referred to as the {\it almost Calabi-Yau form}, as it is the Calabi-Yau form $\O$ if the $SU(n)$-structure is integrable.


\section{$SU(4)$-structure}\label{su(4)}
An eight-dimensional manifold admits an $SU(4)$-structure if and only it admits a globally defined nowhere vanishing pure spinor. In eight dimensions, a Weyl spinor $\eta$ is pure if and only if it satisfies
\al{
\tilde{\eta} \eta =0 \;.
}
As $\eta$ is non-trivial, it follows that
\eq{
\eta &= \frac{1}{\sqrt{2}} \left(\eta_R + i \eta_I \right) \\
\tilde{\eta}_R \eta_R &= \tilde{\eta}_I \eta_I \equiv 1
}
with $\eta_{I,R}$ real-valued and the normalisation chosen for convenience, as it implies $\widetilde{\eta^c} \eta = 1$. Existence of $\eta_{I,R}$ reduces the structure group of the spin structure from $Spin(8)$ to $Spin(6)$, which is isomorphic to $SU(4)$ due to a so-called `accidental isomorphisms'.
Let us define
\eq{
J_{mn} &\equiv - i \widetilde{\eta^c}\g_{mn}\eta\\
\Omega_{mnpq} &\equiv \we\g_{mnpq}\eta\;.
}
Because of the demand that $\eta$ is nowhere vanishing, both of these forms are non-degenerate. Thus, $J$ is an almost symplectic structure.
Using the Fierz identity \eqref{eq:fierz}, one can then explicitly confirm that
\eq{
J_m{}^pJ_p{}^n=-\delta_m^n ~,
}
demonstrating that this is indeed an induced almost complex structure.
In terms of $J$, one can define the projection operators
\al{
\left(\Pi^\pm\right)_m^{\phantom{m}n} \equiv \frac12 (\delta^n_m \mp i J_m^{\phantom{m}n} ) \;.
}
By virtue of the fact that $\delta_m^n =  \left(\Pi^-\right)_m^{\phantom{m}n} + \left(\Pi^+\right)_m^{\phantom{m}n}$, $k$-forms can be decomposed into $(p,q)$-forms as
\al{
\a^{(k)} &= \sum_{p+q= k} \a^{(p,q)} \\
\a^{(p,q)}_{m_1...m_{p+q}} &\equiv \frac{(p+q)!}{p!q!}
 \left(\Pi^+\right)_{[m_1}^{\phantom{m}r_1} ... \left(\Pi^+\right)_{m_p]}^{\phantom{m}r_p}\left(\Pi^-\right)_{[m_{p+1}}^{\phantom{m}s_1} ... \left(\Pi^+\right)_{m_{p+q}]}^{\phantom{m}s_q}
 \a_{r_1...r_p s_1 ... s_q} \;.\nn
}
By again making use of Fierz identities, it can be shown that
\al{
(\Pi^{+})_m{}^r\Omega_{rnpq}=\Omega_{mnpq}~,  \qquad(\Pi^{-})_m{}^r\Omega_{rnpq}=0 ~
}
and that
\al{\label{eq:su4}
\frac{1}{4!} J^4 &= \frac{1} {2^4} \O \wedge \O^* = \text{vol}_{8} \;.
}
Thus $\O = \O^{(4,0)}$ is a volume form on $T^{* (1,0)}$. Using the definition of the projection operators, it immediately follows that $J= J^{(1,1)}$, hence
\al{
 J \wedge \O = 0 \;,
 }
leading to the conclusion that $(J, \O)$ indeed define an $SU(4)$-structure; by a very slight abuse of terminology, we will also refer to $(J, \O)$ as the $SU(4)$-structure.  The exterior derivative acting on the the $SU(4)$-structure can be decomposed in terms of torsion classes as follows:
\eq{\label{eq:su4torsion}
\d J &= W^*_1 \lrcorner \O + W_3 + W_4 \wedge J + \text{c.c.} \\
\d \O &= \frac{8i}{3} W_1 \wedge J \wedge J + W_2 \wedge J + W_5^* \wedge \O \;,
}
with
\eq{
W_{1,4,5} &\sim \bf{4} \\
W_{2,3}  &\sim {\bf 20 } \;,
}
or equivalently,
\eq{
W_{1,4,5} &\in \O^{(1,0)}(M) \\
W_{2,3} &\in \Op^{(2,1)} (M) \quad \implies  W_{2,3} \wedge J \wedge J = 0 \;.
}
The prefactor $8 i/3$ is fixed by the constraint $\d (J \wedge \O) = 0$.
The precise way in which the torsion classes are obstructions to integrability of the various structures are given in the following table, where the first column indicates whether $M$ equipped with $(J, \O)$ defines a certain structure, and the second column indicates the necessary and sufficient condition on the torsion classes. \\
\begin{center}
\begin{tabular}{|l|l|}
\hline
Geometrical structure & Torsion classes \\ \hline
Complex & $W_1 = W_2 = 0$ \\ \hline
Symplectic & $W_1 = W_3 = W_4 = 0$ \\ \hline
K\"{a}hler & $W_1 = W_2 = W_3 = W_4 = 0$ \\\hline
Nearly Calabi-Yau & $W_1 = W_3 = W_4 = W_5 = 0$ \\ \hline
Conformal Calabi-Yau & $W_1 = W_2 = W_3 =0$, $2 W_4 =  W_5$ exact \\\hline
Calabi-Yau & $W_j = 0 $ $\forall j$ \\\hline
\end{tabular}
\end{center}
$\phantom{a}$\\
These conditions are all straightforward to deduce. The almost symplectic form is an integrable almost symplectic form (which tends to be refered to by the less silly name `symplectic form') if $\d J = 0 \iff W_{1,3,4 } = 0$. For integrability of the almost complex structure, necessity of the vanishing of the torsion classes $W_{1,2}$ is clear, while the sufficiency follows from calculating the Nijenhuis tensor. A K\"{a}hler structure is nothing more than a compatible complex and symplectic structure, and a Calabi-Yau structure is a K\"{a}hler structure with the additional requirement that $\O$ is a holomorphic section and hence $\d \O=0$.
Under conformal transformations $g \rightarrow e^{2\chi} g$, it follows from  \eqref{eq:su4} that $J \rightarrow e^{2 \chi} J$, $\O \rightarrow e^{4 \chi} \O$, hence
\eq{
W_1 &\rightarrow W_1 \\
W_2 &\rightarrow e^{2 \chi} W_2 \\
W_3 &\rightarrow e^{2 \chi} W_3 \\
W_4 &\rightarrow W_4 + 2 \p^+ \chi \\
W_5 &\rightarrow W_5 + 4 \p^+ \chi\;,
}
with\footnote{It is necessary to introduce the operator $\p^+ $ as the almost complex structure need not be integrable, hence the Dolbeault operator is ill-defined.} $\p^+ \chi \equiv \Pi^+(\d \chi)$. Using these, the constraint for a conformal Calabi-Yau structure follows.

\subsection*{Tensor decomposition}
The precise decomposition of forms, either seen in terms of representation theory or as Lefshetz and Hodge decomposition, is as follows.
Under the $SO(8)\rightarrow SU(4)$ decomposition the two-form, the three-form, the selfdual and the anti-selfdual four-form of $SO(8)$ decompose respectively as:
\eq{
\bf{ 28}&\rightarrow \bf{ (6\oplus {6})\oplus(1\oplus 15) }\nn\\
\bf{ 56}&\rightarrow \bf{ (4\oplus \bar{{4}})\oplus(4\oplus 20)\oplus(\bar{4}\oplus \bar{20}) }\nn\\
\bf{35^+}&\rightarrow \bf{(1\oplus 1)\oplus (6\oplus {6})\oplus 20'\oplus  1}\nn\\
\bf{35^-}&\rightarrow \bf{(10\oplus \bar{10})\oplus 15}~.
}
Here, $\bf{35^\pm}$ correspond to selfdual and anti-selfdual four-forms respectively.
Explicitly, $k$-forms are decomposed as follows:
\eq{\label{eq:su4decomp}
F_m &= f^{(1,0)}_m + \mathrm{c.c} \\
F_{mn} &=f^{(1,1)}_{2|mn}+f_2J_{mn}+\left(f^{(2,0)}_{2|mn}+\mathrm{c.c.}\right) \\
F_{mnp} &=f^{(2,1)}_{3|mnp}+3f^{(1,0)}_{3|[m}J_{np]}+ \tilde{f}^{(1,0)}_{3|s}\Omega^{s*}{}_{mnp} +\mathrm{c.c.}\\
F^+_{mnpq} &=f^{(2,2)}_{4|mnpq}+6f_{4}J_{[mn}J_{pq]} +\left(6f^{(2,0)}_{4|[mn}J_{pq]}+\tilde{f}_4\Omega_{mnps}+\mathrm{c.c.}\right)\\
F^-_{mnpq}&=6f^{(1,1)}_{4|[mn}J_{pq]}+\left(f^{(3,1)}_{4|mnpq}+\mathrm{c.c.}\right)\;,
}
with $f^{(p,q)} \in \Op^{(p,q)}(M)$ $\forall p,q$. In terms of irreducible $SU(4)$ representations, we have that
\eq{\begin{split}
f                 &\sim \bf{1}    \hskip .32cm;  \\
f_{mn}^{(2,0)}    &\sim \bf{6} \oplus \bf{6}  \;;   \\
f_{mnp}^{(2,1)}   &\sim \bf{20}   \;;    \\
f_{mnpq}^{(3,1)}  &\sim \bf{10}   \;;     \\
\end{split}
\quad
\begin{split}
f_m^{(1,0)}       &\sim \bf{4} \\
f_{mn}^{(1,1)}    &\sim \bf{15} \\
& \\
f_{mnpq}^{(2,2)}  &\sim \bf{20'}  \\
\end{split}
}
Reality of a $k$-form is equivalent to demanding  $\left(f^{(p,q)}\right)^* = f^{(q,p)}$.
\\
\\
Note that complex (2,0)-forms $\varphi^{(2,0)}$ take value in the reducible module $\bf{6}\oplus\bf{6}$, and irreducible representations can be formed by imposing a pseudoreality condition:
\eq{\label{pseudo}
\varphi^{(2,0)}_{mn}=\frac18~\! e^{i\theta}\O_{mn}{}^{pq}\varphi^{(0,2)}_{mn}
~,}
where $\theta\in [0, 2\pi)$ is an arbitrary phase. Similarly for complex $(0,2)$-forms.

\section{$SU(5)$-structure}\label{su(5)}
The construction of an $SU(5)$-structure is almost entirely analogous to that of an $SU(4)$-structure, hence everything should be self-explanatory after the previous section. The main purpose here is to give all formulae necessary for doing computations.
\\
\\
An $SU(5)$-structure on a ten-dimensional manifold exists if and only if it admits a globally nowhere vanishing pure spinor. The accidental isomorphism $Spin(6) = SU(4)$ was convenient to see this for $SU(4)$-structures, here this reasoning cannot be used. Instead, the relation between almost complex structures and pure spinors will be exploited to make this evident. As a pure spinor is annihilated by half the gamma matrices, it is possible to define the annihilator space to consist of antiholomorphic gamma matrices, and the remainder of holomorphic gamma matrices (or vice versa). This identification is independent of the normalisation of the spinor. Thus, a line bundle of pure spinors is equivalent to a choice of almost complex structure, reducing the structure group from $SO(10)$ to $U(5)$. Demanding that there is a globally nowhere vanishing section of the line bundle is then equivalent to demanding a smooth section of the canonical bundle, reducing the structure group further to $SU(5)$. This trick will be worked out in greater detail in section \ref{gcg}, where a similar procedure will be used for generalised almost complex structures of $T \oplus T^*$. For all practical purposes, our starting point will simply be to demand existence of the pure spinor, by means of which the almost complex structure and volume form will be constructed.
\\
\\
Given a (normalised) Weyl spinor $\eta$ of $Spin(10)$, $\eta$ is pure if and only if
\eq{
K^m \equiv \we \g^m \eta = 0\;.
}
Given that $\eta$ is pure, the almost symplectic structure and almost Calabi-Yau form are defined as
\eq{\label{eq:su5etajo}
J_{mn} &\equiv - i \widetilde{\eta^c}\g_{mn}\eta\\
\Omega_{mnpqr} &\equiv \we\g_{mnpqr}\eta\;.
}
Fierzing shows that
\eq{
J_m{}^pJ_p{}^n=-\delta_m^n ~,
}
and hence projection operators can again be defined as
\al{
\left(\Pi^\pm\right)_m^{\phantom{m}n} \equiv \frac12 (\delta^n_m \mp i J_m^{\phantom{m}n} ) \;.
}
with respect to which $\O = \O^{(5,0)}$ and $J = J^{(1,1)}$.
Still using no more than definitions and Fierz identities, it follows that
\eq{
J \wedge \O &= 0 \\
\frac{1}{5!} J^5 &= \frac{1} {2^5} i \O \wedge \O^* = \text{vol}_{10}\;,
}
hence $(J, \O)$ defines an $SU(5)$-structure. Decomposing the exterior derivatives of the $SU(5)$-structure itself leads to
\eq{\label{eq:su5torsion}
\d J&=W^*_1\lrcorner\Omega+W_3+W_4\wedge J+\mathrm{c.c.}\\
\d\Omega&= - \frac{16i}{3} W_1\wedge J\wedge J+W_2\wedge J+W^*_{5}\wedge\Omega
~,}
with $W_1 \sim \bf{10}$ a complex (2,0)-form, $W_2 \sim \bf{40}$  a complex primitive (3,1)-form, $W_3 \sim \bf{45}$  a complex primitive (2,1)-form  and $W_{4,5} \sim\bf{5}$ complex (1,0)-forms\footnote{Of course, $(2,0)$- and $(1,0)$-forms are trivially primitive as well, hence all torsion classes are primitive, as they should be.}.
The torsion class conditions on geometrical structures are given by
\\
\begin{center}\label{table:su5}
\begin{tabular}{|l|l|}
\hline
Geometrical structure & Torsion classes \\ \hline
Complex & $W_1 = W_2 = 0$ \\ \hline
Symplectic & $W_1 = W_3 = W_4 = 0$ \\ \hline
K\"{a}hler & $W_1 = W_2 = W_3 = W_4 = 0$ \\\hline
Conformal K\"{a}hler & $W_1 = W_2 = W_3 =0$, $W_4$   exact\\\hline
Nearly Calabi-Yau & $W_1 = W_3 = W_4 = W_5 = 0$ \\ \hline
Conformal Calabi-Yau & $W_1 = W_2 = W_3 =0$, $3 W_4 =  2 W_5$ exact\\\hline
Calabi-Yau & $W_j = 0 $ $\forall j$ \\\hline
\end{tabular}
\end{center}
$\phantom{a}$\\
As is evident, the only difference compared to the $SU(4)$-structure case is the numerical factor for conformal Calabi-Yau structures, as $W_5$ scales slightly differently under conformal transformations.

\subsection*{Tensor decomposition}\label{sec:tensor}
Under an $SO(10)\rightarrow SU(5)$ decomposition the one-form, the two-form, the three-form, the four-form, and the self-dual five-form of $SO(10)$ decompose respectively as:
\eq{\spl{
\bf{10}   &\rightarrow \bf{ 5\oplus \bar{5} }\\
\bf{45}   &\rightarrow \bf{1 \oplus 10 \oplus \bar{10} \oplus  24}\\
\bf{120}  &\rightarrow \bf{5\oplus \bar{5} \oplus 10 \oplus \bar{10} \oplus 45\oplus \bar{45} }\\
\bf{210}  &\rightarrow \bf{ 1\oplus 5\oplus \bar{5} \oplus 10 \oplus \bar{10} \oplus 24 \oplus  40 \oplus \bar{40} \oplus 75 }\\
\bf{126^+} &\rightarrow 1 \oplus \bf{5}\oplus \bar{10} \oplus \bar{15}  \oplus\bar{45}\oplus   50
~.
}}
Note that by `selfdual', it is understood to mean that the $\bf{126^+}$ satisfies $\star F_5 = i F_5$.
Explicitly, $k$-forms can be decomposed as follows:
\eq{\spl{\label{eq:su5decomp}
F_m        &= f_{1|m}^{(1,0)} + f_{1|m}^{(0,1)}                \\
F_{mn}     &= f_2 J_{mn} + f^{(1,1)} + f_{2|mn}^{(2,0)} + f_{2|mn}^{(0,2)}   \\
F_{mnp}    &= 3 f_{3|[m}^{(1,0)} J_{np]} + f_{3|mnp}^{(2,1)} + \frac{1}{2} {f}^{(0,2)}_{3|qr} \O_{mnp}^{\phantom{mnp}qr} +3 f_{3|[m}^{(0,1)} J_{np]} + f_{3|mnp}^{(1,2)} + \frac{1}{2} {f}^{(2,0)}_{3|qr} \O_{mnp}^{*\phantom{mnp}qr} \\
F_{mnpq}   &= 6 f_4 J_{[mn}J_{pq]} + 6 f_{4|[mn}^{(1,1)}J_{pq]} + f_{4|mnpq}^{(2,2)} \\
            &\phantom{=}+ 6 ( f_{4|[mn}^{(2,0)} + f_{4|[mn}^{(0,2)})  J_{pq]} + f_{4|mnpq}^{(3,1)} + f_{4|mnpq}^{(1,3)}
           + {f}^{(0,1)}_{4|r} \O_{mnpq}^{\phantom{mnpq}r} + {f}^{(1,0)}_{4|r} \O_{mnpq}^{*\phantom{mnpq}r} \\
F^+_{mnpqr}&= 30 f_{5|[m}^{(1,0)} J_{np} J_{qr]} + 5 {f}^{(0,2)}_{5|xy} \O_{[mnp}^{\phantom{mnp}xy} J_{qr]} + 10 f^{(1,2)}_{5|[mnp}J_{qr}] \\
 &\phantom{=}+ {f}_5 \O_{mnpqr} + f^{(1,4)}_{5|mnpqr} + f^{(3,2)}_{5|mnpqr}  \;,
}}
where in terms of irreducible $SU(5)$ representations we have:
\eq{\spl{
f                 &\sim \bf{1}    \hskip .32cm;  \\
f_{mn}^{(2,0)}    &\sim \bf{10}   \;;   \\
f_{mnp}^{(2,1)}   &\sim \bf{45}   \;;    \\
f_{mnpq}^{(3,1)}  &\sim \bf{40}   \;;     \\
f_{mnpqr}^{(4,1)} &\sim \bf{15}   \;;
}
\quad
\spl{
f_m^{(1,0)}       &\sim \bf{5} \\
f_{mn}^{(1,1)}    &\sim \bf{24} \\
& \\
f_{mnpq}^{(2,2)}  &\sim \bf{75}  \\
f_{mnpqr}^{(3,2)} &\sim \bf{50}   \;.
}
}
An $SO(10)$ representation is real if and only if the $SU(4)$ representations in the decomposition satisfy $(f^{(p,q)})^* = f^{(q,p)}$.

Using Hodge duality all $k$-forms with $k > 5$ can be similarly expressed in terms of the expansions above; however, let us mention the following expressions which will be needed in section \ref{gcgsolsu5}
\eq{\spl{
F_6 &= -\frac{1}{3} i f_4J \wedge J \wedge J + \frac12 i (f_4^{(1,1)}- f_4^{(2,0)}- f_4^{(0,2)} )\wedge J \wedge J \\
&+ i (f_4^{(3,1)}+f_4^{(1,3)} -f_4^{(2,2)} ) \wedge J -  \tilde{f}_4^{(0,1)} \wedge \O  +  \tilde{f}_4^{(1,0)} \wedge \O^* \\
F_8 &= i \frac{1}{4!}f_2 J^4 + i\frac{1}{3!}(f_2^{(2,0)}+ f_2^{(0,2)}- f_2^{(1,1)})J^3 \\
F_{10} &= -i \frac{1}{5!}f_0 J^5\;.
}}

\newpage

\chapter{Supergravities}\label{sugra}
This section is about the physical theories that are the subject of this thesis, which are all various guises of low-energy string theories. These are described in terms of supergravities. There is a plethora of different supergravities, in various dimensions, with various supersymmetries and with various field content. These are not all independent, as some of them can be obtained from others by means of Kaluza-Klein reduction or dualities.
We will discuss type II supergravity, which describes low-energy type II (i.e., closed) string theory and $D=11$ supergravity, which describes low-energy M-theory (by definition). As there is an abundance of differing conventions in the literature, this will also give us the opportunity to fix ours, which are mostly equivalent to those of \cite{mmlt}.
\\
\\
We start by discussing type II supergravity in section \ref{typeII}, which will be the primary physical focus of this thesis. Secondly, $D=11$ supergravity is discussed in section \ref{m}, after which we consider the relation between type IIA and $D=11$ supergravity in section \ref{dualities}. Next, the method of obtaining lower-dimensional supergravity theories from solutions to the equations of motion which do not break supersymmetry is considered in section \ref{susyvac}. Finding such supersymmetric vacua entails solving Killing spinor equations, which are easier to solve than the equations of motions themselves. In section \ref{integ}, we discuss the conditions for supersymmetry solutions to furnish supersymmetric vacua, a process which goes by the name of integrability.

\section{Type II supergravity}\label{typeII}
Type II supergravity is the low-energy limit of type II string theory. Just as there are two variants of type II string theory, so there are two variants of type II supergravity, namely type IIA \cite{huq} and type IIB \cite{IIB}. Both of these can be described simultaneously by the democratic formalism \cite{demo} as we will use here. Type II supergravity is a theory in  $D=10$ with maximal\footnote{It is maximal in the sense that representation theory tells us that more supersymmetry would lead to massless spin $>$ 2 partcles.} $\ncal = 2$ supersymmetry.
\\
\\
Our starting point is the bosonic pseudo-action of type II supergravity on a ten-dimensional manifold $\mcal_{10}$.
The bosonic\footnote{The reason why we are only interested in the bosonic terms will be made clear in section \ref{susyvac}. See e.g. \cite{demo} for the fermionic terms up to fourth order if one is interested. } pseudo-action of type II supergravity is given by\footnote{
Our conventions for Hodge duality (see \eqref{hodge}) are such that the pseudo-inner product of (suitable, as $\mcal_{10}$ is not compact) differential forms is given by
\al{
\langle \a, \b \rangle &\equiv \int_{\mcal_{10}} \star \a \wedge \b = \int d^{10} x \sqrt{g} \; \frac{1}{k!} \a_{M_1...M_k} \b^{M_1 ... M_k} \;, \qquad \a, \b \in \O^k (\mcal_{10})
}
rather than by having the star act on the second form, thus staying true to the time-honoured tradition that physicists' conventions for (pseudo-)inner products are exactly the opposite of what mathematicians would expect.}
\eq{ \label{eq:tIIaction}
2 \k_{10}^2 S_{II} &= \int d^{10} x \sqrt{g}  e^{-2 \phi} \left( R + 4 (\p \phi)^2  - \frac{1}{2}H^2 \right)  - \frac{1}{4} \fcal^2 \\
&=\int_{\mcal_{10}}   e^{-2 \phi} \left( \star 1 \wedge R + 4 \star \d \phi \wedge \d \phi  -  \frac{1}{2}\star H \wedge  H \right)  - \frac{1}{4} \star \fcal \wedge \fcal\;,
}
supplemented by the additional relation
\al{\label{eq:selfduality}
\fcal = \star \s (\fcal)\;,
}
which needs to be be enforced on the (classical) equations of motions rather than on the action itself: it is this necessity which makes it a pseudo-action rather than a regular action.
In the first line of \eqref{eq:tIIaction} we use the following convention for index contraction of tensors:
\al{
T^2 \equiv \frac{1}{p!} T_{M_1...M_p} T^{M_1... M_p}\;.
}
Here and from now on, $\s$ is defined as the map acting on a $k$-form $\a$ as
\eq{
\s \a_{M_1... M_k} dx^{M_1} \wedge ... \wedge dx^{M_k} &= \a_{M_k ... M_1} dx^{M_1} \wedge ...\wedge dx^{M_k} \\
&= (-1)^{\frac{1}{2}k(k-1)}\a\;.
}
It goes without saying that we will extend this definition to polyforms by linearity and by abuse of notation to act on coefficients $\a_{M_1...M_k}$.

The symmetries of the theory are given by $\ncal = 2$ local supersymmetry, which includes diffeomorphism invariance. The fermionic supersymmetry transformations are given by
\eq{\label{eq:susyeq}
\delta \lambda^1 &= \left(\underline{\p}\phi + \frac{1}{2} \underline{H} \right) \epsilon_1+ \left( \frac{1}{16} e^{\phi} \G^M \underline{\fcal}\G_M \G_{11} \right) \epsilon_2  \\
\delta \lambda^2 &= \left( \underline{\p }\phi - \frac{1}{2} \underline{H} \right) \epsilon_2 - \left( \frac{1}{16} e^{\phi} \G^M \sigma(\underline{\fcal})\G_M \G_{11} \right) \epsilon_1  \\
\delta \psi^1_M &= \left( \nabla_M + \frac{1}{4} \underline{H}_M \right) \epsilon_1 + \left( \frac{1}{16} e^{\phi} \underline{\fcal}\G_M \G_{11} \right) \epsilon_2  \\
\delta \psi^2_M &= \left( \nabla_M - \frac{1}{4} \underline{H}_M \right) \epsilon_2 - \left( \frac{1}{16} e^{\phi} \sigma(\underline{\fcal})\G_M \G_{11} \right) \epsilon_1  \;,
}
while the bosonic transformations will turn out to be of less interest. The variational parameters $\epsilon_{1,2}$ are Majorana-Weyl spinors of $Spin(1,9)$. For type IIA $\epsilon_{1,2}$ are of opposite chirality, whereas for type IIB, they have the same chirality, taken to be positive without loss of generalisation. The use of a non-canonical curvature term $ e^{-2 \phi} R$ compared to the usual $R$ as occurs in $d=4$ general relativity is referred to as the `string frame'. The canonical form, known as the `Einstein frame', is recovered by making the field redefinition
\al{\label{eq:frame}
g_ E \equiv e^{- \frac12 \phi} g_{\text{str}} \;.
}
Generally, the string frame will be more convenient for our purposes.
Including fermions, the action contains the following fields:
\begin{itemize}
\item A vector-spinor $\psi_M$, the gravitino.
\item A spinor $\l$, the dilatino.
\item A Lorentzian metric $g$.
\item A scalar field $\phi$, the dilaton.
\item A closed three-form field, locally $H= \d B$, the Neveu-Schwarz Neveu-schwarz (NSNS) flux.
\item For type IIA, twisted closed $2k$-forms, locally $\fcal = \d_H C$, the Ramond Ramond (RR) fluxes.
\item For type IIB, twisted closed $(2k+1)$-forms, locally $\fcal = \d_H C$, the Ramond Ramond RR fluxes.
\end{itemize}
More specifically, the RR charge $C$ is either an even or odd polyform depending on the type of supergravity; `$C_{-1}$' should be considered formally as a convenient way to unify $\fcal_0 \in \rbb$, known as the Roman's mass, with the other fluxes. The other RR charges $C$ can always be defined locally by Poincar\'{e}'s lemma, but are globally ill-defined, undergoing gauge transformations $C \rightarrow C + \d_H X$; this is similar to garden variety (abelian) gauge theory, except that here, $X$ is a differential form rather than a scalar. A similar story holds for the NSNS charge $B$.
 The twisted exterior derivative $\d_H : \O^\bullet(\mcal_{10}) \rightarrow \O^\bullet(\mcal_{10})$ is defined as
\al{
\d_H \equiv \d + H \wedge.
}
Due to closure of $H$, it satisfies $\d_H^2=0$, however, it does not satisfy the product rule.

The bosonic equations of motion that follow from this action are given by
\eq{
\begin{array}{ccccl}\label{eq:eom}
E_{MN}            &=& R_{MN} + 2 \nabla_M \nabla_N \phi - \frac12 H_M \cdot H_N - \frac14 e^{2 \phi} \fcal_M \cdot \fcal_N &\stackrel{!}{=}&    0\\
\delta H_{MN}     &=&  e^{2 \phi} \star_{10} \left( \d \left( e^{- 2 \phi} \star_{10} H\right) -
                       \frac12 [\left( \star_{10}\fcal\right) \wedge \fcal]_8\right)_{MN}  &\stackrel{!}{=}& 0\\
 D                &=&2 R - H^2 + 8 \left( \nabla^2 \phi- (\p \phi)^2 \right)  &\stackrel{!}{=}&  0
\end{array}
}
with the quantities $D, \delta H_{MN}, E_{MN}$ defined as
\eq{
E_{MN} &\equiv 2 \k_{10}^2  \frac{e^{2\phi}}{\sqrt{g_{10}}}\left( \frac{\delta S}{\delta g_{MN}} - \frac14 g_{MN} \frac{\delta S}{\delta \phi} \right)\\
\delta H_{MN} & \equiv - 2 \k_{10}^2 e^{2 \phi}\star_{10}  \frac{\delta S} {\delta B_{MN}} \\
D & \equiv - 2 \k_{10}^2  \frac{e^{2\phi}}{\sqrt{g_{10}}} \frac{\delta S} {\delta \phi}\;.
}
The Bianchi identities
\eq{\label{eq:bianchi}
\d H &= 0 \\
\d_H \fcal &= 0
}
are required by definition of the fluxes as field strengths. The equations of motion for the RR charges $C$ are incorporated in the Bianchi identity by virtue of \eqref{eq:selfduality}.


\section{M-theory}\label{m}
The field content of any supergravity theory is determined by representation theory of the Poincar\'{e} superalgebra, which is an extension of the Poincar\'{e} algebra by odd generators, known as supercharges. In fact, there is a wide variety of Poincar\'{e} superalgebras, determined by the dimension of the fundamental representation of its Poincar\'{e} algebra, and the number of supercharges. The former corresponds to the dimension $D$ of spacetime for the supergravity theory, the latter is in some dimension-dependent way enumerated as $\ncal$.
By making use of the Poincar\'{e} algebra, it is possible to decompose irreduble reprentations of the superalgebra into a number of irreducible representations of the algebra, and thus assign spin values to the latter. If the superalgebra contains more than 32 supercharges, a supercharge acting on a representation of spin two (i.e., a graviton) will yield a higher spin field, i.e., a representation of spin greater than two. If we assume that such are unphysical, and note that the minimal number of supercharges is related to the dimension, it follows that there must be a supergravity theory of maximal dimension without higher spin fields. This theory is $D=11$, $\ncal = 1$ supergravity \cite{jcs}. Similar to how type II supergravity is the low-energy limit of type II string theory, it is expected that there is a UV (i.e., high energy) completion to $D=11$ supergravity. This hypothesis is supported by the relation between (massless) type IIA supergravity and $D=11$ supergravity that will be discussed in the next section; since $D=11$ supergravity and IIA supergravity are, in some sense that will be made precise, equivalent, and since IIA has a UV completion, it is expected that the same holds for $D=11$ supergravity. This notion is consistent with the existence of M2- and M5-branes, which translate to $D2$- and $NS5$-branes in type IIA supergravity, for which a string interpretation is clear. In particular, the M5-brane plays a crucial role in the anomaly cancellation of $D=11$ supergravity.
This completion of $D=11$ supergravity is defined to be M-theory. We will abuse this notion and not make a clear distinction between the two; for example, solutions to the equations of motion of $D=11$ supergravity will be referred to as M-theory vacua.
\\
\\
Similar to type II supergravity, the starting point for M-theory is given by the bosonic action. It is given by\footnote{The conventions here differ from those in \cite{pta} by $A_3 \rightarrow - A_3$, $C_3 \rightarrow - C_3$, $B \rightarrow + B$. The conventions here make identification of the democratic type IIA formalism to the result of dimensional reduction of M-theory more convenient. See section \ref{dualities} for more details.}
\al{\label{eq:maction}
2 \k_{11}^2 S_{11} &= \int_{\mcal_{11}} \star 1 \wedge R   - \frac{1}{2} \star G_4 \wedge  G_4 + \frac{1}{6} A_3 \wedge G \wedge G\;.
}
Including fermions, the action depends on the following fields:
\begin{itemize}
\item A vector-spinor $\Psi_M$, the gravitino.
\item A Lorentzian metric $g$.
\item A closed four-form, locally $G = \d A_3$, the flux.
\end{itemize}
The first term is the curvature, the second one the kinetic term for the flux, the last one the Chern-Simons term\footnote{Note that the Chern-Simons term is only sensible in the case where $\mcal_{11}$ is the boundary of some twelvefold on which we extend $G$. See \cite{witten} for details.}.
The Bianchi identity is given by
\al{
\d G = 0 \;.
}
The action is invariant under $\ncal=1$ local supersymmetry, which includes diffeomorphism invariance. The supersymmetry transformations of the gravitino is given by
\al{\label{eq:mkse}
\delta_\epsilon \Psi_M = \nabla_M \epsilon + \frac{1}{288} \left( G_{NPQR} \G_M^{\phantom{M}NPQR} - 8 G_{MNPQ} \G^{NPQ}\right)  \;,
}
with $\e$ a Majorana spinor of $Spin(1,10)$.
The (bosonic) equations of motion are
\eq{\label{eq:meom}
2 \k_{11}^2\frac{\delta S_{11}}{\delta g^{MN}} &= R_{MN} - \frac{1}{12} \left(G_{MQRS} G_N^{\phantom{N}QRS} - \frac{1}{12} g_{MN} G_{PQRS} G^{PQRS}\right)  = 0 \\
2 \k_{11}^2 \delta_{A_3}  S_{11} &= - \d \star G + \frac12 G \wedge G = 0
}
See \cite{jcs} for the full fermionic action.

\section{Dualities}\label{dualities}
There are a number of dualities which connect various string theories with one another. Such dualities descend to the level of supergravities. In particular, type IIA and IIB supergravity are related by means of T-duality, whereas $D=11$ supergravity is related to type IIA by means of dimensional reduction.

A rough sketch of the former is as as follows: dimensional reduction of type II supergravity on an $S^1$ yields $\ncal =2 $, $D=9$ supergravity. The theory one obtains from reducing IIA is the same as the one obtained from IIB, up to field redefinition. Thus, given a IIA vacuum on $\mcal_9 \times S_A^1$, one can reduce it to a $D=9$ supergravity vacuum on $\mcal_9$, then lift it to a IIB vacuum on $\mcal_9 \times S_B^1$ and vice versa. The radii $\r_{A,B}$ of $S_{A,B}^1$ are related by the usual $\r_A \r_B = 1$. This duality will be of no concern to us. On the other hand, the duality between M-theory and type IIA will play a part in what follows. It is also less convoluted, as it purely a dimensional reduction of $D=11$ supergravity which leads to type IIA supergravity. Let us examine this in more detail.
\\
\\
Let $\mcal_{11}$ be an $S^1$-bundle over a base space $\mcal_{10}$, such that the transition functions are independent of the local coordinate describing the $S^1$ fibre. Locally,  the metric can then be written as
\al{\label{eq:mdrmetric}
g_{11} &= e^{-\frac16 \phi} g_{10} + e^{\frac43 \phi} \left(\d z + C_1\right)^2
}
with $\phi \in C^\infty(\mcal_{10}, \rbb)$ the dilaton and $C_1$ the RR one-form charge; $C_1$ is the connection one-form of the fibration and is thus a locally defined one-form of $\mcal_{10}$ which is not invariant with respect to diffeomorphisms of the fibres. Furthermore, the M-theory flux charge $A_3$ can be decomposed as
\al{ \label{eq:Gdr}
A_3 = C_3 - B \wedge \d z \;,
}
with the gauge fields $C_3, B$ local sections taking value in $\bigwedge^\bullet T^* \mcal_{10}$; in other words, in terms of coordinates we have $B = B_{mn}(x,z) \d x^m \wedge \d x^n$ and similarly for the three-form $C_3$\footnote{The fact that $\d z$ does not appear in the expression is occasionally referred to as not having a `leg along the fibre'. We shall not use this terminology.}.
Plugging these into the $D=11$ supergravity action \eqref{eq:maction}, it is found that the three terms correspond respectively to
\al{
\int \d^{10} x \;\d z \sqrt{g_{11}}  R_{11} &= \int \d^{10} x \;\d z \sqrt{g_{10}} e^{- \frac16 \phi}  \left(R_{10} - \frac12 \left(\p \phi\right)^2 - \frac12 e^{\frac32 \phi} F_2^2\right)\nonumber \\
\int \d^{10} x \;\d z \sqrt{g_{11}} \left(-\frac12 G_4^2\right) &=  \int \d^{10} x \;\d z \sqrt{g_{10}} e^{- \frac16 \phi} \left(- \frac12 e^{- \phi} H^2 - \frac12 e^{\frac12 \phi} \tilde{F}_4^2 \right) \nonumber \\
\frac16 \int_{\mcal_{11}}  A_3\wedge G_4 \wedge G_4 &= - \frac12 \int_{\mcal_{11}} \d z \wedge  B \wedge F_4 \wedge F_4 \;,
}
where an exact term has been discarded in the last line\footnote{Note also that the last line is somewhat of an abuse of notation, since of course the fibre coordinate $\d z$ is not globally defined.}, and
\al{
\tilde{F}_4 \equiv F_4 + H \wedge C_1 \;, \qquad F_4 = \d C_3\;.
}
Let us now make the ansatz that the fields are independent of the fibre coordinate $z$, such that all fields are (local) sections of the base space $\mcal_{10}$: this is the dimensional reduction ansatz. As a result, the integral over $z$ can be performed explicitly. After changing to the string frame by using \eqref{eq:frame}, the following action is found
\eq{
2 \k_{10}^2 S_{10}^{ND} =&  \int \d^{10} x  \sqrt{g_{10}} \left(e^{-2 \phi} \left[R + 4 (\d \phi)^2 - \frac12 H^2\right] - \frac12 F_2^2 - \frac12 \tilde{F}_4^2 \right)\\
              &- \frac12 \int_{\mcal_{10}} B \wedge F_4 \wedge F_4
}
with
\al{\label{eq:couplingconstants}
 2 \k_{10}^2  \equiv \frac{2 \k_{11}^2}{\int \d z} \;.
}
As can be deduced from the equations of motion, this non-democratic action is equivalent to the democractic action \eqref{eq:tIIaction} up to the obvious field redefinition, except for the Roman's mass term $\fcal_0$, which is not present here. As a final remark, let us note that this describes only the bosonic reduction. There is a more involved procedure that confirms that the $D=11$ fermionic action also reduces to the type IIA fermionic action on a circle fibration. See \cite{huq} for more details.

\section{Supersymmetric Vacua \& Effective Theories }\label{susyvac}
The theories described so far have been in dimensions greater than four. However, as the correspondence between type IIA and $D=11$ supergravity demonstrated, the manifest dimension of the theory (that is, as appears in the action) is not quite as definitive as it appears at first. In this section, we discuss the relation between higher dimensional theories and lower dimensional ones, a relation that is necessary to understand if one is interested in describing four-dimensional physics by means of string theory.
\\
\\
Dimensional reduction lies at the heart of the correspondence between type IIA and $D=11$ supergravity. Similarly, dimensional reduction can be used to obtain lower dimensional supergravities; for example, $\ncal = 8$, $d=4$ supergravity can be obtained by dimensional reduction of $D=11$ supergravity on $\mcal_4 \times T^7$. This procedure requires as input a theory to reduce (which we will refer to as a `higher theory') and a manifold $\mcal_d \times T^{D-d}$ on which to reduce; the result is a supergravity on $\mcal_d$ after integrating out the torus $T^{D-d}$. One might expect this procedure to generalise by taking $T^{D-d} \rightarrow \mcal_{D-d}$ by simply restricting the fields to be sections of $\mcal_d$ and integrating over $\mcal_{D-d}$. However, this is not the case, as it is not a priori clear that vacua of the lower dimensional theory can be uplifted to vacua of the higher theory; by definition, this is the case only for {\it consistent truncations}.
The reason that toroidal dimensional reduction is a consistent truncation is because there is a deeper mathematical motivation behind it than just the simple ansatz of disregarding  the fibre dependence of the fields, namely Kaluza-Klein reduction.

Kaluza-Klein theory was originally developped for the reduction of (non-supersymmetric) gravity on $\mcal_5 \rightarrow \mcal_4 \times S^1$, but the generalisation from five to $D$ dimensions  is straightforward. Consider a vacuum with everything but the metric trivial. Then the $g_{DD}$ component of the metric corresponds to a scalar, the dilaton $\phi$, which results in a massless particle. That is to say, the equation of motion for the dilaton is given by
\al{
\nabla^2_{(D)} \phi = 0 \;.
}
We now consider the expansion $\phi(x,z) = \sum \phi_n(x) \o_n(z)$, with $\o_n$ eigenfunctions of the operator $\p_z^2$; clearly, these eigenfunctions are given by $\exp\left(( i n \frac{z}{\rho}\right)$, with $\rho$ the circle radius, as dictated by standard Fourier theory. Expanding the Laplacian of  the (trivial) fibration $\mcal_D = \mcal_{D-1} \times S^1$ into the Laplacians of the base and the fibre, it follows that
\eq{
\left(\nabla^2_{(D-1)} + \p_z^2 \right) \sum \phi_n(x) e^{i n \frac{z}{\rho}}  = 0 \implies \left( \nabla_{(D-1)}^2 - \left(\frac{n}{\rho}\right)^2 \right) \phi_n(x) = 0\;.
}
Therefore, the Kaluza-Klein expansion results in a massless particle $\phi_0$ and an infinite number of massive particles with mass $m^2 = n^2 / \rho^2 $, the so-called tower of massive KK modes. All equations depend on just one field, and all fields are irreducible representations of the $U(1)$ isometry group of the $S^1$ fibres.  Therefore, one can consider vacua with all massive fields set to zero; physically, this is interpreted as the massive fields being heavy and therefore not relevant to the low-energy field theory. This is exactly dimensional reduction. The generalisation to vacua on higher dimensional tori is straightforward.
\\
\\
Let us now consider how to generalise this idea to non-toroidal vacua on $\mcal_d \times \mcal_{D-d}$. The manifold $\mcal_d$ will be considered as a model for spacetime and is referred to as the {\it external space}, whereas $\mcal_{D-d}$ will be considered as a microscopical object and will be referred to as the {\it internal manifold}\footnote{This terminology is based on the fact that after KK reduction, diffeomorphism invariance of the internal manifold becomes gauge invariance (or internal symmetries) whereas the diffeomorphism invariance of spacetime is known as external symmetry. This should remind one strongly of the phrasing of the Coleman-Mandula theorem. See the duality between $D=11$ and IIA supergravity for an explicit example.}.
For simplicity, our higher theory will be $D=11$ supergravity, and the internal manifold will be taken to be Calabi-Yau. For Calabi-Yau metrics, it is known that \cite{delaossa} fluctuations of the Calabi-Yau metric are given by the K\"{a}hler and complex structure moduli.

Consider the case where one has a fluxless vacuum and is searching for another vacuum by taking small fluctuations around it. That is, one has
\eq{
A_3 &= \langle A_3 \rangle + \delta A_3 \\
G &= \langle G \rangle + \d \delta A_3 \;.
}
with $\langle G \rangle = 0$. Neglecting terms of order $\ocal((\delta A_3)^2)$, the equation of motion for $G$ yields $\d \star \d \delta A_3 = 0$.
Gauge fixing $\d^\dagger \delta  A_3 = 0$, with $d^\dagger \sim \star \d \star$, it follows that
\eq{\label{eq:aeom}
\left(\d \d^\dagger + \d^\dagger \d\right)  A_3 = \nabla^2 A_3 = 0
}
and hence $A_3$ is harmonic. Let us assume that the metric splits similarly to the topology,  $g_{11} = g_d + g_{11-d}$, such that the Laplacian splits accordingly. Let us take the expansion
\eq{
A_3(x,y) = \sum_n C_n(x) \o_n(y)\;,
}
with $C_n$ forms on the external space, and $\o_n$ the eigenforms (or eigenmodes) on the internal space. Clearly, for internal eigenmode $k$-forms, the external forms are $(3-k)$-forms. By assumption, $\nabla^2_{(11)} = \nabla^2_{(d)} + \nabla^2_{(11-d)}$, and as a consequence, \eqref{eq:aeom} reduces to
\eq{
\sum_n \o_n \left(\nabla^2_{(d)}  - m_n^2\right) C_n(x) = 0 \;.
}
Thus, this procedure yields the equations of motions for a number of massless fields, which can be zero-, one-, two-, or three-forms, the number of each determined by the space of harmonics (or zero-modes) for the internal Laplacian.

Next, let us consider the case where $\langle G \rangle \neq 0$. Generically, it is always possible to expand $\delta A_3$ in terms of eigenforms of the Laplacian (of course, these are not known). Although the equation of motion now reduces to
\eq{
\d \star \d \delta A_3 = \frac12 \langle G \rangle \wedge \d \delta A_3 \;,
}
nevertheless we will discard all terms in the expansion of $\delta A_3$ which are not harmonic. This is justified provided the scale of the internal space is smaller than the order of the vacuum flux, such that the non-harmonic modes can be considered far heavier than the harmonic modes and can thus  be discarded to obtain a low-energy effective action. This procedure yields a consistent truncation, despite the fact that $\delta A_3$ is such that $\langle A_3 \rangle + \delta A_3$ no longer satisfies the equation of motion.

There are a number of complications to the procedure. Firstly, the set of harmonics for an arbitrary geometry is not known. In case Hodge's theorem is applicable, there is an isomorphism between cohomology and harmonic forms, hence the Betti numbers will at least tell us how many harmonic forms there are. The second complication is that due to fluctuations of the metric, $\mcal_{11-d}$ is not fixed. The third one is that in the higher theory, the equations of motions are not always so clean, leading to more complicated mass-operators. Finally, let us also note that in all cases discussed, we have disregarded fermions. The procedure for fermions is, to some extent, analogous to the procedure for fluxes, albeit with a different mass operator than the Laplacian, provided one considers a bosonic vacuum, i.e., a vacuum with all fermionic profiles\footnote{By profile, we mean a specification of the field in question. For example, given a scalar field $\phi$, an example of a profile of the scalar field would be $\phi(x) = x$, or $\phi(x) = x^2$. } trivial. Physically, this is the sensible thing to do, as a non-trivial profile for a fermionic field breaks Lorentz invariance of the vacuum.

In general, the procedure is well-understood for compact (bosonic) Calabi-Yau vacua, whereas beyond that, it is an active field of research. For more details, see for example \cite{theisen, dnp}. We will not go through with any reduction, but will keep this in mind as a guideline to what sort of vacua are of interest.
\\
\\
Consider a higher supergravity theory with fields collectively denoted by $\{ \phi_i ,\psi_j\}$, with $\{\phi_i\}$ the bosonic and $\{\psi_j\}$ the fermionic fields.
\dfn{
Let $\mathcal{V}_S \equiv \{\hat{\phi}_i, \hat{\psi}_j\}$ be a set of profiles for the fields $\{ \phi_i ,\psi_j\}$.
If there are non-trivial supersymmetry transformations under which all elements of $\mathcal{V}_S$ are invariant, $\mathcal{V}_S$ is defined to be a {\it supersymmetric solution}. If
$\mathcal{V}_S$ is a supersymmetric solution that satisfies the equations of motion, $\mathcal{V}_S$ is defined to be a {\it supersymmetric vacuum}.
}
The first constraint can be considered to be the fermionic equivalent of demanding that non-trivial isometries of the metric exist and leave all fields invariants; in fact, a supersymmetric solution implies the existence of and invariance under isometries\footnote{Note that since supersymmetry invariance is a local constraint, the resulting isometries will also be local.}. A spinor $\epsilon$ that generates such a supersymmetry transformations is known as a {\it Killing spinor}. The reason they are known as such is because a Killing spinor bilinear yields a Killing vector. For flat space, this can be shown as follows; since
\al{
[\bar{\e}\bar{Q}, \e Q] \sim \bar{\e} \G^M \e P_M \equiv K^M \p_M,
}
invariance of $\hat{g} \in \mathcal{V}_S$  under such supersymmetry transformations implies
\al{
\lcal_K \hat{g} = 0\;,
}
hence $K$ is a Killing vector generating an isometry. A similar though more involved argument can be invoked for vacua on more general spaces. Thus, $\e$ could be thought of as a fermionic generator of a superisometry in the framework of superspace, or as a `square root' of a Killing vector.
\\
\\
At first glance, it appears that finding supersymmetric vacua is more cumbersome than finding arbitrary vacua, as supersymmetric vacua are vacua with an additional constraint. However, it turns out that this is not the case; if one can find a set of fields satisfying some mild assumptions (the {\it integrability conditions}\footnote{Not to be confused with integrability of $G$-structures, which is unrelated.}) which are invariant under a supersymmetry transformation, these will automatically satisfy the equations of motion and thus furnish a supersymmetric vacuum. This is known as the {\it integrability theorem}. The integrability theorem for type II supergravity will be demonstrated in section \ref{integ}.
\\
\\
The vacua we are interested in are those that (vaguely) resemble physical vacua, and therefore ought to satisfy the following constraints: \\
\begin{itemize}
\item The total space is a direct product space: $\mcal_D = \mcal_d \times \mcal_{D-d}$.
\item The metric splits accordingly as
 \eq{
 \hat{g}_{D} = e^{2 A} \hat{g}_d +\hat{g}_{D-d}
 }
 with $g_d$ a Lorentzian metric on $\mcal_d$, $g_{D-d}$ a Riemannian metric on $\mcal_{D-d}$,  $A \in C^\infty(\mcal_{D-d}, \rbb)$. The function $A$ is known as the {\it warp factor}\footnote{Note that it is vital that the prefactor $e^{2A}$ does not change sign or vanish. For an example why, consider the `warped metric' $g_2 = \d \t^2 + \sin^2( \t) \d \phi^2 $: clearly the underlying space of this metric is $S^2$ rather than $S^1 \times S^1$.}.
\item Since the external manifold $\mcal_d$ is to be interpreted as spacetime, it is required to be maximally symmetric. Maximally symmetric Lorentzian manifolds are conformal to either Minkowski, AdS, or dS space. The fact that in our case the conformal warp factor is a function on the internal manifold does not change this.
\item All fields of the vacuum are invariant under global isometries of the spacetime metric. In particular, this means that all fermions must vanish as $\rbb^{1,d-1}$, $\text{AdS}_d$ and $\text{dS}_d$  are invariant under $SO(1,d-1)$.
\end{itemize}
As a consequence of the last point, all bosonic fields are automatically invariant under any supersymmetry transformation, as the result will be proportional to the vanishing fermions.
Every $k$-form can be split into a term proportional to the volume and a term which will be a section of the internal space only; we will come back to this in more detail in section \ref{susysol}, where we explicitly solve supersymmetric vacua for type II supergravity on $\mathbbm{R}^{1,1} \times \mcal_8$.

\section{Integrability}\label{integ}
As stated, one of the reasons among many that supersymmetric vacua are popular is because they are easier to construct. In this section, we will consider the integrability theorem for type II supergravity, which shows, given a set of fields invariant under certain supersymmetry transformations, which additional constraints these fields need to satisfy in order to form a supersymmetric vacuum. The (sourceless) integrability theorem was first discussed in \cite{integIIA} for type IIA, \cite{integIIB} for type IIB. An equivalent theorem for $D=11$ supergravity will be briefly touched on, which was first discussed in \cite{pakis}.
\\
\\
We look for a set of fields invariant under certain supersymmetry transformations. Because a vacuum ought to respect the external isometries, we set all fermions to zero and thus, supersymmetry transformations of bosonic fields automatically vanish. Therefore, what needs to be imposed is that the supersymmetry transformations of the fermions \eqref{eq:susyeq} vanish:
\eq{\label{kse}
\delta \lambda^1 &= \left(\underline{\p}\phi + \frac{1}{2} \underline{H} \right) \epsilon_1+ \left( \frac{1}{16} e^{\phi} \G^M \underline{\fcal}\G_M \G_{11} \right) \epsilon_2 = 0 \\
\delta \lambda^2 &= \left( \underline{\p }\phi - \frac{1}{2} \underline{H} \right) \epsilon_2 - \left( \frac{1}{16} e^{\phi} \G^M \sigma(\underline{\fcal})\G_M \G_{11} \right) \epsilon_1 = 0 \\
\delta \psi^1_M &= \left( \nabla_M + \frac{1}{4} \underline{H}_M \right) \epsilon_1 + \left( \frac{1}{16} e^{\phi} \underline{\fcal}\G_M \G_{11} \right) \epsilon_2 = 0 \\
\delta \psi^2_M &= \left( \nabla_M - \frac{1}{4} \underline{H}_M \right) \epsilon_2 - \left( \frac{1}{16} e^{\phi} \sigma(\underline{\fcal})\G_M \G_{11} \right) \epsilon_1 = 0 \;.
}
These Killing spinor equations are known as the {\it supersymmetry equations}\footnote{They also go by the name of `BPS equations' or `the Killing spinor equations'; for consistency's sake, neither of these will be used throughout this thesis.}; finding solutions will be one of the main points of this thesis. These equations can be manipulated to imply the following, as detailed in \cite{mmlt}:
\al{\label{eq:integ}
0 &= \left(- E_{MN} \G^N + \frac{1}{2} \left(\delta H_{MN} \G^N + \frac{1}{3!} (\d H)_{MNPQ} \G^{NPQ} \right)\right) \epsilon_1 - \frac{1}{4} e^{\phi} \underline{\d_H \fcal} \G_M \G_{11} \epsilon_2 \nonumber \\
0 &= \left(- E_{MN} \G^N - \frac{1}{2} \left(\delta H_{MN} \G^N + \frac{1}{3!} (\d H)_{MNPQ} \G^{NPQ} \right)\right) \epsilon_2 - \frac{1}{4} e^{\phi} \underline{\sigma \d_H \fcal} \G_M \G_{11} \epsilon_1  \nonumber \\
0 &= \left(-\frac{1}{2} D + \underline{\d H} \right) \epsilon_1 + \frac12 \underline{\d_H \fcal} \epsilon_2 \nonumber \\
0 &= \left(-\frac{1}{2} D - \underline{\d H} \right) \epsilon_2 + \frac12 \underline{\sigma \d_H \fcal} \epsilon_1
}
with $D$, $\delta H_{MN}$, $E_{MN}$ defined in (\ref{eq:eom}). After enforcing the Bianchi identities \eqref{eq:bianchi} the above reduces to
\eq{
\ecal^+_{MN}\G^N \e_ 1 &= 0 \\
\ecal^-_{MN}\G^N \e_ 2 &=0 \\
D &= 0 \;,
}
with $\ecal^\pm_{MN} \equiv - 2E_{MN} \pm \delta H_{MN}$. Consider a fixed index $M = \tilde{M}$; the first equation then yields
\eq{
0 = \ecal^+_{\tilde{M}P}\ecal^+_{\tilde{M}N} \G^P \G^N \epsilon_1 =  \ecal^+_{\tilde{M}P}\ecal^+_{\tilde{M}N}  g^{NP} \e_1 \;.
}
Thus, together with a similar procedure for the second equation, it follows that
\al{\label{eq:ecal}
\sum_{N, P = 0}^9 \ecal^\pm_{\tilde{M}N}\ecal^\pm_{\tilde{M}P} g^{NP} = 0 \;.
}
The summation has been spelled out explicitly for clarity. Let us explicitly seperate space and time indices by setting $N  \in \{0, n\}$, $n \in \{1, ...,9\}$. Consider now the fields to be constrained to satisfy
\al{\label{eq:ansatz}
E_{0n} = \delta H_{0n} = 0 \iff \ecal^\pm_{0n} = 0\;.
}
Note that this requires setting $g_{0n} = 0$ and hence, $g_{np}$ is a Riemannian metric whereas $g_{00}$ is Lorentzian.
After splitting indices, \eqref{eq:ecal} reduces to
\al{
\ecal^+_{\tilde{M}n}\ecal^+_{\tilde{M}p} g^{np} + \ecal^+_{\tilde{M}0}\ecal^+_{\tilde{M}0} g^{00} = 0 \;.
}
Setting $\tilde{M} = \tilde{m}$ and imposing \eqref{eq:ansatz}, one finds that the vector $\ecal^\pm_{\tilde{m}n}$ has trivial norm and thus vanishes, whereas setting $\tilde{M} = 0$ leads to $\ecal^\pm_{00} = 0$, thus leading to
\al{
E_{MN} = \delta H_{MN}= 0\;.
}
As it was already deduced that $D=0$ and the equations of motion for the RR fields $\d_H \star \s \fcal = 0$ are encapsulated in the Bianchi identity, it thus follows that all bosonic equations of motion are satisfied. Of course, the fermionic equations of motion have been trivialised by setting all fermions to zero, so a supersymmetrc vacuum has been found. Thus, the following has been shown:

\thm{ \label{thm1}
Let $\vcal \equiv \{g, \phi, \fcal, H, \psi_M^{1,2}, \l^{1,2} \}$ be such that
\begin{enumerate}
\item The supersymmetry equations \eqref{kse} are satisfied
\item $\psi^{1,2}_M = \l^{1,2} = 0$
\item $\d_H \fcal = \d H = 0$,
\item  $E_{0n} = \delta H_{0n} = 0$ for $n \in \{1,...,9\}$.
\end{enumerate}
Then the bosonic equations of motion
\al{
D = E_{MN} = \delta H_{MN} = \d_H \star \s \fcal  = 0
}
as well as the fermionic equations of motion are satisfied, there are supersymmetry transformations leaving $\vcal$ invariant and thus, $\vcal$ is a supersymmetric vacuum. \hfill $\square$
}
This theorem therefore gives a description of how to ensure that a supersymmetric solution of type II supergravity is a supersymmetric vacuum. The next question to consider is how to construct supersymmetric vacua for M-theory. For M-theory, there are two options. Given an M-theory supersymmetric solution, there is a similar integrability theory as the one for type II discussed.  On the other hand, given a type IIA vacuum, it uplifts to an M-theory vacuum. In case the starting point is a type IIA supersymmetric solution, it is possible to either first apply integrability and then uplift or vice versa; the resulting M-theory vacua are the same. In other words, the diagram below is commutative:
\all{
\xymatrixcolsep{5pc}
\xymatrix{
\text{IIA susy solution}\ar[d]^{\text{integrability}} \ar[r]^{\text{uplift}} &  \text{M-theory susy solution}\ar[d]^{\text{integrability}}\\
\text{IIA susy vacuum} \ar[r]^{\text{uplift}} & \text{M-theory susy vacuum}
}
}
The integrability theorem for $D=11$ supergravity is as follows, as given in e.g. \cite{pakis}:
\thm{
Let $\vcal \equiv \{ g, G, \Psi_M \}$ be such that
\begin{enumerate}
\item The $D=11$ supergravity supersymmetry equation \eqref{eq:mkse} is satisfied
\item $\Psi_M  = 0$
\item The Bianchi identity $\d G = 0$ is satsifed
\item $E_{0n} =  0$ for $n \in \{1,...,10\}$.
\item The equation of motion for the flux  $- \d \star G + \frac12 G\wedge G = 0$ is satisfied
\end{enumerate}
Then the remaining equation of motion for the metric $g$ is automatically satisfied, hence $\vcal$ is a supersymmetric vacuum.
}
In addition to the integrability, it can be shown \cite{witten} that the quantum theory for M-theory requires a quantisation condition on the flux $G$ as follows from the duality between M-theory and heterotic $E_8 \times E_8$ string theory. This condition is given by
\eq{\label{eq:quant}
\frac{[G]}{2 \pi} - \frac{p_1}{4} \in H^4(\mcal_{11}, \zbb)\;,
}
where $[G]$ denotes the cohomology class of $G$. Furthermore, anomaly cancelation requires a higher order correction to the equation of motion of the flux given in \eqref{eq:meom}, namely
\eq{
- \d \star_{11} G +  G \wedge G =  (2 \pi)^2 X_8 \;.
}
See \cite{dlm}, \cite{svw}, \cite{bb}. This will be discussed in more detail in section \ref{nogo}.

\section{Branes}
Type II supergravity describes low-energy type II string theory, whereas $D=11$ supergravity describes low-energy M-theory. Both type II string theory as well as M-theory describe branes.
Branes can be viewed as higher-dimensional analogues of point-particles (i.e., as `extended objects'), and hence are located at (or, they `wrap') a submanifold $\scal \subset \mcal_D$, rather than having a wordline description in spacetime. On the one hand, branes are solutions to the supergravity equations of motion supplemented by a source-term. Thus, by expanding around these solutions, effective actions can be found describing dynamics of the branes themselves. On the other hand, as branes have energy, their presence affects the gravitational field, resulting in a backreaction.
This effect is analogous to what is encountered in classical electrodynamics: on the one hand, electrons are pointlike and can be found as solutions to sourced Maxwell's equations, on the other hand, placing an electron in an electric field backreacts to modify the electric field.

There are various kinds of branes. The ones that will be of relevance to us are $M_2$- and $M_5$-branes in M-theory, D$p$-branes and NS5-branes in type II supergravity. $M_2$-branes are 2+1 dimensional and can act as sources of the charge $A_3$ with fieldstrength $G$. D$p$-branes are $p+1$ dimensional and source the RR charges $C_{(p+1)}$ of the RR fluxes $\fcal_{(p+2)}$. Hence $p$ is even/odd for IIA/IIB. NS5-branes are 5+1 dimensional, and correspond to $M_5$-branes of M-theory by duality. These source the NSNS charge $B$, albeit in a far more convoluted way than the $M_2$- and $D$-branes source their respective charges.

There are a number of situations where these branes will play a part in the following, and more details will be provided as required when they show up. For a far more in-depth description than what is needed, we refer the reader to \cite{johnson} or especially \cite{simon}, or alternatively to the more general texts \cite{bbs} \cite{pol}.

\newpage

\chapter{Supersymmetric vacua on manifolds with $SU(4)$-structure}\label{III}
This section will give explicit constraints for supersymmetric flux vacua. In particular, we will give families of $\ncal = (1,1)$ type IIA vacua on $\rbb^{1,1} \times \mcal_8$, $\ncal = (2,0)$ type IIB vacua on $\rbb^{1,1} \times \mcal_8$ and $\ncal = 1 $ M-theory vacua on $\rbb^{1,2} \times \mcal_8$; these are results of \cite{pta}, \cite{ptb}. In order to do so, it will be assumed that $\mcal_8$ is such that it admits an $SU(4)$-structure, which will aid in the process tremendously. A number of other ans\"{a}tze will also be made, which will be spelled out explicitly, as well as the rationale behind the specific supersymmetry and the external spacetime. The procedure to obtain the type II vacua is to find solutions to the supersymmetry equations of type II supergravity, and then apply the integrability theorem of section \ref{integ}. In order to find M-theory vacua, the type IIA vacua is uplifted. The result is then double checked by doing the computation directly in $D=11$. Explicit examples showcasing existence are presented on $K_3$ surfaces.
\\
\\
In section \ref{setup}, all of the technical details of how to obtain our solutions to the supersymmetry are discussed, folllowed by the implementation and the results in section \ref{susysol}. The integrability theorem is applied in \ref{appliedinteg}, which leads to constraints on specific type II supersymmetric flux vacua. The $D=11$ supergravity analogue is given in section \ref{mtheory}. There are some issues with flux vacua on compact internal manifolds as described by the Maldacena-Nu\~{n}ez no-go theorem, which is adapted to our circumstances in section \ref{nogo}.
In \ref{k3vacua}, explicit examples of M-theory vacua are given on the product of two K3 surfaces; the details of K3 surfaces are discussed, as well as supersymmetry enhancement and an escape from the no-go theorem by means of higher order corrections.

\section{Setting up for solving the type II supersymmetry equations}\label{setup}
Due to the integrability theorem \ref{thm1}, the main task to be done to find supersymmetric vacua of type II supergravity is to solve the susy equations \eqref{kse}. A generic solution to the susy equations is still too much to ask for, however. We will simplify the problem by making a number of ans\"{a}tze. Firstly, the only vacua of concern will be those satisfying the constraints mentioned in section \ref{susyvac}. We consider $d=2$, hence
\al{
\mcal_{10} = \mcal_2 \times \mcal_8
}
with $\mcal_2 \in \{\text{AdS}_2, \text{dS}_2, \rbb^{1,1} \}$ identified as spacetime.
Secondly, we restrict ourselves to the case where $\mcal_8$ admits an $SU(4)$-structure.
Thirdly, we consider only solutions with a `strict $SU(4)$'-ansatz, which is an ansatz about the relation between the $SU(4)$-structure and the supersymmetry that leaves the solution invariant; at this point, this ansatz will be considered purely as a technical ansatz, but a deeper interpretation will be given in section \ref{su5}.
Fourth and finally, we consider vacua that are invariant under the minimal supersymmetry that still admits $\k$-symmetric D-branes\footnote{
The action of a D$p$-brane is invariant under worldvolume supersymmetry in $p+1$ dimensions, which differs from the $D=10$ target space supersymmetry of $\mcal_{10}$. In order to write down an action where the target space supersymmetry is manifest (the Green-Schwarz formalism), it is necessary to make use of gauge degrees of freedom, which are described by $\k$-symmetry. See \cite{koerber-gcg} and references therein or \cite{simon} for more details.};
this means $\ncal = (1,1)$ for type IIA and $\ncal = (2,0)$ for type IIB. As invariance under less supersymmetry implies less restrictions on the solution, the IIA solutions will thus generalise the known case of strict $SU(4)$ IIA, $d=2$, $\ncal = (2,2)$ \cite{gukov}.
\\
\\
Let us discuss these simplifications in more detail. The metric is given by
\al{
g_{10} = e^{2 A} g_2 + g_8 \;,
}
with $g_2$ Lorentzian, $g_8$ Riemannian, and the warp factor $A \in C^\infty(M_8, \rbb)$. To this metric one can thus associate an $SO(1,1) \times SO(8)$-structure\footnote{After choosing an orientation; existence of both orientation and a spin-structure on $\mcal_8$ are assumed.} and hence, all fields can be decomposed into representations of this subgroup of $SO(1,9)$.
The $G$-structure on $\mcal_2$ is a subgroup of the group of global isometries and hence, invariance with respect to the spacetime metric of $\mcal_2$ requires that the fields are $SO(1,1)$ invariant. Thus, the fluxes ought to be decomposable as
\eq{\label{eq:fluxdecomp}
\fcal &= \vol_2   \wedge F^{el} + F \;, \quad F^{el} =  e^{2A} \star_8 \s F \\
H &= \vol_2 \wedge H_1 + H_3
}
with $\vol_2$ unwarped, and as mentioned before, all fermions ought to vanish. $F^{el}$ is known as the `electric' flux, $F$ as the `magnetic' flux. The expression for $F^{el}$ is equivalent to the selfduality constraint $\fcal = \star_{10} \s \fcal$.
\\
\\
The second ansatz is the existence of an $SU(4)$-structure on $\mcal_8$. As explained in section \ref{su(4)}, existence of an $SU(4)$-structure is equivalent to existence of a globally well-defined nowhere vanishing pure spinor. Existence of an $SU(4)$-structure has multiple benefits. First of all, it allows all fluxes to be decomposed according to \eqref{eq:su4decomp}. This allows the supersymmetry equations to be decomposed into equations for each seperate irreducible representation. Secondly, existence of the volume form $\O^{(4,0)}$ allows for further decomposition by means of something akin to holomorphic Hodge duality (despite the fact that the almost complex structure need not be integrable). This is made manifest in identities such as \eqref{eq:su4spinorhodge}. Thirdly, by on the one hand relating $\e_{1,2}$ to the pure spinor $\eta$, and on the other hand relating $\nabla \eta$ to $\d J$ and $\d \O$ through the fact that $(J, \O)$ are spinor bilinears, $\nabla_m \e$ is expressible in terms of torsion classes and the $SU(4)$-structure. More specifically, the relation \eqref{eq:su4torseta} holds. As a result, the supersymmetry equations become purely algebraic, and thus solvable. See appendix \ref{su4der} for more details on this procedure.
\\
\\
As the spin-structure of the spinor bundle splis accordingly into a $Spin(1,1) \times Spin(8)$-structure, it will also be possible to decompose the Killing spinors of the supersymmetry transformations, $\e_{1,2}$. The specifics of this decomposition are rather intricate and have substantial consequences.  Firstly, the number of external degrees of freedom is equivalent to the number of supercharges of the vacuum. In the case of type IIA, the vacua of interest will have $\ncal = (1,1)$. Thus, for type IIA, the spinors decompose\footnote{See appendix \ref{spinors} for details on spinor conventions.} as
\eq{\label{eq:ksiia}
\e_1^A &=  \z_+ \otimes \eta^1_+  + \z_- \otimes \eta^1_-\\
\e_2^A &= \z_- \otimes \eta_+^2 + \z_+ \otimes \eta^2_-  \;,
}
with the free parameters $\z_\pm$  Majorana-Weyl spinors of $Spin(1,1)$ of $\pm$ chirality, and $\eta^{1,2}_\pm$ fixed real Weyl spinors of $Spin(8)$ of $\pm$ chirality.  As a Majorana-Weyl spinor of $Spin(1,1)$ has one real degree of freedom, the vacuum thus indeed has $\ncal = (1,1)$ supersymmetry. On the other hand, for type IIB, the spinors decompose as
\eq{\label{eq:ksiib}
\e_1^B &=  \z_+ \otimes \eta^1_ +  + \z^c_+ \otimes (\eta^1_+)^c\\
\e_2^B &= \z_+ \otimes \eta_+^2 + \z^c_+ \otimes (\eta^2_+)^c  \;,
}
where now, the free parameter $\z_+$ is a Weyl spinor of $Spin(1,1)$ and $\eta^{1,2}_\pm$ are fixed Weyl spinors of $Spin(8)$. Let us stress that this formulation using a complex-valued Weyl spinor of $Spin(1,1)$ is equivalent to using two independent Majorana-Weyl spinors, but more convenient in practice. This is thus the most general $\ncal = (2,0)$ spinor decomposition.
\\
\\
The pure spinor $\eta$ that determines the $SU(4)$-structure can assist in decomposing $\epsilon_{1,2}$ further. The Clifford algebra $Cl(8)$ acts transitively on the fibers of the spin bundle, and hence locally any spinor $\xi_\pm$ of $\pm$ chirality can be expressed as\footnote{A quick degree of freedom counting shows that both LHS and RHS have eight complex dofs. Hodge duality and equations \eqref{eq:su4spinorhodge} have been used to rewrite higher order gamma matrices. }
\eq{\label{eq:spinordecomp}
\xi_+ &= a \eta + b \eta^c + c_{mn} \g^{mn} \eta \\
\xi_- &=  d_m \g^m \eta + e_m \g^m \eta^c
}
hence $\eta_\pm^{1,2}$ can be expressed in terms of $\eta$. The strict $SU(4)$-ansatz now entails that the internal spinors are directly proportional to $\eta$, i.e., $b = c_{mn} = d_m = e_m = 0$.
Let us also note the following. The requirement that the vacuum admits $\k$-symmetric branes comes down to demanding that (see \cite{koerber-gcg} and references therein for details)
\eq{
\e_1 = \G \e_2 \;,
}
with $\G$ known as the worldvolume chiral operator.  Using the fact that $\G^\dagger \G =  \obb$ and our spinor ans\"{a}tze \eqref{eq:ksiia}, \eqref{eq:ksiib}, it follows that the norms of $\eta_{1,2}$ must be the equivalent. This also immediately implies that  $\ncal = (1,0)$ vacua for both IIA and IIB are incompatible with $\k$-symmetric branes.

Enforcing this equivalence and making use of the strict $SU(4)$-ansatz, the parameters determining the supersymmetry transformations are given by
\eq{\label{eq:IIAks}
\e_1^A &=  \frac{\a}{\sqrt{2}} \z_+ \otimes \left(\eta + \eta^c \right) \\
\e_2^A &=  \frac{\a}{\sqrt{2}} \z_- \otimes \left(e^{i \t} \eta + e^{- i \t} \eta^c \right)
}
for type IIA and
\eq{\label{eq:IIBks}
\e_1^B &=  \a \;\z_+ \otimes \eta   + \a\; \z^c_+ \otimes \eta^c  \\
\e_2^B &= \a \;\z_+ \otimes  e^{i \t} \eta + \a \;\z^c_+ \otimes  e^{-i \t} \eta^c  \;,
}
for type IIB, with $\a, \t \in C^\infty (\mcal_8, \rbb)$. Finally, let us remark on the relation between the Killing spinors and the external space.
Consider the Majorana spinor $\z = \z_+ + \z_-$. As it is a Killing spinor, it satisfies
\al{\label{eq:extkse}
\nabla_\m^{(2)} \z = \rho \g_\m \z \implies \nabla_\m^{(2)} \z_+ = \rho \g_\m \z_- \;,
}
with the curvature given by
\al{
R_{(2)} = - 8 \rho^2 \;.
}
As $R_{\text{AdS}} < 0 = R_{\text{Mink}} < R_{\text{dS}}$, the connection acting on the external part of the spinor tells us something about the external spacetime. It follows immediately that, since our IIB Killing spinors satisfy $\z_- = 0$ due to the strict $SU(4)$-ansatz, our IIB solutions will all have $\rbb^{1,1}$ as external spacetime.

\section{Solving the type II supersymmetry equations}\label{susysol}
The previous section consisted of explanations and justifications for the following:
\begin{itemize}
\item We use explicit expressions \eqref{eq:IIAks}, \eqref{eq:IIBks} for the parameters $\e_{1,2}$ in the supersymmetry equations.
\item We decompose gamma matrices into external and external parts according to \eqref{eq:gammadecomp}.
\item We decompose the fluxes into external and internal parts according to \eqref{eq:fluxdecomp}.
\item We decompose the internal fluxes in $SU(4)$ representations using \eqref{eq:su4decomp}.
\end{itemize}
In addition, we do the following:
\begin{itemize}
\item Any term with $>$ 4 (internal) indices will be Hodge dualised to a term with $<$ 4 indices.
\item As we have seen, any gamma matrix with $>$ 2 internal indices can be expressed in terms of the $SU(4)$-structure and gamma matrices with $<$ 2 indices. These identities, given by \eqref{eq:su4spinorhodge}, are inserted wherever possible.
\item The connection acting on the internal Killing spinors can be expressed in terms of torsion classes via \eqref{eq:su4torseta} due to the Killing spinors being expresseable in terms of the $SU(4)$-structure. The connection acting on the external Killing spinors can be expressed in terms of the external scalar curvature via \eqref{eq:extkse}.
\end{itemize}
Putting all this together leads to the following main results of \cite{pta}, \cite{ptb}:
\\
\\
The type IIB $\ncal = (2,0)$ solutions on $\mcal_2 \times \mcal_8$ with an $SU(4)$-structure on $\mcal_8$ and a strict $SU(4)$-ansatz for the Killing spinors are given by
\eq{\spl{\label{eq:iib}
\rho &= 0\\
W_1 &= W_2 = 0 \\
W_3 &= i e^\phi (\cos\t f^{(2,1)}_3  - i \sin\t f^{(2,1)}_5) \\
W_{4} &= \frac{2}{3}\p  (\phi- A )\\
W_{5} &=  \p (\phi - 2 A +  i \t)\\
\a &= e^{ \frac{1}{2} A}\\
\tilde{f}^{(1,0)}_{3} &=  \tilde{f}^{(1,0)}_{5} = \tilde{h}^{(1,0)}_{3} =0 \\
h_{1}^{(1,0)} &= 0\\
h_{3}^{(1,0)} &= \frac{2}{3} \p \t \\
f_{1}^{(1,0)} &=-i \p ( e^{- \phi}\sin\t )\\
f_{3}^{(1,0)} &= - \frac{i}{3}e^{2A} \p ( e^{-2A- \phi}\cos\t )\\
f_{5}^{(1,0)} &=\frac{1}{3}e^{-4A} \p( e^{4A- \phi}\sin\t )\\
f_{7}^{(1,0)} &=e^{-2A} \p( e^{2A- \phi}\cos\t )\\
h^{(2,1)} &= e^\phi (- \cos\t f^{(2,1)}_5  + i \sin\t f^{(2,1)}_3)
~.
}}
The free parameters of the solution are the warp factor, dilaton and an internal Killing spinor parameter, $A, \phi, \t \in C^\infty(\mcal_8, \rbb)$, as well as parts of the RR flux $f_{3,5}^{(2,1)} \in \O^{(2,1)}_{\text{p}}(\mcal_8)$. Although a priori the fluxes are sections of $\mcal_{10}$, vanishing of $\rho$ means the only allowed external spacetime is $\rbb^{1,1}$. As a consequence, invariance under isometries of the external metric requires external translational invariance of the fields, thus $F$ is a section of the internal manifold instead, i.e., $F(x,y) = F(y)$. The internal manifold is necessarily complex due to $W_1 = W_2 = 0$, and thus, the Dolbeault operator $\p$ is well-defined. The NSNS three-form $H$ is purely internal.
\\
\\
The type IIA $\ncal = (1,1)$ solutions on $\mcal_2 \times \mcal_8$ with an $SU(4)$-structure on $\mcal_8$ and a strict $SU(4)$-ansatz for the Killing spinors are given by three branches, depending on $\t$. Generically,
\eq{\label{eq:hint}
\rho &= 0 \\
\alpha &= e^{ \frac12 A}\\
H_1 &= - 2\d A~,}
with the warp factor $A \in C^\infty(\mcal_8, \rbb)$ a free parameter. Note that again, due to the strict $SU(4)$-ansatz, the only allowed external spacetime is $\rbb^{1,1}$ and thus again, the internal fluxes $F$ satisfy $F \in \O^\bullet(\mcal_8)$. Furthermore, the internal fluxes obey a (twisted) self-duality condition:
\eq{\spl{\label{eq:11fluxes1}
F=\star_8\sigma(F)
~.}}
In addition, the RR-fluxes generically obey the following relations:
\eq{  \label{eq:rrconstraints}
f_4&=\frac16 f_0+\frac43 e^{-i\theta}\cos\theta\tilde{f}_4\\
f_2&=2e^{-i\theta}\sin\theta\tilde{f}_4 \\
\sin \t f_{2|mn}^{(2,0)} &= - \cos \t f_{4|mn}^{(2,0)} - \frac{1}{8} e^{i \t} \O_{mn}^{\phantom{mn}pq} f_{4|pq}^{(0,2)} \\
\sin \t f_{4|mn}^{(2,0)} &=   \cos \t f_{2|mn}^{(2,0)} - \frac{1}{8} e^{i \t} \O_{mn}^{\phantom{mn}pq} f_{2|pq}^{(0,2)}
\;,}
where $f_{0,2,4} \in C^\infty (\mcal_8, \rbb)$, while $\tilde{f}_4 \in C^\infty(\mcal_8, \cbb)$. Note that the last two equations are equivalent for $e^{2i \t} \neq 1$, whereas for $e^{2 i \t} = 1$ they become independent pseudo-reality conditions. The remaining NSNS fields satisfy the constraints of one of the following branches, depending on the internal Killing spinor parameter $\t \in C^\infty(\mcal_8, \rbb)$.
\\
\\
$\bullet$ $e^{2i\t} = 1$:
\eq{
\label{eq:iiasol1}
e^\phi &=  g_s e^A\\
h_{3}^{(1,0)} = \tilde{h}_{3}^{(1,0)} &= 0\\
h_{3}^{(2,1)} &=0\\
W_1 &=-\frac{3i}{4}W_4\\
W_3 &=\frac{1}{2}W_2\\
W_5 &=\frac{3}{2}W_4
~,}
with $g_s$ a non-zero integration constant, $W_4^{(1,0)}$, $W_2^{(2,1)}$ unconstrained.\\

$\bullet$ $e^{2i\t} = -1$:
\eq{\label{eq:iiasol2}
h_{3}^{(1,0)} &= 0\\
\tilde{h}_{3}^{(1,0)} &=  \frac{1}{4}\partial^{+}(A-\phi)\\
W_1 &=0\\
W_{2} &=-2i h_{3}^{(2,1)}\\
W_3 &=0\\
W_4 &=\partial^{+}(\phi - A)\\
W_5 &=\frac{3}{2}\partial^{+}(\phi - A)
~,}
with $\phi$,  $h_3^{(2,1)}$ unconstrained; for any scalar $S$, $\partial^{\pm}S$ denotes the projection of the exterior derivative $\d S$ onto its (1,0), (0,1) parts.\\

$\bullet$ $e^{2i\t} \neq \pm 1$:
\eq{\label{eq:iiasol3}
e^\phi&=g_s e^{A}\cos\theta \\
h_{3}^{(1,0)} &= \frac{2}{3}\partial^{+}\t\\
\tilde{h}_{3}^{(1,0)} &= \frac{1}{4}(i+\tan\t)\partial^{+}\t\\
W_1^{(1,0)}&=\frac{1}{4}(1+i\cot\t)\partial^{+}\t\\
W_{2}^{(2,1)} &=2(-i+\cot\t)~\!h_{3}^{(2,1)}\\
W_3^{(2,1)}&=\cot\t ~\!h_{3}^{(2,1)}\\
W_4^{(1,0)}&=-(\tan\t+\frac13\cot\t)\partial^{+}\t\\
W_5^{(1,0)}&=(i-\frac{1}{2}\cot\t-\frac{3}{2}\tan\t)\partial^{+}\t
~,}
with $g_s$ a non-zero integration constant, $h_3^{(2,1)}$ unconstrained.

\section{Integrability of the supersymmetry solutions}\label{appliedinteg}
To recap: section \ref{susysol} contains solutions to the supersymmetry equations, which, according to section \ref{integ}, will furnish vacua of type II supergravity provided all fermions are set to zero, $E_{0n} = \delta H_{0n} = 0$, and the Bianchi identities are satisfied. The latter two conditions bear closer examination. A generic feature of our supersymmetry solutions is that the metric is given by
\al{
g_{10} = e^{2 A} \eta_2 + g_8 \;.
}
Poincar\'{e} invariance of the vacuum is then enough to ensure that, if $\delta H_{01} = 0$, then $E_{0n} = \delta H_{0n} = 0$. The Bianchi identities will also reduce further upon closer inspection, depending on the kind of internal space under consideration.

\subsection{IIA $\ncal = (1,1)$  Calabi-Yau vacua}\label{IIACY}
Consider type IIA with  $W_j = 0$ $\forall j$. As a consequence, the supersymmetry solution reduces as follows.
The metric is given by
\al{\label{eq:iiametric}
g_{10} & = e^{2A} \eta_2 + g_8\;.
}
The NSNS flux, dilaton and warp factor are constrained to be
\eq{\spl{\label{nsf}
e^{\phi} &= g_se^A \\
H &=-\mathrm{vol}_2\wedge\d e^{2A}\;.
}}
The RR fluxes are given by
\eq{\spl{\label{rrf}
F_0&=f_0\\
F_2&=f_2 J + f_2^{(1,1)}+ \left(f_2^{(2,0)} + \text{c.c.}\right)\\
F_4&= \star_8F_4 = f_4^{(2,2)} + f_4 J \wedge J + \left( \tilde{f}_4 \O + f^{(2,0)}_4 \wedge J + \text{c.c} \right)
\\
F_6&=-\star_8F_2\\
F_8&=\star_8F_0
~,
}}
where $J$ and $\O$ are the K\"{a}hler form and holomorphic four-form of the Calabi-Yau fourfold respectively. In addition, the RR fluxes obey the constraint \eqref{eq:rrconstraints} with $\t$ now constant.
\\
\\
The Bianchi identities reduce as follows. Given the constraint on $H$ \eqref{nsf}, the supersymmetry solution automatically satisfies the NSNS Bianchi identity $\d H=0$. The RR Bianchi identity reduces to closure and co-closure of the magnetic RR flux:
\eq{\label{eq:cybianchi}
\d F=\d\star_8 F=0~.
}
As a result,  $f_{0,2,4} \in \mathbbm{R}$, $\tilde{f}_4 \in \cbb$, and $f_2^{(1,1)}$, $f_2^{(2,0)}$, $f_4^{(2,0)}$, $f_4^{(2,2)}$ are closed and co-closed.
\\
\\
Finally, consider the constraint $\delta H_{01} = 0$. Inserting the fields above, this integrability constraint reduces to
\eq{\label{int2}
-\d\star_8\d e^{-2A}+\frac{g_s^2}{2} F\wedge \sigma(F)|_8=0
~.
}

\subsection{IIA $\ncal = (1,1)$ complex non-Calabi-Yau vacua}\label{cplxvac}
Consider type IIA with $W_1 = W_2 = 0$ with some other torsion not vanishing (obviously, the Calabi-Yau vacua described above are all subcases otherwise). From \eqref{eq:iiasol1}, \eqref{eq:iiasol2}, \eqref{eq:iiasol3} it follows that this requires $e^{2 i \t} = -1$. As follows from the discussion on M-theory uplifts in section \ref{mtheory}, it follows that these vacua will not uplift to M-theory vacua on $\rbb^{1,2} \times \mcal_8$, a fact that will become important for section \ref{stnzl}, but can be ignored for now.
\\
\\
Taking the $e^{2 i \t} = -1$ branch of the supersymmetric solution and enforcing $W_1 = W_2 = 0$, the NSNS terms \eqref{eq:iiasol2} reduces as follows
\eq{ \label{eq:susycons}
W_1 &= W_2 = W_3 = h^{(2,1)} =  0 \\
W_4 &= \frac23 W_5 = - 4 \tilde{h}^{(1,0)} = \p ( \phi - A)
}
and hence the NSNS flux is given by
\eq{ \label{eq:h3}
H =  - \vol_2 \wedge \d e^{2A} - \left( \frac14 W_4^* \lrcorner \O  + \text{c.c.}\right) \;.
}
In particular, the internal flux $H_3$ does not vanish, which severely complicates solving the Bianchi identities. The RR Bianchi identity reduces to
\al{
\d_{H_3} F = 0\;.
}
Generically, this will not be satisfied. Depending on the particulars of $\mcal_8$ and its forms, certain specific solutions may be found. A solution that does generically satisfy this constraint, independent of the particularities of forms on $\mcal_8$, is given by
\al{
F = f^{(2,2)} \;, \qquad \d f^{(2,2)} = 0 \;,
}
since $H_3 = H_3^{(3,0)} + H_3^{(0,3)}$. The NSNS Bianchi identity is given by
\al{
\d \left( W_4^* \lrcorner \O \right) = 0 \;.
}
Whether or not this is satisfied depends on the particularities of the $SU(4)$-structure on $\mcal_8$.

\subsection{IIB $\ncal = (2,0)$ conformal Calabi-Yau vacua}\label{IIBCCY}
Consider the type IIB supersymmetry solution and set
\eq{
\t =& \pi\\
e^{\phi} =& g_s e^{-2A}\\
f_3^{(2,1)} =& 0 \;.
}
As a consequence, the torsion classes are given by
\eq{\label{eq:iibtors}
W_1 = W_2 = W_3 &= 0 \\
W_4 = \frac12 W_5 &= - 2 \p A \;,
}
which indeed gives us a conformal Calabi-Yau structure, with the conformal metric $g_8$ related to a Calabi-Yau metric $g_{CY}$ by
\al{\label{eq:ccymetric}
g_8 = e^{-2 A } g_{CY} ;.
}
The NSNS three-form is given by
\al{\label{eq:iibnsfluxes}
H = h^{(2,1)} + h^{(1,2)}\;,
}
in particular, $H$ is internal and primitive.
The non-vanishing RR fluxes are given by
\eq{\spl{\label{eq:iibrrfluxes}
{g_s}\fcal_3&= \mathrm{vol}_2\wedge\d e^{4A}\\
{g_s}\fcal_5&= e^{4A}\mathrm{vol}_2\wedge H- e^{2 A}\star_8 H\\
g_s \fcal_7 &= \star_{10} \s \fcal_3\;.
}}
The integrability condition $\delta H_{01} = 0$ and the Bianchi identities need to be imposed on this supersymmetric solution to obtain a vacuum.
Unlike in the case of IIA in the previous section, $\delta H_{01}$ is trivial in this case and leads to no additional constraints.
Instead, in IIB, the constraint on the warp factor follows from the Bianchi identities.
The non-trivial Bianchi identies are given by
\eq{
\d H  &= 0 \\
\d \fcal_5 + H \wedge \fcal_3 &= 0\\
\d \fcal_7 + H \wedge \fcal_5 &= 0 \;.
}
The second line implies
\al{ \label{eq:iibbianchis}
\d H = \d e^{2 A}\star_8 H = 0
}
while the third implies
\al{
\d \star_8 \d e^{2 A} +\frac{1}{2} H \wedge e^{2 A} \star_8 H = 0 \;.
}
In terms of the Calabi-Yau metric, these can be rewritten as respectively
\eq{
\d H = \d \star_{CY} H = 0
}
and
\al{\label{eq:iibwarp}
- \d \star_{CY} \d e^{-4 A} + H \wedge \star_{CY} H = 0 \;.
}

\section{M-theory supersymmetry solutions}\label{mtheory}
The M-theory vacua of interest are supersymmetric $\ncal = 1$ vacua on $\rbb^{1,2} \times \mcal_8$, where $\mcal_8$ is equipped with an $SU(4)$-structure, with corresponding warped metric
 \al{\label{eq:mmetric}
 g_{11} = e^{\frac43 \tilde{A} }\eta_3 + e^{- \frac23 \tilde{A}} g_8
  }
 with some\footnote{The exact conformal factors can of course be set however one desires through redefinitions of $\tilde{A}$, $g_8$; the convention used here is most suited for our purpose of uplifting the IIA metric, as it will correspond to $\tilde{A} = A$.} warp factor $\tilde{A}$. Such vacua can be constructed in two ways, as discussed in section \ref{integ}. The straightforward option is to repeat the entire procedure for type II vacua: deduce a suitable Killing spinor ansatz, plug this into the M-theory supersymmetry equation, and then check integrability of the supersymmetric solution to a vacuum. Alternatively, the fact that type IIA theory follows from dimensionally reducing M-theory on $S^1$ as discussed in section \ref{dualities} can be exploited to uplift the IIA supersymmetric solution and the integrability conditions. The second option is more convenient, although slightly less transparent. Both methods will be shown below to obtain supersymmetric solutions, after which it will be demonstrated that the integrability theorems for IIA and M-theory are equivalent.
\\
\\
The first approach will be to exploit the duality between type II and M-theory. In order to make use of this, the following points need to be taken into consideration.
Firstly, the duality does not take into account the Romans' mass $\fcal_0 = F_0$, hence we should set
\al{
F_0 = 0\;.
}
Secondly, the external spacetime in M-theory $\rbb^{1,2}$ is identified with the external spacetime $\rbb^{1,1}$ in type II, together with the $S^1$-fibres of the internal space. Although, of course, $\rbb^{1,1} \times S^1 \centernot{\simeq} \rbb^{1,2}$, one can take the radius of the $S^1$ to be large and note that the dependence of the derivation of the duality depended on the compactness of the fibre lies purely in \eqref{eq:couplingconstants}: by absorbing the divergence of the integral into the coupling constant, one can consider $S^1$ with a large radius to be equivalent to $\rbb$. This procedure goes by the name of `decompactification', and although this is mathematically sketchy, the direct computation will demonstrate that the result is valid. $C_1$ is the connection describing the curvature of the $S^1$-fibration. For the procedure to combine $S^1$ with $\rbb^{1,1}$ into $\rbb^{1,2}$ to work, it must be globally doable and thus, the fibration needs to be trivial. Thus, to uplift our vacuum we should set
\al{
C_1=0 \implies \fcal_2 = F_2 = 0 \;.
}
Thirdly, in order for the metric \eqref{eq:mmetric} to be of the form \eqref{eq:mdrmetric},
\al{
A = \phi + \phi_0 \;, \qquad \phi_0 \in \rbb
}
must be imposed on the type IIA metric. This results in the promised identification of warp factors in type IIA and M-theory. The constant $\phi_0$ is absorbed in a redefinition of the coordinates on $\rbb^{1,2}$. Finally, the four-form flux $G$ of M-theory is related to the fluxes of type IIA via (see \eqref{eq:Gdr})
\al{
G = F_4 - H \wedge dz \;.
}
Imposing these four conditions on the supersymmetric solutions of section \ref{susysol}, one finds that all branches are subcases of the case $e^{2 i \t} = 1$. The supersymmetric solution is given by
\eq{\label{eq:msusysol1}
g_{11} &=  e^{\frac43 A }\eta_3 + e^{- \frac23 A} g_8 \\
G &= - \text{vol}_3 \wedge \d e^{2A} + F\\
F & \equiv f_4 \left(J \wedge J + \frac32 \text{Re}\O \right) + \left( f_4^{(2,0)} - \frac14 f_4^{(2,0)} \lrcorner \O^* \right) \wedge J + f_4^{(2,2)} \;,
}
with torsion classes
\eq{\label{eq:msusysol2}
W_5 &= \frac32 W_4 = 2 i W_1 \\
W_3 &= \frac 12 W_2 \;.
}
The free parameters of the supersymmetric solution are $A, ~f_4 \in C^\infty(M_8, \rbb)$, $f_4^{(2,0)} \in \Op^{(2,0)} (\mcal_8)$, $f_4^{(2,2)} \in \Op^{(2,2)} (\mcal_8)$, $W_1 \in \O^{(1,0)} (\mcal_8)$ and $W_2 \in \Op^{(2,1)} (\mcal_8)$.
\\
\\
The second approach is to do the computation directly. In order to conveniently compare these to the literature \cite{bb}, \cite{ms}, \cite{t8}, slightly different conventions will be used. The starting point is the following ansatz for the metric,
\al{
g_{11} = e^{\frac43 A }\left( \eta_3 + \tilde{g}_8 \right)
}
which leads to
\al{
G = F + e^{2A} \text{vol}_3 \wedge f
}
in order to satisfy isometry invariance. The strict $SU(4)$ ansatz for an $\ncal = 1$ vacuum on $\rbb^{1,2} \times \mcal_8$ is given by
\al{\label{eq:ksm}
\e = \zeta \otimes e^{\frac13 A} ( \eta + \eta^c )\;,
}
with the free parameter $\zeta \in Spin(1,2)$ a Majorana spinor, and the normalisation chosen for convenience. Plugging these into \eqref{eq:mkse} and setting this to zero leads to
\eq{
\nabla_m \left( \eta + \eta^c \right) &= 0 \\
\p_m A + \frac12 f_m &= 0 \\
F_{mnpq} \g^{npq} \left(\eta + \eta^c \right) &= 0 \;.
}
Using the identities \eqref{eq:su4spinorhodge}, the result is equivalent to \eqref{eq:msusysol1}, \eqref{eq:msusysol2}.
\\
\\
This result can be compared to the well-known solution of \cite{bb}, where $\ncal = 2 $ supersymmetry solutions of M-theory on $\rbb^{1,2} \times \mcal_8$ are discussed. In this paper, a strict $SU(4)$-Killing spinor ansatz\footnote{The term `strict $SU(4)$' is not used by the authors of \cite{bb}; it is by definition equivalent to their (2.34) however.}  is taken of the form
\al{\label{eq:ksm2}
\epsilon = \zeta \otimes \eta + \zeta^c \otimes \eta^c \;,
}
with $\zeta \in Spin(1,2)$ a Dirac spinor: as a result, it has four rather than two real degrees of freedom, and thus corresponds to $\ncal = 2$ rather than $\ncal = 1$. As a result of the additional symmetry requirement, the internal flux is primitive (2,2) and all torsion classes vanish. On the other hand, $\ncal = 1$ also permits a non-primitive (2,2)- and  a (4,0)-term, as well as a non-primitive (3,1)-term. Furthermore, the internal manifold of $\ncal = 1$ vacua need not be Calabi-Yau.

\subsection{Integrability: M-theory $\ncal = 1$ vacua}\label{minteg}
It has been shown that the $\ncal = 1$ M-theory supersymmetric solutions on $\rbb^{1,2} \times \mcal_8$ can be obtained by uplifting the $\ncal = (1,1)$ type IIA supersymmetric solutions on $\rbb^{1,1} \times \mcal_8$. As discussed in section \ref{integ}, to find vacua we can thus either uplift type IIA vacua or directly apply the integrability theorem on the M-theory side.
Given the uplift of the IIA fluxes $(\fcal, H)$ to the M-theory flux $G$, the integrability conditions are equivalent, as we will now show.
\\
\\
To uplift IIA on  $\mathbbm{R}^{1,1} \times \mcal_8$ to M-theory on $\mathbbm{R}^{1,2} \times \mcal_8$, it is necessary to take $C_1 = \fcal_0 = 0$.
On the IIA side, integrability requires firstly the Bianchi identities to be satisfied, which reduce to
\eq{\label{eq:bianchisinteg}
&\d H = \d F_4 = 0\\
&\d \star \s \fcal_4 + H \wedge \fcal_4 = 0 \;,
}
and secondly, the equation of motion $\delta H_{01}=0$ to be satisfied, which reduces to
\eq{\label{eq:eominteg}
 - \d \star_8 \d e^{2A} + \frac12 F_4 \wedge F_4 =0 \;.
}
Considering the expression for $G$ given in \eqref{eq:msusysol1} and the expression for $H$ that follows from (\ref{eq:hint}-\ref{eq:iiasol3}) after setting $\t= 0$, namely
\eq{
G &= F_4 - \vol_3 \wedge \d e^{2A} \\
H &= - \vol_2 \wedge \d e^{2A} \;,
}
it follows that the first line of \eqref{eq:bianchisinteg} is equivalent to
\al{
\d G = 0
}
and the second line, in combination with \eqref{eq:eominteg}, reduces to
\eq{
- \d \star_{11} G + \frac12 G \wedge G = 0\;.
}
This shows that indeed, both integrability theorems are equivalent in this case.
\\
\\
To summarise, the following M-theory vacuum is found. The fields are given by \eqref{eq:msusysol1}, the torsion is given by \eqref{eq:msusysol2}, and the integrability conditions are closure and co-closure of $F_4$, together with the Poisson equation for the warp factor as given in \eqref{eq:eominteg}. Generically, such vacua are not Calabi-Yau; the non-integrability of the $SU(4)$-structure is independent of the flux however. Finally, there is the quantisation condition \eqref{eq:quant}.

\section{The Maldacena-Nu\~{n}ez no-go theorem}\label{nogo}
The internal manifold $\mcal_8$ can be either compact or non-compact. There are advantages and disadvantages to both. One of the disadvantages to taking the internal space compact is a no-go theorem discussed in \cite{gibbons}, \cite{nunez}, commonly referred to as the Maldacena-Nu\~{n}ez no-go theorem. Applied to our IIA vacua, the argument goes as follows. Consider the Poisson equation for the warp factor,
\eq{
 - \d \star_8 \d e^{2A} + \frac{g_s^2}{2} \left(\star_8 F \wedge F \right)_8 =0 \;.
}
Integrating over $\mcal_8$ then yields
\eq{
\frac{g_s^2}{2} \int_{\mcal_8} \star_8 F \wedge F = \int_{\mcal_8} \d \left( \star_8 \d e^{2A} \right) = 0\;,
}
so Stokes' theorem leads to $F=0$, and as a consequence $A \in \rbb$. The $D=11$ supergravity case is completely analogous. In order to circumvent this no-go theorem, there are two main options. Firstly, note that if $\mcal_8$ is non-compact, Stokes does not apply and hence $F$ need not be trivial. Secondly, type II supergravity and $D=11$ supergravity are merely lowest order approximations to type II string theory and  M-theory respectively. By including higher order corrections, this argument need no longer hold. Let us sketch this for M-theory vacua on compact Calabi-Yau manifolds.
\\
\\
Some corrections to the $D=11$ supergravity action are known, but not all of them. However, it can be argued \cite{peeters} that for internal manifolds with large volume, all but the following can be neglected. The relevant correction is given by
\eq{
S_X =- T_2 \int_{\mcal_{11}} A_3 \wedge X_8 \;,
}
with, for Calabi-Yau manifolds, $X_8$ satisfying
\eq{
\int_{\mcal_8} X_8 =  \frac{1}{4!}\chi(\mcal_8) \;.
}
In addition, taking into account the presence of localized $M_2$-branes yields an additional contribution to the action. The $M_2$-branes couple to both the charge and the metric. Ignoring the latter for now, the contribution of $M_2$-branes related to the flux is given by
\eq{
S_{M_2} =  T_2 \int_{\mcal_3} A_3\;,
}
where $\mcal_3$ is the cycle on which the brane is located. Specifying this to a number $N_{M_2}$ of $M_2$ branes wrapping the cycles $\rbb^{1,2} \times \{y_j\}$ (i.e., spacetime filling and localized to $y_j \in \mcal_8$) , this can be rewritten as
\eq{
S_{M_2} = \frac{(2 \pi)^2}{2 \k_{11}} \sum_{j=1}^{N_{M_2}} \int_{\mcal_{11}} A_3 \wedge \delta^8(y - y_j) \vol_8\;,
}
where we set $T_2 = 1$\footnote{Note that the parameters of the theory $l_P^3$, $\k_{11}^2$, $T_2$  are related to one another as
\all{
T_2 &= \frac{2\pi}{l_P^3} \\
2 \k_{11}^2 &= \frac{l_P^9}{2 \pi} \;,
}
in our conventions. Here, $l_P$ is the Planck length.
}.
Together, these two modify the warp factor equation to
\eq{
 - \d \star_8 \d e^{2A} + \frac12  F \wedge F + (2 \pi)^2 \sum_{j=1}^{N_{M_2}} \delta^8(y - y_j) \vol_8  = (2 \pi)^2 X_8 \;.
}
Integrating then yields the constraint
\eq{\label{eq:topo}
\frac{1}{8 \pi^2} \int F \wedge F + N_{M_2} = \frac{\chi(\mcal_8)}{24}
}
that needs to be satisfied.
\\
\\
As stated, other corrections can be neglected only in the large volume limit. That is to say, take an expansion of the form
\eq{
g_8 = l^2 g_8^{(0)} + g_8^{(2)} + ...
}
with $\text{Vol}_8 \sim l^8$, and consider only the leading order term. Directly solving the equations of motion for $N_{M_2} = 0$ in this case leads to the conclusion that \cite{bb2}: the internal manifold is Ricci-flat, the warp factor is constant, and the flux is internal, closed, selfdual and satisfies \eqref{eq:topo}. Note that this is for any vacuum, supersymmetric or not. This result is consistent with the vacuum that we have found.

Of course, in the presence of such corrections, the integrability theorem of section \ref{integ} needs to be modified. In \cite{sources}, it was shown that sources for D-branes in type IIA, which similarly modify the action and violate the Bianchi identity of the RR fluxes which couple to the D-branes, are exactly such that the integrability theorem still holds. At the level of supergravity, the two integrability theorems are equivalent. Therefore, we shall operate under the assumption that the results of \cite{sources} can be extrapolated to M-theory and that the presence of $M_2$-branes do not alter the integrability theorem in a meaningful way.

\section{M-theory vacua on $K3 \times K3$}\label{k3vacua}
At this point, it is clear what the conditions are for an M-theory $\ncal = 1$ vacuum on $\rbb^{1,2} \times \mcal_8$. The next question then, is whether or not any manifolds $\mcal_8$ exist which satisfy the necessary conditions for such vacua. Furthermore, it needs to be checked that if an example is found, it is not an $\ncal=2$ or $\ncal = 4$ vacuum in disguise, as obviously, these vacua are subsets of $\ncal = 1 $ vacua. This phenomenon goes by the name of {\it supersymmetry enhancement}. We will demonstrate that explicit vacua exist on $\mcal_8 = S \times \tilde{S}$, with $S$, $\tilde{S}$ K3 surfaces. This extends the results of \cite{drs}, where such vacua with $\ncal=2,4$ were first studied. Because K3 surfaces are well-understood, the fluxes and $SU(4)$-structure can be made very precise, and it can be shown explicitly when supersymmetry enhancement does or does not occur. As $S \times \tilde{S}$ is compact, we run into the no-go theorem of section \ref{nogo}, which will be circumnavigated by the addition of higher order corrections as described.

\subsection{K3 Surfaces}\label{k3}
K3 surfaces are objects that appear in a wide variety of contexts. We will briefly point out the properties of relevance to us, skipping over most of the proofs, as these require a vast array sophisticated algebraic-geometric machinery, all of which is beyond the scope of this thesis and a subset of which is beyond the author besides. Mostly, we follow along the lines of \cite{aspinwall}. See  \cite{bhpvdv} or \cite{huybrechtsk3} for a more thorough exposition.
\\
\\
A {\it K3 surface} is a two-dimensional compact complex two-dimensional manifolds with trivial canonical bundle and vanishing first Betti number. Thus, it is simply connected and (proper) Calabi-Yau.
In fact, the only other compact two-dimensional Calabi-Yau manifolds are topologically $T^4$, for which $b_1 \neq 0$.
All K3 surfaces are diffeomorphic and differ only in choice of $SU(2)$-structure. Let us deduce the Hodge diamond of K3 surfaces. There are nine Hodge numbers, but Hodge duality together with complex conjugation reduces the number of independent Hodge numbers down to four. $h^{0,0} = 1$ due to simple connectedness, $h^{1,0}=0$ by definition, and $h^{2,0} = 1$ because of the Calabi-Yau form, leaving only $h^{1,1}$ undetermined. In order to derive $h^{1,1}$, we make use of the Hirzebruch-Riemann-Roch theorem, relating the Euler characteristic of sheaves to characteristic classes, and the Dolbeault theorem, relating the cohomology of sheaves to the cohomology of the Dolbeault operator\footnote{Unfortunately, a thorough exposition of all concepts required here requires a good deal more mathematics than is worthwhile for the results, thus the reader is expected to either  already be well-versed in these concepts or to just nod with a deep thoughtful look of understanding on her or his face.}.Let $S$ be a K3 surface, $E \rightarrow S$ a holomorphic vector bundle and $\fcal \equiv \G_{\text{h}}(S,E)$ the sheaf of holomorphic sections of $E$. The Euler characteristic of $\fcal$ is defined as
\al{
\chi(S, \fcal) \equiv \sum_k (-1)^k \text{dim} H^k(S, \fcal)\;,
}
where $H^k(S, \fcal)$ is the $k$-th \v{C}ech (or sheaf) cohomology group. Let $\O^p_{\text{h}} (M)\subset \O^{(p,0)}(M)$ be defined as the sheaf of holomorphic sections of the holomorphic vector bundle $T^{*(p,0)}$, with $\ocal_M \equiv \O^0_{\text{h}}(M)$ the sheaf of holomorphic sections of the trivial bundle (i.e., the structure sheaf of holomorphic functions). The Dolbeault theorem states that sheaf cohomology relates to $\p$-cohomology as
\al{
H^q(M, \O^p_{\text{h}}(M)) = H^{p,q}(M)\;.
}
The Hirzebruch-Riemann-Roch theorem states that
\al{
\chi(S, \fcal) = \int ch(E) Td(S)\;,
}
with the usual convention for characteristic classes that $Td(S) \equiv Td (T^{(1,0)}S)$. Both the Chern character and the Todd class can be expanded in terms of Chern classes as
\eq{
Ch(E)&= \text{rank}(E) + c_1(E) + \frac12 \left( c_1(E)^2 - 2 c_2(E) \right) + ...\\
Td(E)&= 1 + \frac12 c_1(E) + \frac{1}{12} \left( c_1(E)^2 + c_2(E) \right) + ...\;,
}
where $...$ are $k$-forms with $k > 2$. Taking $\fcal = \ocal_S$, the Hirzebruch-Riemann-Roch theorem yields
\al{
24 = \int c_2 (S)
}
after making use of the fact that  $c_j (\ocal_S) = 0$, the Dolbeault theorem, and the fact that $h^{0,0} = h^{2,0} = 1$, $h^{1,0} = 0$.
Taking $\fcal = \O_{\text{h}}^1$, it is then found that
\al{
- h^{1,1} = -\frac56 \int c_2(S) = - 20 \;.
}
Therefore, the Hodge diamond of a K3 surface is given by
\all{
\begin{array}{ccccc}
&&h^{0,0}&&\\
&h^{1,0} & & h^{0,1}& \\
h^{2,0} & & h^{1,1} & & h^{0,2}\\
&h^{2,1} & & h^{1,2}& \\
&&h^{2,2}&&
\end{array}
=
\begin{array}{ccccc}
&&1&&\\
&0 & & 0& \\
1 & & 20 & & 1\\
&0 & & 0& \\
&&1&&
\end{array}
}
\indent  A {\it lattice} $\left(L, (.,.)\right)$ is defined to be a finitely generated free $\zbb$ module\footnote{
That is to say, given a vector space $V$ with basis $\{e_j \; |\; j \in \{1, n\}\}$, $L = \{\sum_ja_j e_j \; |\; a_j \in \zbb \} \cong \zbb^n$. $n$ is called the {\it rank} of $L$. Technically, lattices can be defined as free modules of more generic rings than $\zbb$, but this will not be necessary for our purposes. } $L$ with a symmetric bilinear form $(.,.) : L \times L \rightarrow \zbb$. A lattice is {\it unimodular} if $\text{det}\left[ (.,.)\right] \in \{ \pm 1\}$ and is {\it even} if $(x, x) \in 2 \zbb$ $ \forall x \in L$.
The bilinear forms $(.,.)$ of indefinitie signature of  unimodular even lattices can be classified: they are all of the form
\al{
(.,.) = U^n \oplus (\pm E_8)^m
}
for some $m, n \in \mathbbm{N}_0$, with
\al{
U = \left(\begin{array}{cc}
0&1 \\
1 & 0
\end{array} \right)
}
and $E_8$ the unique positive definite even lattice of rank 8, which is given by the root lattice of the Lie algebra.

There exists a natural symmetric bilinear form on the second integral cohomology group
\al{
(.,.) : H^2(S, \zbb) \times H^2(S, \zbb) \rightarrow \zbb
}
called {\it the intersection form}. By Poincar\'{e} duality, this is equivalent to an inner product on the second integral homology group.
We claim that this group is given by
\al{
H_2(S, \zbb) = \zbb^{22} \;,
}
i.e., no torsion terms $\zbb_p$ exist. Thus, the pair $\left(H_2 (S, \zbb), (.,.)\right)$ forms a lattice.

Wu's formula states that for any complex surface $\Sigma$ and $\a \in H^2(\Sigma, \zbb)$, the intersection pairing satisfies
$(\a, \a) + (c_1(T^{(1,0)}\Sigma), \a) \in 2 \zbb$. Taking $\Sigma = S$ and noting that $c_1(T^{(1,0)}S) = 0$ by definition of a K3 surface, $\left(H_2 (S, \zbb), (.,.)\right)$ is an even lattice. By Poincar\'{e} duality, the intersection form has determinant $\pm 1$, thus the lattice is unimodular. By Hodge's index theorem, the signature of the intersection form is given by
$(2 h^{2,0} + 1, h^{1,1} - 1)$ and thus, the lattice has signature $(3,19)$. Thus, the cohomology lattice is unimodular indefinite and even, and has to be of the form $U^n \oplus (\pm E_8)^m$. Noting that $U$ has signature $(1,1)$, there is only one possibility for $n,m$. Thus, the final result of all this is the following:
\propn{
$(H^2(S, \zbb), (.,.))$ is an even unimodular lattice of signature $(3,19)$ with intersection form given by
\al{
(.,.) = U^3 \oplus (- E_8)^2\;.
}
}

A priori, by our definition a K3 surface need not be algebraic, i.e., there need not exist any $n$ such that $S$ can be embedded in $\cbb P^n$. However, all examples that will be given will in fact be algebraic. In order to demonstrate this, some more formalism is necessary.

The second real coholomogy group can be split into selfdual and anti-selfdual forms as
\al{
H^2(S, \rbb) = H^+(S, \rbb) \oplus H^-(S, \rbb)\;.
}
As the signature of the intersection form is $(3,19)$ and can be expressed in terms of Hodge duals, it follows that $\text{dim} H^+(S, \rbb) = 3$, $\text{dim} H^-(S, \rbb) = 19$.
Let $(j, \o)$ define the $SU(2)$-structure on $S$, with $j$ the K\"{a}hler form and $\o$ the Calabi-Yau form. Then $(j, \text{Re} \o, \text{Im} \o)$ span $\text{dim} H^+(S, \rbb)$, as they are all selfdual and the $SU(2)$-structure relations $j \wedge \o = \o \wedge \o = 0 $ guarantee linear independence.

The {\it Picard lattice} is defined as
\al{
\text{Pic}(S) \equiv H^2(S, \zbb) \cap H^{1,1}(S)\;,
}
with $\rho \equiv \text{dim Pic}(S) \leq 20$ the Picard number.
The intersection form on $H^2(S, \zbb)$ induces a bilinear symmetric form on $\text{Pic}(S)$, thus $\left(\text{Pic}(S), (.,.) \right)$ is a sublattice of  $\left( H^2(S, \zbb), (.,.) \right)$ with signature $(1, \rho - 1)$. Its complement is given by the {\it transcendental lattice} and is of signature\footnote{There is a typo in \cite{pta}, p. 24. This is the correct formula.} $(2, 20 - \rho)$. A sublattice $\L \subset L$ is called {\it primitive} if $L / \L$ is free\footnote{Note that for any primitive sublattice $\L \subset L$, L unimodular, $\L \oplus \L^\perp = L$ if and only if $\L$ is unimodular.}.

At this point, we can refer to \cite{morrison} for the following

\thm{
Let $L$ be the lattice $\zbb^{22}$ with signature $(3,19)$. Let $\L \subset L$ be a primitive sublattice with signature $(1, \rho - 1)$.  Then there exists an algebraic K3 surface $S$ with $\text{Pic}(S) = \L$.
}
Thus, it will be possible to give explicit expressions for the (cohomology classes of) the $SU(2)$-structure in terms of a basis for the lattice of an algebraic K3 surface
Before we do so, let us return to physics and consider the constraints on a vacuum with the product of K3 surfaces as internal space.

\subsection{M-theory vacua constraints on $S \times \tilde{S}$}
Let $S$, $\tilde{S}$ be K3 surfaces and let $\mcal_8 = S \times \tilde{S}$. The $SU(4)$-structure on $\mcal_8$ can then be further reduced to an $SU(2) \times SU(2)$-structure. This can be made manifest either at the level of the pure spinor or at the level of the forms $(J, \O)$. The pure spinor of $\mcal_8$  $\eta$ can be split as
\eq{\label{eq:su241}
\eta = \eta_S \otimes \eta_{\tilde{S}} \;,
}
with $\eta_{S, \tilde{S}}$ Weyl spinors of $S$, $\tilde{S}$ (which are automatically pure for dimensional reasons). It is then possible to construct spinor bilinears and use Fierz-identities to confirm that these lead to $(j, \o)$ on $S$ and $(\tilde{\jmath}, \tilde{\o})$ on $\tilde{S}$, with $j$, $\tilde{\jmath}$ the K\"{a}hler forms and $\o$, $\tilde{\o}$ the Calabi-Yau forms.
More straightforwardly, though perhaps slightly less insightful, is to do the following for $S$, and completely analogous for $\tilde{S}$.
Let
\eq{
H^+(S,\rbb) &= \text{span}_\rbb \left\{j^a \;|\; a \in \{ 1,2,3\} \right\}\\
H^-(S,\rbb) &= \text{span}_\rbb \left\{l^\a \;|\; \a \in \{ 1,...,19\} \right\}\;,
}
with the bases normalized such that
\eq{
\frac12 \star_4 j^a \wedge j^b &= \delta^{ab} \text{vol}_4 \\
\frac12 \star_4 l^\a \wedge l^\b &= \delta^{\a\b} \text{vol}_4
}
Let $\r^a \in \rbb$, $c^a \in \cbb$ be such that
\eq{
\r^a \r^a &= \frac12 c^a c^{*a}  = 1 \\
\r^a c^a &=  c^a c^a = 0
}
and define
\eq{
j \equiv \r^a j^a \;, \qquad \o \equiv c^a j^a \;.
}
Then
\eq{
\frac{1}{2!} j \wedge j &= \frac{1}{2^2}\o \wedge \o^* = \text{vol}_4 \\
j \wedge \o &= 0
}
and thus $(j, \o)$ forms a (torsion-free) $SU(2)$-structure.
Setting
\eq{\label{eq:su242}
J \equiv j + \tilde{\jmath}\;, \qquad \O \equiv \o \wedge \tilde{\o}
}
it follows that
\eq{
\frac{1}{4!} J^4 &= \frac{1}{2^4} \O \wedge \O^* = \text{vol}_8 \\
J \wedge \O &= 0
}
and in such a way the $SU(4)$-structure can be constructed in terms of the $SU(2)$-structures which have been expressed explicitly in terms of a basis of the second real cohomology groups of the K3 surfaces.
\\
\\
At this point, we can deliver on the promise of physics that this subsection started with. The $\ncal = 1$ supersymmetric solution was given in \eqref{eq:msusysol1}, \eqref{eq:msusysol2}. Imposing the integrability conditions of section \ref{minteg} then comes down to three things:
\begin{itemize}
\item All (internal) flux forms ought to be harmonic and thus expressible in terms of cohomology
\item The quantization condition $\frac{1}{2\pi} \int_{C_4} F \in \zbb$ needs to be satisfied for any cycle $C_4 \in H_4((S \times \tilde{S}, \zbb)$
\item The integral of the warp factor constraint demands\\ $\frac{1}{8 \pi^2}\int  F \wedge F + N_{M_2} = \frac{\chi(S \times \tilde{S})}{24}$
\end{itemize}
As a basis for the cohomology, the intersection form, and the Euler character are all explicitly known, we are now in a position to find explicit vacua.

Consider the four-form flux. The internal part, given by the last line of\eqref{eq:msusysol1} can be expressed in terms of the cohomology basis and the $SU(2) \times SU(2)$-structure as
\eq{\label{eq:k3flux}
F  =& A \;\text{Re} ( \o \wedge \tilde{\o}) + \text{Re}( B \o \wedge \tilde{\o}^*) + C j \wedge \tilde{\jmath} + (4A- 2 C) \left( \text{vol}_4 + \widetilde{\text{vol}}_4 \right) \\
 &+ f^{\a\b} l^\a \wedge l^\b + \left(D \o \wedge \tilde{\jmath} + D^* \tilde{\o}\wedge j + \text{c.c.} \right)
}
with $A, C, f^{\a\b} \in \rbb$, $B, D \in \cbb$. The terms proportional to $D$, $D^*$ are (3,1) non-primitive, the terms proportional to $A$ are (2,2) non-primitive and $(4,0) + (0,4)$, everything else corresponds to the primitive (2,2) term. Therefore, it also follows that setting $A = D= 0$, the solution actually has $\ncal = 2$ supersymmetry. This process goes by the name of {\it supersymmetry enhancement}. Let us investigate this in more detail.

\subsection{Supersymmetry enhancement}
Since $\ncal = 2 $ and $\ncal = 4$ vacua are a subset of the less constrained $\ncal = 1$ vacua, it is relevant to know under which conditions $\ncal = 1 $ vacua are actually masqueraded higher supersymmetric vacua. The supersymmetry solution determines an $SU(4)$-structure. Conversely, if multiple $SU(4)$-structures exist for a single vacuum, this corresponds to the presence of more free parameters in the Killing spinors and thus to supersymmetry enhancement. Let us be more precise.
\\
\\
Consider a single K3 surface $S$. The metric determines an $SO(4)$-structure on $S$. As usual, there are two equivalent ways to proceed: tensorial or spinorial. From a tensor point of view, $SO(3) \subset SO(4)$ leaves the metric invariant, yet acts fundamentally on the vector $\jcal \equiv (\Im \o, \Re \o, j)$; this comes down to stating that it is possible to deform the complex structure without altering the metric.  Generically, there are no non-trivial $SO(3)_S \times SO(3)_{\tilde{S}}$ rotations of $(\jcal, \tilde{\jcal})$ that leave $F$ invariant. However, consider the case $A = D = 0$. In this case, there is a subgroup of $SO(3)_S \times SO(3)_{\tilde{S}}$ which does leave $F$ invariant, namely the  $SO(2)_S \times SO(2)_{\tilde{S}}$ subgroup which acts on $(\jcal, \tilde{\jcal})$ by leaving $j, \tilde{\jmath}$ invariant and rotating $\Re \o, \Im \o$ into one another. Although not immediately obvious, this is made manifest by noting that $SO(2) \simeq U(1)$ and $F$ is indeed manifestly invariant under $(\o, \tilde{\o})\rightarrow e^{i \vartheta} (\o, \tilde{\o})$. Clearly the $SU(2)_S \times SU(2)_{\tilde{S}}$-structure of $S\times \tilde{S}$ is not invariant under this action, yet all fields of the vacuum $(g, G, \Psi_M)$ are\footnote{Clearly, the $SO(2)_S \times SO(2)_{\tilde{S}}$ group acts trivially on $\Psi_M = 0$ and the external part of $G$.}.  Therefore, we have found a vacuum associated to multiple $SU(4)$-structures and supersymmetry enhancement occurs, as is evident from the fact that we know that when setting $A = D = 0$, the flux reduces to $F = f^{(2,2)}$, which is exactly the $\ncal =2$ supersymmetry solution.
\\
\\
The correspondence between supersymmetry enhancement and multiple $SU(4)$-structures related to the same vacuum is made manifest from the spinorial point of view. Existence of spinors and a metric on $S$ defines a $Spin(4)$-structure. There is a so-called `accidental isomorphism', stating that $Spin(4) \simeq Spin(3) \times Spin(3) \simeq SU(2) \times SU(2)$. In order to avoid confusion, let us stress that this $SU(2) \times SU(2)$-structure is a $G$-structure on $S$, {\it not} on $S \times \tilde{S}$. Existence of a globally defined nowhere-vanishing pure spinor $\eta_S$ reduces the $Spin(4)$-structure to a $Spin(3) \simeq SU(2)$-structure. Thus, to be pedantic to stress the point, $SU(2)\times \{\obb\} $ leaves $\eta_S$, and thus $(j, \o)$ invariant, whereas $ \{ \obb \} \times SU(2)$ acts via the fundamental representation on $\eta_S$, inducing the $SO(3)$ action on $\jcal$ by noting that
\eq{
\jcal_{mn} = - \frac12 i \vec{\s}_{ij} \tilde{\eta}_i^c \g_{mn} \eta_j
}
with $\eta_1 \equiv \eta_S$, $\eta_2 \equiv \eta_S^c$. This also explains the relation between the $SO(2)$ and $U(1)$ actions on $\o$ explained before, as the $SO(2)$ action can be seen as a subgroup of the metric-induced $SO(4)$-structure, whereas the $U(1)$-action can be seen as the non-trivial $U(1)$-action on $\eta_S$. The advantage of the latter viewpoint is that the supersymmetry enhancement becomes manifest when investigating the resulting action on the Killing spinor \eqref{eq:ksm}. Explicitly, for $A=D = 0$, the invariance of the vacuum under $\eta_S \rightarrow e^{ \frac12 i \vartheta} \eta_S$ corresponds to
\al{
\e \rightarrow e^{-\frac13 A} \left( e^{\frac12 i \vartheta} \z \right) \otimes \eta + e^{-\frac13 A} \left( e^{-\frac12 i \vartheta} \z \right) \otimes \eta^c \;,
}
which is exactly the $\ncal = 2$ ansatz \eqref{eq:ksm2}. This demonstrates explicitly how invariance of the vacuum results in a more generic Killing spinor and thus supersymmetry enhancement to $\ncal =2$.

The $\ncal = 2$ enhancement obtained by setting $A=D=0$ can be further enhanced to $\ncal =4$ by taking $B= C$, resulting in
\eq{
F &=  C \;\text{Re} ( \o \wedge \tilde{\o}^*) + C j \wedge \tilde{\jmath}  -2 C \left( \text{vol}_4 + \widetilde{\text{vol}}_4 \right) + f^{\a\b} l^\a \wedge l^\b \\
&= C \jcal_m \delta^{mn} \tilde{\jcal}_n  -2 C \left( \text{vol}_4 + \widetilde{\text{vol}}_4 \right) + f^{\a\b} l^\a \wedge l^\b \;.
}
This flux is invariant under an $SO(3)$ symmetry acting as
\eq{
\jcal \rightarrow R \jcal \;, \qquad \tilde{\jcal} \rightarrow R \tilde{\jcal}\;,
}
for $R \in SO(3)$. Corresponding $SU(2)$ actions on the spinors can be deduced, showing that this indeed leads to $\ncal = 4$ as claimed.
However, the way to achieve supersymmetry enhancement is not unique. Another example leading to $\ncal = 4$ would be to take  $A= C$, $B = D = 0$ such that the flux reduces to
\eq{
F &= A \;\text{Re} ( \o \wedge \tilde{\o})+ A j \wedge \tilde{\jmath} + 2A \left( \text{vol}_4 + \widetilde{\text{vol}}_4 \right) + f^{\a\b} l^\a \wedge l^\b \\
 &=  A \jcal_m \eta^{mn} \tilde{\jcal}_n + 2 A \left( \text{vol}_4 + \widetilde{\text{vol}}_4 \right) + f^{\a\b} l^\a \wedge l^\b \;.
}
In this case, the flux is invariant under a different $SO(3)$ symmetry, which acts as
\eq{
\jcal \rightarrow R \jcal \;, \qquad \tilde{\jcal} \rightarrow \eta_3 R \eta_3 \tilde{\jcal}\;,
}
with $\eta_3$ the matrix $\text{diag}(-1,1,1)$.
At the level of the spinors, the corresponding $SU(2)$ action is given by
\eq{
\left(
\begin{array}{c}
\eta_S\\
\eta_S^c
\end{array}
\right) &\rightarrow  \exp\left( -  \frac 12 i \vartheta \hat{n} \cdot \vec{\sigma}\right)
\left(
\begin{array}{c}
\eta_S\\
\eta_S^c
\end{array}
\right) \\
 \left(
\begin{array}{c}
\eta_{\tilde{S}}\\
\eta_{\tilde{S}}^c
\end{array}
\right) &\rightarrow  \exp\left(   \frac 12 i \vartheta (\eta_3 \hat{n}) \cdot \vec{\sigma}\right)
\left(
\begin{array}{c}
\eta_{\tilde{S}}\\
\eta_{\tilde{S}}^c
\end{array}
\right) \;,}
for $\hat{n}$ a unit vector determining the axis of rotation. As a result, the Killing spinor gets mapped to
\al{
\e \rightarrow e^{-\frac13 A} \left( \z_1 \otimes \eta_S \otimes \eta_{\tilde{S}} + \z_2  \otimes \eta_S \otimes \eta_{\tilde{S}}^c + \text{c.c.} \right) \;,
}
with $\zeta_{1,2}$ complex spinors of $Spin(1,2)$, determined in terms of $\zeta$ and $\hat{n}$. This is indeed an $\ncal = 4$ ansatz.
\\
\\
Finally, let us also note that there are non-trivial symmetries of the vacuum which affect the $SU(2)_S$- and $SU(2)_{\tilde{S}}$-structures, but which leave the $SU(4)$-structure invariant: these do not lead to supersymmetry enhancement. As an example,  consider $B=0$. Then the flux is invariant under the $U(1)$ action $(\o, \tilde{\o}) \rightarrow (e^{i \vartheta} \o, e^{-i \vartheta} \tilde{\o})$. However, although both $SU(2)_{S, \tilde{S}}$ are not invariant, the $SU(4)$-structure is, as follows from either $\eqref{eq:su241}$ or $\eqref{eq:su242}$. As a consequence, the Killing spinor $\e$ is invariant, and no supersymmetry enhancement occurs.

\subsection{Examples}
We now have the expression \eqref{eq:k3flux} for the flux in terms of the cohomology basis and an understanding of when $\ncal = 1$ vacua undergo supersymmetry enhancement. All that remains is to fix the parameters $A,B,C,D, f^{\a\b}$ to satisfy the quantization condtion and see what the number of branes $N_{M_2}$ ought to be to satisfy the topological constraint.
\\
\\
In order to solve the quantization condition, we fix an explicit expression for the $SU(2)$-structures in terms of $H^2(S, \zbb)$, $H^2(\tilde{S}, \zbb)$.
Let a basis of  $H^2(S, \zbb)$ be given by  $\{e_{1, I}, e_{2, I}, e_j\}$, $I \in \{1,2,3\}$, $j \in \{7, ..., 22\}$ such that
\eq{
\int_S e_{a, I} \wedge e_{b, J} = U_{ab} \delta_{IJ} \;.
}
Furthermore, we define $e_{\pm I} \equiv e_{1, I} \pm e_{2, I}$ with norm $\pm 2$ with respect to the intersection form, and a dual basis in $H_2(S, \zbb)$ given by $\{C_{1, I}, C_{2, I}, C_j\}$. Since
\eq{
H_4 (S \times \tilde{S}, \zbb) = \text{span}_\rbb \{ S, \tilde{S}, C \times \tilde{C} \;|\; C \in H_2(S, \zbb), \; \tilde{C} \in H_2(\tilde{S}, \zbb) \}
}
by the K\"{u}nneth formula for cohomology on product spaces,
this immediately yields a basis for $H_4 (S \times \tilde{S}, \zbb) $.
The $SU(2)$-structure can then be taken to satisfy
\eq{
j &= \sqrt{2 \pi v} e_{+1}\\
\o &= \sqrt{2 \pi v} \left(e_{+2} + i e_{+3} \right)
}
and analogous for $\tilde{S}$: as a consequence,
\eq{
\int_S \vol_4 = 2 \pi v\;.
}
We then have the following explicit examples of vacua:
\\
\\

$\bullet$ Example 1: $\ncal = 1$, $N_{M_2} = 12$:
\\
\\
Set $B = D =f^{\a\b} = 0$,
\eq{
C = 2A = \pm \frac{2}{\sqrt{v \tilde{v}}}\;.
}
It can then be seen that the quantization condition \eqref{eq:quant} is satisfied, with the only non-zero charges given by
\eq{
\frac{1}{2\pi}\int_{\mathcal{C}_{a,1}\times\mathcal{C}_{b,1}}G=\pm 2~;~~~
\frac{1}{2\pi}\int_{\mathcal{C}_{a,2}\times\mathcal{C}_{b,2}}G=\pm 1~;~~~
\frac{1}{2\pi}\int_{\mathcal{C}_{a,3}\times\mathcal{C}_{b,3}}G=\mp 1~,
}
for $a,b\in \{1,2\}$.
\\
\\

$\bullet$ Example 2: $\ncal = 2$, $N_{M_2} = 0$:
\\
\\
Set $A= B = D =f^{\a\b} = 0$,
\eq{
C =  \pm \frac{2}{\sqrt{v \tilde{v}}}\;.
}
In this case, the non-trivial integrals are given by
\eq{
\frac{1}{2\pi}\int_{\mathcal{C}_{a,1}\times\mathcal{C}_{b,1}}G=\pm2~;~~~
\frac{1}{2\pi}\int_{S}G=\mp4\sqrt{\frac{v}{\tilde{v}}}
~;~~~\frac{1}{2\pi}\int_{\tilde{S}}G=\mp4\sqrt{\frac{\tilde{v}}{v}}
~,}
for $a,b=1,2$. A solution to the charge quantization constraints is thus obtained by setting
\eq{
v = 4^n \tilde{v}\;,
}
for $n\in \{0,\pm1,\pm2\}$.
\\
\\

$\bullet$ Example 3: $\ncal = 4$, $N_{M_2} = 14$:
\\
\\
Set $A= B = D =f^{\a\b} = 0$,
\eq{
C = A = \pm \frac{1}{\sqrt{v \tilde{v}}}\;.
}
The quantization condition \eqref{eq:quant} is then satisfied, with non-trivial charges
\eq{\spl{
\frac{1}{2\pi}\int_{\mathcal{C}_{a,1}\times\mathcal{C}_{b,1}}G=\pm1~&;~~~
\frac{1}{2\pi}\int_{\mathcal{C}_{a,2}\times\mathcal{C}_{b,2}}G=\pm1~;~~~
\frac{1}{2\pi}\int_{\mathcal{C}_{a,3}\times\mathcal{C}_{b,3}}G=\mp1\\
&\frac{1}{2\pi}\int_{S}G=\frac{1}{2\pi}\int_{\tilde{S}}G=\pm2
~.}}

\chapter{Flux vacua on $SU(4)$-deformed Stenzel space}\label{stnzl}
In section \ref{k3}, examples were given of $\ncal =1$, $d=3$ supersymmetric flux vacua for M-theory with $K3 \times K3$ as internal space, a compact Calabi-Yau. In this section, other examples of flux vacua will be given, with as internal manifold either the Stenzel space fourfold, or something that shall be referred to as `$SU(4)$-deformed Stenzel space'. Stenzel spaces  \cite{stenzel} are a class of non-compact Calabi-Yau manifolds; our interest will be solely the fourfold which we will simply refer to as `Stenzel space'. Other members of this class are Eguchi-Hanson space \cite{eguchi} for $d=4$ and the deformed conifold \cite{candelas} for $d=6$. The deformed conifold $\ccal(T^{1,1})$ has been of particular interest, as the (warped) metric of a compactification on $\mathbbm{R}^{1,3} \times \ccal(T^{1,1})$ asymptotes to $\text{AdS}_5 \times T^{1,1}$, thus lending itself to describe dual conformal field theories \cite{kw}, \cite{kt}. Although the conifold has a singularity at the tip of the cone, this may be smoothed out by either deformation (blow-up) or resolution. Smoothness at the tip of the `throat' of the deformed conifold is then associated to color confinement in the dual CFT \cite{ks}.

In \cite{cglp}, an $\ncal=2$ M-theory flux vacuum was constructed on $\mathbbm{R}^{1,2} \times \scal$, where $\scal$ is the Stenzel fourfold (hereafter simply referred to as `Stenzel space'). This M-theory vacuum was reduced to an $\ncal = (2,2)$ vacuum of type IIA theory in \cite{kp}. Similarly to the deformed conifold, Stenzel space is a deformed cone, in this case over the Stiefel manifold $V_{5,2} \simeq SO(5)/SO(3)$, with an $S^4$ bolt at the origin to smooth out the singularity of $\ccal(V_{5,2})$. Hence this vacuum is a higher dimensional (or lower, depending on perspective) analogue of the one found in \cite{ks}. Recently, this vacuum has attracted interest mostly in the context of metastable vacua in M-theory when supplemented with (anti-) $M_2$ branes \cite{kp, bena, massai}. See also \cite{hashimoto, ms3}.

The purpose of this section is three-fold. First of all, we are interested in extending the results of \cite{cglp} to construct IIA, IIB and M-theory vacua with less supersymmetry, by applying the results of \cite{ptb, pta} in case of vanishing torsion. These vacua allow for more RR fluxes, which can be described in terms of certain closed and co-closed $(p,q)$-forms. Thus, this process involves constructing closed and co-closed forms on Stenzel space and considering their effects on the warp factor. For the case of IIA, we find three new possible contributions: scalar terms, a primitive $(1,1)$-form, and a new primitive $(2,2)$-form. Although massive IIA contributes such scalar terms, non-massive IIA may also do so. All of these terms cause divergences in the warp factor, the scalars in the UV, the (1,1) and (2,2)-form in the IR. Thus, in this sense, these vacua are more along the lines of \cite{kw} than that of \cite{ks}. We explain how such fluxes affect the uplift of the IIA vacua to M-theory vacua on $\mathbbm{R}^{1,2} \times \scal$.

Secondly, we are interested in constructing explicit examples of non-Calabi-Yau spaces. Stenzel space comes equipped with a natural $SU(4)$-structure with vanishing intrinsic torsion defined by its Calabi-Yau structure (i.e., its symplectic form, holomorphic four-form, and metric). We use the coset structure of $V_{5,2}$ to construct a family of $SU(4)$-structures which we call `left-invariant', as these are induced from the left-invariant forms on $SO(5)/SO(3)$. We consider a sub-family of these, which we refer to as `$abc$ $SU(4)$-structures', and give explicit formulae for the torsion classes. This reduces the problem of finding moduli spaces, e.g. of integrable almost complex structures, to ODE which we solve. Such spaces which are diffeomorphic to Stenzel space but equipped with a different $SU(4)$-structure we will refer to as `$SU(4)$-deformed Stenzel spaces'.

Our third point of interest is to construct vacua on such $SU(4)$-deformed Stenzel space. Due to the specifics of forms on Stenzel space, non-Calabi-Yau deformations automatically violate the NSNS Bianchi identity. We construct type IIA $\ncal= (1,1)$ vacua on complex non-symplectic (hence, in particular, non-Calabi-Yau) $SU(4)$-deformed Stenzel space, up to subtleties in the integrability theorem. The geometry is smooth and complete, with an $S^4$ bolt at the orgin and conical asymptotics, thus conformal to $\text{AdS}_3$ in the UV; this is similar to IIA on Stenzel space. The RR flux is primitive, (2,2) and satisfies the RR Bianchi identity. The violation of the NSNS Bianchi is sourced by a distribution of NS5-branes, for which we give the contribution to the action. These vacua do not uplift to M-theory vacua on $\mathbb{R}^{1,2} \times \scal$, as such M-theory vacua require certain constraints on the dilaton in terms of the warp factor which are explicitly not satisfied.

The rest of this section is organized as follows. In section \ref{sspace}, we discuss some known results on Stenzel space, $SU(4)$-structures, and vacua on manifolds with $SU(4)$-structures. In section \ref{fluxstenzel}, we describe the combination of the ingredients of the previous section, applying the $SU(4)$-solutions to Stenzel space. In section \ref{deforms} we describe how to construct $SU(4)$-structures on manifolds diffeomorphic to Stenzel space.  We then examine the moduli spaces and geodesical completeness of such spaces. Finally, in section \ref{generalizations} we discuss how to construct IIA vacua on these $SU(4)$-deformed Stenzel spaces. This involves taking the susy solutions of section \ref{susysol}, applying them with torsion classes described in section \ref{deforms}, and then checking the integrability conditions and Bianchi identities, as described in section \ref{integ}.

For convenience, we will slightly alter our notation. Metrics on $\mcal$ will be denoted by $ds^2(\mcal)$, and the warp factor will be denoted by $e^{2A} \equiv \hcal^{-1}$. In particular, the integrability condition $\delta H_{01} = 0$ for type II vacua (see section \ref{integ} can be rewritten as the scalar equation
\al{\label{eq:warp}
\nabla^2 \hcal = -  g_s^2\star_8 \left(  F_0 \wedge \star_8 F_0 + F_2 \wedge \star_8 F_2 + \frac{1}{2}F_4 \wedge \star_8 F_4 \right)\;,
}
with $\nabla^2$ the Laplacian.

\section{Stenzel Space}\label{sspace}
The vacua of interest will either have a Stenzel space or something closely related to a Stenzel space as internal manifold. A Stenzel space can be viewed as a smoothing of a (singular) cone in such a way that it is still a Calabi-Yau (CY) manifold. Stenzel spaces exist for arbitrary even dimensions; the most well-known is the deformed conifold in $d=6$ \cite{candelas}. We will focus purely on the case where $d=8$ and will simply refer to the CY fourfold Stenzel space as `Stenzel space'. We will review its properties below; see \cite{stenzel}, \cite{cglp}, \cite{kp} for more details.

Stenzel space is defined as the CY manifold $(\scal, J, \O)$, where $\scal$ is a smooth manifold, $J$ its K\"{a}hler form, and $\O$ the Calabi-Yau form.
There are numerous ways to describe Stenzel space. As an algebraic variety, it is defined by the set
\al{
\scal = \{ z \in \mathbbm{C}^5 \; | \; z^2 = \epsilon^2 \}
}
with $\epsilon \in \mathbbm{R}$.  Topologically speaking, it is homeomorphic to $T^*S^4$. Alternatively, and for our purposes more conveniently, Stenzel space can be considered as a deformed cone over the Stiefel manifold $V_{5,2} \simeq SO(5) / SO(3)$, with the singularity at the tip of the undeformed cone blown up to $S^4 \simeq SO(5) / SO(4)$. It is thus a smooth non-compact manifold. $SO(5)$ comes equipped with a set of left-invariant forms $L_{AB}$, $A,B \in \{1, ...,5\}$ satisfying the relation
\al{
\d L_{AB} = L_{AB} \wedge L_{BC} \;.
}
Relabeling $A =(1,2, j)$, $j \in \{1,2,3\}$ and defining $L_{1j} = \s_j$, $L_{2j} = \tilde{\s}_j$, $L_{12} = \n$, one has that $\n, \s_j, \tilde{\s}_j$ span a basis for $T^*V_{5,2}$ and satisfy the following relations (summation implied):
\eq{\spl{\label{eq:liforms}
\d \s_j &= \n \wedge \tilde{\s}_j + L_{jk} \wedge \s_k \\
\d \tilde{\s}_j &= - \n \wedge \s_j + L_{jk} \wedge \tilde{\s}_k \\
\d \n &= - \s_j \wedge \tilde{\s}_j \;.
}}
Stenzel space comes equipped with a metric
\al{\label{eq:stenzelmetric}
ds^2(\scal) = c(\tau)^2 \left( \frac14 d \tau^2 + \n^2 \right) + b(\tau)^2 \tilde{\s}_j^2 + a(\tau)^2 \s_j^2
}
with $a,b,c$ defined as
\eq{\spl{\label{eq:stenzelabc}
a^2 &= 3^{-\frac14} \l^2 \e^{\frac32}  x \cosh \left(\frac{\tau}{2}\right) \\
b^2 &= 3^{-\frac14} \l^2 \e^{\frac32}  x \cosh \left(\frac{\tau}{2}\right) \tanh^2 \left(\frac{\tau}{2}\right) \\
c^2 &= 3^{\frac34} \l^2 \e^{\frac32}  x^{-3} \cosh^3 \left(\frac{\tau}{2}\right)\;,
}}
with
\al{
x \equiv (2 + \cosh \tau)^{1/4}
}
defined for convenience. $\tau \in [0, \infty)$ is the radial coordinate of the deformed cone. Such $a,b,c$ satisfy the differential constraints
\eq{\spl{\label{eq:cycondition}
a'&= \frac{1}{4 b} \left(b^2 + c^2 - a^2 \right)  \\
b'&= \frac{1}{4 a} \left(a^2 + c^2 - b^2 \right) \\
c'&=\frac{3 c}{4 a b} \left(a^2 + b^2 - c^2 \right)
}}
which were determined to be the constraints for Ricci-flatness in \cite{cglp}. Note that $\l \in \mathbbm{R}$ is an arbitrary scaling parameter, as Ricci-flatness is invariant under conformal transformations.
We define orthonormal one-forms
\eq{\begin{alignedat}{2}\label{eq:oncoords}
e_0 =& \frac{c(\tau)}{2} \d\tau \;, \quad &e_j = a(\tau) \s_j \\
\tilde{e}_0 =& c(\tau) \n \;, \quad &\tilde{e}_j= b(\tau) \tilde{\s}_j
\end{alignedat}}
and holomorphic one-forms
\eq{\spl{ \label{eq:holocoords}
\z^0 &= - \frac{c}{2}d\tau +i c \n \\
\z^j &= a \s_j + i b \tilde{\s}_j \;.
}}
Using these, the K\"{a}hler form and Calabi-Yau form are given by
\eq{\spl{ \label{eq:stenzelsu}
J &= \frac{i}{2} \zeta^\a \wedge \bar{\z}^\a  \\
\O &= \z^0 \wedge\z^1 \wedge \z^2 \wedge \z^3
}}
such that $(J, \O)$ forms a Calabi-Yau structure on $\scal$. For small $\tau$, one sees that $a \sim c \sim \l \e^{3/4}$, $b \sim \frac{1}{2} \l \e^{3/4} \tau$. Hence at $\tau = 0$,  the CY-structure degenerates to
\eq{\spl{
\left.ds^2(\scal)\right|_{\tau = 0} &= \l^2 \e^{3/2} \left( \n^2 + \s_j^2 \right) \\
\left. J \right|_{\tau = 0} &= 0 \\
\left. \O\right|_{\tau = 0} &= \l^4 \e^3 i \n \wedge \s_1 \wedge \s_2 \wedge \s_3 \;.
}}
In particular, the metric becomes the standard metric on $S^4$, $J$ vanishes as the $S^4$ bolt is a special Lagrangian submaniold, and $\O$ becomes proportional to the volume form of the bolt.

\section{Type IIA \& M-theory on Stenzel Space}\label{fluxstenzel}
We will now consider the case of IIA theory on Stenzel space. Stenzel space possesses an $SU(4)$-structure as it is CY, hence applying the constraint given in subsection \ref{IIACY} yields a solution to the susy equations on Stenzel space. The only obstructions to finding a supersymmetric vacuum come from the fact that the (magnetic) RR fluxes have to be closed and co-closed on the $\scal$ and that the warp factor has to satisfy an inhomogeneous Laplace equation whose source is given by the RR flux. So the strategy in the following will be to a) construct new closed and co-closed RR fluxes on Stenzel space and b) solve the resulting differential equation for the warp factor.

In \cite{cglp}, an $\ncal = 2$ M-theory vacuum was given on Stenzel space. The vacuum was defined by a primitive selfdual closed (2,2)-form
\al{\label{eq:stenzelL2}
f^{(2,2)} &=      3 f_L [\tilde{e}_0\wedge e_1 \wedge e_2 \wedge e_3 + e_0 \wedge \tilde{e}_1 \wedge \tilde{e}_2 \wedge \tilde{e}_3]\nn\\
&+ \frac{1}{2} f_L \e_{ijk}[e_0\wedge e_i \wedge e_j \wedge \tilde{e}_k +  \tilde{e}_0 \wedge e_k \wedge \tilde{e}_i \wedge \tilde{e}_j ]\;,
}
with
\al{
f_L \equiv \left(\e^3 \cosh^4 \frac{\tau}{2}\right)^{-1}
}
and the associated warp factor
\al{\label{eq:stenzeln=2}
\hcal = 2^{3/2} 3^{11/4} \e^{-9/2} \int^\infty_x \frac{dt}{ (t^4 -1)^{5/2} }
}
which satisfies  \eqref{eq:eominteg}. Clearly, such an $\ncal=2$ M-theory vacuum is also an $\ncal=1$ M-theory vacuum as described in section
\ref{mtheory}, and can be dimensionally reduced to an $\ncal = (1,1) $ IIA  vacuum, as it satisfies \eqref{eq:warp} with $F_0 = F_2 = 0$, $F_4 = f^{(2,2)}$.

More generally, for $\ncal = (1,1) $ IIA  vacuum there are  four different irreps to consider: scalars, primitive $(1,1)$-forms, $(2,0)$-forms and primitive $(2,2)$-forms.
An observation that simplifies the analysis is that the contribution of each irrep to the warp factor equation \eqref{eq:warp} can be considered separately. For example,
switching on a scalar flux $f_s$ leads to the following modification of the warp factor equation
\all{
\nabla^2 \hcal &= -\frac{1}{2} \star_8 \left(f^{(2,2)}\wedge  f^{(2,2)} \right) - f_s^2\;.
}
So without loss of generality we can simply set $\hcal= \hcal_{(2,2)} + \hcal_s$, provided that
\all{
\nabla^2 \hcal_s =  -f_s^2
~;~~~\nabla^2 \hcal_{(2,2)} = -\frac{1}{2}\star_8 \left(f^{(2,2)}\wedge  f^{(2,2)} \right)
~.
}
The upshot is that each irreducible component of the flux $f$ can be considered as a separate source to the warp factor equation, leading to a corresponding solution $\hcal_f$ for the warp factor; the complete expresion for the warp factor is then given by $$\hcal = \sum_f \hcal_f~.$$

{\bf Summary of the results:}
We have been able to construct a non-normalizable closed and co-closed primitive $(1,1)$-form and a non-normalizable closed and co-closed primitive $(2,2)$-form on Stenzel space, and for each of them we have calculated the contribution to the warp factor. We find divergences in the IR\footnote{A note on terminology: `IR' for us means $\tau \rightarrow 0$, whereas `UV' is synonymous with $\tau \rightarrow \infty$. For a function $X$, we will also occasionally use `$X_{IR}$', which is defined by $X \rightarrow X_{IR}$ when $\tau \rightarrow 0$,  and similarly for $X_{UV}$ and $\tau \rightarrow \infty$.} which cannot be canceled against each other; in the UV the warp factor goes to zero in exactly the same way as for \eqref{eq:stenzeln=2}.

In addition we have considered closed and co-closed scalar fluxes $f_0$, $f_2$, $f_4$, $\tilde{f}_4$, which appear through the Romans mass ($f_0$), but also through the two form ($F_2\sim f_2 J+\dots$) and the four-form ($F_4\sim f_4 J^2+(\tilde{f}_4\Omega+\mathrm{c.c.})+\dots$). These scalars are constant as follows from closure and co-closure, and hence the corresponding fluxes are  non-normalizable. We have calculated the induced contribution to the warp factor: in the IR the warp factor stays finite; in the UV the geometry becomes  singular at a  finite value of the radial coordinate.

Finally we also consider adding a homogeneous solution $\hcal_0$ to the warp factor:   $\nabla^2\hcal_0=0$. This is finite in the UV but diverges in the IR and for that reason it was discarded in \cite{cglp}. Here we are including divergent modifications to the warp factor, so it is consistent to take  $\hcal_0$  into account.

\subsection{Scalars}
All scalars will contribute in the same manner but with different prefactors. Specifically, we have
\al{
  f^2 &\equiv  \star_8 \left(  F_0 \wedge \star_8 F_0 + F_2 \wedge \star_8 F_2 + \frac{1}{2}F_4 \wedge \star_8 F_4 \right)\nn\\
&=   \left(f_0^2 +4 f_2^2 + 12 f_4^2 + 16|\tilde{f}_4|^2\right)\;,
}
which gives the source entering the warp factor equation (\ref{eq:warp}).
Using the definitions of the metric \eqref{eq:stenzelmetric} and $a,b,c$ \eqref{eq:stenzelabc}, it can be seen that the warp equation reduces to
\all{
\hcal' &= - \varphi \left(x\sinh\left(\frac{\tau}{2}\right) \right)^{-3}\int^\tau dt \sinh^3 \!t\\
 &=  - \varphi \left(x\sinh\left(\frac{\tau}{2}\right) \right)^{-3}\left(\frac{2}{3}-\cosh\tau +\frac{1}{3}\cosh^3 \!\tau\right)\\
 &=  - \varphi \left(x\sinh\left(\frac{\tau}{2}\right) \right)^{-3}\left(\frac{4}{3}-4\cosh^4\left(\frac{\tau}{2}\right) +\frac{8}{3}\cosh^6 \left(\frac{\tau}{2}\right)\right)
~,}
where we fixed the integration constant in the second line by demanding that there should not be any IR divergence and in the third line we used the identity
$\cosh \tau = 2 \cosh^2\left(\frac{\tau}{2}\right) -1$; we have also defined
\all{
\hcal' &= \p_\tau \hcal \\
\varphi &\equiv g_s^2 f^2 (\frac{3^{3/4} }{2} \l^2 \epsilon^{3/2}) > 0\;.
}
By noting that
\all{
\left(\frac{4}{3}-4\cosh^4\left(\frac{\tau}{2}\right) +\frac{8}{3}\cosh^6\left(\frac{\tau}{2}\right)\right) = \frac{4}{3}x^{4}\sinh^4\left(\frac{\tau}{2}\right)
~,}
we find the explicit expression for the warp factor:
\al{
\hcal = - \frac{8\sqrt{2}}{15}\varphi \int_{3^{5/4}}^{x^5}\frac{dt}{(t^{4/5}-1)^{1/2}}+k\;,
}
with $k$ an integration constant.
There are no singularities in the IR since $\tau \rightarrow 0$ corresponds to  $x^5 \rightarrow 3^{5/4}$. In the UV, on the other hand, $\hcal$ vanishes at a finite
value of the radial coordinate and the metric develops a singularity. More specifically the asymptotics are:
\eq{
\hcal\longrightarrow
\left\{\begin{array}{rl}
k~;&~~~\tau \rightarrow 0\\
\frac{3}{4} k(\tau_{\mathrm{UV}}-\tau)~;&~~~\tau \rightarrow \tau_{\mathrm{UV}}
\end{array}
\right.
~,}
where
\eq{
\tau_{\mathrm{UV}}\equiv\frac{4}{3}\log \big(
\frac{2^{\frac14}9k}{8\varphi}
\big)~,
}
and we have assumed $\tau_{\mathrm{UV}} >>1$.

The ten- and eleven-dimensional metrics\footnote{The solution can be uplifted to eleven dimensions provided the Romans mass vanishes.} are given in \eqref{eq:iiametric} and \eqref{eq:mmetric} respectively. In the IR the metric is regular and asymptotes to the standard metric on $\mathbb{R}^{1,1}\times \scal$, or $\mathbb{R}^{1,2}\times \scal$ after uplift to eleven dimensions. In the UV the metric becomes singular at $\tau=\tau_{\mathrm{UV}}$; the singularity remains even after the uplift.

The dilaton and NSNS flux are given in (\ref{nsf}). The RR flux is given in (\ref{rrf}); in the present case of scalar fluxes this gives:
\eq{\spl{\label{rrscalar}
\mathcal{F}_0&=f_0\\
\mathcal{F}_2&=\hcal^{-1}f_0\mathrm{vol}_2+f_2J\\
\mathcal{F}_4&=\hcal^{-1}f_2J\wedge\mathrm{vol}_2+f_4J\wedge J+(\tilde{f}_4\Omega+\mathrm{c.c.})
~,
}}
where $f_0$, $f_2$, $f_4$ are constants.

\subsection{Two-forms}\label{sec:twoforms}
We now consider the possibility of allowing two-forms. There are two cases: $f_2^{(2,0)}$ or $f_2^{(1,1)}$, since  $f_4^{(1,1)}$ vanishes and $f^{(2,0)}_4$ is determined by $f^{(2,0)}_2$. For simplicity we will henceforth set $f^{(2,0)}=0$. The Bianchi identities require that $F_2$ should be closed and co-closed. Since $\star F_2 \sim f^{(1,1)}_2 \wedge J \wedge J$  and since $J$ is closed, closure of $F_2$ also implies co-closure.

In terms of the holomorphic one-forms \eqref{eq:holocoords}, a closed and co-closed primitive $(1,1)$-form is given by
\eq{\label{eq:(1,1)1}
f^{(1,1)} = i f(\tau) ( \z^0\wedge \bar{\z}^{\bar{0}} - \frac{1}{3} \z^j\wedge \bar{\z}^{\bar{\jmath}})\;,
}
provided that
\al{\label{eq:(1,1)2}
2 a b f' + (3 c^2 +2 (ab)') f = 0\;.
}
Using the explicit expressions for $a,b,c$ \eqref{eq:stenzelabc}, we find that
\eq{\label{ft}
f(\tau) = \frac{m}{\left(x\sinh\left(\tau/2\right)\right)^{4}}\;,
}
with $m$ an integration constant.
The norm of $f^{(1,1)}_2$ can be calculated explicitly:
\all{
F_2 \wedge \star F_2 &= -\frac{1}{2}f^{(1,1)}_2 \wedge f^{(1,1)}_2 \wedge J \wedge J \\
&= \frac{1}{3}f^2 \O \wedge \O^* = \frac{16}{3} f^2 \text{vol}_8 \;,
}
with $f$ given in (\ref{ft}). Note that this flux is non-normalizable: one has that
\eq{\spl{
\int_{\mcal_8} F_2 \wedge \star F_2 &= \int d^8 x \sqrt{g} \frac{16}{3} f^2\\
&\sim  \lim_{\e \rightarrow 0} \int_\e^\infty d\tau \frac{\sinh^3 \tau}{x^8 \sinh^8\left(\tau/2\right)}
~,}}
hence the integral diverges in the IR as $\tau^{-4}$.

The warp factor is given by
\all{
\nabla^2 \hcal = - g_s^2\frac{16}{3}f^2\;,
}
which leads to
\eq{\label{hder1}
 \hcal =  \frac{2}{3^{5/4}} g_s^2 \l^2 \e^{3/2} \int_{\tau}^{\infty}\d t\; \frac{k  \left( x \sinh\left(\frac{t}{2}\right)\right)^4 - m^2 }{\left( x \sinh\left(\frac{t}{2}\right)\right)^{7}}\;,
}
where $k$ is an integration constant, $m$ was introduced in (\ref{ft}), and the limits of the integration have been chosen so that $\hcal$ vanishes in the UV, but diverges in the IR.

Explicitly, in the UV one has:
\eq{
\hcal(\tau)\longrightarrow  \frac{128}{6^{1/4}27}g_s^2\l^2 \e^{3/2}  k e^{-\frac{9}{4}\tau}~;~~\tau \rightarrow \infty
~.}
In the IR, one finds that
\al{
\hcal(\tau)\longrightarrow  \frac{4}{1215} g_s^2 \l^2 \e^{3/2} \left( -480 m^2 \tau^{-6} + 420 m^2 \tau^{-4} + \frac12(497 m^2 - 540 k) \tau^{-2} \right)
}
for $\tau \rightarrow  \tau_{IR}$. $\tau_{IR}$ is determined by the approximate vanishing of $\hcal$, which happens for
\all{
\tau_{IR} =  2 \sqrt{ - \frac{105 m^2 \pm m\sqrt{32400 k - 18795 m^2 }} {540 k - 497 m^2}}\;.
}
In order for this approximation to be sensible, this places bounds on $k$ as a function of $m$. Specifically, let us take
$\sqrt{\frac{m^2}{k} } << 1$. Then, one has that
\al{
\tau_{IR} = \frac{2}{\sqrt{3}} \left(\frac{m^2}{k}\right)^{1/4} \left( 1 - \frac{7}{24} \sqrt{ \frac{m^2}{k}} \right) + \mathcal{O} ( (\frac{m^2}{k})^{5/4} )\;,
}
which goes to 0 as it should to justify the approximation used for $\hcal$.
In the IR the ten-dimensional string-frame metric becomes singular.
In the UV the string metric asymptotes to conformal\footnote{
Alternatively, the metric can be rewritten as
\all{
ds^2 = \L^2 \left( \d R^2 + R^6 ds^2(\mathbb{R}^{1,1}) + R^2 ds^2(V_{5,2})\right)\;.
}
For this reason, conformal AdS is equivalent to a domain wall.
} AdS.
\eq{
ds^2= \L^2 e^{2 \rho} \left[ds^2(\mathrm{AdS}_3)+ds^2(V_{5,2}) \right]
~,}
with
\eq{\spl{\label{eq:confads}
ds^2(\mathrm{AdS}_3) &= \d \rho^2 + e^{4 \rho}  ds^2(\mathbb{R}^{1,1}) \\
ds^2(V_{5,2}) &= \frac{9}{16} \n^2 + \frac{3}{8} \left( \s_j^2 + \tilde{\s}_j^2 \right)\;.
}}
Here, we introduced
\eq{\label{eq:confadspara}
\rho \equiv \frac{3}{8} \tau \;, \qquad \L^2 \equiv \frac{8}{6^{5/4}} \l^2 \epsilon^{3/2}
}
and rescaled
\all{
ds^2(\mathbb{R}^{1,1}) \rightarrow \frac{32}{9} \L^{4} g_s^{2} k\; ds^2(\mathbb{R}^{1,1}) \;.
}
The Einstein metric $ds^2_E=e^{-\frac{\phi}{2}}ds^2$ is also conformal to $\mathrm{AdS}_3\times V_{5,2}$ but with a different conformal factor:
\eq{
ds^2_E= \left(\frac{32 k}{9}\right)^{1/4}  \L^{5/2} e^{\frac{1}{2} \rho} \left[ds^2(\mathrm{AdS}_3)+ds^2(V_{5,2})\right]
~.}
After uplift to eleven dimensions the metric asymptotes $\mathrm{AdS}_{4}\times V_{5,2}$, exactly as was the case in \cite{cglp}.

The dilaton and NSNS flux are given in (\ref{nsf}). The RR flux is given in (\ref{rrf}); in the present case of (1,1) fluxes this gives:
\eq{\spl{\label{rr11}
\mathcal{F}_0&=0\\
\mathcal{F}_2&=f^{(1,1)}\\
\mathcal{F}_4&=\hcal^{-1}f^{(1,1)}\wedge\mathrm{vol}_2
~,
}}
where $f^{(1,1)}$ was given in (\ref{eq:(1,1)1}), (\ref{ft}).

\subsection{Four-forms}
Let us now examine possible four-forms on Stenzel space. The $(3,1)$, $(4,0)$ and non-primitive $(2,2)$ forms have been discussed, so let us examine possible $(2,2)$ forms which are primitive, closed (and hence, co-closed due to selfduality). We have found two such forms; they are given by
\al{
f^{(2,2)}_{NL}& =
            3 f_{NL} [e_0\wedge e_1 \wedge e_2 \wedge e_3 + \tilde{e}_1 \wedge \tilde{e}_2 \wedge \tilde{e}_3 \wedge \tilde{e}_0]\nn\\
&+ \frac{1}{2} f_{NL}  \e_{ijk}[e_0\wedge e_i \wedge \tilde{e}_j \wedge \tilde{e}_k + e_j \wedge e_k \wedge \tilde{e}_i \wedge \tilde{e}_0]
\label{eq:(2,2)NL1}\\
f_L^{(2,2)} &=
     3 f_L [\tilde{e}_0\wedge e_1 \wedge e_2 \wedge e_3 + e_0 \wedge \tilde{e}_1 \wedge \tilde{e}_2 \wedge \tilde{e}_3]\nn\\
&+ \frac{1}{2} f_L \e_{ijk}[e_0\wedge e_i \wedge e_j \wedge \tilde{e}_k +  \tilde{e}_0 \wedge e_k \wedge \tilde{e}_i \wedge \tilde{e}_j ]\;, \label{eq:(2,2)L1}
}
with $f_{NL}$ determined by
\eq{\spl{\label{eq:(2,2)NL2}
2 b c f_{NL}' + ( ac + 6 cb' + 2 bc') f_L =& 0\\
2 c f_{NL} \left(a^2 - b^2 + 2 b a'- 2 a b' \right) =& 0
}}
and $f_L$ determined by
\eq{\spl{\label{eq:(2,2)L2}
2 a c f_{L}' + ( bc + 6 ca' + 2 ac') f_L =& 0\\
2 c f_{L} \left(a^2 - b^2 + 2 b a'- 2 a b' \right) =& 0
}}
Notice that \eqref{eq:(2,2)L2} is nothing more than \eqref{eq:(2,2)NL2} under the transformation $a \leftrightarrow b$. The second equation is a consistency condition which is satisfied for the $a,b,c$ of Stenzel space\footnote{In fact, it is equivalent to integrability of the almost complex structure. See section \ref{moduli} for more details.}. Plugging in $a, b, c$ in the first equation, we find that $f^{(2,2)}_L$ is nothing more than the $L^2$ harmonic form found before \eqref{eq:stenzelL2}. Inserting the expressions for $a,b,c$ \eqref{eq:stenzelabc}, the solution to \eqref{eq:(2,2)NL2} is given by
\al{\label{4}
f_{NL} = \frac{m}{\e^3\sinh^4(\tau/2)}\;,
}
where we have chosen a normalization to match the conventions of \cite{kp}.
This flux is non-normalizable as can be seen from
\all{
\int_{\mcal_8} f^{(2,2)} \wedge f^{(2,2)} &\sim  \lim_{\e \rightarrow 0} \int_\e^\infty d\tau \frac{\sinh^3 \tau}{\sinh^8(\tau/2)}\;,
}
which diverges as $\tau\rightarrow 0$ in the same way as $f^{(1,1)}$ of section \ref{sec:twoforms}. Thus we get the same IR divergence for the warp factor to leading order, although the subleading term will have a different coefficient.
More explicitly the warp factor satisfies
\eq{\spl{
\nabla^2 \hcal &= - \frac{g_s^2}{2} 24 f^2_{NL} \\
&=  - 12 g_s^2\frac{ m^2}{\e^6\sinh^8(\tau/2)}\;,
}}
with solution
\al{
\hcal &= \frac{3^{7/4} }{2}g_s^2 \l^2 \e^{-9/2}\int_{\tau}^{\infty}\d t\;  \frac{(k -2 m^2 \sinh^{-2}\left(\frac{t}{2}\right) - m^2 \sinh^{-4}\left(\frac{t}{2}\right))}
{(x \sinh\left(\frac{t}{2}\right))^{3}} \;,
}
with $k$ an integration constant, and we have chosen the limits of the integration so that $H$ vanishes in the UV but diverges in the IR. Explicitly the UV asymptotics are given by
\al{
\hcal(\tau)\longrightarrow \frac{32}{6^{1/4}} g_s^2 \l^2 \e^{-9/2}  k e^{-\frac{9}{4} \tau}~;~~~\tau \rightarrow \infty
}
and the IR asymptotics by
\eq{
\hcal(\tau)\longrightarrow g_s^2 \l^2\e^{-6}\left( - 32 m^2\tau^{-6}  - 4 m^2 \tau^{-4} + (6k + \frac{223}{30} m^2) \tau^{-2}   \right)
}
for $\tau \rightarrow \tau_{IR}$.
Setting this expression to zero we find
\all{
\tau_{IR} = 2 \sqrt{ \frac{15 m^2 + \sqrt{15 m^2}\sqrt{720 k + 907 m^2}}{180 k + 223 m^2}}\;.
}
Again, we take $k$ such that $\sqrt{\frac{m^2}{k}} << 1$ leading to
\al{
\tau_{IR} = \frac{2}{3^{1/4}} \left(\frac{m^2}{k}\right)^{1/4} \left( 1 + \frac{1}{8 \sqrt{3}}\sqrt{\frac{m^2}{k}}\right)+ \mathcal{O} ( (\frac{m^2}{k})^{5/4} )\;.
}
Similarly to the case where the flux was given by a (1,1)-form, the string metric asymptotes to conformal AdS in the UV
\eq{
ds^2= \L^2 e^{2 \rho} \left[ds^2(\mathrm{AdS}_3)+ds^2(V_{5,2}) \right]
~,}
with $\L$, $\rho$ defined in \eqref{eq:confadspara}, the metrics defined in \eqref{eq:confads}, and the metric rescaled
\all{
ds^2(\mathbb{R}^{1,1}) \rightarrow 24 \L^{4} g_s^{2}\epsilon^{-6} k \;ds^2(\mathbb{R}^{1,1}) \;.
}
Thus, we again find that the Einstein metric is conformal $\mathrm{AdS}_3\times V_{5,2}$ but with a different conformal factor
\eq{
ds^2_E=(24 k)^{1/4}\e^{-3/2}  \L^{5/2} e^{\frac{1}{2}\rho}\left[ds^2(\mathrm{AdS}_3)+ds^2(V_{5,2})\right]
~,}
and again, the uplift to eleven dimensions of the metric asymptotes to $\mathrm{AdS}_{4}\times V_{5,2}$.

The dilaton and NSNS flux are given in (\ref{nsf}). The RR flux is given in (\ref{rrf}); in the present case of (2,2)-fluxes this gives:
\eq{\spl{\label{rr22}
\mathcal{F}_0&=0\\
\mathcal{F}_2&=0\\
\mathcal{F}_4&=f^{(2,2)}
~,
}}
where $f^{(2,2)}$ is given by \eqref{eq:(2,2)NL1},  (\ref{4}).

\subsection{Homogeneous solution}\label{sec:homo}
As already  mentioned, the homogeneous solution was discarded in \cite{cglp} due to the IR divergences but for us it is consistent to include it. The warp factor equation becomes
\all{
\p_\tau [\left(x\sinh\left(\frac{\tau}{2}\right)\right)^{3}\p_\tau \hcal] =0 \;,
}
which gives
\al{\label{eq:homo}
\hcal =   k_1 -k_2 \int^\tau \frac{d t}{\left(x(t)\sinh\left(\frac{t}{2}\right)\right)^{3}} \;,
}
with $k_1$, $k_2$ integration constants.
This is finite in the UV but diverges in the IR. Explicitly
the asymptotics are given by
\eq{ \def\arraystretch{1.8}
\hcal \longrightarrow
\left\{\begin{array}{cc}
k_1 +\frac{4}{3^{3/4}} k_2 \tau^{-2} & \phantom{aaaaaa}\tau \rightarrow 0\\
k_1 + \frac{64}{2^{1/4} 9}k_2 e^{-\frac{9}{4}\tau}& \phantom{aaaaaaa}\tau \rightarrow \infty
\end{array}
\right.
~.}

\section{$SU(4)$-structure deformations of Stenzel space}\label{deforms}
So far, we have discussed supersymmetric vacua of type II and M-theory on Stenzel space, by applying the known results for eight-manifolds with $SU(4)$-structure: Stenzel space is a Calabi-Yau fourfold, hence it has such an $SU(4)$-structure, as determined by \eqref{eq:stenzelsu}. In order to go beyond Stenzel space, we will alter the geometry by considering deformations of the $SU(4)$-structure.
\\
\\
Our starting point will be to consider the topological space $\scal \equiv T^* S^4 $, which is homeomorphic to Stenzel space. We will first investigate all $SU(4)$-structures on this space which are `left-invariant'; we will explain what we mean by this. We will then consider a subset of these, which we dub `$abc$ $SU(4)$-structures'. The canonical metric of such $abc$ $SU(4)$-structures is identical to the Stenzel metric \eqref{eq:stenzelmetric}, but with $a,b,c$ generic functions of the radial direction $\tau$ rather than fixed to satisfy \eqref{eq:stenzelabc}.
\\
\\
By taking generic $a,b,c$, we find that $J, \O$ are no longer closed and thus torsion classes have been turned on, going beyond the Calabi-Yau scenario. As the torsion classes determine integrability of geometrical structures associated to the $SU(4)$-structure (such as the almost complex and almost symplectic structure),  we can examine moduli spaces of such geometrical structures. These moduli spaces are  determined by differential equations which we can solve for a number of cases.
\\
\\
Finally, we should take care that by altering the metric, we do not make the space geodesically incomplete. We analyze various possibilities, and conclude that to avoid this, the straightforward thing to do is to ensure that at $\tau =0$, the space maintains its $S^4$ bolt, albeit possibly squashed.

\subsection{Left-invariant $SU(4)$-structures}
As mentioned, our starting point is $\scal = T^* S^4$  equipped with a Riemannian metric $g$. In order to construct $SU(4)$-structures on this space, we will require some knowledge of the cotangent bundle of $\scal$. Due to its conical structure, we will first focus on the forms on $V_{5,2}$, and for that, it is convenient to first discuss left-invariant forms on cosets.
We consider the left-invariant forms of $SO(5)$, descended on the coset $V_{5,2} \simeq SO(5) / SO(3)$. Concretely, this means the following: the left-invariant one-forms of $SO(5)$ are given by $\n, \s_j, \tilde{\s}_j, L_{jk}$, $j,k \in \{1,2,3\}$, with $L_{jk}$ the left-invariant one-forms of the $SO(3)$ subgroup, while $\s_j, \tilde{\s}_j, \n$ lie in the complement. In terms of these, a $p$-form on $V_{5,2}$ is left-invariant if and only if its exterior derivative lies in the complement, i.e., is expressible solely in terms of $\s_j, \tilde{\s}_j, \n$. Given a left-invariant form, any scalar multiple is also a left-invariant form. The exterior derivative maps left-invariant $p$-forms to left-invariant $p+1$-forms. For more general details, see \cite{nieuw}.
\\
\\
A basis of left-invariant forms up to fourth degree for $V_{5,2}$ and their derivatives is given in the following table (see also \cite{varela}):
\begin{center}
\begin{tabular}{ |l|l|l|}
\hline
1-forms & $\n$ &  \\\hline
2-forms & $\d\n = \tilde{\s}^j \wedge \s^j$ &\\\hline
\multirow{5}{*}{3-forms}
& $\a_0=\n \wedge d\n$ &  $\d \a_0 = \b_0$ \\
& $\a_1=\s^1 \wedge \s^2 \wedge \s^3$ & $\d\a_1 = - \b_4$  \\
& $\a_2=\tilde{\s}^1 \wedge \tilde{\s}^2 \wedge \tilde{\s}^3$   & $\d \a_2 = - \b_3$ \\
& $\a_3=\frac{1}{2} \varepsilon_{ijk} \tilde{\s}^i \wedge \s^j \wedge \s^k$ & $\d \a_3 = 2 \b_3 - 3\b_2$ \\
& $\a_4= \frac{1}{2}\varepsilon_{ijk} \s^i \wedge \tilde{\s}^j \wedge \tilde{\s}^k$  & $\d\a_4 = 2\b_4 - 3 \b_1$  \\ \hline
\multirow{5}{*}{4-forms}
& $\b_0=d\n \wedge d\n$ &\\
& $\b_1= - \n \wedge \a_2$ &\\
& $\b_2=\n \wedge \a_1$ &\\
& $\b_3= \n \wedge \a_4 $ &\\
& $\b_4=- \n \wedge \a_3$ &\\ \hline
\end{tabular}\label{li-forms}
\end{center}
Although left-invariant forms only make sense given a specific group action on the space of forms, we will abuse terminology and make the following definition: Let $\tau$ be the radial coordinate on $\scal$ viewed as a (deformed) cone over $V_{5,2}$. Then we define a \textit{left-invariant form on $\scal$} to be a form that restricts to a left-invariant (`LI') form on $V_{5,2}$ at any fixed $\tau$. This is equivalent to demanding that the exterior derivative acting on such forms can be expressed purely in terms of the radial one-form $\d\tau$ and the left-invariant one-forms of $SO(5)$ that lie in the coset $SO(5)/SO(3)$, i.e., $\n, \s_j, \tilde{\s}_j$.
All such LI $p$-forms on $\scal$ can be gotten by taking linear combinations of LI $p$-forms on $V_{5,2}$, and LI $p-1$-forms on $V_{5,2}$ wedged with $\d\tau$, with coefficients that are functions only of $\tau$.

We define a \textit{left-invariant $SU(4)$-structure} to be an $SU(4)$-structure defined by forms $(J, \O)$ such that $J$ and $\O$ are left-invariant forms on $\scal$\footnote{In\cite{varela}, the LI forms on $V_{5,2}$ are used to construct LI $SU(3)$-structures on $V_{5.2}$.}. The most general LI two-form and LI four-form are given by
\all{
J =& n (\tau) \n \wedge \d\tau - m (\tau) d\nu\\
\O =& k_j(\tau)   \d\tau \wedge \a^j + l_j (\tau) \b^j \;,
}
with $j \in \{ 0, ..., 4\}$ summed over. In order for $(J, \O)$ to form an $SU(4)$-structure, it needs to satisfy the following constraints:
\eq{\spl{\label{eq:su4cons}
J\wedge \O =& 0\\
\frac{1}{4!}J^4 =& \text{vol}_8\\
\frac{1}{2^4} \O \wedge \O^* =& \text{vol}_8\\
\star \O =& \O
\;.
}}
In addition, $\O$ needs to be decomposable\footnote{
In $d=6$, this can be checked by means of the Hitchin functional \cite{hitchin2}: The three-form $\O$ determines the complex structure through
\all{
I_j^{\phantom{j}k} = - \epsilon_{j m_1 ... m_5 } \left(\text{Re}\O\right)^{k m_1 m_2} \left(\text{Re}\O\right)^{m_3 m_4 m_5}\;.
}
In $d=8$, this fails, as one can see from a selfconsistency check: one finds instead that
\all{
- \epsilon_{j m_1 ... m_7 } \left(\text{Re}\O\right)^{k m_1 m_2 m_3} \left(\text{Re}\O\right)^{m_4 m_5 m_6 m_7} = \delta_j^k \;.
}
This is a consequence of the fact that $\star_8\O$ = $\O$ compared with $\star_6 \O = i \O$.} and holomorphic. Note that the final constraint is not independent, yet explicitly working with this redundancy is convenient. Let us see how these constrain the free parameters $k_j$, $l_j$, $n$, $m$. We start by introducing normalised one-forms
\eq{\begin{alignedat}{2}
e_0 &= \frac{1}{\sqrt{g^*(\d\tau, \d\tau)}} \d\tau \;, \quad &e_j = \frac{1}{\sqrt{g^*(\s_j, \s_j)}} \s_j \phantom{\;,}\\
\tilde{e}_0 &= \frac{1}{\sqrt{g^*(\n, \n)}} \n \;, \quad &\tilde{e}_j= \frac{1}{\sqrt{g^*(\tilde{\s}_j, \tilde{\s}_j)}} \tilde{\s}_j\;,
\end{alignedat}}
with $g^*$ the induced metric on the cotangent bundle. We shift the coefficients of $J, \O$ to encompass these normalizations so that we can consider the one-forms normalised without loss of generality.
The first constraint of \eqref{eq:su4cons} sets $k_0 = l_0= 0$.
Let us now choose an orientation by taking
\eq{\spl{
\text{vol}_8 &= \sqrt{g} d\tau \wedge \n \wedge \s_1 \wedge \s_2 \wedge \s_3 \wedge \tilde{\s}_1\wedge \tilde{\s}_3\wedge \tilde{\s}_3\\
&= e_0 \wedge  \tilde{e}_0 \wedge  e_1\wedge  e_2\wedge  e_3\wedge  \tilde{e}_1\wedge  \tilde{e}_2\wedge  \tilde{e}_3\;.
}}
Then the selfduality constraint leads to $k_j = l_j$.
Next, we consider decomposability of $\O$. As detailed in \cite{ckkltz} appendix C, a $p$-form $\o$ is decomposable iff $\forall X \in \mathfrak{X}^{p-1}(M)$, $\iota_X \o \wedge \o = 0$. This gives us 56 equations, a number of which are trivial while the rest are linearly dependent. They boil down to
\all{
k_1^2 + k_2^2 =& 0 \\
k_3^2 + k_4^2 =& 0 \\
k_1^2 + k_3^2 =& 0 \\
k_2^2 + k_4^2 =& 0 \\
k_3 ( k_1 + k_4) =& 0\\
k_4 (k_2 + k_3 )=& 0\;,
}
with solution
\al{
\vec{k} = (k_1, \pm i k_1, \mp i k_1, -k_1)\;.
}
As we will see, the choice of $\pm$ determines which one-forms are holomorphic and which are anti-holomorphic.
Using this solution, the volume constraints reduce to
\all{
1 =& nm^3 = |k_1|^2\;,
}
hence $k_1$ is just a phase. Holomorphic one-forms are then given by
\eq{\spl{
\zeta^0 =& e^{ i \varphi_0} \left(- e_0 + i \tilde{e_0}\right)\\
\zeta^j =& e^{ i \varphi_j} \left( e_j + i \tilde{e_j}\right)\;,
}}
with $e^{ i (\varphi_0+ \varphi_1 + \varphi_2 +\varphi_3) }= k_1$ such that
\al{
\O = \zeta^0 \wedge \zeta^1  \wedge \zeta^2  \wedge \zeta^3\;.
}
In terms of these holomorphic one-forms, we have
\al{
J = \frac{i}{2} \left( n \z^0 \wedge \bar{\z}^0 + m \z^j \wedge \bar{\z}^j \right)\;.
}
Since the complex structure is given by $I_a^{\phantom{a}b}= J_{ac}g^{cb}$, $\O$ is holomorphic, and $g^*(\z^\a, \bar{\z}^\a) = 2$, the constraint $I^2 = - \mathbbm{1}$ enforces
$n = m = 1$, leading to
\al{
J = \frac{i}{2} \zeta^\a \wedge \bar{\zeta}^\a
}
and the metric must be given by
\eq{\spl{
ds^2(\scal) =&\frac{1}{2} \left( \zeta^\a \otimes \bar{\zeta}^\a + \bar{\zeta}^\a \otimes \zeta^\a \right) \\
=& e^\a \otimes e^\a + \tilde{e}^\a \otimes \tilde{e}^\a \\
\equiv& \frac{c^2}{4} \d\tau^2 + c^2 \n^2 + a_j \s_j^2 + b_j \tilde{\s}_j^2\;,
}}
with the additional restriction
\al{
a_1 b_1 = a_2 b_2 = a_3 b_3 \;.
}
Here, we have rescaled $\tau$ such that its coefficient is once more $\frac{c^2}{4}$ and renamed the normalization to match the notation in the Stenzel space scenario. The rotations generated by the four angles $\varphi_0, \varphi_j$ leave the $SU(4)$-structure invariant and hence can be set to zero without loss of generality. Thus, LI $SU(4)$-structures are defined by the five free parameters  $a_1(\tau), b_1(\tau), b_2(\tau), b_3 (\tau), c(\tau)$.

\subsection{$abc$ $SU(4)$-structures}
We will restrict ourselves to the case $a_j = a$, $b_j = b$ $\forall j$, as the generalization does not lead to any novel features outside of more cluttered equations. For lack of imagination, we shall refer to such as $abc$ $SU(4)$-structures to distinguish them from the more general LI $SU(4)$-structures.
Let us explicitly spell out such structures. We have orthonormal one-forms
\eq{
\begin{alignedat}{2}
e_0 =& \frac{c(\tau)}{2} \d\tau \;, \quad &e_j = a(\tau) \s_j \\
\tilde{e}_0 =& c(\tau) \n \;, \quad &\tilde{e}_j= b(\tau) \tilde{\s}_j
\end{alignedat}
}
and holomorphic one-forms
\eq{\spl{
\z^0 &= - \frac{c}{2}d\tau +i c \n \\
\z^j &= a \s_j + i b \tilde{\s}_j \;.
}}
The metric and volume form are given by
\al{
ds^2(\scal) =& (e^\a)^2 + (\tilde{e}^\a)^2 \nn \\
\text{vol}_8 =& e^0 \wedge \tilde{e}^0 \wedge e^1 \wedge e^2 \wedge e^3 \wedge \tilde{e}^1 \wedge \tilde{e}^2 \wedge \tilde{e}^3   \;,
}
with $\a \in \{0,1,2,3\}$, and the $SU(4)$-structure is given by
\eq{\spl{\label{j}
J &= \frac{i}{2}\z^\a \wedge \bar{\z}^{\bar{\a}}\\
\O &= \z^0 \wedge \z^1 \wedge \z^2 \wedge \z^3
~.}}
For $a,b,c$ as in \eqref{eq:stenzelabc}, this $SU(4)$-structure is the one on Stenzel space.
As explained in section \ref{su(4)}, the $SU(4)$-structure determines the existence of geometrical structures in terms of torsion classes, defined by
\eq{\spl{
\d J =& W_1 \lrcorner \O^* + W_3 + W_4 \wedge J + \text{c.c.} \\
\d \O =& \frac{8 i}{3} W_1 \wedge J \wedge J + W_2 \wedge J + W_5^* \wedge \O \;.
}}
Using the explicit expressions for $J$ and $\O$, one finds that
\al{\label{eq:torsion}
W_1 &= 0 \nn \\
W_2 &=\frac{1}{4i abc}(a^2 - b^2 + 2 b a'-2 a b') \varepsilon_{\bar{i}jk} \bar{\z}_i \wedge \z_j \wedge \z_k \nn\\
W_3 &= 0 \\
W_4 &= - \frac{1}{abc}\left((ab)'- \frac{1}{2}c^2\right)\z_0\nn\\
W_5 &=-\frac{1}{4 abc^2}\left(-3 (a^2+  b^2) c + 6 (ab)'c +  4 abc'\right)\z_0\;.\nn
}
Note that these equations are only sensible as long as $a,b,c \neq0$. At any point where this does not hold, the $SU(4)$-structure degenerates: as an example, let $p \in \scal$ be described in local coordinates as  $p = (\tau_0,...)$ and let $b$ satisfy  $b(\tau_0) = 0$. Then the metric reduces to
\all{
\left.g\right|_{T_p\scal}  =  c(\tau_0)^2 \left( \frac14 \d \tau^2 + \n^2 \right) +  a(\tau_0)^2 \s_j^2\;.
}
Let the vectors $\tilde{v}_j \in T_p \scal$ be dual to $\left.\tilde{\s}_j\right|_p$. Then
\all{
\left.g\right|_{T_p\scal} (\tilde{v}_j, v) = 0 \quad \forall v \in T_p \scal
}
and thus the metric is degenerate. On Stenzel space, such a situation occurs at the tip, where $b(0) = 0$  leads to an $S^4$ bolt. In particular, this $S^4$ bolt is a special Lagrangian submanifold with $J=0$.

\subsection{Moduli spaces of $abc$ $SU(4)$-structures}\label{moduli}
Let us examine the geometry determined by the torsion classes more carefully.
As a consequence of \eqref{eq:torsion}, we see that whatever functions we choose for $a,b,c$, $W_1 = W_3 = 0$. This leaves us with $W_2, W_4, W_5$ which can all either be zero or non-zero. Let us first consider $W_2$. The almost complex structure determined by the $SU(4)$-structure is integrable if and only if $W_2=0$, which is equivalent to
\al{\label{eq:W2}
a^2 - b^2 + 2 b a'-2 a b' = 0\;.
}
By substituting $r = b/a$ one can find the solution to this equation to be given by
\al{
r \in \left\{\pm 1, \frac{e^{(\tau + \tau_0)/2} \mp  e^{- (\tau+ \tau_0)/2} }{e^{(\tau+ \tau_0) /2} \pm  e^{- (\tau+ \tau_0)/2}}\;|\; \tau_0 \in \mathbbm{R} \right\}
}
where one should consider $r = \pm1$ to be the limiting cases for the integration constant $\tau_0 \rightarrow \pm \infty$. The moduli space of complex structures for $SU(4)$-structures satisfying our ansatz is then given by these solutions, modulo diffeomorphisms (in particular, $\tau \rightarrow \tau - \tau_0$ and $\tau \rightarrow - \tau - \tau_0$):
\al{
\mathfrak{M}_\mathbbm{C}&= \left\{(a,r a,c)\;|\; r \in \{ \pm 1, \tanh^{\pm1} \left(\frac{\tau}{2}\right)\}  \right\}\;.
}
Note that  $ r \rightarrow r^{-1}$ leaves the moduli space invariant, which is a consequence of  \eqref{eq:W2} being invariant under $a \leftrightarrow b$. In essence, we thus see that there are but two possible complex structures: we will refer to $r = \tanh \left(\frac{\tau}{2}\right)$ as the Stenzel complex structure, and to $r =  1 $ as the conical complex structure (we will describe this in more detail in the next subsection).

Next, let us examine when the $SU(4)$-structure determines an integrable almost symplectic structure. The non-degenerate two-form $J$ is closed (and hence, a symplectic form) if and only if $W_4=0$, which is equivalent to
\al{
(ab)'- \frac{1}{2}c^2 = 0\;,
}
hence the moduli space of symplectic structures is given by
\al{
\mathfrak{M}_S &=\left\{(a,b, \sqrt{2(ab)'} ) \right\}
}
Compatibility of the complex and symplectic structure determines a K\"{a}hler structure, hence the K\"{a}hler moduli space is given by
\al{
\mathfrak{M}_K &= \mathfrak{M}_\mathbbm{C} \cap \mathfrak{M}_S \nn\\
&= \left\{(a,r a,\sqrt{2(r a^2)'})\;|\; r \in \{ \pm 1, \tanh^{\pm1} \left(\frac{\tau}{2}\right)\}  \right\}
}
Conformal Calabi-Yau structures are found by demanding $W_2 = 0$, $2W_4 = W_5$. Setting $b = r a$ and using the expressions for $W_4, W_5$, this reduces to the equation
\al{\label{eq:ccy}
\tilde{r}' + \left(r + 2 r^{-1} \right) \tilde{r} - 2 r^{-1} = 0\;,
}
where we have set $\tilde{r} \equiv a^2 / c^2$. This is solved by
\al{
\tilde{r} = \left\{\label{eq:ccy2}
\renewcommand\arraystretch{2}
\begin{array}{cl}
 \frac{2 + \cosh\left(\tau\right) + k \left(\sinh\left(\frac{\tau}{2}\right)\right)^{-4} }{3 \cosh^2\left(\frac{\tau}{2}\right)} & \phantom{abcde} r = \tanh \left(\frac{\tau}{2}\right) \\
\frac{2}{3} + k e^{\mp 3 \tau} &  \phantom{abcde} r = \pm1 \\
\end{array} \right.
}
for $k \in \mathbbm{R}$. As $(r, \tilde{r})$ determine the ratios between $a$ and $c$ and between $a$ and $b$, and one can by conformal transformation set $a=1$, this determines every possible CCY-structure. Some remarks are in order here. Firstly, note that if we consider $\tilde{r}$ as a function of $r$,
\all{
\lim_{\tau \rightarrow \pm \infty} \tilde{r}\left(\tanh \left(\frac{\tau}{2}\right)\right) = \tilde{r}( \pm 1)
}
after rescaling $k$, thus confirming our earlier argument that such points in the moduli space of complex structures should be considered as limiting cases.
Secondly,  note that for $r = \tanh \left(\frac{\tau}{2}\right)$,  $\tilde{r}$ is singular at $\tau = 0$ indicating that the $SU(4)$-structure degenerates as either $c$ approaches $0$ or $a$ blows up. This happens unless one sets the integration constant $k = 0$, which gives the unique regular solution
\al{\label{eq:stenzelccy}
\tilde{r} =  \frac{2 + \cosh\left(\tau\right) }{3 \cosh^2\left(\frac{\tau}{2}\right)} \;.
}
This is exactly the proportionality between $a^2$ and $c^2$ for Stenzel space \eqref{eq:stenzelabc}, so we have found that the only smooth CCY manifolds with the Stenzel space complex structure are in fact conformal to Stenzel space. On the other hand, the conical complex structures lead to a regular $\tilde{r}$ regardless of choice of $k$. Thirdly, note that $r = 1/ \tanh \left(\frac{\tau}{2}\right)$ is uninteresting as the constraint  $2W_4 = W_5$ is invariant under $a \leftrightarrow b$. Thus the solution is as above but with $a \leftrightarrow b$.

Finally, a Calabi-Yau structure is found by demanding $W_2 = W_4 = W_5 = 0$. Note that this is equivalent to \eqref{eq:cycondition} as given in \cite{cglp}.
In case we take $r=  \tanh \left(\frac{\tau}{2}\right)$, we can solve $W_5 = 0$ to find
\all{
r a^2 =  \frac{1}{2} \l^{8/3}  c^{-2/3}\sinh \tau \;, \qquad \l \in \mathbbm{R}
}
and use this to solve $W_4 =0$, leading to
\all{
c^2 &= 3^\frac{3}{4}    \l^2 \frac{\cosh^3\left(\frac{\tau}{2}\right)}{\left(\frac{k}{\sinh^4\left(\frac{\tau}{2}\right)} + \left(2 + \cosh \tau \right)\right)^{3/4}}\\
a^2 &= 3^{-\frac{1}{4}} \l^2 \cosh\left(\frac{\tau}{2}\right)                                   \left(\frac{k}{\sinh^4\left(\frac{\tau}{2}\right)} + \left(2 + \cosh \tau \right)\right)^{1/4}\\
b^2 &= 3^{-\frac{1}{4}} \l^2 \cosh\left(\frac{\tau}{2}\right)\tanh^2\left(\frac{\tau}{2}\right) \left(\frac{k}{\sinh^4\left(\frac{\tau}{2}\right)} + \left(2 + \cosh \tau \right)\right)^{1/4}
}
Compare with \eqref{eq:stenzelabc}. This derivation is completely equivalent to the derivation of the Stenzel metric in \cite{cglp}, except we have explicitly kept an integration constant $k$: however, any non-zero $k$ will lead to a singularity at $\tau = 0$. For our intents and purposes, we are thus only interested in Stenzel space.

In case we take $r = 1$, the solution to $W_5=0$ is given by
\all{
a^2 = \frac{2}{3} \l^{8/3} c^{-2/3} e^\tau
}
and solving $W_4$ then leads to
\eq{\spl{\label{eq:cone2}
c^2 &=\l^2 \frac{e^{3\tau}}{(e^{3\tau}+ k)^{3/4}} \\
a^2 &= b^2 = \frac{2}{3} \l^2 (e^{3\tau}+ k)^{1/4}
}}
If instead we take $r= -1$, the same solution is found but with $\tau \rightarrow - \tau$ as expected.

\subsection{Geodesical Completeness \& Cones} \label{cones}
Before we start constructing vacua using these $abc$ $SU(4)$-structures, let us first consider the construction of CY-structures with $r=1$, i.e., \eqref{eq:cone2}.
In order to illustrate what is happening here we consider the specific case $k=0$, such that this configuration reduces to
\eq{\spl{\label{eq:cone}
a &=  b = \sqrt{\frac{2}{3}} c\\
c &= \l e^{ \frac38 \tau} \;.
}}
In the UV, this solution is identical to Stenzel space. In the IR it behaves quite differently; at $\tau=0$, $b \neq 0$ hence there is no $S^4$ bolt. Indeed, as $a$, $c$ also do not vanish, $\tau=0$ is no special point and we see that this slice is just a copy of $V_{5,2}$, just like slices for any other $\tau$. In fact, we can make a coordinate transformation to find that the metric is globally given by
\al{
ds^2(\scal) &= \l^2 e^{\frac{3}{4} \tau} \left( \frac{1}{4} \d \tau^2 + \n^2 + \frac{2}{3} (\s_j^2 + \tilde{\s}_j^2 )\right)\nn\\
&=  \frac{16}{9} \l^2  \left[\d R^2 +  R^2 ds^2(V_{5,2}) \right]
}
with $ds^2(V_{5,2})$ defined in \eqref{eq:confadspara} and $R \in [1,\infty)$ for $\tau \in [0, \infty)$. We recognize that this is just the undeformed cone with the tip `cut off', as it were. More precisely, the space is geodesically incomplete: one can solve the geodesic equation to find
\all{
\tau(t) = k_2 - \frac{8}{3} \log(3t + k_1)
}
which is not a solution $\forall t \in \mathbbm{R}$, regardless of $k_{1,2}$. Obviously, the tip of the cone is a singularity; smoothing out this singularity was the reason Stenzel space garnered interest in the first place.
\\
\\
We illustrate this because it points us to a potential pitfall: one can consider arbitrary $abc$ $SU(4)$-structures, but without the $S^4$ bolt at the tip it is not guaranteed that the space is geodesically complete. We have the following possibilities at $\tau=0$:
\begin{itemize}
\item $a=b=c=0$ leads to a singularity. $a,b,c \neq 0$ leads to potentially incomplete spaces as above; all spaces of such type that we have examined are incomplete and have a conical singularity in their completion.
\item $a=0$, $b=  c \neq 0$ leads to an $S^4$ bolt, similar to $b=0$, $a= c \neq 0$, on which $\O$ is proportional to the volume form and $J$ vanishes. This can be deduced by noting that the defining equation for the LI forms of $V_{5,2}$ \eqref{eq:liforms} are invariant under $\n \rightarrow - \nu$, $\s_j \leftrightarrow \tilde{\s}_j$. This transformation interchanges the four-forms \eqref{eq:(2,2)NL1} and \eqref{eq:(2,2)L1}, which also explains why $ a\leftrightarrow b$ leads to $f_L \leftrightarrow f_{NL}$ as follows from \eqref{eq:(2,2)NL2}, \eqref{eq:(2,2)L2}. Thus, which (2,2)-form diverges and which does not is interchanged. This conclusion essentially remains the same for vacua on $SU(4)$-deformed spaces. In case one has $b=0$, $a, c \neq 0$, $a \neq c$ (or similarly for $a \leftrightarrow b$), the bolt will be a squashed $S^4$, with squashing parameter $c(0) / a(0) \equiv \a(0)$, as follows from the metric at $\tau = 0$.
\item The remaining possibilities are
\begin{itemize}
\item[-]$ c = 0, a, b \neq 0$
\item[-] $a = c = 0, b\neq 0$ or $b = c = 0, a\neq 0$
\item[-]$a=b=0$, $c \neq 0$.
\end{itemize}
For these cases, it is unclear whether or not the space is singular at $\tau =0$ (i.e., whether or not the curvature blows up); this is purely due to computational difficulties, as one should be able to calculate the Riemann tensor to see whether or not this is the case.
\end{itemize}
Due to these considerations, we will limit ourselves to cases where the metric comes with a (possibly squashed) $S^4$ bolt to ensure that we do not encounter potential incompleteness issues. As both types of $S^4$ bolts are similar up to transformations $a \leftrightarrow b$, we will only consider the case where
\al{\label{eq:geo}
a(0), c(0) \neq 0 \;,\quad b(0) = 0\;.
}
One can consider this a boundary condition on the differential equations determining our vacua.

\section{IIA on $SU(4)$-structure deformed Stenzel Space}\label{generalizations}
We have discussed supersymmetric vacua of type IIA supergravity and M-theory on Stenzel space, and $SU(4)$-structure deformations of Stenzel space. In this section, we will discuss $\ncal= (1,1)$ vacua of type IIA supergravity on  on $SU(4)$-structure deformed Stenzel space. Our parameters consist of $a,b,c$, which determine the $SU(4)$-structure, and the RR fluxes. We would like to choose these in such a way that the following hold:
\begin{itemize}
\item[1)] $a,b,c$ are such that that $\scal$ equipped with such an $abc$ $SU(4)$-structure is geodesically complete and free of singularities.
\item[2)] The torsion classes and fluxes satisfy one of the branches of the IIA susy solutions, as given in \ref{typeII}.
\item[3)] The Bianchi identities are satisfied.
\item[4)] The integrability conditions, which turn a solution to the susy equations into a solution of the equations of motion, are satisfied.
\item[5)] The warp factor is regular and positive.
\item[6)] All the fluxes are $L^2$.
\end{itemize}
Unfortunately, due to the particularities of left-invariant forms on $\scal$, one cannot go beyond the Stenzel space scenario satisfying the first three constraints. A way out is by discarding the constraint that the Bianchi identities are satisfied. Such violations can come about in the presence of sources, which modify the action. The precise source terms needed can be deduced from the integrability equations. More precisely, we will see that the NSNS Bianchi identity will always be violated, thus indicating the presence of NS5-branes. On the other hand, the RR Bianchi identities need not be violated.
\\
\\
In the rest of this section, we will demonstrate the claims in the above paragraph. We then discuss vacua\footnote{There are, however, some subtleties to the integration theorem which we ignore. This will be explained in section \ref{sources}.} on $\scal$ which are complex but not symplectic (and hence, not CY), with primitive $(2,2)$-flux, with $\ncal=(1,1)$ supersymmetry, with external metrics that are asymptotically conformal $\text{AdS}_3$. We also give the appropriate source term. These vacua will not uplift to M-theory vacua on $\mathbbm{R}^{1,2} \times \scal$, as can readily be deduced from the torsion classes. In particular, these vacua will have the following properties:
\begin{itemize}
\item[1)] The metric will have an $S^4$ bolt at the origin and conical asymptotics, similar to Stenzel space, thus ensuring geodesical completeness.
\item[2)] The torsion class constraint imposed by supersymmetry can be explicitly solved, fixing $c$.
\item[3)] We consider the Bianchi identities on non-CY manifolds and deduce that taking solely a primitive (2,2)-form does not violate the RR Bianchi identities, whereas a number of other possibilities do. In all cases, $\d H \neq 0$.
\item[4)] We will explicitly check the integrability conditions in the presence of source terms. These yield the constraint which determines the source term needed.
\item[5)] The warp factor will be regular and positive, the flux $L^2$.
\end{itemize}

\subsection{A no-go for sourceless IIA $\ncal = (1,1)$ on $\scal$ with non-CY $abc$ $SU(4)$-structure}
We start by examining the constraints on the torsion classes imposed by $\mathcal{N}= (1,1)$ supersymmetry. As given by \eqref{eq:iiasol1}, \eqref{eq:iiasol2}, \eqref{eq:iiasol3},  $W_1 = W_3 = 0$ implies $W_2 = W_4 = W_5=0$ for $e^{2 i \t} \neq -1$. On the other hand, $e^{2 i \t} = -1$ allows for non-CY vacua. The remaining torsion classes are then related to $H$ as
\eq{\spl{ \label{eq:susycons}
\tilde{h}^{(1,0)}_3 =& \frac 14 \p^+ ( A-\phi)\\
W_2 =& - 2 i h^{(2,1)} \\
W_4 =& \p^+ ( \phi - A) \\
W_5 =& \frac32 \p^+ ( \phi - A) \;.
}}
Specifically, this implies that there are obstructions to the existence of a complex or symplectic structure if and only if the NS three-form has an internal component. That is to say, we must have that
\eq{\spl{ \label{eq:h3}
H_{3} =& - \frac14 W_4^* \lrcorner \O + \frac12 i W_2 + \text{c.c.} \\
\equiv& f(\tau) \z^1 \wedge \z^2 \wedge \z^3 + g(\tau) \varepsilon_{\bar{\imath}jk} \bar{\z}^{\bar{\imath}} \wedge \z^j \wedge \z^k+ \text{c.c.} \;,
}}
for some $f, g$ determined by $a,b,c$. Recall that $H$ is defined by \eqref{eq:fluxdecomp} and that \eqref{eq:hint} ensures that $\d H = 0$ if and only $\d H_3 = 0$.
It is easy to show that no functions $f,g$ exist that are non-trivial and are such that the Bianchi identity
\all{
\d H = 0
}
is satisfied. In fact, $H_3$ is a primitive left-invariant three-form, and there are no closed primitive left-invariant three forms as can be checked explicitly by using the basis for left-invariant forms on $V_{5,2}$ given in \ref{li-forms}. Thus, these classes of $SU(4)$-structures admit no non-CY vacua satisfying both susy and the Bianchi identities.
\\
\\
Now let us consider possible Calabi-Yau structures. We impose \eqref{eq:geo} to get a bolt at the origin, which means that we can pick only one possible complex structure, namely
\al{
r \equiv \frac{b}{a} = \tanh\left(\frac{\tau}{2}\right)\;.
}
As discussed in section \ref{moduli}, this complex structure allows only for Stenzel space as regular CY-structure.
\\
\\
Let us make some more remarks. First of all, for generic LI $SU(4)$-structures one still has $W_1 = W_3 = 0$.
Thus, the entire argument above goes through, and it can be concluded that there are also no non-CY vacua with LI $SU(4)$-structure satisfying susy and the Bianchi identities simultaneously. This is the primary reason we consider the more generic LI $SU(4)$-structures to be of little more interest than $abc$ $SU(4)$-structures.
Secondly, let us note that for any supersymmetric solution satisfying \eqref{eq:susycons} with non-vanishing torsion classes, one cannot set $e^\phi = g_s e^A$. As this is necessary to lift $d=2$ IIA vacua to M-theory vacua on $\mathbbm{R}^{1,2} \times \scal$, we immediately find that such IIA solutions do not uplift to M-theory vacua with external spacetime $\mathbbm{R}^{1,2}$. This is also evident from the fact that our $\ncal= 1$ M-theory vacua on $\mathbbm{R}^{1,2} \times \mcal_8$ given in section \ref{mtheory} require \eqref{eq:msusysol2}, which immediately rules out non-CY vacua in the current case where $W_1 = W_3 = 0$.

\subsection{Torsion class constraint}
To summarize the previous section, there are no interesting non-Stenzel CY solutions, and non-CY solutions require violating the NSNS Bianchi identity. Accepting this violation for the moment, we consider the susy branch  $e^{2 i \t} = -1$, which, as noted, is the only branch that allows for non-CY with $W_1 = W_3 =0$. In fact, the vacua that we find are those which are complex, non-symplectic. These are exactly those described in section \ref{cplxvac}. Let us examine how the supersymmetry torsion class constraint translates to the space at hand.
\\
\\
Susy implies \eqref{eq:susycons}, which we take as defining equations for $H$, $\phi$. This constrains $a,b,c$ to satisfy
\al{
W_5 = \frac32 W_4 \;.
}
Plugging in \eqref{eq:torsion}
leads to the constraint
\al{\label{eq:iiasusycons}
- 3 ( a^2 + b^2 - c^2)c + 4 a b c' =0 \;.
}
As an aside, notice that this constraint is actually equivalent to the third equation of \eqref{eq:cycondition}.
One can solve this equation by parametrizing
\eq{\spl{\label{eq:para}
a \equiv& \a^{-1}(\tau) c\\
b \equiv& \b (\tau) a \;,
}}
leading to the solution
\al{\label{eq:susysol}
c (\tau) =& \l \exp\left(\frac34 \int^\tau dt \frac{1+ \b^2(t) - \a^2 (t)}{\b(t)} \right)   \;.
}
Generically, the space satisfies susy and is complex iff
\al{
\b = r \;.
}
It satisfies susy and is symplectic (and is in fact `nearly CY', i.e., only $W_2 \neq 0$) iff
\all{
3 \b \a' - \b'\a + 2 \a^5 - \frac32 \b^2 \a^3 - \frac 32 \a^3 =0 \;.
}

\subsection{Vacua on complex manifolds with $f^{(2,2)}$}\label{cplx22}
The vacua are considered are those described in section \ref{cplxvac}. To summarise, we consider vacua where the flux is given by a four-form RR flux, combined with the NSNS flux. The four-form is primitive, (2,2), and selfdual. The RR Bianchi identities are satisfied if and only if the form is closed (which implies both $W_2=0$ and co-closure, two necessary conditions). There are two such possible four-forms, as given by \eqref{eq:(2,2)NL1} and \eqref{eq:(2,2)L1}. The closure  constraint is given by \eqref{eq:(2,2)NL2}, \eqref{eq:(2,2)L2} respectively for each form. Considering that $W_4 \sim W_5$, $W_1 = W_3 = 0$, we find that we can only find non-CY solutions by ensuring $W_4 \neq 0$. Thus, our vacua will be complex but not symplectic.
\\
\\
We set $\b = r$ to satisfy $W_2 = 0$ and $r = \tanh\left(\frac{\tau}{2}\right)$ to satisfy \eqref{eq:geo}. Using this, the closure condition can be solved. Let us first examine the form which was non-$L^2$ on Stenzel space, whose closure condition is given by  \eqref{eq:(2,2)NL2}. Using our parametrization \eqref{eq:para} and the complex structure, we can solve this equation to find
\al{
f_{NL} =& m \exp(- \int \frac{a}{2b} ) \exp(- \int \frac{2b'}{b}) \exp(- \int \frac{(bc)'}{bc} )  \nn\\
=& m \frac{\a^3} {c^4 \tanh^3\left(\frac{\tau}{2} \right)\sinh\left(\frac{\tau}{2} \right) } \;,
}
with $c$ determined by \eqref{eq:susysol}. At this point $\a$ is still a free function, but we insist that $a \neq 0, \infty$ anywhere. Hence we see that regardless of choice of $\a$, supersymmetry, the Bianchi identities and the bolt at the origin do not allow for regular solutions of this flux, and the IR problems that arose for this form in Stenzel space are still there.

Let us now examine the other four-form, which behaved well on Stenzel space. The closure condition is given by \eqref{eq:(2,2)L2}, which is the same as \eqref{eq:(2,2)NL2} but with $a \leftrightarrow b$. Thus, the solution is given by
\al{
f_L =& m \exp(- \int^\tau dt \frac{b}{2a} ) \exp(- \int^\tau dt \frac{2a'}{a}) \exp(- \int^\tau dt \frac{(ac)'}{ac} )  \nn\\
=& m \frac{\a^3} {c^4 \cosh\left(\frac{\tau}{2} \right)} \;.
}
Using this expression, the warp factor can be calculated, which satisfies
\al{
\nabla^2 \hcal = - 12 g_s^2 f_L^2\;.
}
The solution is found to be
\al{
\hcal =  \frac32 g_s^2 m^2 \int_\tau^\infty dt \;  \left(\frac{\a(t)}{c(t)}\right)^6 \tanh\left(\frac{t}{2}\right) \;,
}
where an integration constant has been fixed to ensure regularity of $\hcal$ at $\tau = 0$.
Our sole free function $\a$ now needs to be chosen such that the following are satisfied:
\begin{enumerate}
\item $\a$ is nowhere vanishing nor blows up anywhere to avoid singularities. As a consequence, we immediately find that $f_L$ is regular.
\item $\hcal > 0$ holds everywhere in order to avoid the metric changing sign.
\end{enumerate}
This is a rather small list of demands, and is easily satisfied, and we will give examples below. Before doing so, however, we would like to make some remarks. The asymptotics of $\a$ govern the squashing of the metric: the $S^4$ bolt is unsquashed for $\a_{IR}=1$ and the asymptotics are precisely conical for $\a_{UV} = \sqrt{3/2}$. One might think that it is possible to use this method to also construct asymptotically $\text{AdS}_3$ rather than asymptotically conformal $\text{AdS}_3$ metrics. This is not the case. The reason is that this requires $c_{UV} \sim 1$ rather than  $c_{UV} \sim \exp(k \tau)$. Examining \eqref{eq:susysol} then leads to the conclusion that $\a_{UV} \sim \sqrt{2}$, and hence the warp factor blows up (or becomes negative, depending on the choice of integration boundaries).

Finally, before moving on to the examples, let us also comment on the possibility of a homogeneous solution. The homogeneous solutions obeys
\al{
\hcal \sim \int^\tau \frac{dt}{a^3 b^3 } \;.
}
Imposing \eqref{eq:geo} without any further restrictions due to susy or fluxes, we immediately find that $\hcal_{IR} =\int \frac{dt}{t^3}$ and hence the non-constant homogeneous warp factor retains its IR divergence regardless of choice of $abc$ $SU(4)$-structure.

\textbf{Example 1:}\\
Set
\al{
\a =  \sqrt{1 + \frac12  \tanh^2\left(\frac{\tau}{2} \right)}
}
such that
\eq{\spl{\label{eq:ex1}
a =& \l \frac{\cosh^{3/4}\left(\frac{\tau}{2} \right) }{\sqrt{1 + \frac12  \tanh^2\left(\frac{\tau}{2} \right)}}\\
b =& \l \tanh\left(\frac{\tau}{2} \right) \frac{\cosh^{3/4}\left(\frac{\tau}{2} \right) }{\sqrt{1 + \frac12  \tanh^2\left(\frac{\tau}{2} \right)}}\\
c =& \l \cosh^{3/4}\left(\frac{\tau}{2} \right)
}}
and
\al{
f_L = \frac{m \left(1 + \frac12  \tanh^2\left(\frac{\tau}{2} \right)\right)^{3/2}}{\l^4 \cosh^4\left(\frac{\tau}{2} \right)}\;.
}
Thus, the warp factor is given by
\al{
\hcal &=   \frac{g_s^2 m^2  \left(55802  + 93933 \cosh \tau + 44982 \cosh 2\tau + 13923 \cosh 3\tau\right)}{198016 \l^6\cosh^{21/2} \left(\frac{\tau}{2}\right) } \;.
}
The asymptotics are given by
\al{ \def\arraystretch{1.8}
\hcal \rightarrow \left\{ \begin{array}{cc}
\frac{1630}{1547} \frac{g_s^2 m^2}{  \l^6} - \frac{3 g_s^2 m^2}{8 \l^6}\tau^2 & \phantom{aaaaa} \tau \rightarrow 0 \\
36 \sqrt{2}\frac{ g_s^2 m^2}{\l^6}e^{-\frac94 \tau}  & \phantom{aaaaaaa} \tau \rightarrow \infty \;.
\end{array}\right.
}
We thus find an asymptotically conformal AdS metric, as the scaling is analogous to all UV-finite cases discussed in section \ref{fluxstenzel}. Specifically, we find that the (string frame) metric asymptotes to
\eq{\spl{
ds^2_{UV} &= \L^2 e^{2 \rho} [ds^2(\text{AdS}_3) + ds^2(V_{5,2})] \\
ds^2(\text{AdS}_3) &=   \d\rho^2 + e^{4 \rho} ds^2(\mathbbm{R}^{1,1}) \;,
}}
where we introduced a constant $\L$, rescaled $ ds^2(\mathbbm{R}^{1,1})$,  and $\rho \equiv \frac38 \tau$.
\\
\\
\textbf{Example 2:}\\
By changing the asymptotics of $\a$, we find solutions which asymptote to conformal AdS with a squashed $V_{5,2}$ as internal space. We now give such an example.
Set
\al{
\a = \sqrt{1 + \frac13  \tanh^2\left(\frac{\tau}{2} \right)}
}
such that
\eq{\spl{ \label{eq:ex2}
a &= \l \frac{\cosh\left(\frac{\tau}{2}\right)}{\sqrt{1+ \frac13 \tanh(\frac{\tau}{2})^2 }}\\
b &= \l   \tanh\left(\frac{\tau}{2}\right) \frac{\cosh\left(\frac{\tau}{2}\right)}{\sqrt{1+ \frac13 \tanh(\frac{\tau}{2})^2 }}\\
c &= \l \cosh\left(\frac{\tau}{2}\right) \;.
}}
This leads to a flux defined by
\al{
f_L &=  \frac{m \left(1 + \frac13  \tanh^2\left(\frac{\tau}{2}\right)\right)^{3/2}}{\l^4 \cosh^5\left(\frac{\tau}{2} \right)}\;.
}
The warp factor is given by
\al{
\hcal =  \frac{g_s^2 m^2 \left(96 + 156 \cosh \tau + 75 \cosh 2 \tau + 20 \cosh 3\tau \right) }{540 \l^6 \cosh^{12}\left(\frac{\tau}{2}\right)}
}
with asymptotics
\al{\def\arraystretch{1.8}
\hcal \rightarrow \left\{ \begin{array}{cc}
 \frac{347 g_s^2 m^2}{540 \l^6} - \frac{3 g_s^2 m^2}{8 \l^6} \tau^2 &\phantom{aaaaa} \tau \rightarrow 0\\
 \frac{2048}{27} \frac{m^2 g_s^2}{\l^6}e^{- 3  \tau} & \phantom{aaaaaa} \tau \rightarrow \infty
 \end{array}\right.
}
The metric then asymptotes to the following:
\eq{
ds^2_{UV} = \L^2 e^{\frac32\sqrt{2} \rho} [ds^2(\text{AdS}_3) + \widetilde{ds^2(V_{5,2})}]
}
where we introduced a (different) constant $\L$, the radial direction is now related to $\tau$ as $\rho = \frac{1}{\sqrt{8}} \tau$, we again rescaled $ds^2(\mathbbm{R}^{1,1})$, but the internal and external metrics are now given by
\eq{\spl{
ds^2(\text{AdS}_3) &= \d \rho^2 + e^{3 \sqrt{2} \rho} ds^2(\mathbbm{R}^{1,1})\\
\widetilde{ds^2(V_{5,2})} &= \frac12 \n^2 + \frac{3}{8} \left( \s_j^2 + \tilde{\s}_j^2\right) \;.
}}
Note that the squashed $V_{5,2}$ metric cannot simply be rescaled to the regular $V_{5,2}$ metric, due to the finite range of the (angular) coordinates. Thus, this solution is inequivalent to the first example.
\\
\\
To summarize, these solutions satisfy the following properties:
\begin{itemize}
\item The vacua are $\ncal = (1,1)$ type IIA solutions.  Due to the torsion classes, no supersymmetry enhancement occurs, and the vacua do not uplift to  $\ncal = 1$ on $\mathbbm{R}^{1,2} \times \scal$ M-theory vacua, as is evident from the fact that \eqref{eq:msusysol2} is not satisfied.
\item The RR flux is given by a closed primitive (2,2)-form.
\item $W_1 = W_2=0$ hence the almost complex structure defined by the $SU(4)$-structure is integrable.
\item $W_4 \neq 0$ hence the almost symplectic structure is not integrable. In particular, the solution is not CY.
\item $\scal$ has an $S^4$ bolt at the origin.
\item The external metric is conformally AdS for $\tau \rightarrow \infty$.
\item The RR Bianchi identity is satisfied. The NSNS Bianchi identity is not, indicating the presence of NS5-branes which act as sources.
\end{itemize}
Thus, we have shown how to break the CY structure of Stenzel space to find a vacuum that consists of a complex space and a $(2,2)$-form, at the price of introducing sources that violate the Bianchi identity for the NSNS flux.

\subsection{Source-action for vacua on complex manifolds with $f^{(2,2)}$ }\label{sources}
As mentioned, the supersymmetric vacua of the previous section violate the Bianchi identity for $H$. The source of this violation can be interpreted as a distribution of NS5-branes, as $\d H$ is not localised on $\scal$. In \cite{koerbt}, it was shown that for any violation of the RR Bianchi, an appropriate source-action can be found such that supersymmetry combined with the specific violation of RR Bianchi identity gives the source-modified equations of motions. When the NSNS Bianchi is violated instead, the situation is more complicated. This is due to the fact that the democratic supergravity action is not used to compute equations of motions for the RR charges; they are already incorporated in the Bianchi identities. Thus, the contribution to the equations of motion from the bulk supergravity action does not change, and one can look purely for a source-action to compensate the violated Bianchi identiy. This is no longer the case when $\d H \neq 0$, as will be explained below. We construct a source-action that ensures integrability is satisfied, but disregard the fact that the contribution from the bulk action should be modified.
\\
\\
Let the action be given by $S = S_{\text{bulk}} + S_{NS5}$.
The equations of motion in our conventions are then given by\footnote{See \cite{mmlt}, appendix A.}
\eq{\spl{\label{eq:eom2}
E_{MN} &= -2 \k_{10} \frac{e^{2\phi}} {\sqrt{g_{10}}}\left(           \frac{\delta S_{NS5} }{\delta g^{MN}} -
                                                            - \frac14 \frac{\delta S_{NS5} }{\delta \phi}g_{MN} \right) \\
\delta H_{MN} &= 2 \k_{10} \frac{\delta S_{NS5}}{\delta B^{MN}}\\
D &=  2\k_{10} \frac{e^{2\phi}} {\sqrt{g_{10}}} \frac{\delta S_{NS5} }{\delta \phi}\;,
}}
with (see \eqref{eq:eom}, repeated here for convenience)
\eq{\spl{
E_{MN} &\equiv R_{MN} + 2 \nabla_M \nabla_N \phi - \frac{1}{2} H_{M}\cdot H_N - \frac{1}{4}e^{2 \phi} \fcal_M \cdot \fcal_N \\
\delta H_{MN} &\equiv \star_{10} e^{2 \phi} \left(\d  \left(e^{-2 \phi} \star_{10} H\right) - \frac{1}{2} \left(\star_{10}\fcal \wedge \fcal \right)_8 \right) \\
D &\equiv 2 R - \frac{1}{3!}H_{MNP} H^{MNP} + 8 \left(\nabla^2 \phi - (\p \phi)^2 \right) \;,
}}
The two subtleties that are not taken into account here are as follows. Firstly, the equation of motion for $B$ has been obtained by setting $H = \d B$ in $S_{\text{bulk}}$. When $\d H \equiv j \neq 0$, this cannot be the case. Secondly, without sources, the RR equations of motion $\d_H \star \sigma \fcal = 0$ follow from the Bianchi identity $\d_H \fcal = 0$ and selfduality (see \eqref{eq:selfduality}). In case $\d H \neq 0$, $\d_H^2 \neq 0$ and hence one cannot use $\d_H \fcal = 0  \implies \fcal = \d_H C$.
\\
\\
Manipulating the supersymmetry equations, one can obtain the following integrability equations \cite{mmlt}:
\al{\label{eq:integ}
0 &= \left(- E_{MN} \G^N + \frac{1}{2} \left(\delta H_{MN} \G^N + \frac{1}{3!} (\d H)_{MNPQ} \G^{NPQ} \right)\right) \epsilon_1 - \frac{1}{4} e^{\phi} \underline{\d_H \fcal} \G_M \G_{11} \epsilon_2 \nonumber \\
0 &= \left(- E_{MN} \G^N - \frac{1}{2} \left(\delta H_{MN} \G^N + \frac{1}{3!} (\d H)_{MNPQ} \G^{NPQ} \right)\right) \epsilon_2 - \frac{1}{4} e^{\phi} \underline{\sigma \d_H \fcal} \G_M \G_{11} \epsilon_1  \nonumber \\
0 &= \left(-\frac{1}{2} D + \underline{\d H} \right) \epsilon_1 + \frac12 \underline{\d_H \fcal} \epsilon_2 \nonumber \\
0 &= \left(-\frac{1}{2} D - \underline{\d H} \right) \epsilon_2 + \frac12 \underline{\sigma \d_H \fcal} \epsilon_1
}
In our case, we have $\d_H \fcal= 0$. The spinor ansatz for $\epsilon_{1,2}$  for our solution is given in \eqref{eq:IIAks}.
The violation of the NS Bianchi identity is purely internal, i.e., the only non-vanishing components of $\d H_{MNPQ}$ are given by $\d H_{mnpq}$.
A priori, one can decompose $\d H_{mnpq}$ like any other four-form (see \eqref{eq:su4decomp}) as
\all{
\d H_{mnpq} =& 6 j_0 J_{[mn} J_{pq]} + 6 \left( j^{(1,1)}_{[mn} + j^{(2,0)}_{[mn}  + j^{(0,2)}_{[mn} \right)J_{pq]} \\
& + \tilde{\jmath}\; \O_{mnpq} + \tilde{\jmath}^* \O^*_{mnpq} + j^{(3,1)}_{mnpq} + j^{(1,3)}_{mnpq} \;.
}
Finally, we know that the metric satisfies $g_{m \n} = 0$, and we use the gamma matrix decomposition of \cite{pta}.
Plugging all of the above into the integrability equations, we find
\eq{\spl{\label{eq:integresults}
D &= 32 \tilde{\jmath} = 32  \tilde{\jmath}^* \\
\delta H_{MN} &= 0\\
E_{\m \n} &= 0\\
E_{mn} &= 2 (\tilde{\jmath} + \tilde{\jmath}^* ) - \frac{1}{4!}\left( j^{(3,1)}_{qrsm} \O_n^{*qrs} + j^{(1,3)}_{qrsm} \O_n^{qrs} \right)\\
j_0 &= j^{(1,1)} = j^{(2,0)} = 0\;.
}}
Our task is to figure out a suitable submanifold $M_6$ for the NS5-brane to wrap such that the contribution of the action of the NS5-brane evaluated on $M_6$ as described in \eqref{eq:eom2} is equivalent to what the integrability equations tell us in \eqref{eq:integresults}, i.e.,
\eq{\spl{
2\k^2 \frac{e^{2\phi}} {\sqrt{g_{10}}} \frac{\delta S_{NS5} }{\delta \phi} &= 16 (\tilde{\jmath} + \tilde{\jmath}^* ) \\
\k^2 \frac{e^{2\phi}} {\sqrt{g_{10}}}\left(\frac12 \frac{\delta S_{NS5} }{\delta \phi}g_{mn} - 2 \frac{\delta S_{NS5} }{\delta g^{mn}}\right) &=
2 (\tilde{\jmath} + \tilde{\jmath}^* ) - \frac{1}{4!}\left( j^{(3,1)}_{qrsm} \O_n^{*qrs} + j^{(1,3)}_{qrsm} \O_n^{qrs} \right)
\\
\k^2 \frac{e^{2\phi}} {\sqrt{g_{10}}}\left(\frac12 \frac{\delta S_{NS5} }{\delta \phi}g_{\m\n} - 2 \frac{\delta S_{NS5} }{\delta g^{\m\n}}\right) &= 0 \\
2 \k^2 \frac{\delta S_{NS5}}{\delta B^{MN}} &= 0 \;.
}}
The action of the NS5-brane is known \cite{bns}, but is rather intimidating. Instead, we will simply try to construct a suitable action from scratch satisfying the above. We find that such a suitable action is given by\footnote{Taking the derivative of $\O^{mnpq}$ with respect to $g^{mn}$ may seem somewhat ambiguous, since indices could be raised with either $-i J^{mn}$ or $g^{mn}$. One should either consider all indices raised with $(\Pi^+)^{mn}$, which treats $J$ and $g$ on equal footing, or as raised with vielbeins acting on the spinor bilinear $\tilde{\eta}\g^{abcd} \eta$ with flat indices. Both give the same correct factor.}
\al{
S_{NS5} = \frac{-1 }{4!4 \k^2} \int d^{10} x \sqrt{-g_{10}}\; e^{-2 \phi} dH_{mnpq} \left( \O^{mnpq} + \O^{*mnpq} \right) \;.
}
This action has no clear DBI or WZ terms. Instead, it can be written as
\all{
S \sim \int_{\mcal_{10}} j \wedge  \Psi  \;,
}
with source $j = \d H$ and we have defined a form $\Psi$ satisfying
 \eq{\spl{
 \Psi &= \widetilde{\text{vol}_2} \wedge  \left[\text{Re}(\Omega) + ...\right]\;.
 }}
The dots represent terms that drop out of the action and the volume form $\widetilde{\text{vol}_2}$ is warped. This seems analogous to the description as given in section \ref{caliDbranes}, where calibrated D-branes were discussed. Hence we would conjecture that this action describes a calibrated distribution of NS5-branes. As far as the author is aware, no analysis is known for calibrated NS5-branes. Such an analysis for NS5-branes is more complicated than for D-branes due to a more complicated action and due to the breaking of the generalised complex geometrical framework, for which $\d H=0$ is essential.

\begin{figure}[h]
\centering
\includegraphics[scale=0.5]{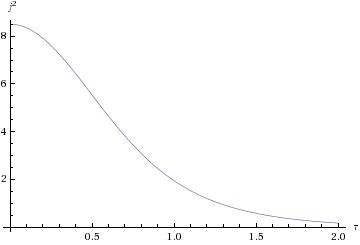}
\caption{The norm $j^2 \equiv \star (\d H \wedge \star \d H)$ as function of $\tau$ with $\l= 1$ for the first example.}
\label{fig1}
\end{figure}

\begin{figure}[h]
\centering
\includegraphics[scale=0.5]{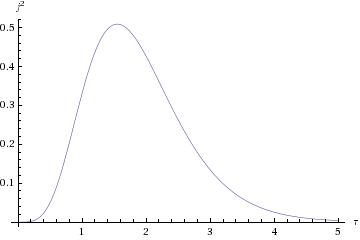}
\caption{The norm $j^2$ as function of $\tau$ with $\l= 1$ for the second example.}
\label{fig2}
\end{figure}

Let us examine the source term in more detail. Generically, it is given by the first line of \eqref{eq:h3}: the solutions we consider in section \ref{cplx22} have $W_2=0$ and thus
\al{
\d H  = \tilde{\jmath} \;\z^{0} \wedge \z^1 \wedge \z^2 \wedge \z^3   +  j_{31} \bar{\zeta}^{\bar{0}} \wedge \z^1 \wedge \z^2 \wedge \z^3 + \text{c.c.} \;,
}
with
\eq{\spl{
\tilde{\jmath} &= - \frac{1}{4 abc} \left(  3 (a^2 + b^2)w + 6 (ab)' w + 4 ab w' \right)\\
j_{31} &= - \frac{1}{4 abc} \left( - 3 (a^2 + b^2)w + 6 (ab)' w + 4 ab w' \right)\\
w &\equiv \frac12 \frac{1}{abc}\left((ab)'- \frac12 c^2 \right) \;.
}}
Thus indeed, $\d H$ is primitive with only $(4,0)$, $(3,1)$, $(1,3)$, $(0,4)$ parts and $\tilde{\jmath}$ is real, as required by \eqref{eq:integresults}.
The norm of the source is given by
\al{
j \wedge \star_8 j = 32 \left(\tilde{\jmath}^2 + j_{31}^2\right) \text{vol}_8 \;.
}
As $W_4 = 0$ if and only if $w =0$, it can be concluded that generically, $j$ will vanish for large $\tau$ due to the conical (and thus, CY) asymptotics. To be more specific, we examine the examples given in the previous section.
\\
\\
\textbf{Example 1:}\\
Imposing \eqref{eq:ex1} leads to
\al{
\star_8 \left(j \wedge \star_8 j\right) =  \frac{8863 + 6816 \cosh \tau + 1308 \cosh 2 \tau + 288 \cosh 3 \tau + 333 \cosh 4 \tau }{8 \l^4 \cosh^7 \left( \frac{\tau}{2}\right) \left(1 + 3 \cosh \tau\right)^4}
\;.
}
Thus the distribution is smooth, does not blow up, is maximal at $\tau = 0$ with norm $17 / 2 \l^4$, and falls off rapidly. The norm of the source is plotted in figure \ref{fig1} for $\l=1$.
 \pagebreak
\\
\\
\textbf{Example 2:}\\
Imposing \eqref{eq:ex2} leads to
\al{
\star\left(j \wedge \star j\right) =
\frac{8 }{9 } \frac{\tanh^4\left(\frac{\tau}{2} \right)\left(553 + 714 \cosh \tau + 278 \cosh 2 \tau + 62 \cosh 3 \tau + 13 \cosh 4\tau\right)}
{\l^4 \cosh^4\left(\frac{\tau}{2} \right)\left(1 + 2 \cosh \tau\right)^4}  \;.
}
Again, the distribution is smooth, does not blow up, and falls off rapidly. It peaks at $\tau \simeq 1.549$ with norm $\simeq 0.5090/ \l^4$. One can thus consider it to be `smeared'. The norm of the source is plotted in figure \ref{fig2} for $\l = 1$.

\chapter{Generalised complex geometry}\label{gcg}
Up to this point, the major mathematical framework that we have used to construct and understand flux vacua is the geometrical language of $G$-structures (with $G = SU(4)$ in particular).
The second major geometrical tool we will make use of is generalised complex geometry. Despite the name, generalised complex geometry simultaneously generalises complex geometry and symplectic geometry. Generalised complex manifolds $M_{2n}$ can be locally trivialised to $\cbb^k \times (\rbb^{2(n-k)}, J)$, with $k$ coordinate-patch dependent and $J$ a symplectic structure.
The focal points of generalised complex geometry are the vector bundle $T\oplus T^*$, the {\it generalised tangent bundle}, and certain endomorphisms of this bundle, {\it generalised (almost) complex structures}, which generalise complex and symplectic structures. Generalised complex geometry came about first in \cite{hitchin}, as a generalisation of Calabi-Yau manifolds, and was expanded upon in \cite{gualtieri} \cite{cavalcanti}. It was quickly recognised as a natural framework for T-duality. Soon it also found other uses in physics due to the natural appearance in the theory of a two-form $B$, which can be interpreted as the NSNS gauge field of the field strength $H$, and of spinors, which can be interpreted as polyforms or as bispinors as is convenient.
Some very few examples of where generalised complex geometry shows up in string theory are in the description of supersymmetry and calibrations of branes \cite{koerbcali} \cite{ms2} \cite{km} \cite{melec}, reformulation of T-duality to construct geometric \cite{pet1} and non-geometric \cite{pet2} vacua, the reformulation of type II and M-theory in terms of generalised objects \cite{wald1} \cite{wald2}, the study of mirror symmetry \cite{wald3},  and, of most concern to us, reformulation of the supersymmetry equations \cite{gmpt} \cite{lpt} \cite{tomasiello} \cite{rosa}. This list is hopelessly incomplete however.

There is wonderfully well-written literature on generalised complex geometry, both from mathematicians \cite{gualtieri} \cite{cavalcanti} as well as physicists \cite{koerber-gcg} \cite{zabzine}. Nevertheless, let us summarise some of the essential parts of this construction, with either self-containment or fixing of conventions as justification. Mostly, we will follow along \cite{cavalcanti}.

\section{Generalised almost complex structures}
Let $M$ be a manifold of dimension $2n$.
\dfn{
The \textit{generalised tangent bundle} is the vector bundle $TM \oplus T^*M$, which we will generally abbreviate to just $T \oplus T^*$.
}
The generalised tangent bundle comes equipped with a natural metric,
\al{
\gcal(X+ \a, Y + \b) \equiv \frac{1}{2} (\a(Y) + \b(X)) \;,
}
with $X,Y \in T$, $\a, \b \in T^*$. This metric is of signature $(n,n)$. We will denote this in matrix notation as
\al{
\gcal = \frac12 \left(
\begin{array}{cc}
0 & \obb \\
\obb & 0
\end{array}
\right)\;,
}
acting on the (composite) vectors of $T\oplus T^*$ ordered as $(X, \a)$.
\\
\\
The central object in generalised complex geometry is the generalised almost complex structure.
\dfn{
A \textit{generalised almost complex structure} $\ical$ is an almost complex structure on $T \oplus T^*$ that is orthogonal with respect to the natural metric. That is to say, $\ical \in \text{End}( T \oplus T^*)$ is fibre-preserving\footnote{Technically, this meas that $\ical \in \G\left(M, \text{End}( T \oplus T^*)\right)$, but this makes notation incredibly inconvenient, as it is usually the image of this section one refers to.}  and satisfies
\all{
\bullet ~~ &\ical^2 = - \obb   \\
\bullet ~~ &\gcal(X+ \a, Y + \b) = \gcal\left(\ical(X+ \a), \ical(Y + \b)\right) \; \qquad \forall (X+ \a),\; (Y+\b) \in T\oplus T^*\;.
}
}
There are two equivalent definitions for almost complex structures. The first makes use of the fact that for ordinary complex geometry, an almost complex structure is equivalent to specifying $T^{(1,0)}$ in the decomposition\\
 $T \otimes \cbb = T^{(1,0)} \oplus T^{(0,1)}$.

\dfn{
An {\it almost Dirac structure} is a subbundle $L \subset (T \oplus T^*) \otimes \cbb$ such that $L$ is isotropic\footnote{That is to say,  $\gcal(X+ \a, Y + \b) = 0$ $\forall (X+ \a),\; (Y+\b) \in L$ .}
 with respect to $\gcal$ and such that, for $\overline{L}$ the complex conjugate of $L$, $L \cap \overline{L} = \{ 0\}$.
}
Specifying an almost Dirac structure is equivalent to specifying a generalized almost complex structure. This comes about as follows.
From now on, let
\eq{
V \equiv T_p M \otimes \cbb\;,
}
which will be made use of for every pointwise investigation. Let $\ical$ be a generalised almost complex structure. Analogously to regular almost complex structures, $\ical$ has eigenvalues $\pm i$ and thus splits $V \oplus V^*$ into eigenspaces: $V \oplus V^* = L^+ \oplus L^-$,   $\text{dim}_\cbb L^\pm= 2n$,
\al{
L^\pm \equiv \{(X + \a) \in V \oplus V^* \;|\; \ical(X + \a) = \pm i ( X+ \a) \}\;.
}
Note that due to orthogonality of the generalised almost complex structure
\al{
\gcal(X+ \a, Y + \b) = - \gcal(X+ \a, Y + \b) \;, \qquad \forall (X+ \a), (Y + \b) \in L^+
 }
hence $L^+$ is isotropic; the same argument holds for $L^-$. Non-degeneracy of $\ical$ implies that $L^+ \cap L^- = \{ 0\}$. Smoothness of $\ical$ then ensures that defining the vector bundle $L$ with fibres $L_p = L^+$ will be a smooth subbundle of $\left(T \oplus T^*\right) \otimes \cbb$ satisfying $ \left(T \oplus T^*\right) \otimes \cbb = L \oplus \overline{L}$.

Conversely, given an almost Dirac structure, $\ical$ is uniquely defined by demanding $\ical \cdot L = i L$.
\\
\\
The third equivalent way to describe generalised almost complex structures is in terms of pure spinors, which will be very convenient in the supergravity context as will be seen later on.
$V \oplus V^*$ has a natural action on the exterior algebra $\bigwedge^{\!\bullet} V^*$, defined by
\al{
(X + \a) \cdot \Phi = X \lrcorner\Phi + \a \wedge \Phi \;, \quad \Phi \in     \bigwedge\nolimits^{\! \bullet} V^*\;.
}
A quick calculation shows that under this action,
\al{
\{X+ \a, Y + \b \}  \cdot \Phi = 2 \gcal(X+ \a, Y + \b) \Phi\;,
}
hence this action is a representation of the Clifford algebra $Cl(V \oplus V^*)$. Thus, we see that in this sense,  we can interpret polyforms as spinors of $Cl(2n,2n)$. We will be somewhat lax in making the distinction between $\O^\bullet (M)$ and $\ea$ and refer to either as `polyforms', similar to how we do not really distinguish clearly between sections  or elements of the spin bundle and refer to either as `spinors'.
\\
\\
Let us consider pure spinors\footnote{See section \ref{su(n)} for a definition.} in this context. A pure spinor $\Psi \in \bigwedge\nolimits^{\! \bullet} V^*$ is annihilated by exactly half the gamma matrices, hence we can define the annihilator space
\al{
L_\Psi = \{X+ \a \in V \oplus V^* \; |\; (X+ \a ) \cdot \Psi =0\} \;.
}
By definition, $L_\Psi$ is a subspace of $V\oplus V^*$ of complex dimension $2n$, and since \\
$\forall (X+ \a), (Y + \b) \in L_\Psi$
 \all{
  0 =(X+ \a)\cdot( Y + \b)\cdot \Psi + ( Y + \b)\cdot(X+ \a)\cdot \Psi =  2\gcal(X+ \a, Y + \b) \Psi\;,
}
$L_\Psi$ is also isotropic. Thus, if the annihilator space additionally satisfies $L_\Psi \cap \overline{L_\Psi} = \{0\}$, it follows that at the point $p$, the pure spinor defines a generalised almost complex structure. Note that due to linearity of the action, at a point $p$, the annihilator space defined by $\Psi$ is the same as that of $\l \Psi$ for any $ \l \in \cbb\backslash\{0\}$. Both purity of $\Psi$ and the orthogonality of $L_\Psi$ to $\overline{L_\Psi}$ can be translated into conditions on the polyform.
\propn{
Let $\Psi \in \bigwedge\nolimits^{\! \bullet} V^*$  be a pure spinor. Then
\eq{\label{eq:purespinor}
\Psi = e^{B - i J} \t^1 \wedge... \wedge \t^k \;,
}
for real two-forms $B$, $J$ and complex linearly independent one-forms $\t^j$. $k$ is defined to be {\it type} of the pure spinor.
}
\\
\\
The {\it Mukai pairing} $\langle .,. \rangle$ is a bilinear map of polyforms, defined as
\eq{
\langle .,. \rangle : \ea \times \ea \rightarrow \bw^{2n} T^* \;,\\
\langle \Phi_1 , \Phi_2 \rangle \equiv \left(\Phi_1 \wedge \s (\Phi_2)\right)_{(2n)} \;.
}
Using the Mukai pairing, we have the following:
\propn{
Let $\Psi_{1,2} \in \bigwedge\nolimits^{\! \bullet} V^*$ be pure. Then
\eq{
L_{\Psi_1} \cap L_{\Psi_2} = \{0 \} \Longleftrightarrow  \langle \Psi_1 , \Psi_2 \rangle \neq 0\;.
}
}
\\
Combining these two, it immediately follows that pointwise, an almost Dirac structure is equivalent to a polyform $\Psi = e^{B - i J} \t^1 \wedge... \wedge \t^k$ satisfying
\eq{\label{eq:ortho}
 J^{n-k} \wedge\left(\t^1 \wedge... \wedge \t^k\right) \wedge \left(\bar{\t^1} \wedge... \wedge \bar{\t^k} \right) \neq 0 \;.
}
Extending these results globally goes as follows. The fact that pointwise, a generalised almost complex structure is equivalent to a form $\Psi$ up to scalar norm means that locally, a generalised almost complex structure is equivalent to $\Psi$ up to multiplication with a smooth nowhere vanishing locally defined function. The global extension is thus that
\propn{
The following are equivalent:
\begin{itemize}
\item A generalised almost complex structure on $M$.
\item A subbundle $L \subset \left(T\oplus T^*\right) \otimes \cbb$ with fibres that are maximal isotropic subspaces, i.e., an almost Dirac structure.
\item A line bundle $U^n \subset \ea \otimes \cbb$ with local frames on open sets $W_j$  given by pure spinor  $\Psi_j = e^{B - i J} \t^1 \wedge... \wedge \t^{k(j)}$ such that \eqref{eq:ortho} is satisfied.
\end{itemize}
}
There are two prime examples of generalised almost complex structures\footnote{Note that due to orthogonality, the perhaps even more trivial case, where one considers $L = T \otimes \cbb$, is excluded.}. Firstly, suppose $M$ is almost complex, with almost complex structure $I$.
Then for $I^* : T^*M \rightarrow T^*M$ the induced almost complex structure on the cotangent bundle,
\al{
\ical_I = \left( \begin{array}{cc}
- I  & 0 \\
0 & I^*
\end{array} \right)
}
is a generalised almost complex structure.

Secondly, suppose $M$ is equipped with an almost symplectic structure $J$ (i.e., a non-degenerate but not necessarily closed two-form). Consider $J$ as a map $J: T \rightarrow T^*$. Then
\al{
\ical_J = \left( \begin{array}{cc}
0  & - J^{-1} \\
J & 0
\end{array} \right)
}
is also a generalised almost complex structure.

Given a generalized almost complex structure $\ical$ and a  two-form $B$, it is possible to construct a new generalized almost complex structure by a process called $B$-{\it twisting}. Considering $B$ as a map $B: T \rightarrow T^*$, $B$ acts on $T \oplus T^*$ as $ B \cdot (X+ \a) = X + \a - X \lrcorner B$. Let
\al{
\ical_B \equiv \left( \begin{array}{cc}
\obb  & 0 \\
B & \obb
\end{array} \right)
\ical
\left( \begin{array}{cc}
\obb  & 0 \\
- B & \obb
\end{array} \right) \;.
}
Then $\ical_B$ is a generalized almost complex structure.
\\
\\
Existence of a generalised almost complex structure is equivalent to a reduction of the structure group of $T \oplus T^*$ to $GL(2n, \cbb)$, which in combination with the natural metric leads to
a $U(n,n)$-structure. Similarly, a $U(n) \times U(n)$-structure is equivalent to existence of two generalised almost complex structures $\ical_1, \ical_2$, which are commutative and are such that
$\tilde{\gcal} \equiv - \ical_1 \ical_2 $ is positive-definite metric on $T \oplus T^*$; specifying such a metric is equivalent to specifying $(g, B)$, a Riemannian metric on $T$ and a two-form. Two such generalised almost complex structures are called {\it compatible}.  A manifold equipped with a torsion-free $U(n) \times U(n)$ structure on the generalised tangent bundle, i.e., equipped with compatible pair of generalised compelx structures, is called {\it generalised K\"{a}hler}. The nomenclature stems from the fact that any K\"{a}hler structure $(g, I, J)$ induces a generalised K\"{a}hler structure $(\ical_I, \ical_J)$.

Given a pair of compatible almost complex structures, the $U(n) \times U(n)$-structure group can be reduced further in the following case. An $SU(n) \times SU(n)$-structure is given by a pair of trivialisations of the almost Dirac structures $L_{\ical_1}$, $L_{\ical_2}$. In other words, this requires that there exists globally defined nowhere vanishing pure spinors $\Psi_{1,2}$ of the form \eqref{eq:purespinor} satisfying \eqref{eq:ortho}.

The reason we spell this out explicitly is that, given an $G$-structure on $T$, the fact that $T \simeq T^*$ implies that $T \oplus T^*$ is equipped with a $G \times G$-structure. Thus, the manifolds that will be relevant for our supergravity vacua will be equipped with $SU(n) \times SU(n)$-structures on the generalised tangent bundle.

\section{Induced grading on the exterior algebra \& the decomposition of forms}\label{gcgdecomp}
Similar to how Lefschetz and Hodge decomposition follow from the existence of almost complex and almost symplectic structures, so too does a generalised almost complex structure induce a grading on the exterior algebra and hence, a decomposition of forms. Perhaps somewhat surprising, in the cases $\ical = \ical_I$, $\ical = \ical_I$, these are not Hodge, Lefschetz decomposition. In order to give a general decomposition, it is first necessary to understand the generic form of $\ical$ and the action on polyforms.
\\
\\
The action of $T \oplus T^*$ on polyforms and the fact that $\ical \in  \text{End}(T \oplus T^*)$ allows us to derive an action of the generalised almost complex structure on polyforms. Using both properties of the generalised almost complex structures, we have that
\eq{
\left.\begin{array}{ccc}
\ical^T \gcal \ical &=& \gcal\\
\ical^{-1} &=& - \ical
\end{array} \right \} \implies \ical^T \gcal + \gcal \ical = 0
}
and hence, $\ical \in so(2n,2n)\subset Cl(2n,2n)$. Working out the definition of $so(2n,2n)$, it follows that pointwise
\eq{
\ical_p = \left(
\begin{array}{cc}
A & \beta \\
B & - A^*
\end{array} \right)
}
with $A \in \text{End}(V)$, $\b \in \bigwedge^2 V$, $B \in \bigwedge^2 V^*$.
We define the actions of these on $V \oplus V^*$ as\footnote{Note that our conventions differ by a sign compared to \cite{gualtieri}, \cite{cavalcanti}.}
\eq{
A \cdot (X+ \a) &= A X  - A^* \a \\
B \cdot (X+ \a) &= - X \lrcorner B \\
\b \cdot (X+ \a) &= \b(\cdot, \a) \ \;,
}
where the last line should be understood as $- \a_m \b^{mn}$. As a consequence, in terms of a basis $\{e_j\}$ for $Cl(2n,2n)$ one can write
\eq{
A &= \frac12 A^i_j (e_i e^j - e^j e_i) \\
B &= \frac12 B_{ij} e^i e^j \\
\b &= \frac12 \b^{ij} e_i e_j
}
and hence the action on polyforms can be deduced. Let  $\Phi \in \bigwedge\nolimits^\bullet V^*$. It then follows that
\eq{\label{eq:jac1}
A \cdot \Phi &= \frac12 \text{Tr}(A) \Phi + A^* \Phi \\
B \cdot \Phi &= B \wedge \Phi \\
\b \cdot \Phi &= - \b \lrcorner \Phi \;,
}
with
\al{\label{eq:jac2}
A^* \Phi_{(k)} (v^1,..., v^k) \equiv \sum_j \Phi_{(k)} (v^1,...,Av^j,..., v^k)\;.
}
Thus, this shows how to calculate $\ical \cdot \Phi$. As a consequence, a grading on the exterior algebra can now be deduced. Suppose $\ical = \ical_I$. As alluded to earlier, the grading of the exterior algebra induced by $I$ is inequivalent to the one induced by $\ical_I$, i.e., the result is not Hodge decomposition into $(p,q)$-forms. Similarly, the grading with respect to $\ical_J$ is not Lefschetz decomposition.

Let $U^n \subset \ea \otimes \cbb$ be the the pure spinor line bundle defining $\ical$ such that
\eq{
L \cdot U^n = \{0\} \;.
}
Then we define
\eq{
U^k \equiv \bigwedge\nolimits^{n-k} \overline{L} \cdot U^n  \;,\qquad k \in \{-n, ...,n\}\;.
}
These are the eigenbundles of $\ical$ as a map on polyforms with eigenvalue $i k$, and a grading on the exterior algebra is given by
\eq{\label{eq:grad}
\ea \otimes \cbb = \bigoplus_{k = -n}^n U^k \;.
}
In turn, this leads to a decomposition of forms. This will be worked out more thoroughly in the examples given in the next section.

\section{A more thorough look at the canonical examples}\label{gcgexamples}
There are three equivalent ways to describe generalised almost complex structure. In this section, we work out all of these, as well as the grading of the exterior algebra, for a number of examples.

\subsection*{The complex case}
Consider
\al{
\ical_I = \left( \begin{array}{cc}
- I  & 0 \\
0 & I^*
\end{array} \right) \;.
}
Making use of the Hodge decomposition with respect to $I$, It follows that
\eq{
L = T^{(0,1)} \oplus T^{*(1,0)}\;.
}
Hence, an associated pure spinor $\Psi$ is given by a trivialisation of the canonical bundle,
\eq{\label{eq:gcgcplx}
\Psi_I = \O \in \O^{(n,0)}(M)\;,
}
since
\eq{
X^{(0,1)} \lrcorner \O^{(n,0)} = \a^{(1,0)} \wedge \O^{(n,0)} = 0\;.
}
The corresponding decomposition of forms is given by acting with $\overline{L} = T^{(1,0)} \oplus T^{*(0,1)}$ on $\O$, and thus it follows that
\al{
U^k = \bigoplus_{p-q = k} T^{* (p,q)} \;.
}
Alternatively, this can be deduced by acting with $\ical_I$ on $T^{* (p,q)}$. The fact that this does not correspond to Hodge decomposition follows from \eqref{eq:jac2}; for $\a \in T^{* (p,q)}$, $\ical_I \a = i (p-q) \a$ due to \eqref{eq:jac1}, \eqref{eq:jac2}. On the other hand, $I \a = \a(I v^1, ..., I v^{p+q})$ and hence,
$I \a = i^{p-q} \a$: the difference is that between the algebra and the group action.

\subsection*{The symplectic case}
Consider
\al{
\ical_J = \left( \begin{array}{cc}
0  & - J^{-1} \\
J & 0
\end{array} \right) \;.
}
Making use of \eqref{eq:jac1} and noting that in indices, $-J^{-1} = J^{mn}$, it follows that
\eq{
L = \{X + i X\lrcorner J \; | \; X \in T \} \;,
}
and thus, it can be shown that
\al{\label{eq:gcgsmplx}
\Psi_J = e^{- i J}
}
satisfies $L \cdot \Psi_J = \{0\}$. The grading of the exterior algebra is given by
\al{
U^k = \{ \e^{-i J} e^{\frac12 i \L} \Phi_{(n-k)} \; | \; \Phi_{(n-k)} \in \bigwedge\nolimits^{n - k} T^* \} \;,
}
with $\L$ defined in \eqref{eq:sympoperators} and we abuse notation by writing $J$ instead of the operator $L$. The proof here is a slightly more involved calculation, for which we refer to \cite{cavalcanti}.

\subsection*{The $B$-twisted case}
Consider
\al{
\ical_B \equiv \left( \begin{array}{cc}
\obb  & 0 \\
B & \obb
\end{array} \right)
\ical
\left( \begin{array}{cc}
\obb  & 0 \\
- B & \obb
\end{array} \right) =  e^{B} \ical e^{-B}\;,
}
where the second equality comes from making use of \eqref{eq:jac1}; one should think of $e^B$ as being an element of a Lie group, as compared to $B$ which belongs to the associated Lie algebra.
Let $L$ be the almost Dirac structure associated to $\ical$. Then
\eq{
L_B = e^B \cdot L = \{ X + \a - X\lrcorner B \; | \; (X + \a) \in L \} \;.
}
Using the explicit expressions and \eqref{eq:jac2}, it follows that if $L \cdot \Psi = 0$, then $L_B \cdot e^B \Psi = 0$,  hence
\eq{
\Psi_B = e^B \Psi
}
is a pure spinor associated to $\ical_B$. By similar arguments, the exterior algebra decomposes as
\eq{
U^k_B = e^B U_k\;.
}

\section{Generalised complex structures}\label{gcginteg}
There is a notion of integrability for generalised almost complex structures, analogous to almost complex and almost symplectic structures. An integrable generalised almost complex structure is called a {\it generalised complex structure}, and a manifold equipped with a generalised complex structure a {\it generalised complex manifold}.
In terms of the almost Dirac structure $L$, integrability is defined as closure of sections of $L$ under the Courant bracket, mimicking the Newlander-Nirenberg theorem for integrability of almost complex structures, but with the Lie bracket replaced by the Courant bracket\footnote{Alternatively, it is possible to work with the Dorfman bracket instead.}. The details are not particularly relevant for us. What is relevant is the following:
\propn{
Let $\Psi$ be a local frame of the pure spinor line bundle $U^n$ associated to $\ical$. Then the $B$-twisted generalised almost complex structure $\ical_B$ is integrable if and only if $\Psi$ there exists $(X+ \a) \in T \oplus T^*$ such that
\eq{
\d_H \Psi = (X + \a) \cdot \Psi  \;.
}
The irrelevance of which local frame one chooses is clear, as the above is satisfied for $\Psi$ if and only if it is satisfisfied for $f \cdot \Psi$ for any nowhere vanishing local function $f$.
}

Given the grading \eqref{eq:grad} of the exterior algebra induced by a generalised almost complex structure, we can define the operators $\p, \bar{\p}$. Let $\a \in \G(M, U^k)$. Then
\eq{ \label{eq:icaldel}
\p^\ical \a &\equiv  \d \a |_{U^{k-1}} \\
\bar{\p}^\ical \a  &\equiv \d \a |_{U^{k+1}}  \;.
}
In other words, $\p^\ical, \bar{\p}^\ical$ project onto a certain subspace, similar to the Dolbeault operators for an almost complex structure. In fact, we have the following:
\propn{
$\ical$ is integrable if and only if $\d = \p^\ical + \bar{\p}^\ical$.
}
\\
\\
For $\ical = \ical_I$, we have that $\p^{\ical_I}= \p$, the Dolbeault operator.
Given integrability of $\ical$, it is also possible to define the operator
\eq{\label{eq:dj}
d^\ical \equiv [d, \ical] = - i (\p^\ical - \bar{\p}^\ical)\;.
}
This operator is better known as $\d^c$ for complex manifolds. See \cite{cavalcanti} for the symplectic analogues.

\section{Generalised complex submanifolds}
Given a generalised complex manifold, it is natural to consider the notion of generalised complex submanifolds. Preferably, the definition of such a submanifold respects the generalised complex structure. In particular, if the generalised complex structure is induced by a complex structure, the generalised complex submanifold ought to be a complex submanifold, whereas if the generalised complex structure is induced by a symplectic structure, the generalised complex submanifold ought to be a Lagrangian submanifold.

\dfn{
Let $M$ be a generalized complex manifold with $B$-twisted generalized complex structure $\ical_B$.
Let $\scal$ be a submanifold of $M$, and let $\fwv$ be a two-form\footnote{The slightly awkward subscript here will become obvious when discussing D-branes.} on $\scal$ satisfying  $\d \fwv = H|_S$.
Then the pair $(\scal, \fwv)$ forms a {\it generalised complex submanifold} with generalised tangent bundle
\eq{
\{ X + \a \;|\; X \in T\scal, \a \in T^*M|_\scal \;, \a|_{T\scal} = - X \lrcorner \fwv \}\;.
}}
The reason it is is necessary to include the two-form $\fwv$ is as follows. Given a generalised complex structure $\ical$ on $M$, the generalised tangent bundle of a generalised complex submanifold is preserved under $\ical$. However, $\ical$ can be $B$-twisted. Thus, it is necessary that the generalised tangent bundle also twists under $B$-field transforms in such a way that the twisted generalised tangent bundle is invariant under the twisted generalised complex structure. This definition accomplishes that, as $e^B \cdot (\scal, \fwv) = (\scal, \fwv + B)$ and $\d (\fwv + B)= H|_\scal$.

\chapter{IIB supersymmetry,  $SU(4)$-structures \\ \& generalised complex geometry }\label{VI}
A number of examples of supersymmetric flux vacua have been discussed. However, these were all ad hoc constructions. In order to find a more systematic way to study supersymmetric flux vacua, existence of an overarching mathematical structure would be most convenient. Generalised complex geometry may offer such a structure. For $\mcal_{10} = \rbb^{1,3} \times \mcal_6$ with warped product metric,
it was shown in \cite{gmpt} that the supersymmetry equations can be geometrically interpreted in the framework of generalised complex geometry both for $\ncal = 1$ and $\ncal =2$. Following this milestone, by making use of the generalised interpretation, it was shown in \cite{ms2} that in fact, another interpretation is possible: the supersymmetry equations are equivalent to the existence of generalised calibrations forms as constructed in \cite{koerbcali}, differential forms that were fine-tuned to determine energy-minimising static magnetic D-branes\footnote{A definition will be given in section \ref{cali}.}. After proving that a similar situation held true for vacua on $\rbb^{1,5} \times \mcal_4$, it was conjectured in \cite{lpt} that the interplay between calibration forms and generalised complex geometry should suffice to also determine supersymmetric flux vacua on $\rbb^{7,1} \times \mcal_2$ and $\rbb^{1,1} \times \mcal_8$. In particular, the supersymmetry equations for $\rbb^{1,1} \times \mcal_8$ should be equivalent to the calibration equations, which are given by
\begin{subequations}\label{eq:gcgeq41}
\al{
\d_H \left(e^{2A - \phi} \Re \Psi_1 \right) &= e^{2A} \star_8 \s (F) \label{eq:gcgeq41a}  \\
\d_H \left(e^{2A - \phi}  \Psi_2 \right) &= 0 \label{eq:gcgeq41b}\;,
}
\end{subequations}
as will be re-derived later on in this section.
\\
\\
It turns out that this is not the case, as was shown in \cite{ptb}. In this section, we will consider $\ncal = (2,0)$ supersymmetric solutions of type IIB on $\rbb^{1,1} \times \mcal_8$, assuming existence of an $SU(4)$-structure on $\mcal_8$ and using a strict Killing spinor ansatz. These supersymmetric solutions are given by \eqref{eq:iib} in section \ref{susysol}. The calibration conditions \eqref{eq:gcgeq41} will be shown to be insufficient to fully capture the supersymmetry equations. Specifically, the supersymmetry equations impose more constraints than the calibration conditions.
As a consequence, the calibration conditions certainly cannot equate to the supersymmetry equations in the far more general case, namely for both type IIA and type IIB on $\rbb^{1,1} \times \mcal_8$ simultaneously, without strict ansatz for the Killing spinor and without $SU(4)$-structure. However, the additional constraints that the supersymmetry equations pose can be captured by one additional equation phrased in the framework of generalised complex geometry. It is given by
\eq{\label{eq:gcgeq42}
\d_H^{\ical_2} \left( e^{- \phi} \Im \Psi_1 \right) = F \;.
}
Currently, we do not have a fully satisfactory (generalised) geometrical interpretation for this equation. From the point of view of calibrations, we would conjecture that this might be the calibration constraint describing branes which are localised in time, i.e., which are instantonic. This interpretation is motivated by the absence of the warp factor and the similarity to codimension 2 calibration equations in spacetimes of dimension greater than two, as described in \cite{lpt}.

Unfortunately, it is not possible to repeat the process of comparing the supersymmetry equations to equations of the form \eqref{eq:gcgeq41}, \eqref{eq:gcgeq42} for type IIA, as the strict $SU(4)$-ansatz for the Killing spinors is too constrictive.
\\
\\
It was conjectured in \cite{ptb} that the calibration conditions \eqref{eq:gcgeq41} together with the additional constraint \eqref{eq:gcgeq42} should capture the supersymmetry equations fully for both type IIA and type IIB on $\rbb^{1,1} \times \mcal_8$ with $SU(4)$-structure but {\it wihtout} strict ansatz, for a minimal amount of supersymmetry admitting $\k$-symmetric branes (specifically, $\ncal = (2,0)$), and that the techniques of \cite{tomasiello} should be useable to prove this statement. In \cite{rosa}, exactly this was done and the conjecture was proved, thus giving a satisfactory formulation, albeit not a full understanding, of the supersymmetry equations on $\rbb^{1,1} \times \mcal_8$ with $SU(4)$-structure.
\\
\\
In section \ref{cali}, generalised calibration forms will be discussed. Using the construction showcase in appendix \ref{calicons}, the calibration will be derived. In section \ref{su4gcg}, the calibration equations will be computed, making use of the $SU(4)$-structure and the strict $SU(4)$-Killing spinor ansatz. It will also be shown why this procedure fails for type IIA. The additional constraint will be shown to complement the calibration conditions in such a way as to equate to the supersymmetry equations. Some interpretations will also be discussed.

\section{D-Branes and calibration forms}\label{cali}
In this section, D$p$-branes are discussed. D-branes are $p+1$ dimensional and arise in type II supergravity, with $p$ even for IIA, odd for IIB.
Consider a D-brane located at a submanifold $\scal$. The effective action describing a D-brane is given by
\al{\label{eq:dbiwz}
S_{Dp} &=  -\mu_p\int_\scal e^{-\phi} \sqrt{-\det( g|_\scal + \fcal_{\text{wv}})}~\!\d^{p+1}\!\sigma
+\mu_p\int_\scal C\wedge e^{\fcal_{\text{wv}}} \;,
}
where the first term is known as the Dirac-Born-Infeld (DBI) and the second as the Wess-Zumino (WZ) action\footnote{The Wess-Zumino action is also referred to as the Chern-Simons action.}. The two-form $\fwv \in \O^2(\scal)$ is known as the {\it worldvolume flux} and satisfies $\d \fwv = H|_\scal$. The metric restricted to $\scal$ is Lorentzian; the coordinates used to parametrise $\scal$ will be denoted by $\left\{\s^j \; | \; j \in \{0, ..., p\}\right\}$ with $\s^0 \equiv \tau$ the temporal direction.

 D-branes are closely related to generalised submanifolds. A D-brane is located on some submanifold of the total space $\mcal_{10}$. Let $\mcal_{10} = \rbb^{1,d-1} \times \mcal_{10-d}$, with warped metric ansatz
\al{
g_{10} = e^{2 A} \eta_d + g_{10-d} \;,  \qquad A \in C^\infty(\mcal_{10-d}, \rbb)\;.
}
and RR-fluxes respecting the isometries of spacetime, i.e., being Poincar\'{e} invariant and thus being of the form
\al{\label{eq:rrbrane}
\fcal \equiv \text{vol}_d \wedge F^{el} + F \;, \qquad F, F^{el} \in \O^{\bullet}(\mcal_{10-d}) \;.
}
We will only consider D-branes that are static and magnetic; the former implies that the brane is constant along the temporal direction, i.e., the submanifold wrapped is of the form $S =  \rbb^{1,q-1} \times \Sigma$, with $\Sigma \subset \mcal_{10-d}$ an $r$-dimensional (boundaryless) submanifold such that $r + q  = p+1$. The latter implies that $\fwv|_{\rbb^{1,d-1}} = 0$, i.e., only the magnetic part of the worldvolume two-form flux is non-trivial on the brane. Then the worldvolume action \eqref{eq:dbiwz} reduces to
\al{
S_{DBI-WZ} &=  -\mu_p \text{Vol}_{q}  \left( \int_{\Sigma} \d^{r}\sigma e^{qA-\phi} \sqrt{\det(\gs+\fcal_{\text{wv}})}
-\int_{\Sigma} \delta_{q,d} C^{el}|_\Sigma \wedge e^{\fcal_{\text{wv}}}\right) \nn \\
 &\equiv - \mu_p \text{Vol}_{q}  \ecal \;,
}
with $C^{el}$ the local gauge field for the electric RR flux satisfying $F^{el} = \d_H C^{el}$ and $\ecal$ the energy density of the D-brane.
\\
\\
Calibration forms \cite{harvey} are differential forms that were introduced as a tool to determine cycles that minimise the volume with respect to a homology class. It turns out that these can be generalised to minimize other functionals as well; in particular, they can be tailored such that they indicate when the energy density of a D-brane is minimised.
\dfn{
Let $(\mcal, g)$ be a Riemannian manifold, $q, d \in \nbb$, $q \leq d$.
A {\it generalised calibration form} is a polyform $\o$ on $\mcal$ satisfying he following two conditions:
\begin{enumerate}
\item Firstly,
\al{
\left(\o \wedge e^{\fcal_{\text{wv}}}\right)_{\tilde{\Sigma}} \leq \d^r \s e^{q A -\phi} \sqrt{g|_{\tilde{\Sigma}} + \tilde{\fcal}_{\text{wv}}}
}
for any generalised submanifold $(\tilde{\Sigma}, \tfwv)$ with coordinates $\{\s^j\}$ on the $r$-dimensional cycle $\tilde{\Sigma}$, and there exists a generalised submanifold $(\Sigma, \fwv)$ that saturates the bound. Such a generalised submanifold is called a {\it (generalised) calibrated submanifold}.
\item Secondly,
\al{
\d_H \o = \delta_{q,d} F^{el}\;.
}
for a polyform $F^{el}$.
\end{enumerate}
}
It will be shown that with this definition, D-branes on calibrated generalised submanifolds minimise the energy density with respect to their generalised homology class. For brevity, we will refer to generalised calibration forms simply as calibration forms.
\\
\\
Given a calibration form, let $(\Sigma, \fwv)$ saturate the bound, and let $(\tilde{\Sigma}, \tfwv)$ be an arbitrary generalised submanifold in the same generalised homology class. That is to say, we require existence of a generalised submanifold $(T, G)$ satisfying
\al{
\p T &= \tilde{\Sigma} - \Sigma \;, \quad \d G = H|_T\;, \quad G|_{\Sigma} = \fwv \;, \quad G|_{ \tilde{\Sigma}} = \tfwv\;.
}
For more on generalised homology, see \cite{em}.
Using the fact that\footnote{The same caveat holds in this case as for the Chern-Simons term of $D=11$ supergravity, namely that
\eq{
\int_\Sigma C^{el} \wedge e^\fwv \equiv \int_T F^{el} \wedge e^G
}
in order to ensure that integration of the non-globally well-defined form $C^{el}$ can be defined properly. }
\eq{
0& = \int_T \left( \d_H \o - \delta_{q,d} F^{el} \right)_T \wedge e^G = \int_T \d \left( \left(\o - \delta_{q,d} C^{el} \right)_T \wedge e^G\right) \\
&=   \int_\Sigma \left(\o - \delta_{q,d} C^{el} \right)_\Sigma \wedge e^\fwv - \int_{\tilde{\Sigma}} \left(\o - \delta_{q,d} C^{el} \right)_{\tilde{\Sigma}} \wedge e^{\tfwv}\;,
}
one finds
\eq{
\ecal(\Sigma, \fwv) &= \int_\Sigma \left( \o - \delta_{q,d} C^{el} \right)_\Sigma \wedge e^{\fwv} =   \int_{\tilde{\Sigma}} \left(\o - \delta_{p,q} C^{el}  \right)|_{\tilde{\Sigma}} \wedge e^{\tfwv} \\
&\leq \ecal(\tilde{\Sigma}, \tfwv)\;.
}
Therefore, $(\Sigma, \fwv)$ minimises the energy-density with respect to its generalised homology class. A D-brane wrapping such a generalised complex submanifold is called a {\it calibrated D-brane}.
\\
\\
The construction of generalised calibration forms for static magnetic branes in our IIB, $d=2$ setup, was done in \cite{lpt} following the procedure of \cite{km}.
Let us define a pair of spinor bilinears,
\eq{\label{eq:su4polyforms}
\underline{\Psi_1}  &= - \frac{2^4}{\a^2} \eta_1 \otimes \tilde{\eta_2^c} \\
\underline{\Psi_2} &= - \frac{2^4}{\a^2} \eta_1 \otimes \tilde{\eta_2}\;.
}
Since $1 \leq q \leq 2$, there are but two possibilities. For $q=1$, the construction as discussed in appendix \ref{calicons} does not yield a calibration form.
For $q = 2$, the calibration form is given by
\eq{
\o = e^{2 A - \phi} \Re \left( \Psi_1 + e^{i \vartheta} \Psi_2\right)
}
with $\vartheta \in \rbb$ arbitrary.  As a consequence, the differential condition on the calibration form is given by
\eq{
\d_H \left(e^{2A - \phi} \Re \Psi_1 \right) &= e^{2A} \star_8 \s (F) \\
\d_H \left(e^{2A - \phi}  \Psi_2 \right) &= 0 \;,
}
which is exactly \eqref{eq:gcgeq41}  as alluded to in the introduction.

\section{Supersymmetry in terms of calibration forms \& generalised complex geometry}\label{su4gcg}
Although rather short, this section, combined with with \eqref{eq:iib}, contains the main result of \cite{ptb}. Expressions have been found for the calibration form in terms of two polyforms, expressed as bilinears in the pure spinor $\eta$. This pure spinor is associated to the $SU(4)$-structure on $\mcal_8$ and hence, the polyforms $\Psi_{1,2}$ can be expressed in terms of $(J, \O)$. Consider \eqref{eq:su4polyforms} and apply the Fierz identity \eqref{eq:fierz} followed by the Clifford map \eqref{eq:cliff}. Then, applying \eqref{eq:su4spinorbilinears}, it follows that
\eq{\label{eq:su4gacs}
\Psi_1 &= - e^{-i \t} e^{- i J} \\
\Psi_2 &= - e^{i \t} \O \;.
}
As discussed in section \ref{gcgexamples}, these are generalised almost complex structures induced by respectively the almost symplectic and almost complex structures defined by the $SU(4)$-structure. Thus, this leads to the conclusion that the strict $SU(4)$-ansatz, which was taken for technical reasons, can alternatively be interpreted as considering the two most obvious generalised almost complex structures, rather than anything more intricate. Taking this point of view, \eqref{eq:gcgeq41b} is nothing more than ($B$-twisted) integrability of $\ical_{I}$.
\\
\\
Explicitly calculating \eqref{eq:gcgeq41} using \eqref{eq:su4gacs} leads to the following result:
\eq{\label{eq:gcgiib1}
W_1 &= W_2 = 0 \\
W_3 &= i e^\phi \left(\cos\t f^{(2,1)}_3  - i \sin\t f^{(2,1)}_5\right) \\
W_{5} &=   \p^+ ( \phi-2A +  i \t)\\
\tilde{f}^{(1,0)}_{3} &=  \tilde{f}^{(1,0)}_{5}  = h^{(1,0)}_1= \tilde{h}^{(1,0)}_{3} =0 \\
f_{1}^{(1,0)} &= ie^{- \phi} \left( \cos\t (\p^+ \t - 3 h^{(1,0)}_{3} ) +  \sin\t( 2 \p^+ A -\p^+ \phi + 3 W_{4})\right)\\
f_{3}^{(1,0)} &= - i e^{- \phi} \left( \cos\t (\p^+ \phi - 2 \p^+ A  - 2 W_{4} ) + \sin\t (\p^+ \t - 2  h^{(1,0)}_{3} )\right)\\
f_{5}^{(1,0)} &= e^{- \phi} \left( \cos\t (\p^+ \t - h_{3}^{(1,0)} ) - \sin\t(\p^+ \phi-2 \p^+ A  - W_{4}) \right)\\
f_{7}^{(1,0)} &= e^{- \phi} \left( \cos\t (2 \p^+A  - \p^+ \phi) - \sin\t( \p^+ \t) \right)\\
h^{(2,1)} &= e^\phi \left(- \cos\t f^{(2,1)}_5  + i \sin\t f^{(2,1)}_3\right)
\;.
}
The vanishing of $W_{1,2}$ and thus the conclusion that $I$ is integrable is no surprise. As a consequence, we can identify $\p^+ = \p$. Note also that the external NSNS three-form vanishes,  $e^{2A} \vol_2 \wedge H_1 = 0$. The fact that the NSNS flux is internal, i.e. $H = H_3$, is important, as $H$ is the form associated to supergravity, whereas $H_3$ is the form that plays a role in the description of the generalised almost complex structures on $\mcal_8$, and thus the interpretation of \eqref{eq:gcgeq41} as twisted integrability is justified.
\\
\\
Comparing \eqref{eq:gcgiib1} with the supersymmetric solution \eqref{eq:iib}, it follows that a  number of the constraints are missing, in the sense that the supersymmetric solution is a subset of the solution to the equations above. Consider
\eq{
\d_H^{\ical_2} \equiv [\d_H, \ical_2] = i \left(\bar{\p}^{\ical_2}_H - \p^{\ical_2}_H \right)\;,
}
as a twisting of \eqref{eq:dj} as defined in section \ref{gcginteg}. Due to the fact that $\ical_2 = \ical_I$, it follows that $\p^{\ical_2} = \p$, i.e., the differential operator induced by the generalised complex structure is just the Dolbeault operator.
Then the equation
\eq{
\d_H^{\ical_2} \left( e^{- \phi } \text{Im} \Psi_1 \right) = F
~,
}
that was refered to in \eqref{eq:gcgeq42} yields the following:
\eq{\label{eq:gcgiib2}
f_{1}^{(1,0)} &= i e^{- \phi} \left(\sin\t \p^+ \phi -  \cos\t (\p^+ \t)\right)\\
f_{3}^{(1,0)} &= i e^{- \phi} \left( \cos\t (\p^+ \phi - W_{4} )
+ \sin\t (\p^+ \t - h^{(1,0)}_{3} )\right)\\
f_{5}^{(1,0)} &= e^{- \phi}
\left(\sin\t( \p^+ \phi - 2 W_{4}) -\cos\t (\p^+ \t - 2
h_{3}^{(1,0)} ) \right)\\
f_{7}^{(1,0)} &= e^{- \phi} \left( \cos\t (\p^+ \phi - 3 W_{4})
+  \sin\t( \p^+ \t - 3  h_{3}^{(1,0)}) \right)~.
}
These are exactly the missing constraints. That is to say, \eqref{eq:gcgiib1} combined with \eqref{eq:gcgiib2} is equivalent to \eqref{eq:iib}. This proves the claim made in the introduction that on $\rbb^{1,1} \times \mcal_8$ with an $SU(4)$-structure on $\mcal_8$ and an $\ncal= (2,0)$ strict $SU(4)$ Killing spinor ansatz, the supersymmetry equations of type IIB are equivalent to \eqref{eq:gcgeq41}, \eqref{eq:gcgeq42}.

\chapter{$SU(5)$-structures and generalised complex geometry}\label{su5}
In the previous section, it has been shown that there is more to the supersymmetry equations than just the calibration conditions. This did not manifest itself until investigating $\mcal_{10} = \rbb^{1,1} \times \mcal_8$, which is a more general manifold than  $\rbb^{1,3} \times \mcal_6$ or $\rbb^{1,5} \times \mcal_4$. The investigation of $\rbb^{1,1} \times \mcal_8$ relied heavily on the existence of an $SU(4)$-structure. In all three cases, a Killing spinor ansatz of the form $\epsilon \sim \zeta \otimes \eta_1 + \text{c.c.}$ was made. Such an ansatz requires the existence of a nowhere vanishing spinor $\eta_1$ on the internal manifold, and as any Weyl spinor is pure in dimension $2n \leq 6$, the ansatz basically already entailed requiring an $SU(n)$-structure for $d=4$ and $d=6$. In the case $d=2$, however, it was necessary to require its existence as an additional constraint on the internal manifold.
\\
\\
In \cite{tomasiello}, the supersymmetry equations were recast into G$\mathbbm{C}$G-inspired equations in terms of a polyform on arbitrary $\mcal_{10}$, with no structure group restrictions other than what is required for the supersymmetry equations to be sensible (global existence of a metric, Killing spinors, etc.). Unfortunately, these equations are not as simple as one might like, and a geometrical interpretation for them has proven to be elusive.
\\
\\
In order to shed light on the situation, a middle ground is sought, on the one hand by imposing some additional ans\"{a}tze on the case of \cite{tomasiello}, on the other hand by extending the succesful $SU(n)$-structure group results. Thus, we consider type II supergravity on $\mcal_{10}$ with an $SU(5)$-structure. Such a setup has two immediate implications. Firstly, an $SU(5)$-structure comes with a ten-dimensional Riemannian metric; as we wish to associate the tensors defined through the $SU(5)$-structure to the physics described by the supersymmetry equations, this necessitates considering Euclidean supergravity in order to account for this non-Lorentzian metric. This is more convoluted than the single Wick rotation that one might a priori expect, as supersymmetry relates the fermionic spinors to the bosonic fields, and requires resorting to a complex action and taking a `real slice' at the end. Unfortunately, no such real slice exists for type IIB. It does for Euclidean type IIA supergravity, which can also be uplifted to Lorentzian M-theory.

Secondly, the machinery of generalised calibrations for static magnetic branes becomes unavailable, as there is no longer a distinguished temporal direction.
\\
\\
The result found is as follows: in case one takes a strict ansatz for the Killing spinors, the supersymmetry equations imply
\begin{subequations}\label{eq:gcgsu5}
\al{
\d_H \left( \a^2 e^{- \phi} \Psi_2 \right) &= 0  \label{eq:gcgsu5a}\\
i \bar{\p}_H^{\ical_2}\left( e^{- \phi} \Im \Psi_1 \right) &= F^-  \label{eq:gcgsu5b}\;,
}
\end{subequations}
with definitions given in section \ref{gcgsolsu5}. Unfortunately, the converse does not hold. \eqref{eq:gcgsu5a} is also implied by the supersymmetry equations in the absence of the strict ansatz.
We will refer to \eqref{eq:gcgsu5} as the {\it G$\cbb$G equations} or {\it generalised equations}.
\\
\\
First, complex supergravity will be discussed in order to see how to construct supergravity on manifolds with $SU(5)$-structure. In section \ref{susysolsu5}, supersymmetric solutions  for complex supergravity on such manifolds will be given for both type IIA and type IIB, using a strict $SU(5)$ Killing spinor ansatz. Next, we will discuss the definitions of and solutions to the the G$\cbb$G equations \eqref{eq:gcgsu5}. We compare these to the solutions to the supersymmetry equations and note the differences. Then the general proof that the supersymmetry equations imply \eqref{eq:gcgsu5a} is given. Finally, the uplift of Euclidean IIA to Lorentzian M-theory is discussed in section \ref{msu5}.

\section{Complex Supergravity}\label{cplxsugra}
All the necessary details for $SU(5)$-structures have been discussed in section \ref{su(5)}. In order to apply this to type II supergravity, the Riemannian metric needs to be accommodated.
To do so, we make use of complex supergravity, via a construction outlined in \cite{bergshoeff}.
The process to obtain complex supergravity goes by the name of `holomorphic complexification': the complex supergravity will be invariant under a complex superalgebra, and by taking various reality conditions (or `real slices'), various real supergravities with various metric signatures can be acquired. In fact, it is possible to construct multiple supergravity actions with the same metric signature. Of course, it is not guaranteed that these are physical, and indeed, cases are known where, for example, some kinetic terms have the wrong sign.
\\
\\
Consider the action for type II supergravity as given in \eqref{eq:tIIaction}, but let all the bosonic fields be complex-valued rather than real. The selfduality constraint on the RR fluxes cannot be sensible generically, as it depends on the signature of the metric. For a Riemannian metric, the sensible selfduality constraint is given by
\eq{\label{eq:cplxsd}
\fcal = - i \star_{10} \s (\fcal)\;.
}
The sign is convention dependent. For the fermionic action\footnote{We have not given the fermionic action because it is irrelevant for our purposes. The only point in this story where fermions show up is when we consider the Killing spinors in the supersymmetry equations.}, first it is necessary to rewrite the action such that no complex conjugates appear anywhere. In other words, the actions needs to be rewritten in terms of Majorana fermions. Having done so, holomorphic complexification consists of ignoring all reality conditions on the Majorana fermions, i.e., replacing them with Dirac fermions. The gamma matrices are taken as a basis of (flat Euclidean) $Cl(10)$, and as per usual, the `indices are curved' by means of vielbeine $\g^m = e^m_a  \g^a$; holomorphic complexification then comes down to taking $e^m_a$ to be complex. The result is a complex action that is invariant under complex supersymmetry transformations; that is to say, the Killing spinors $\e_{1,2}$ are complex Weyl spinors of $Cl(10)$. Thus, the supersymmetry equations for the complex action are still given by \eqref{kse}.
\\
\\
The possible reality conditions one could impose are given in table 1 of \cite{bergshoeff}. In particular, there is no set of signature (10,0) reality conditions for type IIB, which is unfortunate; that is to say, unfortunate from the perspective of the theory being physically sensible. From a technical perspective on the other hand, it is rather irrelevant as there is no problem solving complex supersymmetry equations for complex supergravity. Thus, we will forge on, imposing merely the minimal requirement of a positive definite metric so that it may be identified with the Riemannian metric of the $SU(5)$-sructure.
\\
\\
Unlike type IIB, type IIA does play nice, as there is a set of reality conditions with a (10,0) signature metric. These reality conditions are given by\footnote{We have `decomposed' the flux into electric and magnetic components; obviously the electric component is trivial in the absence of spacetime and thus the magnetic component satisfies $F = \fcal$.}
\eq{\label{eq:reality1}
H &= - H^* \\
F^* &= \s (F)
}
with every other bosonic field real, and the Killing spinors satisfy
\eq{\label{eq:reality2}
\e_2 = - i \e_1^c \;,
}
which is just the Majorana condition.
\\
\\
Imposing reality conditions is only necessary from a physical point of view and will not be necessary to solve the supersymmetry equations, with the exception being that we impose positive-definiteness of the metric from the start. In addition, we take a strict Killing spinor ansatz, which in this case means setting
\eq{\label{eq:strictsu5}
\e_1 = \a \eta \;, \qquad \e_2 = \left\{
\begin{array}{ll}
\a e^{i \t } \eta^c & \quad \text{Type IIA} \\
\a e^{i \t } \eta & \quad \text{Type IIB}
\end{array} \right.
}
where without loss of generality, $\a, \t \in C^{\infty}(\mcal_{10}, \rbb)$. Note that the norm of $\e_2$ has been fixed such that $|\e_2|^2 = |\e_1|^2$; this was necessary to admit $\k$-symmetric branes in the Lorentzian case. Here, it is possible that this constraint could have been relaxed.

\section{Supersymmetric solutions on $SU(5)$-structure manifolds}\label{susysolsu5}
Solving the complex supersymmetry equations on $\mcal_{10}$ equipped with an $SU(5)$-structure and using a strict $SU(5)$ Killing spinor ansatz is entirely analogous to the $\rbb^{1,1} \times \mcal_8$ case as described in section \ref{susysol}. More specifically, the procedure is as follows:
\begin{itemize}
\item We use the explicit expressions \eqref{eq:strictsu5} for the Killing spinors $\e_{1,2}$ in the supersymmetry equations.
\item We decompose the fluxes into $SU(5)$-representations using \eqref{eq:su5decomp}. Note that reality of a $k$-form $\a$ would imply that, after decomposition, $\a^{(p,q)} = \left( \a^{(q,p)}\right)^*$. Taking the forms complex means that this no longer holds.
\item Any term with $>$ 5  indices will be Hodge dualised to a term with $<$ 5 indices.
\item Any gamma matrix with $>$ 2 indices can be expressed in terms of the $SU(5)$-structure and gamma matrices with $\leq$ 2 indices by making use of the identities \eqref{eq:su5spinorhodge}. This leads to the identities \eqref{eq:su5useff1}, \eqref{eq:su5useff2}.
\item The connection acting on the Killing spinors can be expressed in terms of torsion classes as \eqref{eq:su5nablaeta} due to \eqref{eq:strictsu5}, \eqref{eq:su5etajo}, \eqref{eq:su5torsion}.
\end{itemize}
Following these steps and a non-neglible amount of time calculating, these are the resulting solutions.
\\
\\
The solution to the complexified supersymmetry equations of type IIA, using a Riemannian metric defined by an $SU(5)$-structure on $\mcal_{10}$, and using the strict $SU(5)$ Killing spinor ansatz \eqref{eq:su5iia}, is given by
\eq{\label{eq:su5iia}
\t  &\in \rbb\\
\a &= k  e^{\frac12\phi}\\
W_1 &= -i h^{(2,0)}  \\
W_2 &= 4 i e^{i\t} e^{\phi} f_4^{(3,1)} \\
W_3 &= -i h^{(2,1)} \\
W_4 &=  -i h^{(1,0)} \\
W_5 &= \p^+ \phi   \\
h^{(2,0)} &= \frac{3}{16}e^\phi  e^{i \t} \left( f_{2}^{(2,0)} - if_{4}^{(2,0)} \right)\\
f_0 &= 3i f_2 + 4 f_4 \\
f_2^{(1,1)} &=  3i f_4^{(1,1)} \\
f_4^{(1,0)} &= - \frac{1}{2} e^{- \phi} e^{i\t} \left( \p^+ \log\a + i h^{(1,0)} \right)~,
}
in addition to the reality conditions
\al{
\big({f}_{4}^{(1,0)}\big)^*  &= {f}_{4}^{(0,1)} \;, & \big( f_4^{(3,1)} \big)^* &=  f_4^{(1,3)}\;,  & \big(f_{2}^{(2,0)}  - if_{4}^{(2,0)}  \big)^* &= - \big(  f_{2}^{(0,2)} - i f_{2}^{(0,2)}  \big)
\nn \\
\big( h^{(1,0)} \big)^* &= - h^{(0,1)} \;, &\big({h}^{(2,0)} \big)^* &= - {h}^{(0,2)} \;, & \big( h^{(2,1)} \big)^* &= - h^{(1,2)}~.
}
The solution is thus parametrised by
$k, \t \in \rbb$, $\phi, f_2, f_4 \in C^\infty(\mcal_{10}, \cbb)$ and differential forms $h^{(1,0)}$, $f^{(2,0)}_2 $, $f^{(2,0)}_4$, $ h^{(2,1)}$, $f_4^{(3,1)}$. As usual, for a function $f$, $\p^\pm f $ project onto the $(1,0)$ and $(0,1)$ components of $\d f$ with respect to the almost complex structure. Note that the reality conditions enforced by supersymmetry are a proper subset of the necessary reality constraints to fix the action as real, as follows from comparison with \eqref{eq:reality1}, \eqref{eq:reality2}. In particular, what is missing is
\al{ \label{eq:su5reality}
\begin{aligned}
\t &= - \frac12 \pi \;, & ~~~~f_2^* &= - f_2 \;, ~~~~ & \left(f_4^{(1,1)}\right)^* &= f_4^{(1,1)} \\
\phi^* &= \phi \;,      & ~~~~f_4^* &= f_4  \;,  ~~~~ & \left(f_4^{(2,0)}\right)^* &= f_4^{(0,2)} \;.
\end{aligned}
}
In order to ensure that the solution can be identified with the supersymmetry constraints of real Euclidean type IIA supergravity, these need to be enforced by hand.
\\
\\
The solution to the complexified supersymmetry equations of type IIB, using a Riemannian metric defined by an $SU(5)$-structure on $\mcal_{10}$, and using the strict $SU(5)$ Killing spinor ansatz \eqref{eq:strictsu5}, is given by
\eq{\label{eq:su5iib}
W_1&=0\\
W_2&=0\\
W_3^*&= -i e^{\phi} \big( \cos\t f_3^{(1,2)} - \sin\t f_5^{(1,2)} \big)\\
W_4^* &= \frac{1}{2} \p^- \phi \\
W_5^* &= \p^-\big( \phi - 2 \log \a - i \t \big)\\
h^{(0,1)} &= \frac{1}{2} \p^- \t  \\
{h}^{(2,0)} &=  0\\
h^{(1,2)} &= -i e^{\phi}  ( \sin\t f_3^{(1,2)} + \cos\t f_5^{(1,2)}) \\
f_{1}^{(0,1)} &= i \p^- \big(e^{-\phi} \sin\t \big) \\
f_{3}^{(0,1)} &= \frac{i}{2} \p^- \big(e^{-\phi} \cos\t \big) \\
{f}_{3}^{(2,0)} &=0\\
f_5^{(1,4)} &= 0\\
e^\phi e^{i\t} \big( \frac{1}{4} f_{1}^{(1,0)} + i f_{3}^{(1,0)} - \frac{3}{2} f_{5}^{(1,0)} \big)
&= -{\p^+}\big( 2 \log \a  -\frac{i}{2}\t \big)- i h^{(1,0)} \\
e^\phi e^{- i\t} \big(- \frac{1}{4} f_{1}^{(1,0)} + i f_{3}^{(1,0)} + \frac{3}{2} f_{5}^{(1,0)} \big)
&=  -{\p^+}\big( 2 \log \a  +\frac{i}{2}\t \big)+ i h^{(1,0)} ~,
}
The solution is parametrised by $\a, \t \in C^\infty(\mcal_{10}, \rbb)$, $\phi \in C^\infty(\mcal_{10}, \cbb)$, and $f^{(1,2)}_3, f^{(1,2)}_5 \in \Op^{(1,2)}(\mcal_{10})$. Vanishing of $W_{1,2}$ ensures that $\mcal_{10}$ is complex and hence, $\p^\pm$ can be identified with the Dolbeault operator and its complex conjugate.
\\
\\
As mentioned before, unlike type IIA, no consistent set of reality conditions exists for Euclidean type IIB supergravity.

\section{Solutions to the generalised complex geometrical equations}\label{gcgsolsu5}
In order to reformulate the supersymmetry equations in terms of generalised complex geometry, we require a pair of polyforms which correspond to generalised almost complex structures. We define these entirely analogous to the $\mcal_{10} = \rbb^{1,1} \times \mcal_8$ case as given in \eqref{eq:su4polyforms}, by setting
\eq{\label{eq:su5gacs}
\underline{\Psi_1} &\equiv - \frac{2^5}{|\e|^2} \e_1 \otimes \widetilde{\e_2^c} \\
\underline{\Psi_2} &\equiv - \frac{2^5}{|\e|^2} \e_1 \otimes \widetilde{\e_2} \;.
}
Technically, it needs to be checked that these polyforms correspond to pure spinors before they can be identified as generalised almost complex structures. However, as it turns out, it is more convenient to calculate them first, as the result will make it obvious that they are indeed pure.
By making use of the Fierz identity \eqref{eq:fierz} followed by the Clifford map \eqref{eq:cliff}, these are converted into polyform. Plugging in the strict Killing spinor ansatz \eqref{eq:strictsu5}, the following is found.
\\
\\
For type IIA, the generalised complex structures are given by
\eq{\label{eq:su5iiapolyforms}
\Psi_1 &= e^{- i \t} \O \\
\Psi_2 &=  - e^{i \t} e^{- i J} \;.
}
Comparing with \eqref{eq:gcgcplx}, \eqref{eq:gcgsmplx} we see that $\Psi_1$ is associated to $\ical_I$, a generalised complex structure induced by an almost complex structure, whereas $\Psi_2$ is associated to $\ical_J$,  a generalised complex structure induced by an almost symplectic structure.
\\
\\
For type IIB, the generalised complex structures are given by
\eq{\label{eq:su5iibpolyforms}
\Psi_1 &= - e^{- i \t} e^{- iJ} \\
\Psi_2 &=  - e^{i \t} \O \;.
}
Hence, for IIB, it is found that instead $\Psi_1 = \Psi_J$ and $\Psi_2 = \Psi_I$, which is similar to the $SU(4)$-structure manifold case.
\\
\\
From this, two things can be concluded. Firstly, from the point of view of generalised complex geometry, the difference between type IIA and type IIB is equivalent to interchanging almost complex and almost symplectic structure. Secondly, we note again that the strict Killing spinor ansatz ensures that the generalised almost complex structures $\Psi_{1,2}$ correspond to the obvious ones, namely $\Psi_I$, $\Psi_J$.
\\
\\
Let us investigate the generalised equations \eqref{eq:gcgsu5}. The first equation \eqref{eq:gcgsu5a} is the integrability condition on $\ical_2$. The second equation \eqref{eq:gcgsu5b} we have no satisfactory geometrical interpretation for. The projection operator projects onto the eigenbundle $\oplus_{k < 0} U^k$ with respect to the grading induced on the exterior algebra by $\ical_2$, as discussed in section \ref{gcgdecomp}. Comparing the solution of the G$\cbb$G equations to supersymmetry, as we will do in the next section, it follows that this projection is a direct consequence of working with complex supergravity. If we were to ignore this, it follows that the G$\cbb$G equations would be exactly \eqref{eq:gcgeq41b} and \eqref{eq:gcgeq42}. However, as there is no notion of time, there is also no notion of static magnetic branes, hence no interpretation in terms of calibration forms is possible in this case. Also note that the missing supersymmetry constraints are not given by \eqref{eq:gcgeq41a}, which would have been the obvious thing one might have expected.

\subsection{Type IIB}
Although perhaps less interesting due to the fact that the supersymmetry equations do not correspond to a real supergravity theory, the IIB solutions are far less complicated. This is due to the fact that the integrable generalised almost complex structure is induced by the almost complex structure rather than by the almost symplectic structure. As a consequence, $\d^{\ical_2}$ corresponds to $\d^c$, and $\p^{\ical_2}$ reduces to the Dolbeault operator. Furthermore, the eigenbundles of $\ical_2$ are given by
\eq{
U^k = \oplus_{p-q = k} T^{*(p,q)}\;,
}
as explained in section \ref{gcgexamples}. Inserting \eqref{eq:su5iibpolyforms} into \eqref{eq:gcgsu5}, the result is \eqref{eq:su5iib}, with the exception that the equations
\eq{
e^\phi e^{i\t} \big( \frac{1}{4} f_{1}^{(1,0)} + i f_{3}^{(1,0)} - \frac{3}{2} f_{5}^{(1,0)} \big)
&= -{\p^+}\big( 2 \log \a  -\frac{i}{2}\t \big)- i h^{(1,0)} \\
e^\phi e^{- i\t} \big(- \frac{1}{4} f_{1}^{(1,0)} + i f_{3}^{(1,0)} + \frac{3}{2} f_{5}^{(1,0)} \big)
&=  -{\p^+}\big( 2 \log \a  +\frac{i}{2}\t \big)+ i h^{(1,0)} ~,
}
are missing. As with the $SU(4)$-structure case, the solution saves us from an interpretational dilemma: with respect to $\ical_2 = \ical_I$, one may consider
\al{
H^{(0,3)} \wedge : U^k \rightarrow U^{k-3}
}
and hence, it is not clear how to treat this term when making the decomposition $ \d_H = \p_H + \bar{\p}_H$. However, it is found that $h^{(2,0)} = 0$ and thus\footnote{For the convenience of the reader, let us remind that in the decomposition of $H$, $H^{(0,3)} \sim h^{(2,0)} \lrcorner \O^*$.}, $H^{(0,3)}$ vanishes. As a result,
it follows that $\bar{\p}_H = \bar{\p} + H^{(1,2)} \wedge$.

\subsection{Type IIA}
We repeat the process above for type IIA. The integrable generalised almost complex structure is $\ical_2 = \ical_J$. Therefore, its eigenbundle is given by (again, see section \ref{gcgexamples})
\al{
U^k = \{ \e^{-i J} e^{\frac12 i \L} \Phi_{(n-k)} \; | \; \Phi_{(n-k)} \in \bigwedge\nolimits^{n - k} T^* \} \;.
}
Thus, for a polyform $\Phi$ it follows that
\eq{
\Pi_{\ical_2}^- (\Phi) = \sum_{k =-5}^{-1}  e^{-i J } e^{\frac12 i \L} \left(e^{- \frac12 i \L} e^{i J } \Phi \right)_{(5-k)}\;,
}
where the subscript $5-k$ refers to the ordinary grading of the exterior algebra into  differential forms of definitive degree.
Fortunately, we are spared the labour of doing this explicitly, as
\eq{
\Pi_{\ical_2}^- (\Phi) = 0 \implies \left(e^{i J} \Phi\right)_{5-k} = 0\;,
}
which greatly simplifies things. Making use of this understanding of the projection operator, plugging \eqref{eq:su5iiapolyforms} into \eqref{eq:gcgsu5} leads to \eqref{eq:su5iia}, with the exception that the constraint
\eq{
f_4^{(1,0)} &= - \frac{1}{2} e^{- \phi} e^{i\t} \left( \p^+ \log\a + i h^{(1,0)} \right)
}
as well as the reality conditions \eqref{eq:su5reality} are not imposed. The latter one may consider as not so important, as these can be enforced by demanding that the Euclidean IIA has a Lorentzian M-theoretical origin. The former is more problematic, as it ensures that the supersymmetry equations are not equivalent to the generalised equations.

\section{General Proof}
So far, it has been shown that the supersymmetry equations imply the generalized equations for the strict ansatz, by brute force calculating both and comparing the solutions. In this section, some more refined techniques will be used to show that \eqref{eq:gcgsu5a} is implied also without strict ansatz. The proof is similar to the one used in \cite{gmpt} to demonstrate equivalence between supersymmetry and generalised equations for $d=4$. The key point is that the Clifford map can be used to relate on the one hand acting with gamma matrices of $CL(10)$ on $\underline{\Psi_{1,2}}$ as defined in \eqref{eq:su5gacs} to, on the other hand, the action of $T \oplus T^*$ on $\Psi_{1,2}$. We will give the proof for type IIA and note that the proof for type IIB is analogous in method.
 \\
 \\
Let $\ccal$ denote the Clifford map converting polyforms into gamma contractions, i.e,
\eq{
\ccal : \ea \rightarrow Cl(10)\\
\ccal (\Phi) \equiv \underline{\Phi}
}
and let us stress again that it is bijective. Then in terms of local coordinates
\eq{\label{eq:gammaconv}
\ccal \left(dx^m \wedge \Phi + \p_m \lrcorner \Phi\right)  &=   \g^m \ccal(\Phi)\\
\ccal \left(dx^m \wedge \Phi - \p_m \lrcorner \Phi \right) &=  (-1)^{|\Phi|} \ccal(\Phi) \g^m \;,
}
where $|\Phi_{(k)}| = k$; in case $\Phi$ consists of only even- or odd-form terms, the sign is consistent. Let $\a^{(k)}$ be a $k$-form, and let $[.,.]$ be a graded bracket on $Cl(10)$ satisfying
\eq{
[\underline{\a}^{(1)}, \underline{\Phi}] \equiv& \underline{\a}^{(1)} \underline{\Phi} + (-1)^{|\Phi|}  \underline{\Phi} \underline{\a}^{(1)} \\
[\underline{\a}^{(3)}, \underline{\Phi}] \equiv& \underline{\a}^{(3)} \underline{\Phi} + (-1)^{|\Phi|}  \underline{\Phi} \underline{\a}^{(3)}
                                                 +   \g^m \underline{\Phi} \underline{\a}^{(1)}_m + (-1)^{|\Phi|}   \underline{\a}^{(1)}_m \underline{\Phi} \g^m   \;.
}
Then it can be shown that
\eq{\label{eq:cliffforms}
\ccal \left( 2 \a^{(1)} \wedge \Phi \right) &= [\underline{\a}^{(1)}, \underline{\Phi}] \\
\ccal \left( 8 \a^{(3)} \wedge \Phi \right) &= [\underline{\a}^{(3)}, \underline{\Phi}] \;.
}
Let us now apply the above to the generalised almost complex structure $\underline{\Psi_2}$. First, $\eqref{eq:gammaconv}$ implies that
\eq{
\ccal \left( \d \Psi \right) = 16 \left(
\gamma^m\nabla_m\eta_1\otimes\widetilde{\eta^c_2}+
\gamma^m\eta_1\otimes\nabla_m\widetilde{\eta^c_2}+
\nabla_m\eta_1\otimes\widetilde{\eta^c_2}\gamma^m+
\eta_1\otimes\nabla_m\widetilde{\eta^c_2}\gamma^m
\right)
}
The supersymmetry equations can be rewritten as
\eq{\spl{\label{int3}
\frac{1}{2}\underline{H}\eta_1&=
-\underline{\partial\phi}~\!\eta_1+
\frac{1}{16}e^\phi\gamma^m\underline{F}\gamma_m\eta_2^c\\
\frac{1}{2}\widetilde{\eta^c_2}\underline{H}&=
-\widetilde{\eta^c_2}\underline{\partial\phi}-
\frac{1}{16}e^\phi\widetilde{\eta_1}\gamma^m\underline{F}\gamma_m
\\
\nabla_m\eta_1&=
-\eta_1\partial_m\log\alpha-\frac{1}{4}\underline{H}_m\eta_1
+\frac{1}{16}e^\phi\underline{F}\gamma_m\eta_2^c\\
\nabla_m\widetilde{\eta^c_2}&=-\widetilde{\eta^c_2}\partial_m\log\alpha
-\frac{1}{4}\widetilde{\eta^c_2}\underline{H}_m
-\frac{1}{16}e^\phi\widetilde{\eta}_1\gamma_m\underline{F}
~.}}
Thus, using \eqref{int3} and \eqref{eq:cliffforms},
\eq{
\ccal \left( \d \Psi \right) = \frac12 [\underline{\d \phi} - 2 \d \log \a, \underline{\Psi_2}] - \frac18 [ \underline{H}, \underline{\Psi_2}]
}
Using $\ccal^{-1}$ to rewrite the right-hand side in terms of polyforms, we can conclude that
\eq{
\ccal \left( \d_H \left( \a^2 e^{- \phi} \Psi_2 \right) \right) = 0
}
and thus \eqref{eq:gcgsu5a} follows due to injectivity of $\ccal$.

\section{Lorentzian M-theory from Euclidean IIA}\label{msu5}
Real type IIA supergravity with a Riemannian metric is a consistent truncation of Lorentzian M-theory. Hence, the supersymmetric solution described in  \eqref{eq:su5iia} can be uplifted to a supersymmetric solution of M-theory on $S^1$-fibrations over $\mcal_{10}$. We shall describe the reduction of M-theory to Riemannian type IIA, the integrability of the supersymmetric solution to a supersymmetric vacuum, and give an example of such a vacuum on $\rbb \times \mcal_{10}$.
\\
\\
The dimensional reduction of M-theory to Riemannian type IIA supergravity goes as follows. Let $\mcal_{11}$ be a circle fibration, with metric and four-flux reductions
\eq{\label{eq:su5uplift}
g_{11} &= -e^{\frac43 \phi} \left( \d t + C_1 \right) + e^{- \frac23 \phi} g_{10} \\
G &= F_4 - i H \wedge (\d t + C_1) \;.
}
As $H$ is imaginary, $G$ is real.
It is clear that this reduction leads exactly to the bosonic massless Euclidean type IIA supergravity action, as this can be considered a double Wick rotation compared to the usual reduction as described in section \ref{dualities}.
We fix
\al{
F_2 = i \d C_1\;,
}
thus ensuring that the connection one-form $C_1$ is real and $F_2$ is imaginary.
The $Spin(10)$ Killing spinors $\e_{1.2}$ of IIA uplift to the $Spin(1,10)$ Killing spinor $\e$ of M-theory as
\al{
\e = \e_1 + \e_2 \;.
}
The result of these definitions is that the (Euclidean) IIA equations of motion are equivalent to those of M-theory.
Specifically, the eleven dimensional Bianchi identity
\eq{
\d G = 0
}
is equivalent to
\eq{
d F_4 + H \wedge F_2 &= 0 \\
  \d H &= 0
}
whereas the flux equation of motion
\eq{
- \d \star_{11} G + \frac12 G \wedge G =0
}
is equivalent to
\eq{\label{eq:su5reduc}
\d F_6 + H \wedge F_4 &= 0\\
\d \left( e^{- 2 \phi} \star_{10} H \right) + \frac12 \left.\star_{10} F \wedge F\right|_8 &= 0 \;,
}
hence indeed\footnote{
When comparing the sign of the $H$ equation of motion in \eqref{eq:eom} to \eqref{eq:su5reduc}, there is a sign discrepancy. This comes about as follows: the action is a pseudo-action and is supplemented with the additional RR self-duality constraint, \eqref{eq:selfduality} for Lorentzian, \eqref{eq:cplxsd} for Riemannian signature. The sign in \eqref{eq:cplxsd} is convention-dependent.
When we fixed $J^5 \sim + \text{Vol}_{10}$, this determined \eqref{eq:gamhodge}, and thus the selfduality sign as $\underline{F_0}+\underline{F_2} +\underline{F_4}  \stackrel{!}{=}  + \left(\underline{F_6}+\underline{F_8} +\underline{F_{10}}\right)$ (otherwise $F$ drops out of the supersymmetry equations). Taking $F =  + i \star_{10} \s F$ instead, the result is $\star_{10} F \rightarrow - \star_{10} F $ in \eqref{eq:su5reduc} which matches \eqref{eq:eom}.}
, the equations of motion and Bianchi identities of $G$ for M-theory and of $F$ for Euclidean IIA are equivalent\footnote{The Bianchi identity for $F_2$ follows from the equation of motion for $g$ instead.}. Hence the integrability theorem of section \ref{integ} can be applied to find Lorentzian M-theory vacua on $S^1$ fibrations over a base space $\mcal_{10}$ with $SU(5)$-structure. By the usual decompactification procedure, this also yields vacua on $\mcal_{11} = \rbb \times \mcal_{10}$.

\subsection*{Example: Conformal K\"{a}hler on $\rbb \times \mcal_{10}$}
Let us construct an M-theory vacuum on $\mcal_{11} = \rbb \times \mcal_{10}$. As per usual, this involves setting $F_0  = C_1 = 0$.
Taking
\eq{
g_{11} &= -e^{\frac43 \phi} \d t + e^{- \frac23 \phi} g_{10} \\
G &=  - \frac12 \d t \wedge \d \phi \wedge J \;,
}
and making use of \eqref{eq:su5uplift}, it follows that the supersymmetry equations \eqref{eq:su5iia} are satisfied with
\eq{\label{eq:cktorsion}
W_{1,2,3} &= 0 \\
W_4 &= \frac12 W_5 = \frac12 \p \phi\;.
}
Comparison with table \ref{table:su5} leads to the conclusion that $\mcal_{10}$ is therefore a conformal K\"{a}hler manifold for this supersymmetric solution. In order to apply the integrability theorem to ensure that the supersymmetric solution is a supersymmetric vacuum, the equation of motion and Bianchi identity for $G$ need to be satisfied. Due to \eqref{eq:cktorsion}, $G = - \d t \wedge \d J$ and hence is closed. The equation of motion reduces to the co-closure condition $\d \star_{11} G = 0$. Making use of the fact that $\star_{10} G \sim J^3 \wedge \d^c \phi$, it follows that
the equation of motion is satisfied if and only if
\eq{
\p \bar{\p} e^{- \frac12 \phi} = 0 \implies e^{- \frac12 \phi} = f(z) + \overline{f(z)}
}
for an arbitrary holomorphic function $f$. As usual, we run into the Maldacena-Nunez no-go theorem: on compact manifolds, all globally well-defined holomorphic functions are constant, so if we take $\mcal_{10}$ compact and we do not take higher-order corrections into account, $\d \phi = 0$, leading to a fluxless Calabi-Yau vacuum.

\setcounter{secnumdepth}{-1}

\chapter{Conclusion}
\pagestyle{conc}
A number of classes of supersymmetric flux vacua have been discussed. Considering an internal eight-dimensional manifold with $SU(4)$-structure, we have examined $d=2$, $\ncal = (1,1)$ IIA, $d=2$, $\ncal = (2,0)$ IIB, and $d=3$, $\ncal = 1$ M-theory. In all cases, we have considered a strict $SU(4)$ ansatz for the Killing spinors. As a consequence, the external space is constricted to be Minkowski, excluding AdS (and dS). On ten-dimensional manifolds with $SU(5)$-structure, we have examined complex and Euclidean IIA and IIB theory; IIA uplifts to real Lorentzian $d=1$, $\ncal = 1$ M-theory. Again, this was done by making use of a strict Killing spinor ansatz. For the $SU(4)$-structure vacua, explicit examples were discussed on the non-compact CY Stenzel space. Furthermore, by making use of the coset structure of the base space, we have been able to construct families of $SU(4)$-structures on the underlying space. Using these, IIA non-symplectic, and hence non-CY, vacua were constructed. These vacua violate the NSNS Bianchi identity and are thus sourced by NS5-branes. We have been able to construct a source-action which we conjectured to represent calibrated NS5-branes, although we did not take into account some subtleties in this procedure. For the IIB supersymmetry solutions, we have examined the conjecture of \cite{lpt} that the D-brane calibration conditions are equivalent to the supersymmetry conditions and concluded that an additional equation, formulated in the language of generalised complex geometry, is needed for the equivalence in the strict case, whereas for IIA the strict ansatz precludes the habitual method of constructing calibration forms and thus cannot be used to check the equivalence. However, our conjecture that in the non-strict case, our equations are equivalent to supersymmetry and that this can be checked by making use of the equations of \cite{tomasiello} was proven to be true, exactly by making use of the suggested procedure, in \cite{rosa}. We have then checked whether or not something similar holds for the supersymmetric solutions on Euclidean $SU(5)$-structure manifolds. We have found two generalised equations that reproduce part of the supersymmetric solution for the strict ansatz, but do not give all constraints, i.e., the set of solutions to supersymmetry equations is a subset of the set of soltutions to the generalised equations. We have shown that supersymmetry implies one of these two generalised equations generically, strict ansatz or no. Some questions that arise from this work are as follows.
\\
\\
The $\ncal = 1$ M-theory solution on $\rbb^{1,2} \times \mcal_8$ is a generalisation of the well-known $\ncal = 2$ M-theory solution on $\rbb^{1,2} \times \mcal_8$ of \cite{bb2}. For $\mcal_8$ a compact CY, it is known that the latter reduces to $d=3$, $\ncal = 2$ gauged supergravity by KK reduction, which can be lifted to $d=4$, $\ncal=1$ F-theory. It would be interesting to see what our vacua reduce to on compact CY, which is currently work in progress; the answer seems to be $d=3$, $\ncal =1$ ungauged supergravity, with the additional fluxes breaking the Poincar\'{e} invariance of $\rbb^{1,3}$ when lifted to F-theory.
\\
\\
The question of reformulating supersymmetry in terms of G$\cbb$G has not been answered satisfactorily. Firstly, it seems clear now that calibrations cannot be the whole of the story, but this then leaves us to wonder how to explain why the generalised equations are what they are for each dimension. Furthermore, whereas equations of the form $\d_H \Psi \sim F$ can be interpreted as the RR flux being the obstruction to integrability of an almost complex structure, the newly introduced equation of the form $\d_H^\ical \Psi \sim  F$ does not seem to have such a geometric interpretation. It would be interesting to know what the correct geometric interpretation is, if it exists.
\\
\\
The source-action found for the $SU(4)$-deformed Stenzel space seems to hint at the correct form for calibrated NS5-branes. It would be interesting to see if an analogous construction to that of \cite{melec} could be used to construct calibration (poly)forms for NS5-branes. This is made difficult due to the the fact that $\d H \neq 0$ messes with the generalised geometry framework. In addition, it would be interesting to see if the action here, which holds for $d=2$, $\ncal = (1,1)$ IIA vacua given a strict $SU(4)$ Killing spinor ansatz, can be generalised to more general cases, akin to what was done in \cite{koerbt}.

\setcounter{secnumdepth}{2}

\appendix

\chapter{Conventions, notation, identities}
\pagestyle{main}
Lorentzian metrics are taken to be mostly plus.\\
Hodge duality of a $k$-form on $M_n$ is defined by
\eq{\label{hodge}
\left(\star_n \a \right)_{m_1 ... m_{n-k}} \equiv \frac{1}{k!(n-k)!} \e_{m_1...m_n} \a^{m_{n-k+1} ... m_n}
}
Generically, a subscript is used to denote the (dimension of the) space with respect to which the Hodge dual is taken.

The Levi-Civita symbol on a Lorentzian manifold $M_{n+1}$ is taken to satisfy
\eq{
\varepsilon_{0...n} = - \varepsilon^{0...n} = + 1\;.
}
We define the contraction between a $p$-form $\varphi$ and a $q$-form $\chi$, $p\leq q$, by
\eq{
\varphi\lrcorner\chi = \frac{1}{p!(q-p)!} \varphi^{m_1\dots m_p} \chi_{m_1\dots m_p n_1\dots n_{q-p}}\d x^{n_1}\wedge\dots\wedge\d x^{n_{q-p}}
~.}
The volume form is given by
\eq{
\vol_n \equiv \d^n x \sqrt{g_n} = \star_n  1 = \frac{1}{n!} \e_{m_1... m_n} \d x^{m_1} \wedge ...\d x^{m_n} \;.
}
Any conventions not mentioned are most likely to be similar to those of \cite{mmlt}.

\section*{Notation}
\begin{tabular}{l p{9cm}}
$\simeq \quad$ & Either 1) ``Is diffeomorphic to" or 2) `` Is isomorphic to", depending on context.\\
$\equiv$ & Is defined as \\
$M_n$ & A manifold of dimension $n$. \\
$T = TM$ & The tangent bundle of $M$. \\
$T^* = T^*M$ & The cotangent bundle of $M$. \\
$T^{(1,0)}\equiv T^+$ & The +i-eigenbundle of an almost complex structure. The holomorphic tangent bundle in case the almost complex structure is integrable. \\
$ \bw^k T^*$  & The bundle of $k$-forms\\
$ \ea \equiv \oplus_{k=0}^n \bw^k T^* $ & The exterior algebra on $M_n$\\
$T^{*(p,q)} \equiv \left(\bigwedge\nolimits^p T^{*(1,0)}\right) \wedge \left(\bigwedge\nolimits^q T^{*(0,1)}\right)$ & The bundle of $(p,q)$-forms \\
$\tcal^{(k,l)} =  T^{\otimes k} \otimes \left(T^*\right)^{\otimes l}$  & The tensor bundle of type $(k, l)$   \\
$\G(M, E) \equiv \G(E)$ & Sections of the vector bundle $E \rightarrow M$\\
$\mathfrak{X}(M)\quad$ & The sheaf of smooth sections of $T$, i.e.,  vector fields \\
$\O^{(p,q)})(M)$ & The sheaf of smooth sections of  $T^{*(p,q)}$\\
$\O^{\bullet}(M)$ & The sheaf of smooth sections of  $\ea$\\
$\Op^{(p,q)}(M)$ & The sheaf of primitive smooth sections of $T^{*(p,q)}$ \\
$\O^p_{\text{h}}(M)$ & The sheaf of holomorphic smooth sections of  $T^{*(p,0)}$ \\
$\mathcal{O} \equiv \O^0_h(M)$ & The sheaf of holomorphic smooth sections of the trivial bundle, i.e., locally-defined holomorphic functions, i.e. the structure sheaf.\\
$\hookrightarrow$ & Embedding \\
$ \{\cdot,\cdot\}$ & Anticommutator\\
$X \lrcorner \equiv \iota_X $ & Interior product with a vector(field) $X$\\
$g_n \equiv ds^2(M_n)$ & Metric of the $n$-dimensional manifold $M_n$\\
$\eta_n$ & The Minkowski metric diag$(-1, .., 1)$ on $\rbb^n$\\
$\rbb^{1, d-1}$ &  Minkowski space  $(\rbb^d, \eta_d)$\\
$A^*$ & Given a map $A$ acting on a vector space $V$, $A^*$ is either 1) the complex conjugate, or  2) the induced operator on the dual space $V^*$. Given a morphism $A$ between manifolds, $A^*$ denotes the pushforward. \\
$\text{Ad}_g$ and $\text{ad}_X$ & Repsectively, the standard representations of $G$ and $\mathfrak{g}$ on $\mathfrak{g}$.\\
$\mathfrak{g}$  & The Lie algebra associated to a Lie group $G$.
\end{tabular}

\section{Spinor and gamma matrix conventions}\label{spinors}

For a spinor $\psi$ in any dimension we define:
\eq{\widetilde{\psi}\equiv\psi^{T}C^{-1}~,}
where $C$ is the charge conjugation matrix. In Lorentzian signatures, we also define
\eq{\overline{\psi}\equiv\psi^{\dagger}\G^{0}~.}
In all dimensions the gamma matrices are taken to obey
\eq{
(\G^M)^{\dagger}=\G^0\G^M\G^0~.
}
Antisymmetric products of gamma matrices are defined by
\eq{
\G^{(n)}_{M_1\dots M_n}\equiv\G_{[M_1}\dots\G_{M_n]}~.
}
Of crucial importance to any computation involving spinors are the Fierz identity and the Clifford map.
The Fierz identity expresses the tensor product of two $n$-dimensional Weyl spinors, viewed as a map acting on spinors, in terms of the Clifford algebra:
\eq{\label{eq:fierz}
\chi_\a \otimes \psi_\b = \frac{1}{2^n} \sum_{k=0}^{2n} \frac{1}{k!} \tilde{\psi} \g_{m_1...m_k} \chi \g^{m_k...m_1}_{\a\b}\;.
}
The Clifford map is an isomorphism relating the Clifford algebra to polyforms via
\eq{\label{eq:cliff}
\frac{1}{2^n} \sum_{k=0}^{2n} \frac{1}{k!} \tilde{\psi} \g_{m_1...m_k} \chi \g^{m_k...m_1}_{\a\b} \mapsto
\frac{1}{2^n} \sum_{k=0}^{2n} \frac{1}{k!} \tilde{\psi} \g_{m_1...m_k} \chi e^{m_k} \wedge ... \wedge e^{m_1} \;.
}

\subsection*{Two Lorentzian dimensions}

The charge conjugation matrix in $1+1$ dimensions satisfies
\eq{
C^{T}=-C; ~~~~~~ (C\g^\mu)^{T}=C\g^\mu; ~~~~~~ C^*=-C^{-1}~.
}
The spinors we consider are in the fundamental representation; complex-valued chiral (or Weyl). They posses one (complex) degree of freedom. We define:
\eq{
\zeta^c\equiv\g_0C\zeta^*~.
}
The representation is `real' in the sense that:
\eq{
\left(\zeta^c\right)^c = \zeta
}
The chirality matrix is defined by
\eq{
\g_3\equiv{{-}}\g_0\g_1 ~.
}
The Hodge-dual of an antisymmetric product of
gamma matrices is given by
\eq{
\star\g_{(n)}\g_3={{- (-1)^{\frac{1}{2}n(n+1)}}}\g_{(2-n)}~.
\label{hodge2}
}

\subsection*{Eight Euclidean dimensions}

The charge conjugation matrix in $8$ dimensions satisfies
\eq{
C^{T}=C; ~~~~~~ (C\g^\mu)^{T}=C\g^\mu; ~~~~~~ C^*=C^{-1}~.
}
The fundamental (eight-dimensional, chiral) spinor representation is real.
We work with a complexified chiral spinor $\eta$ (i.e. eight complex degrees of freedom). We define:
\eq{
\eta^c\equiv C\eta^*~.
}
The chirality matrix is defined by
\eq{
\g_9\equiv\g_1\dots\g_8 ~.
}
The Hodge-dual of an antisymmetric product of
gamma matrices is given by
\eq{
\star\g_{(n)}\g_9=(-)^{\frac{1}{2}n(n+1)}\g_{(8-n)}~.
\label{hodge8}
}

\subsection*{Ten Lorentzian dimensions}

The charge conjugation matrix in $1+9$ dimensions satisfies
\eq{
C^{T}=-C; ~~~~~~ (C\G^M)^{T}=C\G^M; ~~~~~~ C^{*}=-C^{-1}~.
}
The fundamental (16-dimensional, chiral) spinor representation
$\e$ is real, where we define the reality condition by
\eq{
\overline{\e}=\widetilde{\e}~.
}
The chirality matrix is defined by
\eq{
\G_{11}\equiv{{-}}\G_0\dots\G_9 ~.
}
When considering a 2+8 split of the ten-dimensional Lorentzian space, we decompose the ten-dimensional Gamma matrices as
\eq{\label{eq:gammadecomp}
\left\{ \begin{array}{ll}
\G^{\mu}=\g^\mu\otimes \obb &, ~~~~~\mu=0, 1\nn\\
\G^m=\g_3\otimes \g^{m-1} &, ~~~~~ m=2\dots 9
\end{array} \right.
~.}
It follows that
\eq{
C_{10}=C_2\otimes C_8;~~~~~~ \G_{11}=\g_3\otimes\g_9~.
}
The Hodge-dual of an antisymmetric product of
gamma matrices is given by
\eq{
\star \G_{(n)} \G_{11}= {{- (-1)^{\frac{1}{2}n(n+1)}}}      \G_{(10-n)}~.
}

\subsection*{Ten Euclidean dimensions}

The charge conjugation matrix obeys:
\begin{align}
C^T = - C; \quad C^\dagger=  C^{-1} ; \quad C^* = - C^{-1}
\end{align}
The complex conjugate $\eta^c$ of a spinor $\eta$ is given by
\eq{\eta^c=C\eta^*~.}
The chirality operator is defined by:
\begin{align}
\g_{11} = i \g_1 ... \g_{10}~.
\end{align}
The irreducible spinor representations of $Spin(10)$ are given by  sixteen-dimensional  Weyl
spinors which are complex, in the sense that $\eta^c$ and
$\eta$ have opposite chiralities.
The Hodge dual of an antisymmetric product of $k$ gamma matrices is given by:
\begin{align}\label{eq:gamhodge}
\star \g_k = - i (-1)^{\frac{1}{2} k (k+1)} \g_{10-k} \g_{11}~.
\end{align}

\subsection*{Eleven Lorentzian dimensions}

Given a set gamma-matrices $\{\gamma_a\}$, $a=1,\dots 10$, generating the Clifford algebra in ten Euclidean dimensions the
eleven-dimensional Lorentzian gamma matrices are given by:
\eq{\label{gammauplift}
\Gamma_a=\left\{\begin{array}{rl}
i\gamma_{11}~;&~~~ a=0\\
\gamma_a~;&~~~ a=1,\dots 10\\
\end{array}\right.
~.}
In our conventions the charge conjugation matrix $C$ in eleven Lorentzian dimensions is the same as the one in ten Euclidean dimensions.

Consider a Dirac spinor $\epsilon$ of $Spin(1,10)$ (with 32 complex components).
Under $$Spin(1,10)\rightarrow Spin(10)~,$$
the spinor $\epsilon$ decomposes as $${\bf 32}\rightarrow{\bf 16_+}\oplus {\bf 16_-}~,$$ where ${\bf 16_{\pm}}$ are the positive-, negative-chirality Weyl spinors of $Spin(10)$ (with 16 complex components each). Explicitly we have:
\eq{\epsilon=\epsilon_1+\epsilon_2
~,}
where $\epsilon\sim{\bf 32}$ of $Spin(1,10)$, $\epsilon_1\sim{\bf 16_+}$ of $Spin(10)$ and $\epsilon_2\sim{\bf 16_-}$ of $Spin(10)$.
Imposing the Majorana condition on $\epsilon$,
\eq{\bar{\epsilon}=\widetilde{\epsilon}
~,}
is equivalent to
\eq{\label{real}\epsilon_2=-i\epsilon_1^c~,}
where we have taken (\ref{gammauplift}) into account.
\\
\\
Similarly, the decompostion  $Spin(1,10)\rightarrow Spin(1,9)$ also leads to
\eq{
{\bf 32}&\rightarrow{\bf 16_+}\oplus {\bf 16_-}\\
\e &= \epsilon_1 + \epsilon_2 \;,
}
with $\epsilon_{1,2}$ $\pm$-chirality Majorana-Weyl spinors of $Spin(1,9)$ in this case.

\section{$SU(4)$-structure identities}

\subsubsection{Torsion classes and the covariance of the pure spinor}\label{su4der}
As will be explained in appendix \ref{torsion}, the intrinsic torsion $T_{\text{int}}$ for an $SU(4)$-structure takes value in $T^*M \otimes \text{ad}Q_{SU(4)}$.
This module of $SU(4)$ can be decomposed into irreducibles as
\eq{\label{eq:su4torsdecomp}
\wcal_{\text{int}} &\in(\bf{4}\oplus\bf{\bar{4}})\otimes(\bf{1}\oplus \bf{6}\oplus\bf{6})\\
&\simeq (\bf{4}\oplus\bf{\bar{4}})\oplus(\bf{20}\oplus\bf{\bar{20}})
\oplus(\bf{20}\oplus\bf{\bar{20}})
\oplus(\bf{4}\oplus\bf{\bar{4}})\oplus(\bf{4}\oplus\bf{\bar{4}})
~.}
In other words, the torsion classes are three (complex) (1,0)-forms and two (complex) primitive (2,1)-forms. This is exactly what has been described in  section \ref{su(4)}, where the defining equations for the torsion classes as obstructions to closure of $(J, \O)$ were given as
\eq{\label{eq:su4tors}
\d J&=W_1\lrcorner\Omega^*+W_3+W_4\wedge J+\mathrm{c.c.}\\
\d\Omega&=\frac{8i}{3}W_1\wedge J\wedge J+W_2\wedge J+W^*_{5}\wedge\Omega\;.
}
Since the $SU(4)$-structure can be defined just as well in terms of the pure spinor $\eta$, the obstruction to covariant constancy of $\eta$ with respect to the Levi-Civita tensor (i.e., the failure of the Levi-Civita connection to be compatible with the $SU(4)$-structure) should equally well be expressible in terms of $W_{1, ..,5}$.

Explicitly we have:
\eq{\spl{\label{eq:su4torseta}
\nabla_m \eta
&= \left( \frac{3}{4}W_{4m} - \frac{1}{2} W_{5m} - \mathrm{c.c.}\right) \eta +\frac{i}{24} \O^*_{mnkl}  W_{1}^{n} \g^{kl}\eta\\
& +\left(
-\frac{i}{16} W_{2mkl}
- \frac{1}{32}  \O_{mnkl} W_{4}^{n*}
+ \frac{i}{64} W^*_{3mnp} \O^{np}{}_{kl}
 \right) \g^{kl}\eta^c~.
}}
This can be seen as follows. As follows from the discussion around \eqref{eq:spinordecomp} and the fact that $\nabla_m\eta$ is an $\bf{8}\otimes\bf{8^+}$ module of $\mathfrak{so}(8)$, it can be expand as
\eq{\label{toretaprov}
\nabla_m\eta= \phi_m \eta + \vartheta_m \eta^c + \varphi_{m,pq} \Omega^{pqrs} \gamma_{rs} \eta^c~,}
for some complex coefficients $\phi_m$, $\vartheta_m\sim{\bf 4}\oplus{\bf \bar{4}}$, $\varphi_{m,pq}\sim({\bf 4}\oplus{\bf \bar{4}})\otimes{\bf 6}$. Furthermore we decompose:
\begin{align}
\varphi_{m,pq} &= \Omega^*_{mpqr}A^r + (\Pi^+)_{m[p}B^*_{q]} + (\Pi^+)_{m}^{\phantom{m}n} C^*_{npq} +  \Omega_{pq}^{*\phantom{pq}rs} D_{rsm}
\end{align}
where $A,B \sim{\bf 4}$ are complex (1,0)-forms and $C,D\sim{\bf 20}$ are complex traceless (2,1)-forms.
Multiplying (\ref{toretaprov}) on the left with $\widetilde{\eta^c}\g_{ij}$ and $\we\g_{ijk}$, antisymmetrising in all indices  in order to form $\d J$ and $\d \Omega$ respectively as spinor bilinears and  comparing with \eqref{eq:su4tors} then leads to \eqref{eq:su4torseta}.

\subsection{Identities}
The following useful identities can be proved  by Fierzing \cite{5brane}:
\eq{
\frac{1}{4!\times 2^4}~&\Omega_{rstu}\Omega^{*rstu}=1\\
\frac{1}{6\times 2^4}~&\Omega_{irst}\Omega^{*mrst}
=(\Pi^+)_{i}{}^{m}\\
\frac{1}{4\times 2^4}~&\Omega_{ijrs}\Omega^{*mnrs}
=(\Pi^+)_{[i}{}^{m}(\Pi^+)_{j]}{}^{n}\\
\frac{1}{6\times 2^4}~&\Omega_{ijkr}\Omega^{*mnpr}
=(\Pi^+)_{[i}{}^{m}(\Pi^+)_{j}{}^{n}(\Pi^+)_{k]}{}^{p}\\
\frac{1}{4!\times 2^4}~&\Omega_{ijkl}\Omega^{*mnpq}
=(\Pi^+)_{[i}{}^{m}(\Pi^+)_{j}{}^{n}(\Pi^+)_{k}{}^{p}(\Pi^+)_{l]}{}^{q}~,
\label{bfive}
}
Moreover, we have
\eq{\label{eq:su4spinorbilinears}
\widetilde{\eta^c}\eta=1; &~~~~~\we\eta=0\\
\widetilde{\eta^c}\g_{mn}\eta=iJ_{mn}; &~~~~~\we\g_{mn}\eta=0\\
 \widetilde{\eta^c}\g_{mnpq}\eta=-3J_{[mn}J_{pq]}; &~~~~~\we\g_{mnpq}\eta=\Omega_{mnpq}\\
\widetilde{\eta^c}\g_{mnpqrs}\eta=-15iJ_{[mn}J_{pq}J_{rs]}; &~~~~~\we\g_{mnpqrs}\eta=0\\
\widetilde{\eta^c}\g_{mnpqrstu}\eta=105J_{[mn}J_{pq}J_{rs}J_{tu]}; &~~~~~\we\g_{mnpqrstu}\eta=0 ~,
}
where we have made use of the identities
\eq{\label{jvol}
\sqrt{g}\; \varepsilon_{mnpqrstu}J^{rs}J^{tu}&=24J_{[mn}J_{pq]}\\
\sqrt{g} \;\varepsilon_{mnpqrstu}J^{tu}&=30J_{[mn}J_{pq}J_{rs]}\\
\sqrt{g} \;\varepsilon_{mnpqrstu}&=105J_{[mn}J_{pq}J_{rs}J_{tu]}
~.
}
Note that the bilinears
$\we\g_{(p)}\eta$, ~$\widetilde{\eta^c}\g_{(p)}\eta$, vanish for $p$ odd.
The last line of equation (\ref{bfive}) together with the last line of the
equation above imply
\eq{\label{omvol}
\Omega_{[ijkl}\Omega^*_{mnpq]}=\frac{8}{35}\sqrt{g}\; \varepsilon_{ijklmnpq}~.
}
%
%
%
Finally, the following relations are useful in the analysis of the Killing spinor equations:
\eq{\label{eq:su4spinorhodge}
\g_m\eta&=(\Pi^+)_{m}{}^{n}\g_n\eta\\
\g_{mn}\eta&=iJ_{mn}\eta -\frac{1}{8}\Omega_{mnpq}\g^{pq}\eta^c   \\
\g_{mnp}\eta&=3iJ_{[mn}\g_{p]}\eta
-\frac{1}{2}\Omega_{mnpq}\g^q\eta^c\\
\g_{mnpq}\eta&=-3J_{[mn}J_{pq]}\eta -\frac{3i}{4}J_{[mn}\Omega_{pq]ij}\g^{ij}\eta^c
+\Omega_{mnpq}\eta^c
~.
}
The action of $\gamma_{m_1\dots m_p}$, $p\geq 5$, on $\eta$ can be related to
the above formul{\ae}, using the Hodge property of gamma matrices \eqref{hodge8}.

\subsection{Useful formul\ae{}}
In order to solve the dilatino equations, the following are used:
\eq{\label{c4}
{{F}}_0 \eta &= f_0 \eta \\
\underline{F_1} \eta &= h_{1|m}^{(0,1)} \g^m \eta \\
\underline{F_2} \eta &= 4 i f_2 \eta - \frac{1}{16} f^{(0,2)}_{2|mn} \Omega^{mnpq}\gamma_{pq} \eta^c \\
\underline{F_3} \eta &= 3i h_{3|m}^{(0,1)} \g^m \eta+8 \tilde{h}_{3|m}^{(1,0)} \g_m \eta^c \\
\underline{F_4} \eta &= -12 f_4 \eta + 16\tilde{f}^{*}_{4} \eta^c - \frac{i}{8} f{^{(0,2)}_{4|mn}} \Omega^{mnpq} \gamma_{pq} \eta^c~,
}
and similarly for the Hodge duals of $k$-forms $F_k$ with $k>4$. Obviously, the decomposed NSNS flux $H_{1,3}$ satisfies similar relations as $F_{1,3}$.
\\
\\
For the gravitino equations with $M=m$, we require
\eq{
\underline{H_{3|m}}  \eta =\;& 3i ( h^{(1,0)}_{3|m} + h^{(0,1)}_{3|m}) \eta -
\left(\frac{i}{8}h^{(0,1)}_{3|n}\O^{\phantom{m}nrs}_m + \frac{1}{16} h^{(1,2)}_{3|mpq} \O^{pqrs}\right)\g_{rs} \eta^c\\
&-\frac{1}{2}\tilde{h}^{(1,0)}_{3|n}\O^{*\phantom{m}npq}_m\g_{pq} \eta
}
and
\eq{\label{c7}
{F}_0 \g_m \eta &= f_0   \g_m \eta\\
\underline{F_1} \g_m \eta &= 2 f_{1|m}^{(1,0)}  \eta + \frac{1}{8}f_{1|n}^{(0,1)} \O_m^{\phantom{m}npq}\g_{pq} \eta^c \\
\underline{F_2} \g_m \eta &=  \left(2i f_2 \g_m - 2 f_{2|mn}^{(1,1)}\g^n \right) \eta - \frac{1}{4}   f_{2|np}^{(0,2)} \g_q \O_m^{\phantom{m}npq} \eta^c\\
\underline{F_3} \g_m \eta &=   6i f_{3|m}^{(1,0)}\eta - 16 \tilde{f}_{3|m}^{(1,0)} \eta^c
                            + \left(\frac{1}{8}i f_{3|n}^{(0,1)}  \O_m^{\phantom{m}nrs}- \frac{1}{8} f_{3|mpq}^{(1,2)}  \O^{pqrs} \right)\g_{rs} \eta^c\\
\underline{F_4} \g_m \eta &=- 4i f_{4|mn}^{(1,1)}\g^n  \eta + \frac{1}{6}  f_{4|mnpq}^{(1,3)} \g_r \O^{npqr} \eta^c~,
}
and similarly for the Hodge duals of $k$-forms $F_k$ with $k>4$.

\section{$SU(5)$-structure identities}

\subsubsection{Torsion classes and the covariance of the pure spinor}
Similar to the $SU(4)$-structure case, the obstruction to covariant constancy of $\eta$ with respect to the Levi-Civita connection is expressible in terms of the torsion classes. Let us first note that the intrinsic torsion decomposes as
\eq{\label{eq:su5torsdecomp}
\wcal_{\text{int}} &\in(\bf{5}\oplus\bf{\bar{5}})\otimes(\bf{1}\oplus \bf{10}\oplus\bf{\bar{10}})\\
&\simeq(\bf{10}\oplus\bf{\bar{10}})\oplus(\bf{40}\oplus\bf{\bar{40}})
\oplus(\bf{45}\oplus\bf{\bar{45}})
\oplus(\bf{5}\oplus\bf{\bar{5}})\oplus(\bf{5}\oplus\bf{\bar{5}})
~,}
and hence all torsion classes are present in the decomposition of $(\d J, \d \O)$ as given in section \ref{su(5)}, namely
\eq{\label{eq:su5tors}
\d J&=W^*_1\lrcorner\Omega+W_3+W_4\wedge J+\mathrm{c.c.}\\
\d\Omega&= - \frac{16i}{3} W_1\wedge J\wedge J+W_2\wedge J+W^*_{5}\wedge\Omega \;.
}
Using this, it is found that
\al{\label{eq:su5nablaeta}
\nabla_m \eta =& \left(W_4 - \frac{1}{2}W_5 - \mathrm{c.c.}\right) \eta \\
&+ \left( - \frac{i}{48} \O_{mnpqr}^*W_1^{qr} + \frac{1}{4}
\Pi^+_{m[n}W_{4p]}^* - \frac{i}{8}  W_{3mnp}^* + \frac{i}{4!2^5}  W_{2mxyz} \O_{np}^{*\phantom{np}xyz} \right) \g^{np} \eta ~. \nn
}
The covariant derivative of $\eta$ can be parametrised as
\eq{\label{b10}
\nabla_m \eta = \phi_m \eta + \vartheta_{m,n} \g^n \eta^c + \varphi_{m,np} \g^{np} \eta \;,
}
for some complex coefficients $\phi_m\sim (\bf{5} \oplus \bf{\bar{5}} )$, $\vartheta_{m,n}\sim(\bf{5} \oplus \bf{\bar{5}} )\otimes\bf{5}$,
$\varphi_{m,np}\sim (\bf{5} \oplus \bf{\bar{5}} )\otimes\bf{\bar{10}}$.
Moreover, the purity of $\eta$ implies $\widetilde{\eta}\gamma_m\nabla_n\eta=0$ and thus $\vartheta_{m,n}=0$. Similarly the constancy of the norm of $\eta$
implies that $\phi_m$ is imaginary: $\phi^{(1,0)*}_m = - \phi^{(1,0)}_m$.
We can further decompose
\eq{
(\bf{5} \oplus \bf{\bar{5}} )\otimes \bar{10}= \bf{\bar{5}} \oplus  \bf{10} \oplus \bf{40} \oplus \bf{45}~,
}
which explicitly amounts to parameterising
\eq{\label{b11}
\varphi_{m,pq} = \O_{mpq}^{*\phantom{mpq}ab}E_{ab}^{(2,0)} + \Pi^+_{m[p}F^{(0,1)}_{q]} + G_{mpq}^{(1,2)} + H^{(3,1)}_{mabc}\O^{*abc}_{\phantom{*abc}pq} \;,
}
where now all coefficients on the right-hand side above are in irreducible $SU(5)$ modules. Taking the above into account we can now multiply (\ref{b10}) on the left with $\widetilde{\eta^c}\g_{ij}$ and $\we\g_{ijklr}$, and antisymmetrise in all free indices  in order to form $\d J$ and $\d \Omega$ respectively as spinor bilinears. Comparing with \eqref{eq:su5tors} then leads to \eqref{eq:su5nablaeta}.

\subsection{Identities}

The spinor bilinears are given by:
\eq{\spl{\label{eq:su5spinorbilinears}
\widetilde{\eta^c}\eta=1; &~~~~~\we \g_m \eta= 0 \\
\widetilde{\eta^c}\g_{mn}\eta=iJ_{mn}; &~~~~~\we\g_{mnp}\eta=0\\
 \widetilde{\eta^c}\g_{mnpq}\eta=-3J_{[mn}J_{pq]}; &~~~~~\we\g_{mnpqr}\eta=  \O_{mnpqr}\\
\widetilde{\eta^c}\g_{mnpqrs}\eta=-15iJ_{[mn}J_{pq}J_{rs]}; &~~~~~\we\g_{mnpqrst}\eta=0\\
\widetilde{\eta^c}\g_{mnpqrstu}\eta=105J_{[mn}J_{pq}J_{rs}J_{tu]}; &~~~~~\we\g_{mnpqrstuv}\eta=0 ~\\
\widetilde{\eta^c}\g_{mnpqrstuvw}\eta= 945 i J_{[mn}J_{pq}J_{rs}J_{tu}J_{vw]} ~,
}}
whereas the bilinears $\we\g_{(2p)}\eta$, ~$\widetilde{\eta^c}\g_{(2p-1)}\eta$, vanish. The following useful identities can be proved  by Fierzing
\eq{\spl{ \label{eq:omegas}
\frac{1}{5!\times 2^5}~&\Omega_{vwxyz}\Omega^{*vwxyz}=1\\
\frac{1}{4!\times 2^5}~&\Omega_{awxyz}\Omega^{*mwxyz}
=(\Pi^+)_{a}{}^{m}\\
\frac{1}{12\times 2^5}~&\Omega_{abxyz}\Omega^{*mnxyz}
=(\Pi^+)_{[a}{}^{m}(\Pi^+)_{b]}{}^{n}\\
\frac{1}{12\times 2^5}~&\Omega_{abcyz}\Omega^{*mnpyz}
=(\Pi^+)_{[a}{}^{m}(\Pi^+)_{b}{}^{n}(\Pi^+)_{c]}{}^{p}\\
\frac{1}{4!\times 2^5}~&\Omega_{abcdz}\Omega^{*mnpqz}
=(\Pi^+)_{[a}{}^{m}(\Pi^+)_{b}{}^{n}(\Pi^+)_{c}{}^{p}(\Pi^+)_{d]}{}^{q}\\
\frac{1}{5!\times 2^5}~&\Omega_{abcde}\Omega^{*mnpqr}
=(\Pi^+)_{[a}{}^{m}(\Pi^+)_{b}{}^{n}(\Pi^+)_{c}{}^{p}(\Pi^+)_{d}{}^{q}(\Pi^+)_{e]}{}^{r}\\
}}
Moreover:
\eq{\spl{\label{jvol}
\varepsilon_{mnpqrstuvw}J^{mn}J^{pq}J^{rs}J^{tu}J^{vw} &=3840\\
\varepsilon_{mnpqrstuvw}J^{pq}J^{rs}J^{tu}J^{vw}       &=384 J_{mn}\\
\varepsilon_{mnpqrstuvw}J^{rs}J^{tu}J^{vw}             &=144 J_{[mn}J_{pq]}\\
\varepsilon_{mnpqrstuvw}J^{tu}J^{vw}                   &=120  J_{[mn}J_{pq}J_{rs]}\\
\varepsilon_{mnpqrstuvw}J^{vw}                         &=210 J_{[mn}J_{pq}J_{rs}J_{tu]}\\
\varepsilon_{mnpqrstuvw}                               &=945 J_{[mn}J_{pq}J_{rs}J_{tu}J_{vw]}
~.
}}
The last line of the above equation together with the last line of \eqref{eq:omegas}
imply
\eq{\label{omvol}
\Omega_{[a_1\dots a_5}\Omega^*_{a_6\dots a_{10}]}=- \frac{8i}{63}\varepsilon_{a_1\dots a_{10}}~;~~~
\Omega_{a_1\dots a_5}=-  \frac{i}{5!}\varepsilon_{a_1\dots a_{10}}
\Omega^{a_6\dots a_{10}}
~.
}
%
%
%
Finally, the following relations are useful in the analysis of the Killing spinor equations.
\eq{\spl{\label{eq:su5spinorhodge}
\g_m\eta      &= (\Pi^+)_{m}{}^{n}\g_n\eta\\
\g_{mn}\eta   &= iJ_{mn}\eta + (\Pi^+)_{[m}^{\phantom{[m}p}(\Pi^+)_{n]}^{\phantom{[m}q} \g_{pq}   \eta  \\
\g_{mnp}\eta  &= 3iJ_{[mn}\g_{p]}\eta  +\frac{1}{8}\Omega_{mnpqr}\g^{qr}\eta^c\\
\g_{mnpq}\eta &= -3J_{[mn}J_{pq]}\eta + 6i J_{[mn}(\Pi^+)_{p}^{\phantom{[m}r}(\Pi^+)_{q]}^{\phantom{[m}s} \g_{rs}   \eta
                 - \frac{1}{2} \O_{mnpqr} \g^r \eta^c \\
\g_{mnpqr}\eta &=  - \O_{mnpqr} \eta^c + \frac{5i}{4} J_{[mn}\O_{pqrst}\g^{st} \eta^c - 15 J_{[mn}J_{pq}\g_{r]} \eta
~.
}}

\subsection{Useful formul\ae{}}

\subsection*{Formul\ae{} needed for the dilatino equations}
In order to solve the dilatino equations, the following are used:
\eq{\spl{\label{eq:su5useff1}
\underline{F_1}\eta &= f_{1|m}^{(0,1)} \g^m \eta \\
\underline{F_2} \eta &= 5 i f_2 \eta + \frac{1}{2}f_{2|mn}^{(0,2)} \g^{mn} \eta \\
\underline{F_3}\eta &= 4i f_{3|m}^{(0,1)} \g^m \eta + 4 {f}_{3|mn}^{(2,0)} \g^{mn} \eta^c\\
\underline{F_4} \eta &= - 20 f_4 \eta +  \frac{3i}{2} f_{4|mn}^{(0,2)} \g^{mn} \eta  - 16 \tilde{f}_{4|m}^{(1,0)} \g^m \eta^c\\
\underline{F_5}\eta &= 0
~.}}
The necessary expression for $\underline{H}\eta$ is analogous to the decomposition of $F_3$.
\\
\\
Intermediary results used to solve the gravitino equations are given by
\al{\label{eq:su5useff2}
\underline{F_1} \g_m \eta &= 2f_{1|m}^{(1,0)}\eta - \Pi^+_{mn} f_{1|p}^{(0,1)} \g^{np} \eta              \nn \\
\underline{F_2}\g_m \eta &= 3 i f_2 \g_m \eta - 2 f_{2|mn}^{(1,1)} \g^n \eta + \frac{1}{16} f_{2|qr}^{(0,2)} \O_{mnp}^{\phantom{mnp}qr} \g^{np} \eta^c \nn\\
\underline{F_3} \g_m \eta &= 8i f_{3|m}^{(1,0)}\eta    -2i \Pi^+_{mn}f_{3|p}^{(0,1)} \g^{np} \eta
                           - 16  {f}_{3|mn}^{(2,0)} \g^{n} \eta^c + f_{3|mnp}^{(1,2)} \g^{np} \eta  \nn\\
\underline{F_4} \g_m \eta &= - 4 f_4 \g_m \eta - 6 i  f_{4|mn}^{(1,1)} \g^n \eta - 32 {f}_{4|m}^{(1,0)} \eta^c
                             - \frac{1}{24}  f_{4|mqrs}^{(1,3)} \O_{np}^{\phantom{np}qrs} \g^{np} \eta^c
                                                  + \frac{i}{16}  f^{(0,2)}_{4|qr} \O_{mnp}^{\phantom{mnp}qr} \g^{np} \eta^c\nn\\
\underline{F_5} \g_m \eta &= - 24 f_{5|m}^{(1,0)} \eta
                             + 2i  f_{5|mnp}^{(1,2)} \g^{np} \eta
                             - \frac{1}{4!} f_{5|mnpqr}^{(1,4)} \O^{npqrs} \g_s \eta^c
~.}
Furthermore, the NSNS flux satisfies
\eq{
\underline{H}_m \eta &= 4 i \left( h_m^{(0,1)} +h_m^{(1,0)} \right)\eta + i (\Pi^+)_{mn} h_p^{(0,1)}
                        \g^{np}\eta + \frac{1}{2} h_{mnp}^{(1,2)}\g^{np} \eta
                        + \frac{1}{4} {h}^{(2,0)}_{qr} \O_{mnp}^{*\phantom{mnp}qr}\g^{np} \eta ;.
}
In addition, when doing the generalised geometric IIA computation, the identities
\eq{\spl{\label{for}
(h^{(0,2)}\lrcorner \Omega)\wedge\Omega^*&=-\frac{16i}{3}h^{(0,2)}\wedge J^3\\
(\tilde{f}_4^{(1,0)} \lrcorner \O^* )\wedge J &=  i \tilde{f}_4^{(0,1)} \wedge \O^*
~.}}
are convenient.

\chapter{Intrinsic torsion}\label{torsion}
As seen in section \ref{III}, a major reason why existence of a $G$-structure is beneficial to solve the supersymmetry equations is due to the presence of torsion classes. From a practical point of view, torsion classes allow for an algebraic description of the Killing spinor equations. Furthermore, they characterise the geometry of the internal manifold. In this appendix, intrinsic torsion and torsion classes will be investigated in some more depth. As is often the case in physics, a thorough understanding of the underlying geometrical concepts is not really necessary to be able to do the computations, yet it certainly might assist in figuring out how to move forward from this point on. At the very least, such an understanding is gratifying. As usual, proofs will be sketched at best; see for example \cite{nakahara} or \cite{salamon}  for a more thorough approach. Mostly, we follow the approach of \cite{waldram}, \cite{joyce}.

In section \ref{connections}, connections on vector bundles as well as principal bundles are discussed. In section \ref{torsion2}, we apply this to give a definition for intrinsic torsion and characterisation in terms of connections. In section \ref{holo}, the holonomy of a connection is defined, and the relation of intrinisic torsion to holonomy is described.

\section{Connections}\label{connections}
Differentiation of sections of fibre bundles is characterised in terms of connections. It will be necessary to understand in some detail both connections on vector bundles as well as on principal bundles, both in terms of differentiation as well as in terms of subbundles, in order to understand intrinsic torsion.
\dfn{
Let $E$ be a vector bundle. A \textit{connection} on $E$ is a map $\nabla : \G(M, E) \rightarrow \G(M, E \otimes T^* M )$ such that $\nabla$ is a linear differential operator:
\eq{
\nabla (f \s ) &= f \nabla \s + \d f \otimes  \s\;, \quad \s \in \G(M, E),\; f \in C^\infty (M)\\
\nabla (\s_1 + \s_2) &= \nabla \s_1 + \nabla \s_2 \;.
}
We will use the standard notation $\nabla_X \s \equiv  X \lrcorner\nabla \s$ such that one could also think of $\nabla$ as a map $\nabla : \G(M, E) \otimes \mathfrak{X}(M) \rightarrow \G(M, E)$, $(X, \s) \mapsto \nabla_X \s$.
}
Given a connection on $T^*M$, a unique connection on the tensor bundle $\tcal^{(k,l)}M$ can be induced, which will also be denoted $\nabla$.

A completely equivalent way of thinking about connections is in terms of horizontal and vertical subbundles. The advantage is that these concepts generalise straightforwardly to principal bundles as well. Let us first consider such subbundle connections in the context of vector bundles. Given a vector bundle $E$, one can consider the map $\d \pi : TE \rightarrow TM$. Since for any point $x\in E$ with $\pi(x) = p$, $\text{Ker}(\d \pi_x)$ is a vector subspace of $TE$, and since $\d \pi$ is surjective,  one can define the vector subbundle $V \rightarrow E$ with fibres $V_x \equiv \text{Ker}(\d \pi_x)$. This subbundle is called the \textit{vertical subbundle}. Staring at definitions for a while one can convince oneself that $V_x \simeq T_x E_p$ with isomorphism $\d_x \iota_p$, with $\iota_p : E_p \hookrightarrow E$ the inclusion map.

\dfn{
Let $E$ be a vector bundle. A \textit{horizontal subbundle} $H$ is a subbundle of $TE$ that is complementary to the vertical subbundle, in the sense that $\forall x \in E$, $V_x \oplus H_x = T_x E$ and $V_x \cap H_x = \{0\}$.
}
Note that since $V_x = \text{Ker}(\d \pi_x)$ and $\d \pi_x : V_x \oplus H_x \rightarrow T_p M$ is surjective and linear, $H_x \simeq T_p M$.

Let us now consider connections on principal bundles; they are identical to the horizontal subbundles defined for vector bundles, with the additional demand that the horizontal subbundle is invariant under the group action.
\dfn{
Let $(P, G, M, \pi)$ be a principal bundle. Then $\d \pi : TP \rightarrow TM$ defines the \textit{vertical subbundle} $V$ with $V_x \equiv \text{Ker}(\d \pi_x)$ $\forall x \in P$.     A \textit{connection} on a principal bundle $P$ is a horizontal subbundle $H$ that is complementary to the vertical subbundle and additionally, is invariant under the action of $G$ on $P$.
}
Given a Lie algebra element $X \in \mathfrak{g}$, it can be associated to a vertical vector field $\hat{X}^V \in \mathfrak{X}(V)$ satisfying
\eq{
\hat{X}^V_x = \frac{d}{dt} \left. x \exp(t X) \right|_{t=0}.
}
In fact, this defines a isomorphism between $V_x$ and  $\mathfrak{g}$ (with the Lie bracket given by the vector field bracket restricted to the fibre) and thus, $H_x \simeq T_p M$.

Another way to define the connection is to view the horizontal subbundle as the kernel of a {\it connection one-form}. Given a horizontal bundle, the connection one-form is constructed as follows.
Let $\hat{X} \equiv \hat{X}^V + \hat{X}^H$ be the decomposition of a  vector field on $P$ in terms of its horizontal and vertical component. Then the connection one-form $\o  : \mathfrak{X}(P) \rightarrow \mathfrak{g}$ can be defined as
\eq{
\o_x (\hat{X}_x) = X \;.
}
In other words, $\o \in \O^1(P ; \mathfrak{g})$ is a Lie algebra-valued one-form on $P$. Note that $\o$ can be seen as composition between projection onto the vertical vector fields, followed by the isomorphism identifying the vertical fibres with the Lie algebra.
This shows that existence of a horizontal subbundle implies existence of a connection one-form. Conversely, given a connection one-form, its kernel can be defined to be the horizontal subbundle. Hence connection one-forms and horizontal subbundles are equivalent.
\\
\\
The connection one-form is a globally well-defined form on the principal bundle $P$. However, generally, it is more convenient not to deal with forms on $P$ but with forms on the base space $M$ instead. It is possible to relate the connection one-form $\o$ to a one-form on $M$, but the price to pay is that this form is not globally well-defined.

Let $U_\a \subset M$, and let $\tau_\a : \pi^{-1}(U_\a) \rightarrow U_\a \times G$ be a local trivialisation of the principal bundle. To this local trivialisation, it is possible to uniquely associate a canonical local section $s_\a \in \G(U_\a, \pi^{-1}(U_a))$ defined by
\eq{
\tau_\a \circ s_\a (p) = (p, e) \;.
}
In other words, such a (local) section is the principal bundle equivalent of the (global) zero section of a vector bundle.
Given a local section as such, the connection one-form on $M$ can be defined as the pullback of $\o$ by means of $s_\a$:
\eq{
A_\a \equiv \left(s_\a\right)^* \o \in \G\left(U_\a, T^*M \otimes \mathfrak{g}\right) \;,
}
and is perhaps better known under the name {\it gauge field}. The transformation rules can be deduced by noting that
$\o$ is $G$-equivariant, in the sense that
\eq{
R^*_g \o = \text{ad}_{g^{-1}} \o  \;,
}
with $R^*_g$ the pullback of right multiplication with $g$ on $P$.
Let
\eq{
\tau_{\a\b} (p, h) = (p, g_{\a\b}(p) h)
}
such that $g_{\a\b}$ are the transition functions of the principal bundle. Then the gauge field transforms\footnote{We only consider principal bundles satisfying $G \subset GL(n, \rbb)$.} as
\eq{
A_\a = \text{ad}_{g_{\a\b}} A_\b  +  (\d g_{\a\b})  g_{\a\b}^{-1}
}
on overlaps. Although this is not the proper transformation rule for a globally defined section, the difference between two gauge fields is in fact globally defined.
\dfn{
Let $P$ be a principal bundle with structure group $G$, with $k$-dimensional Lie algebra $\mathfrak{g}$. Then the {\it adjoint bundle} is the associated vector bundle $P \times \mathfrak{g} / \sim$, with the representation of $G$ on $\mathfrak{g}$ given by the adjoint representation.
}
Consider two gauge fields $A_\a, A_\a'$ associated to two connections $\o, \o'$. Then
\eq{
\acal_\a \equiv A_\a - A_\a' = \text{ad}_{g_{\a\b}} \acal_\b
}
and hence indeed, $\acal \in \G(M, T^*M \otimes \text{ad}P)$ is globally well-defined.
\\
\\
Finally, let us describe the relation between connections on vector bundles and connections on the corresponding frame bundles.
\prop{
Connections $\nabla$ on $E$ can be mapped bijectively onto connections on the corresponding frame bundle $FE$.
}
See the discussion around Definition 2.18 in \cite{joyce} for a precise construction of the map. The story is more complicated for generic principal bundles, but these are not relevant for our purposes. Let us end the discussion on connections with two key notions with respect to intrinsic torsion, the first related to connections on subbundles, the second the torsion of a connection.

\defn{
Let $Q \subset FTM$ be a $G$-structure on $M$ and let $H \subset TFTM$ be a horizontal subbundle on the frame bundle of $M$. Then $H$ {\it reduces to} $Q$ if $H \subset T Q$. Let $\nabla$ be the differential operator acting on tensors of $M$ induced by $H$\footnote{To be precise, $H$ induces a connection one-form $\o$ on $FTM$, which induces a local connection one-form $A$ on $M$, which induces a differential operator $\nabla$ acting on sections of $TM$, which induces a differential operator $\nabla$ acting on sections of the tensor bundle.}. Then $\nabla$ is said to be {\it compatible} with $Q$ if $H$ reduces to $Q$.
}

\defn{
Let $\nabla$ be a connection on $TM$. Then the {\it torsion} of the connection $T(\nabla) \in \G(M, TM \otimes \bw^2 T^*M)$ is defined by
\eq{
T(\nabla)(X, Y) \equiv \nabla_X Y - \nabla_Y X - [X,Y]
}
for $X,Y \in \mathfrak{X}(M)$.
}

\section{Intrinsic torsion}\label{torsion2}
Given the knowledge of connections, we are now in a position to define the intrinsic torsion of a $G$-structure.

First, there is the following relation between $G$-structures and connections.
\prop{
Let $Q$ be a $G$-structure defined by a tensor $\phi$, i.e., $G$ is the stabiliser of $\phi$. Let $\nabla$ be the connection acting on tensors of $M$ associated with a horizontal subbundle $H \subset TFTM$. Then
\eq{
\nabla \phi = 0
}
if and only if  $\nabla$ is compatible with $Q$. There is no topological obstruction to the existence of such a connection.
}
Consider now two connections $\tilde{\nabla}_{1,2}$ which are compatible with $Q$, such that
\eq{
\acal(\tilde{\nabla}_1, \tilde{\nabla}_2) \equiv \tilde{\nabla}_{1} - \tilde{\nabla}_2 \in \G( \text{ad} Q \otimes T^*M) \;.
}
Let $\tau$ be defined by
\eq{\label{eq:injection}
\iota &: \text{ad} Q \otimes T^*M \hookrightarrow \text{ad}FTM \otimes T^* M \simeq TM \otimes T^*M \otimes T^*M \\
\text{ASym} &:  TM \otimes T^*M \otimes T^*M \rightarrow TM \otimes \bw^2 T^*M \\
\tau &\equiv \text{ASym} \circ \iota \;,
}
that is, $\tau$  the composition of inclusion and antisymmetrising, with the diffeomorphism on the first line following from the fact that for $M$ $n$-dimensional, $\mathfrak{gl}(n) = \text{End}(T_p M) \simeq \left(TM \otimes T^*M\right)_p$.
Then
\eq{
\tau (\acal) = T(\tilde{\nabla}_{1}) - T(\tilde{\nabla}_{2})\;,
}
with $T$ the torsion. Define the space
\eq{
\wcal_{\text{int}} \equiv \left(TM \otimes \bw^2 T^*M\right) / \Im (\tau) \;,
}
which depends purely on the topology of $Q$. Thus, we find the following:
\dfn{
Let $\Pi : TM \otimes \bw^2 T^*M \rightarrow \wcal_{\text{int}}$ be the projection operator. Then the {\it intrinsic torsion} of a $G$-structure is given by
\eq{
T_{\text{int}}(Q) = \Pi \circ T (\nabla)\;,
}
for any connection $\nabla$ on $Q$: the intrinsic torsion is independent of this choice of connection.
}
The intrinsic torsion is therefore an obstruction to finding connections which are 1) torsion-free and 2) compatible with the $G$-structure $Q$.
Let us consider the application of this concept to $O(n)$-structures, which are associated to Riemannian metrics. Since by using the metric $\mathfrak{so}(n) \simeq \bw^2 T^*_p M$, the map $\tau$ defined in \eqref{eq:injection} is a diffeomorphism and hence $\wcal_{\text{int}}$ is trivial. Therefore, a connection exists that is compatible with the Riemannian metric and is torsion-free, namely the Levi-Civita connection $\nabla$. On the other hand, consider a $G$-structure $Q_G$ with $G \subset O(n)$ and a connection $\tilde{\nabla}_1$ that is compatible with $Q_G$, but not necessarily torsion-free. Since $\tilde{\nabla}_1$ is compatible with $Q_G$,  clearly it will also be compatible with the $O(n)$-structure $Q_{O(n)}$ and hence
\eq{
\acal(\nabla, \tilde{\nabla}_1) = \nabla - \tilde{\nabla}_1 \in \G\left(M, \text{ad} Q_{O(n)} \otimes T^*M \right) \;.
}
Setting $\mathfrak{so}(n) = \mathfrak{g} \oplus \mathfrak{g}^\perp$ and denoting $\text{ad} Q_{G}^\perp \equiv \text{ad}Q_{O(n)} / \text{ad} Q_G  $, it follows that
\eq{
\acal = \acal^{\mathfrak{g}} + \acal^{\mathfrak{g}^\perp} \;.
}
Since the connection one-forms compatible with $Q_G$ are $\mathfrak{g}$-valued, the best one can do to `gauge away' the torsion is to define the $Q_G$-compatible connection $\tilde{\nabla}_2 \equiv \tilde{\nabla}_1 - \acal^{\mathfrak{g}}$. This leads to the conclusion that
\eq{
\wcal_{\text{int}} = \text{ad}Q_G^\perp \otimes T^*M\;.
}
Thus, the intrinsic torsion is given precisely by $\nabla \phi$, i.e., by the failure of the torsion-free Levi-Civita connection to be compatible with $Q_G$.
\\
\\
A perhaps more practical way of viewing intrinsic torsion is to note the following. Assume that $\phi \in \O^k(M)$. Then $\d \phi$ and $\d \star \phi$ are expressible in terms of $\nabla \phi$, and hence in terms of the intrinsic torsion. By the existence of the $G$-structure, the transition functions take value in $G$ and hence, $\d \phi$, $\d \star \phi$ can be globally decomposed into irreducible representations of $G$. These are the {\it torsion classes}, which take value in the (decomposition of) $\wcal_{\text{int}}$. The fact that the torsion classes fully encode the intrinsic torsion can be checked by verifying that all components of $\wcal_{\text{int}} = \text{ad}Q_G^\perp\otimes T^* M $, decomposed in irreducible representations of $G$, are present in the decompositions of $\d \phi$, $\d \star \phi$. This has been done explicitly for $SU(4)$ and $SU(5)$\footnote{Note that due to the fact that for $SU(n)$-structures one has $J^n \sim \O \wedge \O^* \sim \vol_n$, it follows that $\star J \sim J^{n-1}$, $\star \O \sim \O^*$ and hence the derivative of the Hodge dual contains no new information.} in \eqref{eq:su4torsdecomp}, \eqref{eq:su5torsdecomp}; a more detailed proof, albeit for $G_2$ instead, is given in \cite{fernandez}.

\section{Holonomy}\label{holo}
A concept that sheds a different light on intrinsic torsion is holonomy. To properly define holonomy, let us first consider the following reminder:
\dfn{
Let $\g : [0,1] \rightarrow M$ be a piecewise smooth curve, and let $\s \in \G(M, E)$ be a section of a vector bundle $E$ of rank $k$. Then $\s$ is \textit{parallel} if $\nabla_{\dot{\g}(t)} \s =0$ $\forall t \in [0,1]$. Let $v \in E_p$ be a vector at $p \equiv \g(0)$. Then there exists a unique parallel section $s \in \G(M,E)$ such that $s(p) = v$. The \textit{parallel transport of $v$ along $\g$} is defined as $P_\g (v) = \s\left(\g(1)\right)$.
}
Given the fact that parallel transport is a map $P_\g : E_{\g(0)} \rightarrow E_{\g(1)}$ for a given curve, one might wonder what sort of transformations of vectors one can attain in the same fibre.
\dfn{
Let $\ccal(p)$ be the set of piecewise smooth curves $\g : [0,1] \rightarrow M$ with $\g(0) = \g(1) = p$.
The \textit{holonomy} of $M$ with connection $\nabla$ at  $p$ is given by
\al{
\text{Hol}_p(\nabla) = \{ P_\g  \;|\; \g \in \ccal(p)\} \;.
}
}
Some relevant properties of the holonomy are as follows:
\begin{itemize}
\item The holonomy is a subgroup of $GL(k, \rbb)$.
\item Up to conjugation with elements in $GL(k, \rbb)$, the holonomy is independent of the base point and can be denoted $\text{Hol}(\nabla)$.
\item If $M$ is simply connected, $\text{Hol}(\nabla)$ is connected. In case $M$ is not, one can restrict the holonomy group to only parallel transports along null-homotopic loops such that that this restricted holonomy group $\text{Hol}^0(\nabla) \subset \text{Hol}(\nabla)$  is the connected component of the holonomy containing $\obb$.
\end{itemize}
One of the most important results with respect to holonomy is that there is a classification theorem with regards to the possible holonomy groups of the Levi-Civita connection.
\defn{
A Riemannian manifold $(M, g)$ is {\it locally reducible} if $\forall p \in M$, there is an open neighborhood $U_p \ni p$ such that there exists opens with metrics $(U_1, g_1)$, $(U_2, g_2)$ with $U_p$ isometric to $U_1 \times U_2$ and the metric satisfying  $g|_{U_p} = g_1 + g_2$.
}

\defn{
A Riemannian manifold is {\it locally symmetric} if and only if $\nabla R = 0$, where $R$ is the Riemann curvature tensor and $\nabla$ is the Levi-Civita connection.
}

\begin{ber}
Let $(M_n, g)$ be a Riemannian manifold that is simply connected, irreducible, and nonsymmetric. Then $\text{Hol}(\nabla)$ is one of the following:
\begin{enumerate}
\item $SO(n)$
\item $U(n)$ for $2n \geq 2$
\item $SU(n)$ for $2n \geq 2$
\item $Sp(n)$ for $4n \geq 2$
\item $Sp(n)Sp(1)$ for $4n \geq 2$
\item $G_2$ for $n = 7$
\item $Spin(7)$ for $n=8$
\end{enumerate}
\end{ber}
The relation between holonomy and intrinsic torsion is then as follows.

\prop{
Let $Q$ be a $G$-structure on $(M, g)$. Then  $\text{Hol}(\nabla) \subset G$, and $\text{Hol}(\nabla) = G$ if and only if all torsion classes vanish.
}
Thus, the intrinsic torsion can also be considered as the obstruction for the manifold to have special holonomy. This interpretation is especially convenient when considering $SU(n)$-structure manifolds to be generalisations of proper Calabi-Yau manifolds, which by definition have holonomy group $SU(n)$.

\chapter{Construction of calibration forms}\label{calicons}
In this appendix, it will be demonstrated explicitly how to construct generalised calibration forms for static magnetic branes in type IIB supergravity vacua on $\rbb^{1,1} \times \mcal_8$ with Killing spinor decomposition as given by \eqref{eq:IIBks}, following the procedure of \cite{melec}. In order to ensure that a polyform is a generalised calibration form, an algebraic and a differential condition need to be satisfied. The algebraic inequality follows from the characterisation of $\k$-symmetric branes, the differential condition from the supersymmetry equations. We will describe the general case, plug in our specific setup, and deduce the calibration form for branes on $\scal = \rbb^{1,1} \times \Sigma$.
\\
\\
Let us define the so-called `background structures', which are given in terms of the Killing spinors as
\eq{\label{eq:bgs}
\Psi &\equiv - \sum_{k=0}^{10} \frac{1}{k!} \tilde{\e}_1 \G_{M_1 ... M_k} \e_2 \d X^{M_1} \wedge ... \wedge \d X^{M_k} \\
K &\equiv  - \frac{1}{2} \left( \tilde{\e}_1 \G^{M} \e_1 + \tilde{\e}_2 \G^{M} \e_2 \right)\p_{M}\\
\tilde{K} &\equiv  - \frac{1}{2} \left( \tilde{\e}_1 \G_{M} \e_1 - \tilde{\e}_2 \G_{M} \e_2 \right)\d X^{M} \;.
}
Let us also define the generalised vector fields $\pcal, \kcal \in \G(T \oplus T^*)$ satisfying
\eq{
\pcal &\equiv - \m_p e^{-\phi} \sqrt{ - \det( g|_{\scal} + \fwv)} \p^{\tau} X^M \p_M \\
\kcal &\equiv K + \tilde{K} \;,
}
where for $\pcal$, use has been made of the magneticity of the worldvolume flux, i.e., $\p_{\tau} \lrcorner \fwv =0$.
Finally, let us define a generalised metric $\tilde{\gcal} \equiv \text{diag}(g, g^{-1})$.

In \cite{tomasiello}, it was shown that the supersymmetry equations imply
\eq{\label{eq:tomasusy}
\d_H (e^{-\phi} \Psi ) = - \kcal \cdot \fcal\;,
}
with $\fcal$ the total RR flux. In \cite{melec}, it was shown that
\eq{\label{eq:martucali}
- \tilde{\gcal}(\pcal, \kcal) \geq \left(e^{-\phi}\Psi \wedge e^\fwv\right)_{\Sigma'} \;,
}
for a D-brane on $\scal = \rbb \times \Sigma' \subset \mcal_{10}$ and with equivalence if and only the D-brane on the generalised submanifold $(\Sigma', \fwv)$ is $\k$-symmetric.

Using our Killing spinor ansatz \eqref{eq:ksiib}, the background structures reduce to
\eq{
\Psi &= \a^2 (\d t + \d x)\wedge  \Re \left( \Psi_1 + e^{i \vartheta} \Psi_2 \right) \\
K &= \a^2 \left( \p_t - \p_x \right) \\
\tilde{K} &= 0\;,
}
with $\e^{i \vartheta} \equiv \tilde{\z} \g_0 \z$ and $\Psi_{1,2}$ defined by acting with the Clifford map on \eqref{eq:su4polyforms}.
Plugging these into \eqref{eq:tomasusy} together with the RR flux decomposition \eqref{eq:fluxdecomp} and noting that $\a^2 = e^A$ leads to
\eq{
\d_H \left( e^{2A- \phi} \Re  \left( \Psi_1 + e^{i \vartheta} \Psi_2 \right) \right) = F^{el} \;.
}
Using a parametrisation  $t = \tau$, $x = \s^1$ on our spacetime-filling D-brane on $\scal = \rbb^{1,1} \times \Sigma$, \eqref{eq:martucali} reduces to
\eq{
\left(e^{2A - \phi}\Re \left( \Psi_1 + e^{i \vartheta} \Psi_2 \right) \wedge e^{\fwv}\right)_\Sigma  \leq d^{p-1} \s \; e^{2A- \phi} \sqrt{ \det (\gs + \fwv)}\;.
}
Note that this inequality would have been trivialised for branes of spacetime codimension one, i.e., wrapping $\rbb \times \Sigma$, $\Sigma \subset \mcal_8$.
Comparing with the definition, it follows that
\eq{
\o \equiv e^{2A - \phi}\Re \left( \Psi_1 + e^{i \vartheta} \Psi_2 \right)
}
is a calibration form for spacetime filling D-branes.

\chapter{Calibrated probe branes on Stenzel Space}\label{caliDbranes}

The purpose of this section is to determine which are the admissible calibrated probe branes on the $\ncal=(1,1)$ IIA vacuum geometry discussed in section \ref{fluxstenzel}.

Our setup will be as follows. Consider the background structures as given by \eqref{eq:bgs}, and a probe  Dp-brane with $(p+1)$-dimensional worldvolume $\Sigma$\footnote{Unfortunately, we have been using $\scal$ for both brane worldvolumes and Stenzel space throughout. In this section, $\Sigma$ will refer to the the total worldvolume of the D-brane rather than just the internal cycle.}

and worldvolume electromagnetic fieldstrength $\fwv$ whose Bianchi identity reads:
\eq{\label{wvbianchi}
\d\fcal_{\text{wv}}=\left.H\right|_{\Sigma}
~.}
The necessary and sufficient conditions for the $Dp$-brane to be calibrated can then be formulated as follows \cite{melec}:
\eq{\spl{\label{nscal}
\left(\d x^M\wedge\Psi\wedge e^{\fcal_{\text{wv}}}\right)\left.\right|_{\Sigma}&=K^M\sqrt{-\det(\gs+ \fcal_{\text{wv}})}~\!\d^{p+1}\!\sigma\\
\left(\p_M \lrcorner \Psi\wedge e^{\fcal_{\text{wv}}}\right)\left.\right|_{\Sigma}&=\tilde{K}_M\sqrt{-\det(\gs+ \fcal_{\text{wv}})}~\!\d^{p+1}\!\sigma
~,}}
where $\left.g\right|_{\Sigma}$ is the induced metric on the worldvolume and $\sigma^{a}$, $a=1,\dots,p+1$, are coordinates of  $\Sigma$; in  both  left-hand sides above it is understood that we only keep the top $(p+1)$-form.

Let us note two important corollaries that follow from (\ref{nscal}). Firstly it can be shown that the electric worldvolume field defined by $\mathcal{E}\equiv\iota_K \fcal_{\text{wv}}$ is constrained to satisfy:
\eq{\label{cor1}
\mathcal{E}=\tilde{K} \bs
~.}
Secondly it can be shown that the vector $K$ belongs to the tangent space of $\Sigma$:
\eq{\label{cor2}
K\in T\Sigma
~.}

\section*{Worldvolume equations of motion}

Let the worldvolume be parameterized by $\sigma^{\alpha}$, $\alpha=1,\dots,p+1$.
The dynamical fields on the worldvolume of the branes are the embedding coordinates $X^M(\sigma)$ and the gauge field $A_{\alpha}$. Varying the D$p$-brane action,
\eq{\label{dp}
S_{Dp}= -\mu_p\int_{\Sigma} e^{-\phi}
\sqrt{-\det(\gs+\fcal_{\text{wv}})}~\!\d^{p+1}\!\sigma
+\mu_p\int_{\Sigma}C\wedge e^{\fcal_{\text{wv}}}
~,
}
with respect to $X^M(\sigma)$ we obtain
\eq{\label{xeom}
\partial_{\beta}P_M^{\beta}=0~,}
where we have defined
\eq{\spl{\label{defp}
P_M^{\beta}&\equiv e^{-\phi}\sqrt{-G} \left(G^{(\alpha\beta)}g_{M\alpha} + G^{[\alpha\beta]}B_{M\alpha} \right)
-\frac{1}{p!}\varepsilon^{\beta\alpha_1\dots\alpha_p}\left[\iota_M(C\wedge e^B)\wedge e^{\fcal_{\text{wv}}-B} \right]_{\alpha_1\dots\alpha_p}\\
G_{\alpha\beta}&\equiv \left(g\bs\right)_{\alpha\beta} + \left(\fcal_{\text{wv}}\right)_{\alpha\beta}
~;~~~g_{M\alpha}\equiv g_{MN}\partial_{\alpha}X^N~;~~~
B_{M\alpha}\equiv B_{MN}\partial_{\alpha}X^N
~.}}
Varying (\ref{dp}) with respect to $A_{\alpha}$ we obtain
\eq{\label{aeom}
\partial_{\beta}\Pi^{\alpha\beta}=0~,}
where we have defined
\eq{\label{defpi}
\Pi^{\alpha\beta}\equiv e^{-\phi}\sqrt{-G}
~\!G^{[\alpha\beta]}-\frac{1}{p!}~\!\varepsilon^{\alpha\beta\gamma_1\dots\gamma_p}\left(
C\wedge e^{\fcal_{\text{wv}}}\right)_{\gamma_1\dots\gamma_p}
~.}

\subsection*{$\ncal=(1,1)$ IIA CY vacua}
For the  $\ncal=(1,1)$ IIA CY vacua of section \ref{IIACY} the two ten-dimensional Majorana-Weyl Killing spinors are given by \eqref{eq:IIAks},
where $\eta$ is a unimodular covariantly-constant spinor of $\mcal_8$  and $\alpha^2=e^A=\hcal^{-1/2}$. It is then straightforward to compute:
\eq{\label{data}
\Psi = \hcal^{-1/2}\left( 1+\hcal^{-1}\d t\wedge\d x \right)\wedge \varphi~;~~~
K =\frac{\partial}{\partial t} ~;~~~
\tilde{K}=-\hcal^{-1}\d x
~,
}
where we have defined
\eq{
\varphi\equiv\mathrm{Re}\left[e^{i\theta}(\Omega+e^{iJ})\right]
~.}
Provided a $Dp$-brane is calibrated, i.e. (\ref{nscal}) is satisfied,
its action takes the form:
\eq{\label{dpaction}
S_{Dp}=-\mu_p\int_{\Sigma}\Big(\frac{1}{g_s}~\!\d t\wedge\varphi-C\Big)\wedge e^{\fcal_{\text{wv}}}
~,}
where $\mu_p=(2\pi)^{2p}(\alpha')^{-\frac{p+1}{2}}$ and we have used the fact that  the DBI part of the action saturates the BPS bound
\eq{\label{dbi}
S^{\mathrm{DBI}}_{Dp}= -\mu_p\int_{\Sigma} e^{-\phi}
\sqrt{-\det(\gs+\fcal_{\text{wv}})}~\!\d^{p+1}\!\sigma
=-\mu_p\int_{\Sigma} \frac{1}{g_s}~\!\d t\wedge\varphi \wedge e^{\fcal_{\text{wv}}}
~,}
with $e^{\phi}=g_s \hcal^{-1/2}$.

From (\ref{cor2}) and (\ref{data}) it follows that a calibrated $D$-brane
must extend along the time direction $t$.
We will further distinguish two different subcases according to whether
the $Dp$-brane extends along the spatial non-compact direction $x$ (in which case it is spacetime-filling) or not (in which case it is a domain wall).

\section{Spacetime-filling $Dp$-branes}
Let us  assume that $\Sigma$ wraps $t$, $x$ and an odd $(p-1)$-cycle inside $\mcal_8$. Explicitly let $\Sigma$ be parameterized by coordinates $(t,x,\sigma^a)$, $a=1,\dots,p-1$, so that
\eq{
x^m=x^m(\sigma)
~,}
where $x^m$ are coordinates of $\mcal_8$. The condition (\ref{cor1}) implies that the worldvolume fieldstrength is of the form:
\eq{\label{313}\fcal_{\text{wv}}=-\hcal^{-1}\d t\wedge\d x+ \d x\wedge f+\hat{\fcal}_{\text{wv}}~,}
where we have defined
\eq{f=f_a~\!\d\sigma^a~;~~~\hat{\fcal}_{\text{wv}}=\frac{1}{2}\left(\hat{\fcal}_{\text{wv}}\right)_{ab}~\!\d\sigma^a\wedge\d\sigma^b~.}
Moreover it follows from the form of the NSNS three-form in (\ref{nsf}) that for the worldvolume Bianchi identity (\ref{wvbianchi}) to be satisfied,the last two terms on the right-hand side of (\ref{313})  must be closed. Note that the electric worldvolume field is necessarily  non-vanishing.

Taking (\ref{data}) into account, equations (\ref{nscal}) can be seen to reduce to the following condition
\eq{\label{reducedcalib}\left[\hcal^{-1/2}f\wedge\varphi\bs\wedge e^{\hat{\fcal}_{\text{wv}}}\right]_{p-1}
=\sqrt{-\det(\gs+\fcal_{\text{wv}})}~\!\d^{p-1}\!\sigma
~,}
where on the left-hand side above it is understood that we keep only the  $(p-1)$-form.

Let us now describe explicitly some calibrated spacetime-filling branes.

\subsection*{D2}

Consider the case of a D2 brane $\Sigma$ extending along $(t,x)$ and wrapping an internal direction parametrized by $\psi$. We take $\psi$ such that
\al{
\n|_\Sigma = - \d \psi \;, \quad \s_j|_\Sigma = \tilde{\s}_j|_\Sigma = 0
}
for the left-invariant forms\footnote{See \cite{kp} for an explicit parametrization of the left-invariant forms in terms of coordinates on $V_{5,2}$.
The left-invariant forms $\s_j$ are not to be confused with the coordinates $\s^a$ on $\Sigma$. We do not refer to $\s_j$ elsewhere in this section.}  $\n$, $\s_j$, $\tilde{\s}_j$ on $V_{5,2}$.

Specializing (\ref{313}) to the case at hand, the worldvolume fieldstrength reads:
\eq{\fcal_{\text{wv}}=-\hcal^{-1}\d t\wedge\d x+\d x\wedge f~;~~~f=f_{\psi}\d\psi~;~~~\hat{\fcal}_{\text{wv}}=0~,}
and automatically satisfies (\ref{wvbianchi}) for $f_{\psi}$ an arbitrary function of $x$, $\psi$. Moreover:
\eq{\left[f\wedge\varphi\bs\wedge e^{\hat{\fcal}_{\text{wv}}}\right]_{1}=f_{\psi}\d\psi~,}
and
\eq{\label{matrix1}
\gs+\fcal_{\text{wv}}
=\left(
\begin{array}{ccc}
-\hcal^{-1} & -\hcal^{-1} &0\\
\hcal^{-1} & \hcal^{-1} & f_{\psi}\\
0& -f_{\psi}& c^2\\
\end{array}
\right)
\Longrightarrow -\det(\gs+\fcal_{\text{wv}})=H^{-1}f_{\psi}^2
~,}
so that the calibration condition (\ref{reducedcalib}) is satisfied provided $f_{\psi}$ is everywhere positive. Note that the DBI part of the action evaluates to:
\eq{\label{dbid2}
S^{\mathrm{DBI}}_{D2}= -\mu_2\int_{\Sigma} e^{-\phi}
\sqrt{-\det(\gs+\fcal_{\text{wv}})}~\!\d^{3}\!\sigma
=-\mu_2\int_{\Sigma} \frac{1}{g_s}\hcal^{1/2}\hcal^{-1/2}f_{\psi}\d t\d x\d\psi
~.}
The above is in agreement with the BPS bound (\ref{dbi}), as it should.

Note that the fieldstrength $\fcal_{\text{wv}}$ is not closed, in accordance with (\ref{wvbianchi}). To identify the wotldvolume electromagnetic field $F$ in $\fcal_{\text{wv}}$ we should split:
\eq{\fcal_{\text{wv}}=B\bs+2\pi\alpha' F~,}
with $F$ closed. The form of $B$ will impose constraints on the quantization of Page charges in the RR sector.

We should finally check the worldvolume equations of motion. Consider (\ref{defp}), (\ref{defpi}). In order to calculate these explicitly, the RR flux needs to be specified explicitly. For convenience, we consider only massless IIA, i.e., we set the RR flux $\fcal_0 = F_0 = 0$. Then regardless of which other RR fluxes we turn on (other scalar terms, (1,1)-forms, (2,2)-forms as discussed in section \ref{fluxstenzel}), the Wess-Zumino term reduces to
\al{
\left.C\wedge e^{\fcal_{wv}}\right|_{\Sigma} =0 \;.
}
This results in:
\eq{
P_M^{\beta}=\frac{1}{g_s}\left(
\begin{array}{ccc}
f_{\psi}&0&0\\
0&0&0\\
c^2&0&0\\
\end{array}
\right)~;~~~
\Pi^{\alpha\beta}=\frac{1}{g_s}\left(
\begin{array}{ccc}
0&-\frac{c^2}{f_{\psi}}&0\\
\frac{c^2}{f_{\psi}}&0&-1\\
0&1&0\\
\end{array}
\right)
~,}
where the indices $M$, $\alpha$ should be understood
as enumerating the rows, while the index $\beta$ enumerates the columns.
It can then be checked that the $X^M$-eoms (\ref{xeom}) are automatically satisfied. The $A_{\alpha}$-eoms (\ref{aeom}) are also automatically satisfied except for the Gauss law constraint $\partial_{\alpha}\Pi^{t\alpha}=0$, which imposes that $f_{\psi}$ should only depend on the coordinate $\psi$.

\section{Domain wall $Dp$-branes}
Let us now assume that $\Sigma$ wraps $t$ and an even $p$-cycle inside $\mcal_8$. Explicitly let $\Sigma$ be parameterized by coordinates $(t,\sigma^a)$, $a=1,\dots,p$, so that
\eq{
x^m=x^m(\sigma)
~,}
where $x^m$ are coordinates of $\mcal_8$. The condition (\ref{cor1}) implies that the worldvolume fieldstrength is of the form:
\eq{\fcal_{\text{wv}}=\frac{1}{2}\left(\fcal_{\text{wv}}\right)_{ab}~\!\d\sigma^a\wedge\d\sigma^b~,}
i.e. contrary to the spacetime-filling case here there can be no electric worldvolume field.
Moreover it follows from the form of the NSNS three-form in \eqref{nsf} that for the worldvolume Bianchi identity (\ref{wvbianchi}) to be satisfied  ${\fcal_{\text{wv}}}$ must be closed.

Taking (\ref{data}) into account, equations (\ref{nscal}) can be seen to reduce to the following condition
\eq{\label{reducedcalib2}\left[\hcal^{-1/2}\varphi\bs\wedge e^{\fcal_{\text{wv}}}\right]_p
=\sqrt{-\det(\gs+\fcal_{\text{wv}})}~\!\d^{p}\!\sigma
~,}
where on the left-hand side above it is understood that we keep only the  $p$-form.

\chapter{More on $SU(4)$-deformed Stenzel space vacua}

\section{$\ncal=(2,0)$ IIB on Stenzel Space and $SU(4)$-structure deformed Stenzel Space}\label{sec:iib}
In section \ref{fluxstenzel} an analysis is given of $\ncal = (1,1)$ IIA supergravity on Stenzel space. We wish to give a similar analysis of $\ncal = (2,0)$ IIB supergravity on Stenzel space. The supersymmetric solutions are given by \eqref{eq:iib}. Classes of IIB conformal Calabi-Yau vacua satisfying the integrability conditions are given in section \ref{IIBCCY}.

\subsection{Vacua on Stenzel Space}\label{sec:iibstenzel}
Let us specialize the solution of section \ref{IIBCCY} to the case where $\mcal_8$ is the Stenzel space $\scal$ with the appropriate CY metric and trivial conformal factor.
We have been unable to construct closed and co-closed primitive three-forms on Stenzel space so we will set $H=0$ and thus $\fcal_5=0$ as follows from the second line of (\ref{eq:iibrrfluxes}). The torsion classes all vanish for Stenzel space, hence \eqref{eq:iibtors} implies
\al{
\d A = 0\;,
}
leading to a constant warp factor. As a result, all RR fluxes vanish as well, leading to a fluxless vacuum.
In the UV, the metric asymptotes to
\al{
ds^2_{UV} = \L^2 e^{ 2 \rho} \left[ \left(e^{-2 \rho} ds^2 (\mathbbm{R}^{1,1} ) + \d\rho^2 \right) + ds^2 (V_{5,2})\right]\;,
}
with $\rho, \L$ defined in \eqref{eq:confadspara}.

\subsection{Vacua on $SU(4)$-structure deformed of Stenzel space}\label{sourcediib}
We now consider $\scal$ with a different $SU(4)$-structure. The IIB solution that we have discussed is CCY, which, together with the requirement that we have an $S^4$ bolt at the origin, leads to the conclusion that we can only consider $\scal$ as a CCY conformal to Stenzel space, as discussed in section \ref{moduli} around \eqref{eq:stenzelccy}. Specifically, we have $b, c $ fixed in terms of $a$ as
\eq{\spl{
b &=  \tanh\left(\frac{\tau}{2}\right) a \\
c &= \sqrt{\frac{2 + \cosh\tau }{3 \cosh^2\left(\frac{\tau}{2}\right)}}a \;.
}}
As $W_{4,5} \neq 0$, the warp factor is no longer constant. Thus, from \eqref{eq:iibrrfluxes}, we find that $\fcal_{3,7} \neq 0$. Therefore, the torsion constraint \eqref{eq:iibtors} and the last line of the Bianchi identities \eqref{eq:iibbianchis} impose
\eq{\spl{
- \d A &= \text{Re} W_4\\
\nabla^2 e^{-4 A} &= 0 \;.
}}
Spelled out in terms of  $a$ and $A$, these constraints are given by
\eq{\spl{
A' &=  - \frac{1}{2\tanh\left(\frac{\tau}{2}\right) a^2 } \left( (\tanh\left(\frac{\tau}{2}\right) a^2)' - \frac12 \frac{2 + \cosh\left(\tau\right) }{3 \cosh^2\left(\frac{\tau}{2}\right)} a^2 \right)\\
\p_\tau \left(\tanh^3\left(\frac{\tau}{2}\right) a^6 e^{-4 A} A'\right) &= 0 \;.
}}
The first equation can be solved explicitly to find that
\all{
a^2 = e^{-2A} \l^2 \left(2 + \cosh\tau\right)^{1/4}  \cosh\left(\frac{\tau}{2}\right) \;,
}
which is a consistency check of the fact that the metric under consideration is indeed CCY, i.e., it confirms  \eqref{eq:ccymetric} combined with \eqref{eq:stenzelabc}.
Inserting this into the second equation yields an equation for the warp factor, solved by
\al{
e^{-10 A} &=   k_1 - k_2 \int^\tau \frac{dt}{\left(x \sinh\left(\frac{t}{2}\right)\right)^3} \;,
}
with $k_{1,2}$ integration constants. Comparison with \eqref{eq:homo} leads to the conclusion that $\hcal^5$ for the IIB CCY vacuum is equivalent to the homogeneous warp factor for the IIA Stenzel space vacuum and is thus singular in the IR unless trivialized.

\section{RR-sourced IIA solutions on $SU(4)$-deformed Stenzel space}
We have been unable to find flux configurations with non-zero scalars or two-form on an $SU(4)$-deformed Stenzel space  that satisfy the RR Bianchi identities. We can violate the RR Bianchis at the cost of introducing more sources. These sources can be determined analogous to section \ref{sources}, but now with both RR and NSNS sources. The benefit is that, in this case, there is no other constraint than susy on the RR fluxes at all. A choice of RR fluxes then determines the warp factor. As the RR fluxes are then independent of $a,b,c$, one also has the freedom to tailor the geometry to one's wishes. We will consider this scenario here briefly to illuminate some more possibilities for the geometry.
\\
\\
\\
The IIA susy constraint is the torsion constraint \eqref{eq:iiasusycons}, solved by \eqref{eq:susysol}. This leaves two free functions $a,b$. In particular, these determine two interesting features to consider: the torsion and the boundary.

Let us first consider the boundary conditions we wish to impose. As before, we wish to have a (squashed) $S^4$ bolt at the origin, which means that either $(\a(0), \b(0)) = (a_0,0)$ or $(a(0),b(0)) = (0, b_0)$, $a_0, b_0 \neq 0$ and the squashing determined by the proportionality of $c(0)$ with respect to $a_0$ or $b_0$. Before, we have considered boundary conditions such that the space is asymptotically conical. Another geodesically complete option is to put another bolt at $\tau \rightarrow \infty$. We can have either similar bolts at the origin and at infinity, or different bolts, and with possibly different squashing. Our first example will interpolate between these two options, with trivial squashing on both. Our second example will have similar bolts, with one bolt with a fixed squashing and the other squashing determined by a free parameter.

The second point to consider is the torsion of the $SU(4)$-structure. There are four possibilities for the torsion classes: $W_2$ and $W_4 \sim W_5$ can both be either zero or non-zero. $W_2 = 0$ has been considered when discussing the four-form solution in section \ref{cplx22}, while $W_2 = W_4 = W_5 = 0$ is the CY case, which is necessarily Stenzel space after imposing \eqref{eq:geo}. Our first example below has $W_{2,4,5} \neq 0$ whereas our second example has $W_4 = W_5 = 0$, $W_2 \neq0$: manifolds with $W_1=W_3=W_4=W_5 = 0, W_2 \neq 0$ are also referred to as `nearly Calabi-Yau'. We reiterate that $W_2 = 0$ with a bolt at the origin if and only if $ \left(\frac{a}{b}\right)^{\pm1} = \tanh\left(\frac{\tau}{2}\right)$. On the other hand, the moduli space of nearly CY spaces is more difficult to deduce. We will only consider the case where $c$ is fixed by the susy constraint \eqref{eq:iiasusycons}.  $W_4=0$ is equivalent to
\al{
(a b)' - \frac12 c^2 = 0\;,
}
which, after imposing \eqref{eq:susysol}, is equivalent to
\al{
\check{r}' + \frac32 ( \b + \frac{1}{\b} ) \check{r} - 2 =0 \;,\qquad \check{r} \equiv \frac{\a}{\b^2}
}
where we made use of the parametrization \eqref{eq:para}. Solving this equation leads to the conclusion that, for this specific $c$, the space is nearly CY if and only if
\al{\label{eq:ncy}
\a^2 = \frac{\b \exp \left(\int^\tau dt \b + \frac{1}{\b}\right) }{k + 2 \int^\tau dt \exp \left(\int^\tau dt \b + \frac{1}{\b}\right)}\;,
}
with $k$ an integration constant.

\textbf{Example 1:}
Let us consider the case where we want $W_{2,4,5} \neq 0$ with bolts at  $\tau = 0, \infty$. In this case, it will be easiest to forego the parametrization \eqref{eq:para}.
Set
\eq{\spl{
c &= \l\\
a &= \l \cos(h(\tau)) \\
b &= \l \sin(h(\tau)) \;.
}}
It can easily be verified that this satisfies \eqref{eq:iiasusycons} and that $W_{2,4,5} \neq 0$,  for any function $h(\tau)$. Furthermore, if $h(0) = 0$, \eqref{eq:geo} is satisfied, thus leading to the usual $S^4$ bolt at the origin.
Let us set
\al{
h(\tau) = k \arctan(\tau) + (1-k) N \tau e^{- \tau} \;.
}
$k \in [0,1]$ interpolates between the solution where $c_{UV} = a_{UV} = \l, b_{UV} = 0$ for $k= 0$ and  $a_{UV} = 0, c_{UV} = b_{UV} = \l$ for $k=1$. Hence we can choose which of the two possible (non-squashed) $S^4$ bolts we have at $\tau \rightarrow \infty$. Of course, to swap bolt type at $\tau = 0$, we need simply swap $ a\leftrightarrow b$.
We choose the constant $N$ to be suitably small such that $h(\tau) \in (0, \frac{\pi}{2})$  $\forall \tau \in (0, \infty)$, hence ensuring that $a,b$ do not vanish or blow up at any other point.
By choosing a suitable warp factor, the external metric can be taken to be $\text{AdS}_3$, globally rather than just asymptotically.

\textbf{Example 2:} \\
Let us now construct a nearly CY with two similar squashed $S^4$ bolts, i.e.,
\eq{\spl{
a(0) &= a_0 \;, \quad b(0) = 0 \\
\lim_{\tau \rightarrow \infty} a(\tau) &\equiv a_{UV} \;, \quad \lim_{\tau -> \infty} b(\tau) = 0 \;.
}}
Such a solution is given by
\al{
\b = \frac{k_1 \tau}{k_2 \tau^{1 + k_3 } + 1} \;, \qquad k_{1,2,3} \in (0, \infty) \;.
}
Clearly this satisfies $b(0) = \lim\limits_{\tau -> \infty} b(\tau) = 0$. Defining $\a$ as in \eqref{eq:ncy} with $k=0$, we find that
\eq{\spl{
\a(0) &= \sqrt{\frac{3}{4} + \frac{k_1}{2}}\\
\lim_{\tau \rightarrow \infty} \a &=  \sqrt{3/4}\;,
}}
thus satisfying all boundary conditions. As the squashing of the $S^4$ is non-trivial for $\a \neq 1$, the bolt at $\tau = \infty$ has fixed non-trivial squashing whereas the squashing of the $S^4$ at the origin is determined by $k_1$.

\end{document}